\documentclass[twoside,12pt]{article}
\usepackage{epsfig}
\usepackage{amssymb}
\usepackage[english]{babel}

\usepackage{rotate}
\usepackage{graphicx}

\def\Journal#1#2#3#4{{#1} {#2} (#4) #3 }

\def\NPB{{\em Nucl. Phys.} B}

\def\PLB{{\em Phys. Lett.} B}

\def\PRL{\em Phys. Rev. Lett.}

\def\PREP{\em Phys. Rep.}

\def\PRD{{\em Phys. Rev.} D}

\def\ZPC{{\em Z. Phys.} C}

\def\ANNP{\em Ann. Phys. (N.Y.)}

\def\JHEP{{\em JHEP}\,}

\newcommand{\beq}[1]{
\begin{equation}\label{#1}}
\newcommand{\eeq}{\end{equation}}
\newcommand{\bea}[1]{
\begin{eqnarray}\label{#1}}
\newcommand{\eea}{\end{eqnarray}}

\newcommand\lr[1]{{\left({#1}\right)}}

\newcommand{\insertfig}[2]{\mbox{\epsfxsize=#1cm
\epsfbox{#2.eps}}}

\newcommand{\lsc}{\mbox{$\ell $}}

\newcommand{\JJ}{L}
\newcommand{\xxi}{u}

\newcommand{\ccal}{\mathbb}

\begin{document}

\thispagestyle{empty}
\date{}
\title{
\vspace*{1cm}
The Uses of Conformal Symmetry in QCD}

\author{V.\ M.\ Braun,$^a$ G.\ P.\ Korchemsky,$^b$ and  D.\ M{\"u}ller\,$^c$ \\
\\
$^a${\small \it Institut f{\"u}r Theoretische Physik, Universit{\"a}t Regensburg}\\
            \vspace*{-0.8cm}\\{\small\it D-93040 Regensburg, Germany}\\
$^b${\small\it Laboratoire de Physique Theorique, Universite Paris Sud,}\\
            \vspace*{-0.8cm}\\{\small\it 91405 Orsay Cedex, France}\\
$^c${\small\it Bergische Universit{\"a}t Wuppertal,  D-42097 Wuppertal, Germany}
       }
\maketitle

\vspace*{-9cm}
\begin{flushright}
 hep-ph/0306057
\end{flushright}
\vspace*{8.8cm}

\begin{abstract}
The Lagrangian of Quantum Chromodynamics is invariant
under conformal transformations. Although this symmetry is
broken by quantum corrections, it has important consequences for strong
interactions at short distances and provides one with powerful tools
in practical calculations. In this review we give a short exposition
of the relevant ideas, techniques and applications of conformal symmetry
to various problems of interest.
\end{abstract}

\newpage

\tableofcontents


\section{Introduction}

Conformal symmetry has a long history. It has been known for more than a century
that equations of electromagnetism are invariant with respect to the inversion of
coordinates $x^\mu \to x^\mu/x^2$. This symmetry is useful for solution of many
nontrivial problems and has numerous applications in electrostatics. Later it was
realized that the inversion symmetry is related to the scale invariance of the
underlying $U(1)$ gauge theory, at least at the classical level. The scale
transformations and the inversion together with the usual translations and
Lorentz rotations form the conformal group, which proves to be the maximal
extension of the Poincar\'e group that leaves the light-cone invariant. The
conformal symmetry of a quantum field theory was intensively studied in the end
of sixties and in the seventies, with the motivation coming independently from
the Bjorken scaling hypothesis (see \cite{Bj}) and the theory of second order
phase transitions. The theoretical description of critical behavior in many cases
can be obtained within the framework of conformal invariant theories. It turns
out that conformal symmetry of a two-dimensional quantum field theory implies
existence of infinitely many conservation laws and strong constraints on the
correlation functions. In particular, spectacular results were obtained in
statistical mechanical models in two dimensions. The subject of two-dimensional
conformal field theories (CFT) has become very fashionable because of connections
with string theory and quantum gravity (see, e.g., \cite{Polyakov}), and has
grown into a separate branch of mathematical physics with its own methods and own
language. In some 4-dimensional theories with supersymmetry the $\beta$-function
vanishes (to lowest order or to all orders) and some featurs reminicent of
two-dimensional CFT emerge. This subject attracted a lot of interest, enhanced in
recent years in connection with the conjecture about AdS/CFT correspondence, see
\cite{AGMOO} for a review.

On the other hand,
the impact of conformal symmetry on the development of  Quantum
Chromodynamics (QCD) has been rather modest. In the QCD community, a
widespread scepticism exists to the use of methods based on conformal
invariance, for several reasons. As we will see below, the conformal
symmetry of a quantum theory implies that its $\beta$-function must
vanish. In QCD, this is only the case for the free theory, $\alpha_s\to
0$, and in this limit it essentially reduces to the parton model as far
as ``hard'' processes at large energies and large momentum transfers are
concerned. The scaling laws of the parton model (e.g., quark counting
rules etc.) can indeed be derived directly from the conformal symmetry
of the classical QCD Lagrangian, but this formal connection is usually
not what a QCD theorist is interested in. The proximity to the conformal
limit is determined by the value of the strong coupling and can be
controlled in hard processes. It is usually thought, however, that the
study of this behavior, QCD at small coupling, is not really interesting
for a ``hep-th'' theorist, while using specific techniques based on
conformal symmetry in the QCD phenomenology is complicated and is not
worth the effort.

This review is an attempt to overcome this prejudice. We will restrict ourselves
to the perturbation theory and demonstrate that the structure of perturbative
predictions for light-cone dominated processes reveals the underlying conformal
symmetry of the QCD Lagrangian. Exposing the symmetry allows one to simplify
calculations, obtain new results and new insight. It is worthwhile to mention
that although the subject of conformal symmetry is old, the applications
considered in this review are rather recent. Loosely speaking, the subject of
perturbative QCD is calculation of the scale dependence of physical observables,
which is governed by evolution  equations. The known evolution equations for
structure functions,  distribution amplitides and generalized parton
distributions can be understood as the renormalization group equations for the
light-cone operators and have much in common. Using conformal symmetry in order
to understand the structure of these equations presents our main topic. In
addition, we will consider the so-called BFKL equation for the perturbative
resummation of large logarithms of energy, which is relevant in Regge limit.

The outline of the review is as follows. Sect.~2 is introductory and is
meant to provide basic information about the conformal group, conformal
invariance in a field theory and its violation. In particular, the
construction of conformal operators, the conformal operator product
expansion and conformal Ward identities are important for QCD
applications. Our presentation is necessarily heuristic and incomplete.
More details can be found in numerous reviews and in particular we
recommend \cite{Jackiw,Coleman} for the first reading.

Sect.~3 is devoted to the conformal partial wave expansion of hadron
distribution amplitudes. The approach is very similar to the partial
wave expansion of scattering amplitudes in nonrelativistic quantum
mechanics and aims at the separation of variables. In quantum mechanics,
$O(3)$ symmetry of the scattering potential allows one to separate the
dependence of the wave function on angular coordinates in terms of
spherical harmonics which realize an irreducible representation of the
symmetry group $O(3)$. In QCD the corresponding symmetry group is
$SL(2,{\mathbb R})$ and it allows one to separate the dependence of hadron
wave functions on transverse and longitudinal distances with respect to
the scattering plane. The dependence on longitudinal degrees of freedom
(momentum fractions) is included in ``spherical harmonics'' of
$SL(2,{\mathbb R})$ while the dependence on transverse coordinates is
traded for the scale dependence and is governed by simple
renormalization group equations. Historically this expansion was first
suggested around 1980 \cite{ER80,LB80} and was instrumental for the
proof of the QCD factorization (see \cite{exclusive} for a review) for
the elastic and transition form factors. This, essentially algebraic,
program has been then extended to general multi-particle distributions
\cite{BF90} and is still in progress. For higher twists, it is important
that conformal expansion is fully consistent with QCD equations of
motion.

In Sect.~4 we consider a general approach to QCD evolution equations which is
based on the representation of evolution kernels as formal Hamilton operators
depending on Casimir operators of the $SL(2,{\mathbb R})$ group
\cite{BukFroKurLip85}. In this way the $SL(2,{\mathbb R})$ symmetry of the
leading-order evolution equations is made manifest. Using this technique it has
recently become possible to solve evolution equations for three-parton operators
which govern, e.g., the scale dependence of baryon distribution amplitudes and
the structure function $g_2(x,Q^2)$ in polarized deep inelastic scattering. It
turns out that some three-particle evolution equations are completely integrable
\cite{BDM98}, since they possess a ``hidden'' symmetry alias a new quantum
number. Complete integrability is a new and unexpected feature which comes on the
top of conformal symmetry and is not seen at the classical level from the QCD
Lagrangian. The same methods turn out to be applicable in a different, Regge
kinematics in which case conformal symmetry allows one to solve the so-called
BFKL equation for arbitrary momentum transfers \cite{Lipatov85}. Miraculously,
the complete integrability appears here as well \cite{Lipatov94,FK95} and allows
us to find the Regge intercepts of the system of interacting reggeized gluons in
the large $N_c$ limit.

Sect.~5 is devoted to applications of conformal symmetry to QCD calculations
beyond the leading order, in which case change of scaling dimensions for
elementary fields and composite operators and the running of the QCD coupling
have to be taken into account. It was conjectured long ago
\cite{Nie77AdlColDun77ColDunJog77Min76} that the symmetry breaking corrections
must be proportional to the QCD $\beta$-function and have a rather simple form.
The concrete predictions \cite{BroFriLepSac80} based on conformal symmetry seemed
to be in contradiction with explicit calculations
\cite{DitRad84Sar84Kat85MikRad85}, however, and the resolution of this paradox
was only found much later \cite{Mue91a,Mue94}. About the same time a pre-QCD
conformal prediction, the so-called Crewther relation \cite{Cre72}, was verified
within perturbative QCD calculations at next-to-next-to-next-to leading order
\cite{BroKat93}. We explain these developments and present the calculation of
conformal anomalies that are used to reconstruct anomalous dimensions and
evolution kernels at next-to-leading order (NLO) \cite{BelMue98a,BelMue98c}. Last
but not least, we demonstrate the predictive power of the conformal operator
product expansion for the perturbative expansion of light-cone dominated
two-photon processes \cite{Mue97a,BelMue97a}. As a concrete application, the
photon-to-pion transition form factor at next-to-next-to-leading order (NNLO)  is
considered \cite{MelMuePas02}.

Finally, in Sect.~6 we summarize. The review includes two Appendices:
Appendix A is devoted to the construction of the orthonormal basis of
three-particle conformal operators and in Appendix B we sketch the
derivation of conformal Ward identities, including ghosts and equation
of motion terms.

\section{Preliminaries}
\subsection{\it Conformal Group and its Collinear Subgroup}

Among the general coordinate transformations of the 4-dimensional
Minkowski space that conserve the interval $ds^2 = g_{\mu\nu}(x) dx^\mu
dx^\nu$ there are transformations that change only the scale of the
metric:
\beq{scale1}
  g'_{\mu\nu}(x') = \omega(x) g_{\mu\nu}(x) \,
\eeq
and, consequently, preserve the angles and leave the light-cone
invariant.
 All transformations
belonging to this subclass form, by definition, the conformal group. It
is obvious that conformal transformations correspond to a generalization
of the usual Poincar\'e group, since the Minkowski metric is not changed
by translations and Lorentz rotations. Examples of specific conformal
transformations are the dilatation (global scale transformation) and
inversion
\beq{scale2}
  x^\mu\to x'^\mu = \lambda x^\mu, \qquad
  x^\mu \to x'^\mu  = \frac{x^\mu}{x^2}\,,
\eeq
with real $\lambda$.  Another important example is the so-called
special conformal transformation
\beq{special}
    x^\mu \to x'^\mu = \frac{x^\mu + a^\mu x^2}{1+2 a\cdot x + a^2 x^2}\,,
\eeq
which corresponds to the sequential inversion, translation by an arbitrary
constant vector $a_\mu$ and one more inversion with the same center.

The full conformal algebra in 4 dimensions includes fifteen generators
\cite{MS69,Jackiw}:

\vskip0.3cm

\hspace*{4cm} \parbox[t]{10cm}{${\bf P}_\mu$ (4 translations)\\
                               ${\bf M}_{\mu\nu}$ (6 Lorentz rotations)\\
                               ${\bf D}$~~\,(dilatation) \\
 ${\bf K}_\mu$ (4 special conformal transformations)
                                    }

\vskip0.5cm

\noindent
and is a generalization of the familiar 10-parameter Lie algebra of
the Poincar\'e group generated
by ${\bf P}_\mu$ and ${\bf M}_{\mu\nu}$:
\bea{Poin}
   &&  i[{\bf P}_\mu,{\bf P}_\nu] = 0\,,\quad
     i[{\bf M}_{\alpha\beta}, {\bf P}_{\mu}] =
     g_{\alpha\mu} {\bf P}_\beta - g_{\beta\mu} {\bf P}_\alpha\,,
\nonumber\\
   &&  i[{\bf M}_{\alpha\beta},{\bf M}_{\mu\nu}] =
g_{\alpha\mu} {\bf M}_{\beta\nu}  - g_{\beta\mu} {\bf M}_{\alpha\nu}
- g_{\alpha\nu} {\bf M}_{\beta\mu}  + g_{\beta\nu} {\bf M}_{\alpha\mu}\,.
\eea
The remaining  commutation relations that specify the conformal algebra are:
\bea{confalgebra}
  &&  i[{\bf D},{\bf P}_\mu] = {\bf P}_\mu\,, \quad
    i[{\bf D},{\bf K}_\mu] = -{\bf K}_\mu\,,
\nonumber\\
  &&  i[{\bf M}_{\alpha\beta}, {\bf K}_{\mu}] = g_{\alpha\mu} {\bf K}_\beta
- g_{\beta\mu} {\bf K}_\alpha\,,\quad
    i[{\bf P}_\mu,{\bf K}_\nu] = -2 g_{\mu\nu}{\bf D} + 2 {\bf M}_{\mu\nu}\,,
\nonumber\\
  &&  i[{\bf D},{\bf M}_{\mu\nu}] =  i[{\bf K}_\mu,{\bf K}_\nu] \, = \, 0\,.
\eea
The generators act on a generic fundamental field $\Phi(x)$ (primary field,
in the language of conformal field theories) with an arbitrary spin as
\bea{trafo}
 \delta_P^{\mu} \Phi(x) &\equiv&  i[{\bf P}^\mu, \Phi(x)] \,
=\,  \partial^\mu \Phi(x)\,,
\nonumber\\
 \delta_M^{\mu\nu}\Phi(x) &\equiv&  i[{\bf M}^{\mu\nu},\Phi(x)] =
\left(x^\mu\partial^\nu-x^\nu\partial^\mu-
                         \Sigma^{\mu\nu}\right)\Phi(x)\,,
\nonumber\\
  \delta_D\Phi(x) &\equiv&  i[{\bf D},\Phi(x)] =
\left(x\cdot\partial + \ell\right)\Phi(x)\,,
\nonumber\\
  \delta^\mu_K\Phi(x) &\equiv&  i[{\bf K}^\mu,\Phi(x)] =
\left(2 x^\mu x\cdot\partial -x^2 \partial^\mu+2 \ell x^\mu
   - 2 x_\nu\Sigma^{\mu\nu}\right)\Phi(x)\,.
\eea
For example, if we consider infinitesimal translations $x_\mu\to x'_\mu =
x_\mu+\epsilon_\mu$, the field $\Phi(x)$ transforms as $\Phi'(x) = [1
+\epsilon_\mu \delta_P^\mu] \Phi(x)$, etc. $\Sigma^{\mu\nu}$ is the
generator of spin rotations of the
field $\Phi$.
For scalar, Dirac spinor (quark) and vector (gluon) fields
\beq{spin1}
 \Sigma^{\mu\nu}\phi(x) =0, \qquad \Sigma^{\mu\nu}\psi =
\frac{i}{2}\sigma^{\mu\nu}\psi, \qquad
 \Sigma^{\mu\nu}A^{\alpha} =
g^{\nu\alpha}A^{\mu}- g^{\mu\alpha}A^{\nu}\,,
\eeq
respectively, where $\sigma^{\mu\nu} = i[\gamma^\mu,\gamma^\nu]/2$. The parameter
$\ell$ is called the scaling dimension and it specifies the field transformation
under the dilatations. In a free theory (i.e.,\ at the classical level) the
scaling dimension coincides with the canonical dimension $\ell^{\rm can}$ which
is fixed by the requirement that the action of the theory is dimensionless. In
the quantum theory $\ell \not= \ell^{\rm can}$, in general, and the difference is
called the anomalous dimension.

An ultra-relativistic particle (quark or gluon) propagates close to the light-cone so that one
may envisage that separation of transverse and longitudinal coordinates will prove to be
essential. For further use, we introduce shorthand notations for the projections on the two
independent light-like vectors. For arbitrary four-vector $A_\mu$ we define
\beq{dot}
                        A_+ \equiv A_\mu n^\mu\,,\qquad A_- \equiv A_\mu \bar n^\mu\,,
                  \qquad n^2=\bar n^2 =0\,, \qquad n\cdot \bar n =1\,
\eeq
and the metric tensor in the directions orthogonal to the light-cone
\beq{gperp}
       g_{\mu\nu}^\perp = g_{\mu\nu} - n_\mu \bar n_\nu - n_\nu \bar n_\mu\,.
\eeq
We will also use the notation $A_\perp$ for a generic transverse projection and
$A_\perp^\mu$ for the vector that only has transverse components.
For example
\beq{sudakov}
          x^\mu = x_- n^\mu + x_+ \bar n^\mu + x^\mu_\perp\,,\qquad x^\mu_\perp \equiv g^{\mu\nu}_\perp x_\nu
\eeq
and, therefore, $x^2 = 2 x_+ x_- - x_\perp^2$.

Consider a special case of the conformal transformation in Eq.\ (\ref{special}) with $a_\mu$ being a
light-like vector $a_\mu=a \bar n_\mu$. One finds
\beq{lc1}
     x_- \to x_-' = \frac{x_-}{1+ 2 a\, x_-}
\eeq
so that  these transformations map the light-ray in the $x_-$-direction into
itself. Together with the translations and dilatations along the same direction,
$x_-\to x_-+c$ and $x_-\to \lambda x_-$, they form a so-called collinear subgroup
of the full conformal group. This subgroup is nothing else than the familiar
$SL(2,{\mathbb R})$ group \cite{lang}. It will play a central role in our
analysis.

In the parton model, hadron states are replaced by a bunch of partons
that are collinear, that is all move in the same direction, say $\bar
n_\mu$.
 Whenever
this picture applies, one typically only needs to consider quantum fields
``living'' on the light-ray
\beq{lray}\Phi(x) \to \Phi(\alpha n)\,,\eeq
where $\alpha$ is a real number. In what follows we will often use a
shorthand notation
$\Phi(\alpha)\equiv \Phi(\alpha n )$ in order not to overload the formulas.
We will further assume that the field $\Phi$ is chosen to be
an eigenstate of the spin operator $\Sigma_{+-}$
and so it has fixed projection $s$ on the ``+'' direction:
\beq{spin}
          \Sigma_{+-}\Phi(\alpha) = s \,\Phi(\alpha)\,.
\eeq
This property is automatically satisfied for leading twist
operators \cite{Ohrndorf}. In the general case one should use suitable projection operators
to separate different spin components. Several examples will be considered below.

 Within these specifications, four-dimensional conformal transformations are reduced to the
 collinear subgroup that generates projective transformations on a line:
\bea{project}
         \alpha \to \alpha' &=& \frac{a\alpha+b}{c\alpha + d}\,,~~ ad-bc =1\,,
\nonumber\\
 \Phi(\alpha) \to \Phi'(\alpha) &=& (c\alpha +d)^{-2j} \Phi\left(\frac{a\alpha+b}{c\alpha+d}\right),
\eea
where $a,b,c,d$  are real numbers and
\beq{cspin}
      j= (\ell+s)/2\,.
\eeq
These transformations are generated by the four generators ${\bf P}_{+}$, ${\bf
M}_{-+}$, ${\bf D}$ and ${\bf K}_{-}$ which form a collinear subalgebra of the
conformal algebra (\ref{Poin}) and (\ref{confalgebra}). In order to bring the
commutation relations to the standard form, it is convenient to introduce the
following linear combinations \cite{Ohrndorf,BDKM}:
\bea{J}
  {\bf \JJ}_+ = {\bf \JJ}_1 + i {\bf \JJ}_2  = - i {\bf P}_+\,, &~~~~~~&
  {\bf \JJ}_- = {\bf \JJ}_1 - i {\bf \JJ}_2  = ({i}/{2}) {\bf K}_-\,,
\nonumber\\
 {\bf \JJ}_0 = ({i}/{2}) ({\bf D}+{\bf M}_{-+})\,,  &~~~~~~&
 {\bf E} =   ({i}/{2})({\bf D}-{\bf M}_{-+})\,.
\eea
We obtain
\bea{SL2Rop}
  [{\bf \JJ}_0,{\bf \JJ}_\mp] =\mp  {\bf \JJ}_\mp\,, &~~&  [{\bf \JJ}_-,{\bf \JJ}_+] = -2 {\bf \JJ}_0 \,,
\eea
which is the algebra of $SL(2,{\mathbb R}) \sim O(2,1)$.
Action of the generators defined in Eq.\ (\ref{J}) on quantum fields can easily be derived from
Eq.\ (\ref{trafo}), and it can be traded for the algebra of differential
operators acting on the field coordinates
\bea{SL2R2}
 {}[{\bf \JJ}_+,\Phi(\alpha)] &=& - \partial_\alpha\Phi(\alpha) \equiv \JJ_+\Phi(\alpha)\,,
\nonumber\\
 {}[{\bf \JJ}_-,\Phi(\alpha)] &=&  \left(\alpha^2\partial_\alpha+2 j \alpha \right)\Phi(\alpha)
          \equiv \JJ_-\Phi(\alpha)\,,
\nonumber\\
 {}[{\bf \JJ}_0,\Phi(\alpha)] &=&  \left(\alpha\partial_\alpha+ j\right)\Phi(\alpha) \equiv \JJ_0\Phi(\alpha)\,,
\eea
where $\partial_\alpha = d/d\alpha$. They satisfy the similar $SL(2)$
commutation relations:
\bea{SL2R}
  [{ \JJ}_0,{ \JJ}_\mp] = \pm { \JJ}_\mp\,, &~~&  [{ \JJ}_-,{ \JJ}_+] = 2 { \JJ}_0 \,.
\eea
Note that we use boldface letters for the generators acting on quantum fields to distinguish
them from the corresponding differential operators acting  on the field coordinates.
The remaining generator ${\bf E}$ counts the {\it twist} $t=\ell-s$ of the field $\Phi$:%
\beq{E}
  [{\bf E},\Phi(\alpha)] = \frac12(\ell-s)\Phi(\alpha)\,.
\eeq
It commutes with all ${\bf \JJ}_i$ and is not relevant for most of our
discussions.
{The definition in (\ref{E}) corresponds to the so-called
{\it collinear twist} (``dimension - spin projection on the
plus-direction''), which differs from {\it geometric twist} (``dimension
- spin'') \cite{GroTre71} that refers to the full conformal group instead of the
collinear subgroup. Relations between operators, distribution
functions etc. of definite collinear twist are well understood by the
classification in terms of geometric twist since the latter respects
full Lorentz symmetry, see e.g. \cite{JJ91,BBKT98,GL01}.}

A local field operator $\Phi(\alpha)$ that has fixed spin projection (\ref{spin}) on the light-cone
is an eigenstate of the quadratic Casimir operator
\beq{casim1}
    \sum_{i=0,1,2}[{\bf \JJ}_i,[{\bf \JJ}_i,\Phi(\alpha)]] = j(j-1) \Phi(\alpha) = \JJ^2 \Phi(\alpha)
\eeq
with the operator $L^2$ defined as
\beq{casimir}
   {\JJ}^2 = {\JJ}_0^2+{\JJ}_1^2+{\JJ}_2^2 = \JJ_0^2-\JJ_0+ \JJ_-\JJ_+\,,
   \qquad [\JJ^2,\JJ_i]=0\,.
\eeq
Eqs.~(\ref{project}) and (\ref{casim1}) imply that the field $\Phi(\alpha)$ is
transformed under the projective transformations according to a representation of
the $SL(2,\mathbb{R})$ group specified by the parameter $j$, to which we shall
refer as the {\it conformal spin} of the field. As follows from its definition,
Eq.~(\ref{cspin}), $j$ is positive and takes (half)integer values. We will see below that
we are dealing with infinite-dimensional representations of the collinear conformal group.

Besides the collinear subgroup just described, one can consider another subgroup
corresponding to transformations of the two-dimensional transverse plane
$x_\perp^\mu=(0,x_1,x_2,0)$ introduced in (\ref{sudakov}). This ``transverse''
subgroup involves six generators $\mathbf{P}_\perp^\mu$,
$\mathbf{M}_\perp^{\mu\nu}$, $\mathbf{D}$ and $\mathbf{K}_\perp^\mu$ and is
isomorphic to $SL(2,\mathbb{C})$. It is convenient to introduce complex
coordinates
\beq{z}
z=x_1+ix_2\,,\qquad \bar z=x_1-ix_2=z^*
\eeq
in terms of which the group transformations become
\beq{z-trans}
z\to \frac{az+b}{cz+d}\,,\qquad \bar z\to \frac{\bar a\bar z+\bar b}{\bar c\bar
z+\bar d}\,,
\eeq
where $a,b,c,d$ are complex numbers and $ad-bc=1$. For the fundamental fields
living on the transverse plane, $\Phi=\Phi(z,\bar z)$, the corresponding
transformation laws are
\beq{laws}
\Phi(z,\bar z) \to \Phi'(z,\bar z)=(cz+d)^{-2h} (\bar c \bar z + \bar d)^{-2\bar
h} \Phi\left(\frac{az+b}{cz+d},\frac{\bar a\bar z+\bar b}{\bar c\bar z+\bar d}
\right)\,,
\eeq
where $h=(\ell+\lambda)/2$ and $\bar h=(\ell-\lambda)/2$, and $\lambda$ is the
helicity of the field defined as $\Sigma^{z\bar z} \Phi=\lambda \Phi$. This
subgroup turns out to be relevant for the QCD description of the high energy
scattering in the Regge kinematics, see Sect.~4.5. Note that the collinear and
the transverse subgroups share a common generator of dilations $\mathbf{D}$ and,
therefore, are not independent.

\subsection{\it Conformal Towers}

The algebra of the collinear conformal group coincides with the algebra $O(2,1)$
of hyperbolic rotations which are simply Lorentz transformations in 2+1
dimensional space-time. Similar as for usual spin rotations, it is convenient to
introduce a complete set  of states that in addition to the conformal spin $j$
also have fixed conformal spin projection on a certain axis in this internal
space. The standard way to construct such a conformal basis is in terms of local
composite operators built of fundamental (primary) fields and their derivatives.

As easily seen from Eq.\ (\ref{SL2R2}), the field at the origin of the light-cone
$\Phi(0)$ is an eigenstate of ${\bf \JJ}_0$ with the eigenvalue $j_0=j$ and is
annihilated by ${\bf \JJ}_-$:
\beq{highest}
[{\bf \JJ}_-,\Phi(0)]=0\,,\qquad [{\bf \JJ}_0,\Phi(0)]=j\, \Phi(0)\,.
\eeq
It defines, therefore, the so-called highest weight vector on the
$SL(2,\mathbb{R})$ representation space. A complete basis of operators on this
space can be obtained from $\Phi(0)$ by applying $k$-times the ``raising''
operator ${\bf \JJ}_+$:
\beq{oper}
           {\cal O}_{k} =  [{\bf \JJ}_+,\ldots,[{\bf \JJ}_+,
    [{\bf \JJ}_+,\Phi(0)]]] = [(-\partial_+)^k \Phi(\alpha)]\bigg|_{\alpha=0}
 \,,\qquad {\cal O}_{0} \equiv \Phi(0)\,.
\eeq
{}From the commutation relations it follows that
\bea{oper1}
&&{}[{\bf \JJ}_0, {\cal O}_{k}] = (k+j) {\cal O}_{k}\,,\qquad
     {}  [{\bf \JJ}_+, {\cal O}_{k}] = {\cal O}_{k+1}\,,\qquad
     {}  [{\bf \JJ}_-, {\cal O}_{k}] = -k(k+2j-1) {\cal O}_{k-1}\,.
\eea
The primary field operator at an arbitrary position on the light-cone $\Phi(\alpha)$
can formally be expanded in a Taylor series over local conformal operators
\beq{Phi-exp}
\Phi(\alpha)= \sum_{k=0}^\infty \frac{(-\alpha)^k}{k!} {\cal O}_{k}\,,
\eeq
which can be interpreted as  the relation between  two different
complete sets of states, $\Phi(x)$ for arbitrary $x$ and ${\cal O}_{k}$
for arbitrary $k$.

This construction presents the simplest example of what is called
the {\it conformal tower}. The lowest operator in a conformal tower is
the highest weight vector in the space of representations. Higher operators
are obtained by adding total derivatives,
each of which adds one unit to the conformal spin projection on the
``zero'' axis.
Aim of this  Section is to
construct conformal towers for a general situation of
local composite operators built of several fundamental
fields and (covariant) derivatives. Before doing this, we have
to give some more definitions.

Any local composite operator can be specified by a polynomial that details
the structure and the number of derivatives acting on the fields. For example, the tower of
local operators introduced in Eq.\ (\ref{oper}) can be presented as
\beq{cf1}
   {\cal O}_{k} = [{\cal P}_{k}(\partial_\alpha) \Phi(\alpha)]\bigg|_{\alpha=0}\,,
   \qquad {\cal P}_{k}(\xxi) = (-\xxi)^k\,,
\eeq
where $\partial_\alpha \equiv d/d\alpha$.
The algebra of generators acting on the composite operators (\ref{oper1}) can equivalently
be rewritten as algebra of differential operators acting on the space of characteristic
polynomials. Requiring
\bea{cf2}
 {}[\widetilde{\JJ}_{0} {\cal P}(\partial_\alpha) \Phi(\alpha)]\bigg|_{\alpha=0} &=&
 {\cal P}(\partial_\alpha)[\JJ_{0}\Phi(\alpha)]\bigg|_{\alpha=0}\,,
\nonumber\\
 {}[\widetilde{\JJ}_{\pm} {\cal P}(\partial_\alpha) \Phi(\alpha)]\bigg|_{\alpha=0} &=&
 {\cal P}(\partial_\alpha) [\JJ_{\mp}\Phi(\alpha)]\bigg|_{\alpha=0}
\eea
one finds the following `adjoint' representation of the generators
acting on this space:
\bea{adjoint}
\widetilde{\JJ}_{0}{\cal P}(\xxi) &=& \left(\xxi\partial_\xxi +j\right) {\cal P}(\xxi)\,,
\nonumber\\
\widetilde{\JJ}_{-}{\cal P}(\xxi) &=& -\xxi \,{\cal P}(\xxi)\,,\quad
\nonumber\\
\widetilde{\JJ}_{+}{\cal P}(\xxi) &=&
\left(\xxi\partial^2_\xxi +2 j \partial_\xxi\right){\cal P}(\xxi)\,,
\end{eqnarray}
where $\partial^2_\xxi =d^2/d\xxi^2$ and in  order to maintain the same
commutation relations (\ref{SL2R}), we have defined $\widetilde{\JJ}_-$ as the
adjoint to $\JJ_+$, and vice versa. Note that the generators in the adjoint
representation are more complicated compared to the generators (\ref{SL2R2}), in
particular $\widetilde{\JJ}_+$ contains the second derivative. The following
trick \cite{Ohrndorf,BDKM} allows us to avoid this problem and proves to
be very convenient in applications.

 Let us introduce a new variable $\kappa$ instead of $\xxi$ by the following rule
\beq{T}
  \frac{\xxi^n}{\Gamma(n+2j)} \to \kappa^n\,,\qquad n=0,1,2,\ldots
\eeq
In this way a characteristic polynomial ${\cal P}_n(\xxi)$ is mapped onto a
polynomial in $\kappa$:
\beq{T2}
   {\cal P}_n(\xxi) \to \widetilde{\cal P}_n(\kappa)\,.
\eeq
In what follows we refer to $\widetilde{\cal P}_n(\kappa)$ as the
characteristic polynomial in the adjoint representation. The
rationale for this name is that the generators of projective
transformations become in this space
\bea{T3}
     \JJ_0 \widetilde{\cal P}_n(\kappa) &=&\left(
  \kappa {\partial_\kappa}
   +j\right)\widetilde{\cal P}_n(\kappa)\,,
\nonumber\\
    \JJ_- \widetilde{\cal P}_n(\kappa) &=&
   - {\partial_\kappa}\widetilde{\cal P}_n(\kappa)\,,
\nonumber\\
    \JJ_+ \widetilde{\cal P}_n(\kappa) &=&\left(
   \kappa^2 {\partial_\kappa}
   +2j\kappa \right)\widetilde{\cal P}_n(\kappa)\,,
\eea
i.e., they are given by the same differential operators as the
original generators in Eq.~(\ref{SL2R2}) acting on the field coordinate.
At this point, the construction of the characteristic polynomial in
the adjoint representation  can look very formal.
We will later see, however, that the transformation (\ref{T}) follows naturally
from the conformal transformation properties of correlation functions.

Let us now consider the product of two fields ``living'' on the light-cone,
\beq{co-0}
  O(\alpha_1,\alpha_2) = \Phi_{j_1}(\alpha_1)\,\Phi_{j_2}(\alpha_2)
\eeq
with $\alpha_1\neq \alpha_2$. Here we
inserted subscripts to indicate that the fields may have different conformal
spins. Expanding this product at short distances $|\alpha_1-\alpha_2| \to0$ one encounters the
set of local composite operators%
\footnote{Throughout this review we adopt the following notations: $j$ (or $J$) always stands
 for the conformal spin; $n$ (or $N$) denotes the degree of the characteristic polynomial
(the number of derivatives of the highest weight (conformal) operator), $l$ is the total
number of derivatives and
and $k=l-n$ (the number of total derivatives) denotes the $k$-th term climbing
up the conformal tower, so that $|j_0| = j+k$.}
\beq{co-1}
              {\cal O}_n(0)= {\cal P}_n(\partial_1,\partial_2)
\Phi_{j_1}(\alpha_1)\,\Phi_{j_2}(\alpha_2)\bigg|_{\alpha_1=\alpha_2=0}\,,
\eeq
where $\partial_1 \equiv \partial/\partial \alpha_1$ etc.,  and with ${\cal
P}_n(\xxi_1,\xxi_2)$ being a homogeneous polynomial of degree $n$. Generally
speaking the operators ${\cal O}_n$ do not have any simple properties under
conformal transformations. Our aim is, first,  to construct a complete basis of
local operators that form a conformal tower, and, second, to work out the
operator product expansion (OPE) in this basis --- the conformal operator product
expansion (COPE).

The collinear conformal transformations of two-particle operators defined in Eq.\
(\ref{co-0}) or (\ref{co-1}) correspond to independent transformations
of the fields; the group generators are given, therefore,  by the sum of one-particle
generators
\beq{co-2}
          {\bf \JJ}_a = {\bf \JJ}_{1,a} + {\bf \JJ}_{2,a}\,,\qquad a=0,1,2
\eeq
and the two-particle Casimir operator is
\beq{co-3}
       {\bf \JJ}^2 = \sum_{a=0,1,2} \left( {\bf \JJ}_{1,a} + {\bf \JJ}_{2,a}\right)^2.
\eeq
Our task is to work out the conformal tower of local operators with the given conformal spin
and the conformal spin projection on a given axis, which is the classical problem of spin summation
in nonrelativistic quantum mechanics, albeit for a different group, $SL(2,\mathbb{R})$.

We define a local {\it conformal operator} ${\mathbb O}_n$ by requiring
that its transformation properties under the collinear conformal
subgroup are the same as for the fundamental field (\ref{project}).
This is equivalent to imposing the following three
requirements
\bea{co-4}
{} [{\bf \JJ}^2,{\mathbb O}_n] &=& j(j-1) {\mathbb O}_n\,,
\nonumber\\
{} [{\bf \JJ}_0,{\mathbb O}_n] &=& (j_1+j_2+n)  {\mathbb O}_n\,,
\nonumber \\
{} [{\bf \JJ}_-,{\mathbb O}_n] &=& 0\,.
\eea
The first equation in (\ref{co-4}) states that ${\mathbb O}_n$ is an eigenstate of the quadratic
Casimir operator (\ref{co-3}), i.e., it has conformal spin $j$. The second equation requires
that  ${\mathbb O}_n$ has a given spin projection $j_1+j_2+n$. This condition is satisfied trivially once
 ${\mathbb O}_n$ is a homogeneous polynomial in derivatives of degree $n$, cf.\ Eq.\ (\ref{co-1}).
Finally, the third equation in (\ref{co-4}) is that ${\mathbb O}_n$ is
annihilated by the ``lowering'' operator ${\bf \JJ}_-$, i.e., it defines the
highest weight on the $SL(2,\mathbb{R})$ representation space, corresponding to
the minimum value of the spin projection $j_0$ for a given spin $j$. In other
words, the conformal operator is defined as the lowest state of an irreducible
conformal tower. Since ${\bf \JJ}^2 = {\bf \JJ}_0^2 - {\bf \JJ}_0 + {\bf
\JJ}_+{\bf \JJ}_-$, the highest weight condition together with the second equation
in (\ref{co-4}) make the first equation redundant and imply
\beq{co-5}
                           j= j_1+j_2+n
\eeq
which is the reason that we do not need the second label $j$ for the conformal operator.

Once the conformal operator (highest weight) is known, there are two
possibilities to construct a complete operator basis. The first possibility
is to introduce conformal operators for each space-time point ${\mathbb
O}_n(0)\to {\mathbb O}_n(x)$ with the transformation laws
({\ref{trafo}}) and (\ref{project}), the same as for the fundamental fields.
The second possibility is to build the conformal tower of
operators with higher spin projections that can be constructed by a repeated application
 of the ``raising'' operator ${\bf \JJ}_+$:
\beq{co-6}
           {\mathbb O}_{n,n+k} =
    \underbrace{[{\bf \JJ}_+,\ldots,[{\bf \JJ}_+,[{\bf \JJ}_+}_{k},
    \Phi(0)]]] = (-\partial_+)^k {\mathbb O}_{n} \,,\qquad {\mathbb O}_{n,n}
     \equiv {\mathbb O}_{n}\,,
\eeq
in full analogy with the one-particle case. Here $\partial_+$ is the total derivative.

In order to construct two-particle conformal operators explicitly one has to
find the corresponding characteristic polynomials (\ref{co-1}). This task can be
simplified going over to the adjoint representation ${\cal P}_n(\xxi_1,\xxi_2)
\to
\widetilde{\cal P}(\kappa_1,\kappa_2)$ at the intermediate step \cite{Ohrndorf}.
For two variables, the corresponding rule is
\beq{co-7}
  \frac{\xxi_1^{n_1} \xxi_2^{n_2}}{\Gamma(n_1+2j_1)\Gamma(n_2+2j_2)} \to \kappa_1^{n_1}\kappa_2^{n_2}
\eeq
and in this space
\bea{co-8}
   \JJ_0 \widetilde{\cal P}_n(\kappa_1,\kappa_2) &=&\left(
  \kappa_1 \partial_1+\kappa_2 \partial_2
   +j_1+j_2 \right)\widetilde{\cal P}_n(\kappa_1,\kappa_2)\,,
\nonumber\\
    \JJ_- \widetilde{\cal P}_n(\kappa_1,\kappa_2) &=& -\left(
   \partial_1+ \partial_2
             \right)\widetilde{\cal P}_n(\kappa_1,\kappa_2)\,,
\nonumber\\
   \JJ_+ \widetilde{\cal P}_n(\kappa_1,\kappa_2) &=&\left(
   \kappa_1^2 \partial_1+\kappa_2^2 \partial_2
   +2j_1\kappa_1+2j_2\kappa_2 \right)\widetilde{\cal P}_n(\kappa_1,\kappa_2)\,,
\eea
where $\partial_1 = \partial/\partial\kappa_1$ and
$\partial_2 = \partial/\partial\kappa_2$.
In particular, if $\widetilde{\mathbb P}_n(\kappa_1,\kappa_2)$ defines
the characteristic polynomial of
a conformal operator (highest weight), it has to satisfy
\beq{co-9}
   \JJ_- \widetilde{\mathbb P}_n(\kappa_1,\kappa_2) = -\left(
   \partial_1+ \partial_2
       \right)\widetilde{\mathbb P}_n(\kappa_1,\kappa_2)=0\,,
\eeq
which is solved by
\beq{co-10}
    \widetilde{\mathbb P}_n(\kappa_1,\kappa_2) \propto (\kappa_2-\kappa_1)^n.
\eeq
The coefficient of proportionality is irrelevant. Going over to the
original variables $u_1,u_2$ we obtain
\beq{co-11}
  {\mathbb P}_n^{j_1,j_2}(\xxi_1,\xxi_2) = \sum_{n_1+n_2=n}
\left(
      \begin{array}{c}
       n \\ n_1
      \end{array}
 \right)
 \frac{(-\xxi_1)^{n_1} \xxi_2^{n_2}}{\Gamma(n_1+2j_1)\Gamma(n_2+2j_2)} = (\xxi_1+\xxi_2)^n P^{(2j_1-1,2j_2-1)}_n
 \left(\frac{\xxi_2-\xxi_1}{\xxi_1+\xxi_2}\right),
\eeq
where $P^{(a,b)}_n(x)$ are the Jacobi polynomials \cite{BE} and we have added the
superscripts to indicate the conformal spins of the constituent fields. Stated
differently, we have proven that the coefficients of the Jacobi polynomials are
the Clebsch-Gordan coefficients of the collinear conformal group. The
corresponding local conformal operator is
\beq{co-12}
  {\mathbb O} _n^{j_1,j_2}(x) = \partial_+^n \left[\Phi_{j_1}(x) P^{(2j_1-1,2j_2-1)}_n
\left(\frac{\stackrel{\rightarrow}{\partial}_+ - \stackrel{\leftarrow}{\partial}_+}
           {\stackrel{\rightarrow}{\partial}_+ + \stackrel{\leftarrow}{\partial}_+}
\right)
\Phi_{j_2}(x)\right],
\eeq
where we have restored $x$ as the coordinate.
The expression in Eq.~(\ref{co-12}) was first obtained by Makeenko \cite{Makeenko81}
using a different method.

Note that the characteristic polynomials corresponding to the conformal
operators with different conformal
spin are mutually orthogonal with an appropriate weight function:
\beq{co-13}
    \langle {\mathbb P}_n | {\mathbb P}_m\rangle \equiv
\int_0^1\! d\xxi_1\, d\xxi_2\, \delta(1-\xxi_1-\xxi_2)\,
     \xxi_1^{2j_1-1} \xxi_2^{2j_2-1} {\mathbb P}_n^{j_1,j_2}(\xxi_1,\xxi_2)\,
     {\mathbb P}_m^{j_1,j_2}(\xxi_1,\xxi_2) = {\cal N} \delta_{mn}\,,
\eeq
where ${\cal N}$ is a normalization constant. This property
is related to the fact that the Casimir operator ${\bf \JJ}^2$ is
self-adjoint with respect to the so-called conformal scalar product, see Sec.~2.4.2.
For the operators (polynomials) that satisfy the highest weight condition, the
conformal scalar product reduces to Eq.~(\ref{co-13}). The orthogonality of
characteristic polynomials for the conformal operators with different conformal
spin follows from hermiticity of ${\bf \JJ}^2$ in the same way as the
orthogonality of wave functions with different energy follows from hermiticity of
the Hamiltonian in  usual quantum mechanics.

The above construction can easily be generalized for the products of three and more fields
in which case characteristic polynomials becomes functions of three and more independent
variables. For instance, for three fields Eq.~(\ref{co-9}) will be modified to
\beq{co-90}
   \JJ_- \widetilde{\mathbb P}_n(\kappa_1,\kappa_2,\kappa_3) = -\left(
   \partial_1+ \partial_2+\partial_3
                     \right)\widetilde{\mathbb P}_n(\kappa_1,\kappa_2,\kappa_3)=0\,,
\eeq
which is solved by any linear combination of terms
\beq{co-91}
    \widetilde{\mathbb P}^{n_1,n_2}_n(\kappa_1,\kappa_2,\kappa_3)
 \propto (\kappa_3-\kappa_2)^{n_1}
    (\kappa_2-\kappa_1)^{n_2},
\eeq
where $n_1+n_2=n$. Existence of multiple solutions means that the representation
is degenerate: there exist several ways to sum up the three conformal spins
$j_1$,  $j_2$,  $j_3$ to an overall spin $j=j_1+j_2+j_3+n$. Accepting Eq.\
(\ref{co-91}) as it stands and going over to original variables $\xxi_i$, one
obtains a particular basis of solutions that are known as Appell's polynomials
\cite{BE}. A disadvantage of using this basis is that the Appell's polynomials
corresponding to the different choice of $n_1,n_2$ are not mutually orthogonal. A
better, orthonormal basis, can be constructed if in addition to the total
conformal spin $j$ one requires a definite value of the conformal spin in a given
two-particle channel, see \cite{BDKM,BKM01} and Appendix A.

The conformal towers formed by local composite operators are the
standard, but not the only possible choice of the operator basis. In
fact, going over to local operators in coordinate space is not always
convenient.  Another option is to integrate the field $\Phi(\alpha)$
over the light-ray with a certain coefficient function \cite{BB89}
\beq{coh-1}
               {\cal S} = \int \!d\alpha\, \phi(\alpha) \Phi(\alpha)\, ,
\eeq
where it is assumed that the field decreases sufficiently fast at infinity so that the
integral is well defined, and fix $\phi(\alpha)$ by requiring that the nonlocal operator
${\cal S}$ has desired properties under conformal transformations, for arbitrary $\Phi$.
In particular, one can define the {\it conformal coherent state} as an
eigenstate of the
``raising'' operator
\beq{coh-2}
         [{\bf \JJ}_+, {\cal S}_p] = i p \,{\cal S}_p\,.
\eeq
Using Eq.~(\ref{SL2R2}) and integrating by parts one finds $\phi_p(\alpha) = \exp(ip\alpha)$,
so that Eq.~(\ref{coh-1}) becomes the usual Fourier transformation
and the eigenvalue $p$ can be identified with the momentum $p_+$ along the light-cone:
\beq{coh-3}
     {\cal S}_p \equiv \Phi(p_+) = \int \!d\alpha\, {\rm e}^{ip_+\alpha} \Phi(\alpha)\,.
\eeq
One can reexpress $\Phi(\alpha)$ in terms of the momentum components
(alias coherent states)
\beq{coh-4}
    \Phi(\alpha) = \int\frac{dp_+}{2\pi} \, {\rm e}^{-ip_+\alpha} \Phi(p_+)\,.
\eeq
Note similarity with Eq.~(\ref{Phi-exp}): In the both cases we expand a certain
vector in the Hilbert space over the complete basis of functions.

Any given momentum component $\Phi(p_+)$  (eigenstate of ${\bf \JJ}_+$) can be
decomposed into the eigenstates of ${\bf \JJ}_0$. Similar as in Eq.\
(\ref{SL2R2}) the action of generators on the field in the momentum
representation is reduced to differential operators acting on $p_+$.
{}For the
discussion of hadron distribution amplitudes in Sect.\ 3 it is convenient to use
polynomial solutions in momentum space, which can be constructed as follows.
Consider the complete set of functions \cite{BF90}
\beq{coh-55}
             \langle p_+|j,k\rangle = \frac{1}{\Gamma(2j+k)} (i p_+)^{k+2j-1}\,.
\eeq
As it is easy to see, the action of the generators (\ref{SL2R2}) in this space yields
\bea{coh-5}
            \JJ_+ \langle p_+|j,k\rangle &=& (k+2j)\langle p_+|j,k+1\rangle\,,
\nonumber\\
            \JJ_- \langle p_+|j,k\rangle &=& k \langle p_+|j,k-1\rangle\,,
\nonumber\\
            \JJ_0 \langle p_+|j,k\rangle &=& - (k+j)\langle p_+|j,k\rangle\,.
\eea
Thus $\JJ_+$ acts as the step up and $\JJ_-$ as the step down operator; the state
with $k=0$ corresponds to the lowest $|j_0|= j$ and it is annihilated by $\JJ_-$
so that it defines the highest weight in the momentum  representation. Any field
can formally be expanded in this basis
\beq{coh-NN}
      \Phi_j(p_+) = \sum_k  \varphi_k \cdot \langle p_+|j,k\rangle\, ,
\eeq
where $\varphi_k$ are operator coefficients that do not depend on
momenta.

\subsubsection{\it Conformal Operators in QCD: Examples}

Our discussion so far was completely general, let us now consider some examples.
In many QCD applications to hard processes one encounters a nonlocal operator
built of the quark and the antiquark at light-like separation
\beq{co-14}
     {\cal Q}_\mu(\alpha_1,\alpha_2) = \bar \psi(\alpha_1) \gamma_\mu [\alpha_1,\alpha_2]
     \psi(\alpha_2)\,,
\eeq
where
\beq{co-15}
     [\alpha_1, \alpha_2 ] = \mbox{\rm Pexp}
    \left[ig \int_{\alpha_2}^{\alpha_1} \! dt A_+(t)\right]
\eeq
is the Wilson line. Expanding ${\cal Q}_\mu(\alpha_1,\alpha_2)$ at short distances gives rise to local
operators built of the quark and the antiquark field and covariant derivatives, of the type
$
 \bar\psi (\stackrel{\leftarrow}{D}_+)^{n_1} \gamma_\mu (\stackrel{\rightarrow}{D}_+)^{n_2} \psi\,.
$
Our task is to find the corresponding conformal operators.

The starting observation is that we cannot identify the quark field operator $\psi$
directly with the fundamental
field $\Phi$, but first have to separate contributions with different spin projections. To this end
we introduce the spin projection operators
\beq{co-16}
                     \Pi_+ = \frac12 \gamma_-\gamma_+\,, \qquad
                     \Pi_- = \frac12 \gamma_+\gamma_-\,, \qquad
                     \Pi_+ + \Pi_- = 1
\eeq
and define ``plus'' and ``minus'' components of the quark field as
\beq{co-17}
         \psi_+ = \Pi_+\psi\,, \qquad \psi_- = \Pi_-\psi\,,\qquad \psi = \psi_+ +\psi_-\,,
\eeq
which is the same decomposition as is made in light-cone quantization, although the motivation
is in our case different. Using Eq.\ (\ref{spin1}) one easily finds that $\psi_+$ and $\psi_-$
correspond to the spin projections (\ref{spin}) $s=+1/2$ and $s=-1/2$, respectively.
Remembering that the canonical dimension of a spinor field equals
$\ell_\psi =3/2$, one finds for $\psi_+$ ($\psi_-$)
the values of the conformal spin (\ref{cspin}) $j=1$ ($j=1/2$) and twist
(\ref{E}) $t=1$ ($t=2$), respectively. Now we see that different Lorentz projections of the
operator ${\cal Q}_\mu(\alpha_1,\alpha_2)$ correspond to taking into account different quark field
components and have different properties under conformal transformations:
\bea{co-18}
 \mbox{\rm twist-2:}~~~~~
    {\cal Q}_+ &=& \bar \psi_+\gamma_+\psi_+ \equiv  {\cal Q}^{1,1}\,,
\nonumber\\
\mbox{\rm twist-3:}~~~~~
     {\cal Q}_\perp &=& \bar\psi_+\gamma_\perp\psi_-
                                     +\bar\psi_-\gamma_\perp\psi_+ \equiv  {\cal Q}^{1,1/2} + {\cal Q}^{1/2,1}\,
\nonumber\\
\mbox{\rm twist-4:}~~~~~
     {\cal Q}_- &=& \bar \psi_-\gamma_-\psi_- \equiv  {\cal Q}^{1/2,1/2}\,,
\eea
where the superscripts stand for the conformal spins of the quark and the antiquark.
The corresponding local conformal operators are
\bea{co-19}
  {}{\mathbb Q} _n^{1,1}(x) &=& (i\partial_+)^n \left[\bar\psi(x)
 \,\gamma_+\,C^{3/2}_n
\left(\stackrel{\leftrightarrow}{D}_+/\partial_+\right)
\psi(x)\right],
\nonumber\\
  {}{\mathbb Q} _n^{1,1/2}(x) &=& (i\partial_+)^n \left[\bar\psi(x)
  \,\gamma_+\gamma_\perp\gamma_-\,P^{(1,0)}_n
\left(\stackrel{\leftrightarrow}{D}_+/\partial_+\right)
\psi(x)\right],
\nonumber\\
  {}{\mathbb Q} _n^{1/2,1/2}(x) &=& (i\partial_+)^n \left[\bar\psi(x)
  \,\gamma_-\,C^{1/2}_n
\left(\stackrel{\leftrightarrow}{D}_+/\partial_+\right)
\psi(x)\right],
\eea
where $\stackrel{\leftrightarrow}{D}_+ = \stackrel{\rightarrow}{D}_+ -
\stackrel{\leftarrow}{D}_+$, $\partial_+= \stackrel{\rightarrow}{D}_+ +
\stackrel{\leftarrow}{D}_+$  and
we replaced Jacobi polynomials with two equal superscripts by the Gegenbauer
polynomials $P_n^{(1,1)} \sim C^{3/2}_n$, $P_n^{(0,0)} \sim C^{1/2}_n$ \cite{BE}. We
also added an overall factor $i^n$ which is convenient for applications. The
conformal operators ${\mathbb Q} _n^{1/2,1}$ are given by the same expression as
${\mathbb Q} _n^{1,1/2}$ with an obvious substitution $P_n^{(1,0)}\to
P_n^{(0,1)}$.

The similar analysis can be carried out for gluons. In this case one finds that
the $G_{+\perp}$ component of the gluon field strength tensor corresponds to
$s=1$ and therefore a ``legitimate'' fundamental field with the conformal spin
$j=\frac12(2+1) = 3/2$ and twist $t=2-1=1$, $G_{\perp\perp}$ and $G_{+-}$ both
have $s=0$, $j=1$ and $t=2$, and finally $G_{-\perp}$ has $s=-1$, $j=1/2$ and
twist $t=3$.  Local conformal operators built of two gluon fields of leading
twist are, therefore
\bea{co-20}
  {}{\mathbb G} _n^{3/2,3/2}(x) &=& (i\partial_+)^n \left[G^a_{+\perp}(x)\,C^{5/2}_n
\left(\stackrel{\leftrightarrow}{D}_+/\partial_+\right)
G^a_{+\perp}(x)\right].
\eea
Depending on the particular combination of transverse indices there
exist three  towers of gluon operators corresponding to the two-gluon
states with different helicity.

\subsection{\it The Conformal Operator Product Expansion}
\label{SubSec-COPE}

It has been known for a long time that the conformal symmetry provides powerful
constraints for the Wilson coefficients that appear in the operator product
expansion (OPE) of currents at light-like distances. Using conformal operators as
a basis suggests itself and  was studied in a number of papers, starting in the
seventies with the work of Ferrara, Grillo and Gatto
\cite{FerGriGat71,FerGriGat71a,FerGriGat72a,BonSarTon72,CraDobTod85}. It has been
employed in QCD to find the eigenfunctions of the flavor non-singlet evolution
kernel  beyond the leading logarithms \cite{BroFriLepSac80} and to relate the
Wilson coefficients and hard-scattering amplitudes that appear in a number of
exclusive two-photon processes
\cite{MuePHD,MueRobGeyDitHor94,Ji96Rad96,DieGouPirTer98}, to those known from
deeply inelastic scattering \cite{Mue97a,BelMue97a,MelMuePas02}.

Consider the operator product $A(x) B(0)$ of two local conformal
operators with twist $t_A$ ($t_B$) and spin projection on the
``+'' direction $s_A$ ($s_B$), respectively. We want to expand
this product in the light-cone limit $x_+,x_\perp \to 0$ and
$x_-$\ fixed, i.e., $x^2\to 0$, over a given  tower of local
conformal operators and their derivatives ${\mathbb O}_{n,n+k}^{j_1,j_2}$
which, as we assume,  forms a complete operator basis.
For simplicity, we will restrict ourselves to the leading twist
contributions and also neglect (the most singular) contribution of
the unit operator which is most often irrelevant.
In a free field theory, for dimensional reasons, this expansion
looks as follows:
\beq{OPE-def}
A(x) B(0) =
        \sum_{n=0}^\infty\sum_{k=0}^\infty C_{n,k}
        \left(\frac{1}{x^2}\right)^{\frac{t_A+t_B-t_n}{2}}
   \;  x_-^{n+k+\Delta}\, {\mathbb O}_{n,n+k}^{j_1,j_2}(0)
+ \mbox{higher twist}\, .
\eeq
Here $\Delta = s_1+s_2-s_A-s_B$, where $s_1$ and $s_2$ are the spin
projections of the constituent fields in the local operators
${\mathbb O}_{n,n+k}^{j_1,j_2}$ and $C_{n,k}$ are the corresponding
Wilson coefficients.
The singular behavior for $x^2\to 0$ is governed
by the twist of the operators: $t_A+t_B-t$, where $t_n=
\ell_n-n-s_1-s_2= \ell_1+\ell_2-s_1-s_2$ is the twist of the conformal
operators ${\mathbb O}_n^{j_1,j_2}$. We remind that the twist of all operators in the
conformal tower and thus the $1/x^2$ behavior of all contributions are
the same. In the interacting theory the scaling dimensions
of the operators $\ell_n$ will be modified by the anomalous dimensions and
this universal behavior no longer holds.

The power behavior in $x_-$ is
determined by the angular momentum conservation. This can be established
in the following way. We have assumed that both  $A$ and $B$ have
definite spin projection
\bea{OPE-SpiProj}
\Sigma_{+-} A(0) = s_A A(0)\, , \qquad \Sigma_{+-} B(0) = s_B B(0)\,
\eea
on the ``+'' direction. On the other hand
\bea{OPE-10}
i \left[{\mathbf M}_{-+},{\mathbb O}_{n,n+k}^{j_1,j_2}(0)\right] &=&
(s_1+s_2 + n+k){\mathbb O}_{n,n+k}^{j_1,j_2}(0)\,,
\nonumber\\
i \left[{\mathbf M}_{-+},A(x) B(0)\right] &=&  \left[x_-  n\cdot\partial - x_+
\bar n\cdot\partial + s_A+s_B\right]A(x) B(0)\,.
\eea
Insert the OPE (\ref{OPE-def}) in the second equation. On the l.h.s., evaluate
the commutators \\ $i \left[{\mathbf M}_{-+},{\mathbb
O}_{n,n+k}^{j_1,j_2}(0)\right]$, which gives  $s_1+s_2+n+k$ for a generic term,
see Eq.~(\ref{OPE-10}). On the r.h.s., take the derivatives which gives a factor
$s_A+s_B+\Delta+n+k$. Comparing the both expressions we conclude that $\Delta =
s_1+s_2-s_A-s_B$, as anticipated in Eq.\ (\ref{OPE-def}).

The conformal symmetry allows us to make one further step and obtain
Wilson coefficients $C_{n,k}$  with $ k=1,2,\ldots $ corresponding to operators
with total derivatives in terms of the coefficient of the highest weight
operator $C_{n,k=0} \equiv C_{n}$. One way to see this is to act with the
generator ${\bf \JJ}_+$ on both sides of the OPE (\ref{OPE-def})
and compare the two results in the leading twist
sector, see e.g. \cite{FerGriGat71a}. This is completely analogous to the above derivation of the constraint
arising from Lorentz invariance. Consider an infinitesimal special conformal
transformation
\bea{cOPE-act-K-cur}
[{\bf \JJ}_-,A(x)B(0)]  = \left\{
                x_- \left(2 j_A+x\cdot\partial_x\right) A(x)  -
\frac{1}{2} x^2\, \bar{n}\cdot \partial_x
           A(x)\right\}B(0) + \cdots\, ,
\eea
Here $j_A= (s_A+\ell_A)/2$ is the conformal spin of the operator $A$ and
ellipsis stand for higher twist contributions arising from the spin
rotation, i.e., proportional to $x_\perp\Sigma_{-\perp}A(x) $. Inserting
the OPE (\ref{OPE-def}) in Eq.\ (\ref{cOPE-act-K-cur})
and taking into account that
\bea{traOnk}
\left[ {\bf \JJ}_-,{\mathbb O}_{n,n+k}^{j_1,j_2}(0) \right] =
- k (k+2j_n -1) {\mathbb O}_{n,n+k-1}^{j_1,j_2}(0)
\quad\mbox{with}\quad  j_n=j_1+j_2 +n \, ,
\eea
cf. Eq.~(\ref{oper1}),
results in  a first-order recurrence relation
for the Wilson coefficients
\beq{cOPE-recrel1}
C_{n, k+1}=
        -\frac{j_A-j_B+j_n+k}{(k+1)(k+2j_n)} C_{n, k}\,,
\eeq
which is solved by
\bea{cOPE-recrel}
C_{n,k} =(-1)^k
        \frac{(j_A -j_B+j_n)_{k}}{k!
        (2j_n)_{k}} C_{n}\,,\qquad C_{n}\equiv C_{n,0}\,,
\eea
where $(a)_k = \Gamma(a+k)/\Gamma(a)$ denotes the Pochhammer symbol.
Inserting this solution in the OPE (\ref{OPE-def}) allows us to sum
contributions of operators containing total derivatives
and yields the representation:
\bea{cOPE-def}
A(x) B(0) &\simeq&
        \sum_{n=0}^\infty
        C_n
        \left(\frac{1}{x^2}\right)^{\frac{t_A\!+\!t_B\!-\!t_n}{2}}
\!\!\!\!
\frac{x_-^{n+s_1+s_2-s_A-s_B}}{{\rm B}(j_A\!-\!j_B\!+\!j_n,j_B\!-\!j_A\!+\!j_n)}
\nonumber\\&&{}\times
\int_0^1\!\! du\, u^{j_A\!-\!j_B\!+\!j_n\!-\!1} (1-u)^{j_B\!-\!j_A\!+\!j_n\!-\!1}
        {\mathbb O}_{n}^{j_1,j_2}(u x_- )\, ,
\eea
where ${\rm B}(a,b)= \Gamma(a)\Gamma(b)/\Gamma(a+b)$ is the Euler Beta
function. The total derivatives get absorbed in a translation of the
conformal operator along the light-cone with an appropriate weight function.
The case of an interacting theory will be considered in detail in
Sect.\ 5. We will find that the scaling dimensions of operators are
modified by quantum corrections so that the twist and the conformal
spin have to be modified accordingly: $j_n = (\ell_n+s_1+s_2+n)/2$ and
$t_n= (\ell_n-s_1-s_2-n)$.

To give an example, consider the operator product expansion of two
electromagnetic currents $J^\mu =
\sum_{i} e_i \bar\psi\gamma^\mu \psi$, where $e_i$ is the electric
charge and the sum runs over (light) quark flavors
$i=\{u,d,s\}$. At tree level only the
transverse components of the currents are of interest, which have the spin
projections $s_J=0$. Since the current has dimension three, it possesses
twist $t_J=3$. The
relevant operator basis is provided by the quark operators ${\mathbb
Q}_{n}^{1,1}$ (\ref{co-19}) with
conformal spin $j_n=(\ell_n+ 1 +n)/2=2+n$ and twist $t_n=(\ell_n-1
-n)/2=2$. Moreover, $\Delta=1$ and, therefore,  we find from
Eq.~(\ref{cOPE-def})
\bea{COPE-EleCur}
J^\perp(x)J^\perp(0) \simeq
\sum_{n=0}^\infty
C_n   \left(\frac{1}{x^2}\right)^\frac{6-t_n}{2} (-i x_-)^{n+1}
\frac{1}{{\rm B}(j_n,j_n)} \int_0^1 du\, u^{j_n-1} (1-u)^{j_n-1}
{\mathbb Q}_{n}^{1,1}(u x_-)\,.
\eea
Note that the remaining coefficients $C_n$ are those
that appear in the forward kinematics:
\bea{COPE-EleCur-ForCas}
\langle P| J^\perp(x)J^\perp(0) |P \rangle \simeq
\sum_{n=0}^\infty
        C_n   \left(\frac{1}{x^2}\right)^\frac{6-t_n}{2} (-ix_-)^{n+1}
          \langle P| {\mathbb Q}_{n}^{1,1}(0)|P \rangle \, .
\eea
Up to a trivial difference in normalization that arises from the definition of
conformal operators they coincide with the Wilson coefficients known
from deep inelastic scattering.

As a further example, consider the short-distance expansion of a bilocal
light-ray operator ${\cal Q}_+$ in Eq.\ (\ref{co-14}). In this case we have to
set $s_A=s_B=s_1=s_2=1/2$ so that $\Delta=0$ and
$\ell_A=\ell_B=\ell_1=\ell_2=3/2$. We find
\bea{COPE-QuaOpe}
{\cal Q}_+(\alpha_1,\alpha_2) \simeq
\sum_{n=0}^\infty
\widetilde{C}_n  (-i)^n \left({\alpha_1-\alpha_2}\right)^n
 \int_0^1 du\, u^{n+1} (1-u)^{n+1}
{\mathbb Q}_{n}^{1,1}(u \alpha_1+(1-u) \alpha_2)\, ,
\eea
where we set $\widetilde{C}_n= C_n/{\rm B}(n+2,n+2)$. The simplest way to
evaluate the coefficients $\widetilde{C}_n$ is to form forward matrix elements.
The result reads $\widetilde{C}_n=2(2n+3)/(n+1)!$, in agreement with Ref.\
\cite{BB89,BroFriLepSac80}.

Before closing this section, we want to mention an alternative approach
suggested in \cite{BB89}, which is based on the introduction of the
complete basis of multi-particle coherent states. For instance, for two particles,
one can define the coherent state
\beq{co-23}
      {}{\cal S}_{j,p} = \int \! d\alpha\, d \beta \,\phi_{j,p}(\alpha,\beta)\,
      \Phi_{j_1}\left(\frac12(\alpha+\beta)\right)\,
      \Phi_{j_2}\left(\frac12(\alpha-\beta)\right)
\eeq
by imposing the two conditions:
\beq{co-24}
 [{\bf \JJ}^2, {\cal S}_{j,p}] = j(j-1) \,{\cal S}_{j,p}\,, \qquad
  [{\bf \JJ}_+, {\cal S}_{j,p}] = i p \,{\cal S}_{j,p}\,,
\eeq
 c.f. Eqs.~(\ref{coh-1})--(\ref{coh-3}).
These conditions can be translated to a second-order differential equation for
the function $\phi_{j,p}(\alpha,\beta)$, which can be solved by separation of
variables and turns out to be related to the Bessel equation, see \cite{BB89} for
the details. An important difference with the expansion in local operators is
that the eigenvalue $j$ is not quantized in this case, so that we are tacitly
going over to a different series of representations of the $SL(2,{\mathbb R})$
group \cite{lang}. For the leading twist quark-antiquark operator (\ref{co-14})
one obtains, in particular
\beq{co-25}
 ({\cal Q}_+)_{j,p} = \int_{-\infty}^{\infty} \! d\alpha\, {\rm e}^{ip\alpha}
  \int_{0}^\infty\! \beta^{1/2} d \beta \,H^2_{j-1/2}(\beta p) \,
 \bar \psi\left(\frac12(\alpha+\beta)\right)\gamma_+
 \psi\left(\frac12(\alpha-\beta)\right)\,,
\eeq
where $H^2_{j-1/2}(z) = J_{j-1/2}(z) -iN_{j-1/2}(z) $ is the Bessel
function. The bilocal field product on the light-cone can be expanded in
contributions of coherent states making use of the Kantorovich-Lebedev
transformation
\beq{co-26}
     \beta \delta(\beta-\beta') = -\frac12 \int_{-i\infty}^{i\infty}\! jdj\, J_j(\beta)\,H^2_j(\beta')\,.
\eeq
One obtains, for quarks:
\beq{co-27}
   \bar \psi\left(\frac12(\alpha+\beta)\right)\gamma_+
        \psi\left(\frac12(\alpha-\beta)\right)
      = -\int_{-\infty}^\infty \!\frac{dp}{4\pi}
      \int_{1/2-i\infty}^{1/2+i\infty}\!dj\, (j-1/2)\beta^{-3/2}\, J_{j-1/2}(p\beta) \,
        ({\cal Q}_+)_{j,p}\,.
\eeq
Similar representations exist for gluon operators \cite{BB89}.

Note that a bilocal quark operator on the light cone involves conformal
coherent states with an imaginary conformal spin for which the weight
function $H^2_{j-1/2}$ in Eq.\ (\ref{co-25}) decreases exponentially at large
quark-antiquark separations. Such states can be interpreted as
quasiparticles (wave packets) moving along the ``plus'' direction on the
light-cone with momentum $p$ and having ``mass'' $j$ in some internal
space. We will argue in what follows that operators with different
conformal spin do not mix under renormalization in one loop. This
implies that the shape of the conformal wave packets remains the same at
all scales, to this accuracy.

\subsection{\it Conformal Invariance of a Field Theory and its Violation}

Up to this point our discussion has been entirely about the group-theoretical
aspects of the conformal symmetry and its collinear subgroup: we discussed the
field transformations and learned how to construct composite objects with simple
properties, which is the task of the representation theory, in mathematical
sense. We are now going to address the field-theoretical aspects of conformal
symmetry and the first question to ask is: Under which conditions the field
theory is conformally invariant at the classical level? We demonstrate that the
necessary and the sufficient condition for conformal invariance is the
possibility to introduce the traceless, symmetric and  divergenceless
energy-momentum tensor. As well known, symmetries of a classical field theory can
be broken by quantum corrections, in which case one speaks about quantum
anomalies. This indeed turns out to be the case for QCD. We introduce the
conformal anomaly and show why it is related to the Gell-Mann--Low
$\beta$-function. A powerful tool to obtain constraints on the correlation
functions that follow from the conformal invariance is provided by conformal Ward
identities (CWIs) \cite{Par72,Nie73,Sar74,Nie75}. We will see that the dilatation
Ward identity is nothing else but the familiar Callan-Symanzik equation
\cite{Cal70Sym70} while the Ward identity corresponding to special conformal
transformation is the genuine new feature of a conformal theory and has important
consequences. Discussion of the conformal invariance in a gauge theory like QCD
involves subtleties because the symmetry can be broken spuriously by the gauge
fixing procedure. These difficulties are of technical nature and are discussed in
detail in Appendix B, where we present a complete set of QCD conformal Ward
identities including the contributions of BRST exact and equation of motion (EOM)
operators.

\subsubsection{\it Energy-Momentum Tensor}
\label{SubSec-EneMomTen}

According to the Noether theorem, the invariance of a
four-dimensional theory under conformal transformations implies
the existence of fifteen conserved currents.
The corresponding conserved charges are the fifteen generators of the conformal
group: $\mathbf{P}_\mu$, $\mathbf{M}_{\mu\nu}$, $\mathbf{D}$, and $\mathbf{K}_\mu$.
We will show that these generators can be
expressed in terms of the so-called improved energy-momentum tensor
${\Theta}_{\mu\nu}(x)$ which is divergenceless, symmetric and traceless:
\bea{Theta-def}
{\Theta}_{\mu\nu}(x)={\Theta}_{\nu\mu}(x)\,,\qquad
\partial^\mu{\Theta}_{\mu\nu}(x)=0\,, \qquad
g^{\mu\nu}{\Theta}_{\mu\nu}(x)=0\,.
\eea
We will also see that the possibility to define the energy-momentum
tensor with such properties is closely related to the conformal
symmetry of the Lagrangian.

{}For simplicity let us consider a generic Lagrangian ${\cal
L}(\Phi(x),\partial_\mu\Phi(x))$ that depends on a single fundamental field
$\Phi(x)$ and its first derivative $\partial_\mu \Phi(x)$. Invariance of the
action under a \emph{global} continuous transformation of the field, $\Phi\to
\Phi+\delta_\varepsilon \Phi$, implies that its variation under the similar \emph{local}
infinitesimal transformation characterized by the parameter
$\varepsilon_\alpha(x)$ takes the form
\bea {trans-S}
\delta_\varepsilon S=\int d^4x\, \delta_\varepsilon{\cal L}
=
\int d^4 x\left[
\frac{\partial{\cal L}}{\partial(\partial^\mu\Phi(x))}
 \, \delta_\varepsilon \left(\partial^\mu\Phi(x)\right)
+
\frac{\partial {\cal L}}{\partial\Phi}\, \delta_\varepsilon \Phi(x) \right]
\equiv
  \int d^4x\, J_\alpha^\mu(x)\,
 \frac{\partial \varepsilon^\alpha(x)}{\partial x^\mu}\,.
\eea
On the other hand, any variation of the action must vanish at the
stationary point, i.e., as soon as the fields satisfy the Euler-Lagrange
equations of motion (EOM)
\bea{E-L}
\partial^\mu \Pi_\mu(x) =  \frac{\partial {\cal L}}{\partial\Phi(x)}\,,
\qquad \Pi_\mu(x)=\frac{\partial{\cal L}}
{\partial(\partial^\mu\Phi(x))}\,.
\eea
As a consequence, the current $J_\alpha^\mu(x)$ must be conserved,
$\partial_\mu J_\alpha^\mu(x) = 0$
and the corresponding conserved charge is given by
$Q_a=\int d^3x\,J^0_\alpha(x)$, $d Q_\alpha/{dt}=0$.

In this way, the invariance of the action under space and time
translations leads to conservation of  momentum and energy,
respectively. Substituting
$\delta_\varepsilon\Phi(x)=\varepsilon_\alpha(x) \delta_P^\alpha
\Phi(x)=\varepsilon_\alpha(x)
\partial^\alpha \Phi(x)$ (see Eq.~(\ref{trafo})) in Eq.~(\ref{trans-S}) one
identifies the corresponding conserved current $J_{P,\alpha}^\mu$ as the
canonical energy-momentum tensor
\bea{Def-CanEneMomTen}
J_{P,\alpha}^\mu= T^\mu{}_{\alpha}=\Pi^{\mu} \partial_\alpha \Phi -
g^{\mu}{}_{\alpha} {\cal L}\,,\qquad
\partial_\mu {T}^\mu{}_{\alpha} \stackrel{\mbox {\tiny  EOM}}{=}0\, ,
\eea
where the superscript ${\rm \scriptstyle EOM}$ indicates that the fields
satisfy the equations of motion (\ref{E-L}).
The corresponding conserved charge is the operator of the four-momentum
\bea{Def-ChaTra}
\mathbf{P}_\alpha= \int d^3x\,  {T}^{0}{}_\alpha(x)\,.
\eea
The invariance of the action under Lorentz rotations,
$\delta_\varepsilon\Phi(x)=\varepsilon_{\alpha\beta}\delta_M^{\alpha\beta}\Phi(x)$,
(see Eq.~(\ref{trafo})) implies further six conserved currents which are
contained in the angular momentum tensor:
\bea{ConCurLorInv}
J_{M,\alpha\beta}^{\mu} = x_\alpha  {T}^{\mu}{}_{\beta}(x) - x_\beta
{T}^{\mu}{}_{\alpha}(x) - \Pi^\mu(x) \Sigma_{\alpha\beta}\Phi(x)\,.
\eea
It is given in terms of the canonical energy-momentum tensor
(\ref{Def-CanEneMomTen}), plus the additional term $\Pi^\mu
\Sigma_{\alpha\beta}\Phi$ that arises from the transformation
of the spin degrees of freedom. The tensor ${T}^{\mu}{}_{\beta}(x)$
has no definite symmetry with respect to its Lorentz indices, in
general.  However, conservation of the angular momentum
(\ref{ConCurLorInv}) implies that the antisymmetric part
${T}_{\alpha\beta}(x) -  {T}_{\alpha\beta}(x)$  is given by a total
divergence, since
\bea{antisym}
\partial_\mu J_{M,\alpha\beta}^{\mu}
= {T}_{\alpha\beta}(x) -  {T}_{\beta\alpha}(x) -
\partial_\mu \left( \Pi^\mu(x) \Sigma_{\alpha\beta}\Phi(x) \right)
\stackrel{\mbox{\tiny EOM}}{=}  0\,.
\eea
This suggests to define the symmetric, so-called Belinfante, form of the
energy-momentum tensor \cite{Bel40}:
\bea{DefBelEneMomTen}
T_{\mu\nu}^{\rm B}= {T}_{\mu\nu}  +
\frac{1}{2} \partial^\lambda \left( \Pi_\lambda \Sigma_{\mu\nu} \Phi -
\Pi_\mu \Sigma_{\lambda\nu}\Phi - \Pi_\nu \Sigma_{\lambda\mu} \Phi\right)\,.
\eea
The symmetry $T_{\mu\nu}^{\rm B}= T_{\nu\mu}^{\rm B}$ follows readily from Eq.\
(\ref{antisym}). Since the expression in the parenthesis in the r.h.s.\ of Eq.\
(\ref{DefBelEneMomTen}) is antisymmetric with respect to  $\lambda$ and $\mu$,
the redefined energy-momentum tensor is conserved, $\partial^\mu T_{\mu\nu}^{\rm
B}=0$, so that we can equally well use it instead of the original Noether current
$T_{\mu\nu}$ as the conserved  current associated with the translation
invariance. Note that the operator of four-momentum (\ref{Def-ChaTra}) is not
affected by this redefinition. We can also use the Belinfante form of the
energy-momentum tensor to redefine the  angular momentum tensor density as
\bea{Def-CurLorInv}
J_{M,\alpha\beta}^{\mu} = x_\alpha T_{\beta}^{{\rm B},\, \mu} - x_\beta
T_{\alpha}^{{\rm B},\, \mu}
\, ,
\qquad
\partial_\mu J_{M,\alpha\beta}^{\mu} \stackrel{\mbox{\tiny EOM}}{=} 0\, .
\eea
The corresponding conserved charges define the operator of angular momentum
\bea{M}
\mathbf{M}_{\alpha\beta}=\int d^3x\,J_{M,\alpha\beta}^{\,0}(x)\,.
\eea
We stress that the possibility to define the symmetric energy-momentum
tensor reflects the Lorentz--invariance of the theory.

Next, consider scale transformations $x_\mu \to\lambda x_\mu$ and $\Phi(x) \to
\lambda^{-\ell_\Phi}\,\Phi(\lambda x)$ with $\ell_\Phi$ being the scaling
dimension of the field. The action of a four-dimensional theory is scale
invariant provided that ${\cal L}\to\lambda^{-4}{\cal L}$, i.e.,  the scaling
dimension of the Lagrangian is equal to its canonical dimension (four). {}For the
usual Lagrangians that are given by polynomials in the fields and their
derivatives, this condition translates to the requirement that all couplings have
to be dimensionless and the scaling dimensions of the fields are uniquely fixed
and are equal to their canonical values. The conserved current corresponding to
the scale transformations can be  constructed by calculating the variation of the
action (\ref{trans-S}) under dilatations $\delta_\varepsilon\Phi\to
\varepsilon(x)\delta_D\Phi(x)$ (see Eq.~(\ref{trafo}))
\bea{D-variation}
\delta_D S=\int d^4x\,\varepsilon(x)\left[-\partial_\mu D^\mu +
\Delta_D(x)\right]
\,.
\eea
The Noether current $J_D^\mu = D^\mu$ is called the dilatation current. It is written
in terms of the canonical energy-momentum tensor
\bea{D-can}
D^\mu=x_\alpha T^{\mu\alpha}+\ell_\Phi\Pi^\mu\Phi\,,
\eea
and $\Delta_D(x)$ is defined as
\bea{Delta-D}
\Delta_D =
  (\ell_\Phi + 1) \, \frac{\partial{\cal L}}{\partial(\partial_\mu \Phi)}
\partial_\mu \Phi + \ell_\Phi \, \frac{\partial{\cal L}}{\partial\Phi}\Phi
-d \, {\cal L}\,,
\eea
where $d=4$ is the space-time dimension. Invariance of the action under
dilatations implies that $\Delta_D(x)$ is a total divergence. If the Lagrangian
${\cal L}(\Phi,\partial\Phi)$ is a polynomial in $\partial_\mu\Phi$ and $\Phi$,
the only possibility is  $\Delta_D(x)=0$ by itself, or, equivalently,
$\mathcal{L}(\Phi/\lambda^{\ell_\Phi},\partial\Phi/\lambda^{\ell_\Phi+1})
=\mathcal{L}(\Phi,\partial\Phi)/\lambda^{4}$, as anticipated.

Last but not least, examining the transformation of the action
under the special conformal transformations
$\delta_\varepsilon\Phi=\varepsilon_\alpha(x)\delta_K^\alpha\Phi(x)$,
see Eq.~(\ref{trafo}), one gets
\bea{C-variation}
\delta_K S=\int d^4x\,\varepsilon_\alpha(x)\left[-\partial_\mu K^{\alpha\mu}
+2x^\alpha \Delta_D(x) +
\Delta_K^\alpha(x)\right]
\, ,
\eea
where
\beq{kkk}
K^{\alpha\mu}=(2x^\alpha x_\nu-g_{\nu}^\alpha
x^2)T^{\mu\nu}+2x_\nu\Pi^\mu(\ell_{\Phi}g^{\nu\alpha}-\Sigma^{\nu\alpha})
\Phi
\,,
\eeq
$\Delta_D(x)$ is defined Eq.~(\ref{Delta-D}) and
\beq{DeltaK}
\Delta_K^\alpha(x) =
2 \Pi_\nu\left(\ell_\Phi \,g^{\alpha\nu}+
\Sigma^{\alpha\nu} \right)\Phi \, .
\eeq
Invariance of the action (\ref{C-variation}) under the special conformal
transformations requires that $\Delta_D(x)=0$ and, in addition,
$\Delta_K^\alpha(x)$ has to be a total divergence
\bea{Con-SpeConInv}
\Pi_\nu\left(\ell_\Phi \,g^{\alpha\nu}+
\Sigma^{\alpha\nu} \right)\Phi = \partial^\nu \sigma_{\alpha\nu}(x)\,,
\eea
where $\sigma_{\mu\nu}$ is some tensor. Under these conditions the conserved
current corresponding to special conformal transformations is given by
$J_{K,\alpha}^\mu=K_\alpha{}^{\mu}-2\sigma_\alpha{}^{\mu}$.
Explicit calculation shows that for physical spin$-1/2$ and spin$-1$ fields
with the canonical dimension $\ell=3/2$ and $\ell=1$,
respectively, the field $\sigma_{\alpha\nu}(x)$ is identically zero.
In other words, in this case invariance under special conformal
transformations proves to be a consequence of
the dilatation (scale) invariance.

Using the condition (\ref{Con-SpeConInv}), one can again redefine
the energy-momentum tensor to make it traceless, and so
the expressions for the currents $J_{D}^\mu$ and $J_{K,\alpha}^\mu$
are significantly simplified. One defines \cite{CalColJac70}:
\bea{Def-ImpEneMomTen}
\Theta_{\mu\nu} = T_{\mu \nu}^{\rm B} + \frac{1}{2} \partial^\lambda \partial^\rho  X_{\lambda\rho\mu\nu}\,
\, ,
\eea
where $T_{\mu \nu}^{\rm B}$ is the Belinfante tensor
and $X_{\lambda\rho\mu\nu}$ is chosen in such a way that
$\Theta_{\mu\nu}$ remains symmetric and divergenceless,
and, in addition, $g^{\mu\nu}\partial^\lambda
\partial^\rho X_{\lambda\rho\mu\nu} =
2\partial^\alpha \partial^\lambda \sigma_{\alpha\lambda}= -2g^{\mu\nu}T_{\mu
\nu}^{\rm B}$.
The latter condition ensures that $\Theta_{\mu\nu}$ is traceless, as it is easy
to check using Eqs.\ (\ref{Def-CanEneMomTen}) and (\ref{Con-SpeConInv}). Since
the difference $\Theta_{\mu\nu}-T_{\mu\nu}^{\rm B}$ is given by the total
divergence, the angular momentum tensor density (\ref{Def-CurLorInv}) can be
redefined in terms of $\Theta_{\mu\nu}$ with no effect on the conserved charges
$\mathbf{P}_\mu$ and $\mathbf{M}_{\mu\nu}$. Moreover, the currents associated
with the dilatation and special conformal transformations take a remarkably
simple form:
\bea{Con-Cur}
J_{D}^\mu = x_\nu \Theta^{\mu\nu}\,, \quad\quad J_{K,\alpha}^\mu =
\left(2 x_\nu x_\alpha - x^2 g_{\nu\alpha}\right)
\Theta^{\mu\nu}\, ,
\eea
leading to the conserved charges
\bea{101}
\mathbf{D}=\int d^3 x \,J_{D}^0(x)\,,\qquad
\mathbf{K}_\alpha =\int d^3 x \,J_{K,\alpha}^0\,.
\eea
{}From the representation in Eq.\ (\ref{Con-Cur}) it is obvious that
the dilatation and special conformal currents are conserved
if and only if the improved energy-momentum tensor $\Theta_{\mu\nu}$
is divergenceless and traceless.

In a quantum field theory with a running coupling constant these two conditions
cannot be preserved simultaneously. Due to the appearance of ultraviolet (UV)
divergencies, one has to define the theory with an UV cutoff. Any regularization
that preserves Poincar\'e covariance breaks inevitably the dilatation and
consequently the special conformal symmetry. This implies that the renormalized
improved energy momentum tensor $\Theta_{\mu\nu}$ remains divergenceless and
symmetric but its trace does not vanish \cite{Nie77AdlColDun77ColDunJog77Min76}.
Breaking of a classical symmetry due to quantum effects is called an anomaly, the
conformal anomaly for the case at hand.

The loss of the scale invariance in the quantum theory is rather obvious, because
a dimensionfull parameter appears that plays the r\^ole of  the ultraviolet
cutoff. Consider QCD with massless quarks
\beq{QCD-Lag}
 \mathcal{L}_{\rm QCD}
 = -\frac14 G_{\mu\nu}^aG^{a\mu\nu} + \bar \psi i\!\not\!\!D\, \psi\,,
\eeq
where $D_\mu=\partial_\mu-ig^{(0)} t^a A_\mu^a(x)$ is the covariant derivative,
and divide the quark and gluons fields in ``fast'' and ``slow'', with frequencies above and
below a certain scale $\mu$, respectively. The renormalized QCD action can be thought of as
an effective action, integrating out contributions of fields with large frequencies
(see e.g., \cite{Peskin}) in the background ``slow'' fields. In the one-loop approximation
the effective action can be written, schematically, as
\beq{EffAction}
  S_{\rm eff} = -\frac14 \int\! d^4x
   \left( \frac{1}{(g^{\mbox{\tiny (0)}})^2}-\frac{\beta_0}{16\pi^2} \ln M^2/\mu^2 \right)
   \left[G_{\mu\nu}^aG^{a\mu\nu}\right]_{\rm slow}(x)+\ldots\,,
\eeq
 where $M^2$ is the UV cutoff,
 $\beta_0=(11/3) N_c-(2/3)N_f$ is the first coefficient in the Gell-Mann-Low function and
 the ellipses stand for the quark terms. For the purpose of the argument we rescaled
 temporarily the gluon field $g^{(0)} A_\mu \to A_\mu$ so that the coupling only enters as
 an overall factor in front of $G_{\mu\nu}^aG^{a\mu\nu}$.
 Making the scale transformation $x^\mu \to \lambda x^\mu$,
 $A_\mu(x) \to \lambda A_\mu(\lambda x)$, $\psi(x) \to \lambda^{3/2}\psi(\lambda x)$
 and $\mu \to \mu/\lambda$ with the
 fixed cutoff results in the variation of the effective action
\beq{EffAction2}
 \delta S = -\frac{1}{32\pi^2}\beta_0 \ln \lambda \int\! d^4x\, G^a_{\mu\nu}G^{a\mu\nu}(x)\,,
\eeq
which implies that (cf.\ Eq.\ (\ref{trans-S}))
\beq{anomaly-1}
    \partial_\mu J_{D}^\mu(x) = -\frac{\beta_0}{32\pi^2} G^a_{\mu\nu}G^{a\mu\nu}(x)\,.
\eeq
Restoring the standard normalization of the gluon field and introducing $\alpha_s = g^2/(4\pi)$
we obtain
\beq{anomaly-2}
   \partial_\mu J_{D}^\mu(x)  =  g^{\mu\nu}\Theta_{\mu\nu}^{\mbox{\rm\tiny QCD}}(x)
\stackrel{\mbox{\tiny EOM}}{=}
-\frac{\beta_0\alpha_s}{8\pi} G^a_{\mu\nu}G^{a\mu\nu}(x) + {\mathcal{O}}(\alpha_s^2)\,.
\eeq
This is the desired result. The full answer reads \cite{Nie77AdlColDun77ColDunJog77Min76}
\beq{anomaly-3}
   \partial_\mu J_{D}^\mu(x)  =
   g^{\mu\nu}\Theta_{\mu\nu}^{\mbox{\rm\tiny QCD}}(x) \stackrel{\mbox{\tiny EOM}}{=}
   \Delta_D (x)
    \stackrel{\mbox{\tiny EOM}}{=}  \frac{\beta(g)}{2 g}
   G_{\mu\nu}^a G^{a\mu\nu} (x)\,,
\eeq
where
\beq{beta-f}
    \mu\frac{d}{d\mu} g(\mu)  = \beta(g(\mu))\,,
 \qquad \beta(g)/g  = -\beta_0 \frac{\alpha_s}{4\pi} +\ldots
\eeq
and this is correct in operator sense, for an insertion in arbitrary Green functions.
{Note that $\Delta_D(x)$, which defines the variation of the action
(\ref{D-variation}), is the anomalous term in the trace of
$\Theta_\mu^{\ \mu}$ and both differ by extra EOM terms.}

In practical calculations one usually  employs the dimensional regularization extending
the number of space-time dimensions to $d = 4 -2 \epsilon$ with $\epsilon>0$ and subtracting poles in
$\epsilon$ within the minimal subtraction (MS) scheme. This regularization respects gauge
invariance and also it is explicitly Lorentz covariant so that both the
energy-momentum $\mathbf{P}_\mu$ and the angular momentum tensor
$\mathbf{M}_{\mu\nu}$ remain conserved.
The modified energy momentum tensor in QCD is equal to
\bea{T-QCD}
\Theta_{\mu\nu}^{\mbox{\rm\tiny QCD}}(x)
=-g_{\mu\nu}\mathcal{L}_{\rm QCD}-G_{\mu\lambda}^a G_{\nu\lambda}^a
+\frac{i}4\left[\bar\psi\gamma_\mu \stackrel{\rightarrow}{D}_\nu\psi-
\bar\psi\stackrel{\leftarrow}{D}_\nu\gamma_\mu \psi+(\mu \leftrightarrow\nu)
\right]+\ldots\, ,
\eea
where we do not show  the gauge-fixing terms as well as the terms depending on Faddeev-Popov
ghost fields \cite{Nie77AdlColDun77ColDunJog77Min76}.
One can verify that this expression satisfies the defining relations (\ref{Theta-def}),
 so far as the EOM are employed.
In particular, taking the trace one finds
\bea{Tr-Theta}
g^{\mu\nu}\Theta_{\mu\nu}^{\mbox{\rm\tiny QCD}}(x) \stackrel{\mbox{\tiny EOM}}{=}
(d-4)\frac14 G_{\mu\nu}^aG^{a\mu\nu}(x) +\ldots\, ,
\eea
which would vanish for $d\to 4$ if the operator on the r.h.s.\ were finite.
The renormalization adds $Z-$factors which
have poles in $(d-4)$ and hence a non vanishing contribution
to Eq.\ (\ref{Tr-Theta}) in the limit $d\to4$, reproducing the result in Eq.\ (\ref{anomaly-3}).

There are several subtleties, however.
{}First, since the regularized action (in dimension $d$) has to remain dimensionless,
the quark and the gluon fields must have the canonical (mass) dimensions
$\ell^{\rm can}_\psi = 3/2-\epsilon$,
$ \ell^{\rm can}_A = 1-\epsilon$,
respectively. These dimensions are fixed by the kinetic terms in the QCD
Lagrangian and this implies,
e.g., from the $g^{(0)} \bar\psi \slash\!\!\!\! A\psi$
term, that the bare coupling $g^{(0)}$ has to carry the dimension $[g^{(0)}]= -\epsilon$,
$g^{(0)} \propto \mu^{-\epsilon}$ where $\mu$ is a certain mass scale.

Note that the invariance with respect to special conformal transformations is also
necessarily broken and the corresponding anomaly potentially
contains a new contribution, $\Delta^\alpha_K(x)$,  in addition to  $\Delta_D(x)$, see
Eq.\ (\ref{C-variation}). To avoid extra terms and have a simple pattern of conformal symmetry
breaking governed by $\Delta_D$ alone,  the condition (\ref{Con-SpeConInv}) should be satisfied in the
regularized theory. This forces us to set the scaling dimensions of
physical fields in $d$ dimensions to be the same as in four dimensions and,
therefore, different from their canonical dimensions:
$\ell_\psi = 3/2$ and $\ell_A = 1$.
In other words, the anomalous dimensions $\gamma = \ell-\ell^{\rm can}$ acquire a term in $\epsilon$,
similar as it happens in the $\beta$-function.

Second, there are subtleties related to the gauge fixing procedure.
Fixing the gauge may lead to spurious symmetry breaking already in the
classical theory. For example, the Schwinger gauge $x^\mu A_\mu =0$
respects classical scale invariance and inversion, but breaks the
translational symmetry which is then restored in gauge invariant
quantities. In turn,  the Lorentz invariant gauge
condition $\partial^\mu A_\mu =0$ breaks special
conformal invariance. Hence, one may require an additional gauge
transformation in order to have invariance under the action of group
generators in the simple form (\ref{trafo}) \cite{Jac78a}.
Alternatively, without this additional gauge transformation one finds
that the gauge fixing and ghost terms imply the non vanishing fields
$\sigma_{\alpha\nu}$, defined by Eq.\ (\ref{Con-SpeConInv}). Since the ghost
field $\omega$ plays the role of the gauge parameter, it is convenient
to choose $\ell_\omega = \ell^{\rm can}_\omega =0$ and in this case it does not
contribute to $\sigma_{\alpha\nu}$.  The scaling
dimension of the antighost field $\bar \omega$ can  be chosen as  $\ell_{\bar
\omega} = \ell^{\rm can}_{\bar \omega} = d-2$. Within these settings,
the complete result for the trace anomaly of the energy-momentum tensor reads
\bea{TraAno}
\Delta(x)=
\Delta_D(x)  - (d-2) \partial^\lambda {\cal O}_{B\lambda}(x)\, ,
\eea
where $\Delta_D(x) = (\beta/2g) [G_{\mu\nu}^a G^{a\mu\nu}](x) + \ldots$ is
defined in terms of renormalized operator insertions in
Eq.~(\ref{trace-anomaly}). The additional term here is a total divergence of the
operator ${\cal O}_{B\lambda}(x)$, given in Eq.~(\ref{Def-OpeInsAB}). It stems
from the special conformal variation of the gauge fixing and ghost terms,
$\Delta_K^\alpha =  2(d-2) {\cal O}_B^\alpha$, and is  BRST exact. Thus, it does
not contribute to the physical QCD sector. The expression (\ref{TraAno})
simultaneously determines the (global) dilatation and the special conformal
variations  of the action
\bea{ConAnoCon}
\delta_D S = \varepsilon \int d^d\! x\, \Delta(x)\, , \qquad
\delta_K S = \varepsilon_\alpha \int d^d\! x\, 2x^\alpha\, \Delta(x)\, .
\eea

\subsubsection{\it Conformal Ward identities}
\label{SubSec-ConWarIde}

Symmetries in quantum field theory imply constraints for Green functions, which
are called Ward identities. A generic Green function is given as the vacuum
expectation of the time ordered product of Heisenberg operators and can be
evaluated from the following functional integral:
\bea{Def-GreFun}
\langle  {\cal X}_N \rangle \equiv
\langle 0| {\rm T} \Phi(x_1) \dots \Phi (x_N) |0 \rangle
 \equiv {\cal N}^{-1}\int {\cal D}\Phi\, \Phi(x_1) \dots \Phi (x_N)\, {\rm e}^{iS(\Phi)} \, .
\eea
The Ward identities follow from the invariance of the functional integral under
the change of variables $\Phi(x) \to
\Phi'(x) = \Phi(x)+\delta\Phi(x)$. One obtains~\cite{Sar74,Nie75}%
\footnote{For the situation that we are considering
the Jacobi determinant det$[{\cal D}\Phi'/{\cal D}\Phi]$
is a C-number and, thus, irrelevant.}
\beq{CWI-1}
 0 =
\langle 0| {\rm T} \delta\Phi(x_1) \dots \Phi (x_N) |0\rangle
  + \ldots +
\langle 0| {\rm T}  \Phi(x_1) \dots \delta\Phi (x_N)|0\rangle +
\langle 0| {\rm T} i\delta S\,\Phi(x_1) \dots \Phi (x_N) |0\rangle\,.
\eeq
{}For example, the common wisdom that Green functions in a
Lorentz--invariant theory only depend on the set of invariants
$x^2_{ik}\equiv(x_i-x_k)^2$ is in fact the consequence of the Ward identities for
translations and Lorentz rotations, which leave the action
invariant, of course.  For the dilatations one obtains the Ward identity
\beq{CWI-D}
\sum_{i=1}^{N} \left( {\ell}^{\rm can}_\Phi+x_i\cdot \partial_i\right)
\langle0| {\rm T} \Phi(x_1) \dots \Phi (x_N)|0\rangle
= -i\int\!d^4x \langle0|{\rm T} \Delta_D(x)\,\Phi(x_1)
\dots \Phi (x_N)|0\rangle,
\eeq
where $\partial_i = \partial/\partial x_i$,
which tells that scale transformations of the fields have to be
compensated by the variation of the action. The latter is given by the
zero-momentum insertion of the trace anomaly of the energy-momentum tensor including
EOM terms that are proportional to the canonical dimension and count
the number of the fields.
The Ward identity (\ref{CWI-D}) is nothing else as the Callan-Symanzik
renormalization group equation \cite{Cal70Sym70}, but in order to see
this we have to rewrite it in a somewhat different form.
{}First, the Lorentz invariance and simple dimension counting suggest
that the renormalized Green function in four dimensions can only
be of the form
\beq{CWI-2}
  \langle  {\cal X}_N \rangle(x_1,\ldots,x_N; \mu,g(\mu)) =
     \mu^{N\ell^{\rm can}_\Phi}G\left(x_{ik}^2 \mu^2; g(\mu)\right)\,.
\eeq
As a consequence, the differential operator on the l.h.s.\ of
Eq.\ (\ref{CWI-D}) can be equivalently  written as
\beq{CWI-3}
 \sum_{i=1}^{N} \left( {\ell}^{\rm can}_\Phi+x_i\cdot \partial_i\right)
 \langle  {\cal X}_N \rangle = \mu
 \frac{\partial}{\partial\mu} \langle {\cal X}_N \rangle.
\eeq
Next, consider the operator insertion of the dilatation anomaly
$\partial_\mu J^\mu_D \stackrel{\mbox{\tiny EOM}}{=} \Delta_D$ on the
r.h.s.\ of Eq.\ (\ref{CWI-D}). From our discussion of the one-loop QCD example in
Eq.~(\ref{EffAction}) it is clear that the same result (\ref{anomaly-2})
can be obtained by taking a derivative of the effective Lagrangian over the
UV cutoff,
\beq{CWI-4}
\Delta_D(x) = - M\frac{\partial}{\partial M}\, {\cal L}_{\rm eff}(x),
\eeq
and it is easy to understand that this relation is quite general,
provided the theory does not include explicit dimensionfull parameters
like masses. It follows that
\beq{CWI-5eff}
  i\int d^4 x\,  \Delta_D(x)\,  {\rm e}^{iS_{\rm eff}[\Phi]} = -
  M\frac{\partial}{\partial M} {\rm e}^{iS_{\rm eff}[\Phi]}
\eeq
and the derivative $M\frac{\partial}{\partial M}$ can be taken out of
the functional integral since the fields $\Phi$ are just integration
variables and do not have any $M$-dependence.
We obtain, therefore
\beq{CWI-5}
i \int\!d^4x\, \langle0|{\rm T} \Delta_D(x)\,  \Phi(x_1)
\dots \Phi (x_N)|0\rangle = -  M\frac{\partial}{\partial M} \langle  {\cal X}_N \rangle.
\eeq
On the other hand, renormalizability of the theory implies that the
Green function $\langle  {\cal X}_N \rangle$ depends on the cutoff
through $Z_\Phi$-factors that take into account the field
renormalization
\beq{CWI-6}
   \Phi^{\rm unr}(x) = \sqrt{Z_\Phi}\Phi(x)\,, \qquad
   \gamma_\Phi=\mu\frac{d}{d\mu} \ln \sqrt{Z_\phi} = - M\frac{d}{d M} \ln \sqrt{Z_\Phi}
\eeq
and through the coupling constant. In other words
\beq{CWI-7}
i \int\!d^4x \langle0|{\rm T} \Delta_D(x)\,  \Phi(x_1)
\dots \Phi (x_N)|0\rangle =
   -  M\frac{\partial}{\partial M} \langle  {\cal X}_N \rangle =
    \left[ \beta(g)\frac{\partial}{\partial g} + \sum_{i=1}^N
\gamma_{\Phi_i}\right]
     \langle  {\cal X}_N \rangle.
\eeq
Combining this with Eqs.~(\ref{CWI-D}) and (\ref{CWI-3}), we indeed obtain the Callan-Symanzik equation:
\beq{CalSym}
  \left[ \mu \frac{\partial}{\partial\mu}+
\beta(g)\frac{\partial}{\partial g} + \sum_{i=1}^N
\gamma_{\Phi_i}\right]\langle {\cal X}_N \rangle =0\,.
\eeq

Let us assume for a moment that the $\beta$ function has a fixed point
$g^\ast= g(\mu^\ast)$, such that $\beta(g^\ast)=0$.
In this case by the shift of the scaling dimension
\bea{ShiCanDim}
\ell= \ell^{\rm can}\quad \Rightarrow \quad
\ell(g^\ast) =  \ell^{\rm can} + \gamma(g^\ast) \,
\eea
one can bring the dilatation Ward identity (\ref{CWI-D}) to the
simple form that it has in a free theory:
\beq{CWI-Dfree}
\sum_{i=1}^{N} \left( {\ell}_\Phi(g^\ast)+x_i\cdot \partial_i\right)
\langle0| {\rm T} \Phi(x_1) \dots \Phi (x_N) |0\rangle
= 0\,.
\eeq
In this sense the dilation symmetry is only broken by the coupling
renormalization. Eq.~(\ref{CWI-Dfree}) is a strong constraint which fixes the
structure of two-point Green functions at the fixed point $\beta(g^\ast)=0$ up to
an overall normalization factor $N(g^*)$, e.g., for two scalar fields
\bea{2-point}
\langle 0| \phi(x_1) \phi(x_2) |0\rangle  &\!\!\! = &\!\!\!
 N_2(g^\ast) (\mu^\ast)^{-2\gamma(g^\ast)}
\left(\frac{1}{(x_1-x_2)^2}\right)^{ \ell^{\rm can} + \gamma(g^\ast)}
\eea
where the additional explicit scale dependence  has been introduced for
dimensional reasons. For particles with spin and fixed spin projection $s$ on the
light-cone (\ref{spin}) this is generalized to
\beq{22-point}
\langle 0| \Phi(x_1) \Phi(x_2) |0\rangle   =
 N_2(g^\ast) (\mu^\ast)^{-2\gamma(g^\ast)}
\left(\frac{1}{2(x_1-x_2)_-(x_1-x_2)_+}\right)^{ \ell^{\rm can} +
\gamma(g^\ast)}
\left(\frac{(x_1-x_2)_+}{(x_1-x_2)_-}\right)^s,
\eeq
where it is assumed that $(x_1-x_2)_\perp  = 0$. It follows that the
dependence of the correlation function on the light-cone separation
$(x_1-x_2)_-$
(in the light-cone limit $(x_1-x_2)_+\to 0$, $(x_1-x_2)_- =$~const)
is determined by the conformal spin $j=(\ell(g^\ast)+s)/2$, see Eq.\ (\ref{cspin}).

As explained in the previous Section, a field theory that is invariant under
dilatations is usually also invariant under the inversion of coordinates, $x^\mu
\to x^\mu/x^2$. This implies the existence of  more Ward identities that are
generated by special conformal transformations:
\beq{CWI-K}
\sum_{i=1}^{N}
\left( 2 x_{i}^\mu (\ell^{\rm can}_\Phi+x_i\cdot \partial_i) -
       2 \Sigma_{\ \nu}^{\mu} x_i^{\nu}-x^2_i \partial^{\mu}_i\right)
\langle{\cal X}_N\rangle
= -i\int\!d^4x\, 2x^\mu\, \langle0|{\rm T}
\Delta_D(x)\,\Phi(x_1) \dots \Phi (x_N)|0\rangle\, .
\eeq
The expression on the r.h.s.\ has, up to factor $2 x^\mu$ and subtleties
with the  gauge fixing
procedure, the same form
as in Eq.\ (\ref{CWI-7}) and can be written as
\bea{CWI-K2}
{}\hspace*{-0.5cm}
  i\!\int\!d^4\! x\, 2x^\mu\, \langle0|{\rm T} \Delta_D(x)\,\Phi(x_1)
   \dots \Phi (x_N)|0\rangle
   &\!\!\! =&\!\!\! \sum_{i=1}^N 2x_i^\mu \gamma_{\Phi_i}
\langle0| {\rm T} \Phi(x_1) \dots \Phi (x_N)|0\rangle
\\
&&\hspace*{-0.4cm}{}+ \frac{\beta(g)}{g}\! \int\!d^4\! x\,2x^\mu
\langle0| {\rm T}
g\frac{\partial}{\partial g}{\cal L}(x) \Phi(x_1) \dots \Phi (x_N)|0\rangle\, .
\nonumber
\eea
To be precise, we mention that the renormalization of
$\partial {\cal L}(x)/\partial g$ might require total divergence
counterterms that do not contribute for dilatation.
At the fixed point $\beta(g^\ast)=0$ one obtains
the same form of the special conformal Ward identities as in a free theory
\beq{CWI-Kfree}
\sum_{i=1}^{N}
\left( 2 x_{i}^\mu (\ell_\Phi(g^\ast)+x_i\cdot \partial_i) -
       2 \Sigma_{\ \nu}^{\mu} x_i^{\nu}-x^2_i \partial^{\mu}_i\right)
\langle0| {\rm T} \Phi(x_1) \dots \Phi (x_N)|0\rangle
   = 0\, ,
\eeq
but with the shifted scaling dimension (\ref{ShiCanDim}). The power of conformal
symmetry is to a large extent due to this identity. If combined with the
dilatation Ward identity (\ref{CWI-Dfree}) and Lorentz invariance,
Eq.~(\ref{CWI-Kfree}) is solved by the requirement that the Green functions only
depend on the so-called anharmonic ratios $x_{ik}^2 x_{lm}^2/(x_{il}^2 x_{km}^2)$
which are invariant under the inversion. Most remarkably, conformal invariance
fixes the  three-point Green functions up to an overall normalization factor
\cite{Pol70}, e.g., for three scalar fields
\beq{3-point}
\langle0| \phi(x_1) \phi(x_2) \phi(x_3)|0\rangle  =
 N_3(g^\ast) (\mu^\ast)^{-3\gamma(g^\ast)}
\left(\frac{1}{(x_1-x_2)^2 (x_1-x_3)^2 (x_2-x_3)^2}\right)^{3[l^{\rm can}
+ \gamma(g\ast)]/2}\,.
\eeq

Everything said in this Section remains true if instead of the Green functions
written in terms of fundamental fields $\Phi(x)$ (\ref{Def-GreFun}) one considers
correlation functions of arbitrary local conformal operators provided that their
transformation properties under the conformal group, albeit with different spin
and scaling dimensions, are preserved on quantum level. We will discuss this
point in detail later in Sect.~5.2. An important consequence of the conformal
symmetry, encoded in the special conformal Ward identity (\ref{CWI-Kfree}), is
that the correlation function of two conformal operators vanishes in the
conformal invariant theory, unless they have the same conformal spin:
\beq{CWI-10}
  \langle0|{\rm T}\,{\mathbb O}_{1}(x) {\mathbb O}_{2}(y)|0\rangle =
\delta_{j_1,j_2}\, F(x-y)\,,
\eeq
where we indicated that this can only be a function of $x-y$ thanks to
the translational invariance. Using this
and choosing $y=-x$ the Ward identity (\ref{CWI-Kfree}) for the
correlation function (\ref{CWI-10}) becomes
\beq{CWI-11}
  2\left[x^\mu(\ell_1-\ell_2)-(\Sigma^{(1)}- \Sigma^{(2)})^\mu{}_\nu\,
x^\nu\right]
\langle0|{\rm T}\,{\mathbb O}_{1}(x) {\mathbb O}_{2}(-x)|0\rangle =0\,.
\eeq
This is a vector equation. Multiplying it with $x_\mu$ we conclude
that necessarily $\ell_1=\ell_2$. On the other hand, choosing the
coordinates such  that $x_\perp=0$ and taking the ``minus'' projection
$x^\mu \to x_-\, n^\mu$, we are led to $\ell_1+s_1 = \ell_2+s_2$, or
$j_1=j_2$, as promised. In particular, for the two-particle conformal
operators defined in Eq.~(\ref{co-12}) we have
\bea{2-part}
\langle 0|{\rm T}\,{\mathbb O} _n^{j_1,j_2}(x){\mathbb O} _m^{j_1,j_2}(0)|0\rangle \sim
\delta_{nm}
\eea
since the conformal spins of two operators are different for $n\neq m$.
This equation looks like (and is indeed) the orthogonality relation,
but its actual evaluation is a nontrivial task in an interacting
conformal field theory.
In the free theory, however, we can evaluate the l.h.s.\ rather easily,
writing the conformal operators in terms of their characteristic
polynomials (\ref{co-1}) and replacing the four-point function that
arises by the product of two propagators (\ref{22-point}).
Going over to the light-cone limit $x^2=2x_+x_-$, $x_\perp=0$,
 $x_+\to 0$ with $x_-$ fixed, one obtains from Eq.\ (\ref{2-part})
\bea{ortho-1}
{\mathbb
 P}_n(\partial_{\alpha_1},\partial_{\alpha_2})\,
 {\mathbb
 P}_m(\partial_{\beta_1},\partial_{\beta_2})\,(\alpha_1-\beta_1)^{-2j_1}
 (\alpha_2-\beta_2)^{-2j_2}\bigg|_{{\alpha_1=\alpha_2=0}\atop{\beta_1=\beta_2=0} }
 \sim \delta_{nm}\,.
\eea
This can be simplified using the representation
\bea{102}
(\alpha_k-\beta_k)^{-2j_k}=\int_0^\infty ds_k\,\frac{s_k^{2j_k-1}}{\Gamma(2j_k)}
\exp(-s_k(\alpha_k-\beta_k))
\eea
and changing the integration variables to $s_1= s \xxi_1$ and $s_2=s
\xxi_2$ with $0\le s <\infty$, $0\le \xxi_k \le 1$ and
$\xxi_1+\xxi_2=1$. The integration over $s$ gets factored out, and as the result
one finds the familiar orthogonality relation (\ref{co-13}) for the
characteristic polynomials of conformal operators
\bea{sc-prod}
\langle \mathbb{P}_n|\mathbb{P}_m\rangle \equiv \int_0^1 [d\xxi]\,
\xxi_1^{2j_1-1}
\xxi_2^{2j_2-1}\mathbb{P}_n(\xxi_1,\xxi_2)\mathbb{P}_m(\xxi_1,\xxi_2)\sim \delta_{mn}\,,
\eea
where $[d\xxi]=d\xxi_1 d\xxi_2\, \delta(\xxi_1+\xxi_2-1)$. We see now that its
existence  is a direct consequence of conformal symmetry and in particular of the
Ward identity for the special conformal transformation. Notice that this
orthogonality property alone allows one to identify the two-particle
characteristic polynomials as the Jacobi orthogonal polynomials,
Eq.~(\ref{co-11}), so that their explicit calculation is not necessary.

We recall that the derivation of Eq.\ (\ref{co-11}) was based on the transformation
(\ref{co-7}) which led to particular simple form of the polynomials $\widetilde
{\mathbb{P}}_n(\kappa_1,\kappa_2)$ in the adjoint representation,
Eqs.~(\ref{co-9})--(\ref{co-11}). Using the above construction we can
elucidate the meaning of these polynomials. Consider the
three-point correlation function $x_1^{2j_1}x_2^{2j_2}\langle
0|\Phi_{j_1}(x_1)\Phi_{j_2}(x_2){\mathbb O} _n^{j_1,j_2}(0) |0\rangle$
of the conformal operator ${\mathbb O} _n^{j_1,j_2}$ and two fundamental
fields. Taking the light-cone limit in both coordinates $x_1$ and $x_2$
and denoting $(x_1)_-=1/\kappa_1$,
$(x_2)_-=1/\kappa_2$, one finds that the dependence
of the correlation function on $\kappa_1,\kappa_2$ is factorized into
\bea{dual11}
\widetilde
{\mathbb{P}}_n(\kappa_1,\kappa_2)&=&{\mathbb
P}_n(\partial_{\alpha_1},\partial_{\alpha_2})\,(1-\kappa_1\alpha_1)^
{-2j_1}(1-\kappa_2\alpha_2)^{-2j_2}
\bigg|_{\alpha_1=\alpha_2=0}
\nonumber\\
&=&
\int_0^\infty \prod_{k=1,2}\frac{ds_k\,s_k^{2j_k-1}}{\Gamma(2j_k)}
{\rm e}^{-s_k}{\mathbb P}_n(\kappa_1s_1,\kappa_2s_2)\,.
\eea
It is easy to see that this transformation maps the characteristic polynomial
${\mathbb P}_n$ into another homogeneous polynomial $\widetilde
{\mathbb{P}}_n(\kappa_1,\kappa_2)$ of the same degree and with the expansion
coefficients related to each other according to Eq.\ (\ref{co-7}).
Eq.~(\ref{dual11}) defines, therefore, the characteristic polynomial in the
adjoint representation.


\section{Conformal Partial Wave Expansion }

\subsection{\it Meson Distribution Amplitudes of Leading Twist}

Hadron distribution amplitudes have been introduced for the QCD
description of hard exclusive processes \cite{CZ77,ER80,LB80,CSZ77},
e.g., form factors with large momentum transfer, where they play a
r\^ole similar to that of parton distributions in inclusive reactions.
Much of the discussion in the literature has been concentrated on
the pion distribution amplitude of leading twist which is the simplest
case. In fact,  this is the only distribution for which
sufficient direct experimental evidence is available. It is, therefore,
natural to consider this example.

In an attempt to calculate hard exclusive reactions including a large momentum
transfer to a pion, one is led naturally to the matrix element of a bilocal
quark-antiquark operator between the vacuum and the pion state, of the type
\beq{lt-1}
 \langle 0| \bar d (0) [0,\infty n]\gamma_+\gamma_5[\infty n+x,x]u(x)
 |\pi^+(p)\rangle = i f_\pi p_+ \int_0^1\!du\, {\rm e}^{-iu(p\cdot x)}
 f(u, \ln x^2) + {\cal O}(x^2)\,,
\eeq
where $f_\pi \simeq 132$~MeV is the pion decay constant. We will neglect the mass
of the pion and choose the frame of reference such that $p_\mu =p_+ \bar n_\mu$
and
 $x^\mu = x_- n^\mu + x_\perp^\mu$, $x_+=0$, so
that $x^2 = -x_\perp^2$, cf.\ Eq.\ (\ref{sudakov}). In the light-cone gauge
$A_+=0$ the Wilson lines disappear and the matrix element in Eq.\ (\ref{lt-1})
can be interpreted as the probability amplitude to find the pion in the valence
state consisting of the $u$-quark and $d$-antiquark carrying the momentum
fractions $u$ and $\bar u \equiv 1-u$, respectively, and have the transverse
separation $|x_\perp|$. It has been proven \cite{ER80,LB80} that this amplitude
enters the QCD calculations of the pion electromagnetic form factor for large
momentum transfers $Q^2$ and relevant transverse separations are small, of the
order of $x_\perp^2 \sim 1/Q^2$.

In order to study this limit, we set $x_\perp^\mu =0$ and consider a new
object, the pion distribution amplitude, defined by the bilocal matrix
element similar to Eq.\ (\ref{lt-1}) but taken at exactly light-like separations
\beq{lt-2}
  \langle 0|\bar d (0) [0, \alpha]\gamma_+\gamma_5u(\alpha)
 |\pi^+(p)\rangle = i f_\pi p_+ \int_0^1\!du\, {\rm e}^{-iu \alpha p_+}
 \phi_\pi(u,\mu)\,,
\eeq
where we have replaced $x_-\to \alpha$ in order to recover the notations
used in Sect.\ 2. As always in a field theory, extraction of the
asymptotic behavior creates divergences and the nonlocal operator
appearing on the l.h.s.\ of Eq.\ (\ref{lt-2}) has to be renormalized.
In turn, the distribution amplitude $\phi_\pi(u,\mu)$ is
renormalization scale- and scheme-dependent. The original amplitude
$f(u,\ln x^2)$ is given by the convolution of $\phi_\pi(u,\mu)$
with a certain coefficient function $c(u,x^2\mu^2)$ which can be
calculated order by order in perturbation theory and contains no
large logarithms provided the scale $\mu^2$ is chosen to be of the
order of the transverse separation $\mu^2\sim 1/x^2$. In this way
the small transverse distance behavior of the valence component
of the pion wave function can be traded for the scale dependence of the
distribution amplitude, which is the usual trick of the renormalization
group.
Calculation of this scale dependence is the goal of our analysis
and we will see that conformal symmetry provides one with a very elegant
solution to the leading logarithmic accuracy, with almost no effort.

The renormalization group equation for $\phi_\pi(u,\mu)$ is known as
the Efremov-Radyushkin--Brodsky-Lepage (ER-BL) evolution equation \cite{ER80,LB80}
\beq{ERBL}
  \mu^2  \frac{d}{d\mu^2} \phi_\pi(u,\mu) = \int_0^1 \! dv\,
   V(u,v;\alpha_s(\mu))\, \phi_\pi(v,\mu)\,.
\eeq
The integral kernel $V(u,v)$
is to  leading order in $\alpha_s$  given by
\beq{BLkernel}
 V_0(u,v) = C_F\frac{\alpha_s}{2\pi}
      \left[\frac{1-u}{1-v}\left(1+\frac{1}{u-v}\right)\theta(u-v)
   + \frac{u}{v}\left(1+\frac{1}{v-u}\right)\theta(v-u)\right]_+\, ,
\eeq
where the ``+'' subtraction is defined as
\beq{+}
  [V(u,v)]_+ = V(u,v)-\delta(u-v)\int_0^1\!dt\,\, V(t,v)\,.
\eeq
Solving Eq.~(\ref{ERBL}),
i.e., finding the eigenfunctions
of the  kernel (\ref{BLkernel}), does not look simple
at first sight.

Another way to address the same problem is to expand the both sides of Eq.\
(\ref{lt-2}) in powers of $\alpha$, so that moments of the distribution amplitude
are expressed as matrix elements of renormalized local operators:
\beq{lt-OPE}
 \langle 0|\bar d (0)\gamma_+\gamma_5\,
  \big( i\! \stackrel{\leftrightarrow}{D}_+\!\big)^n u(0)|\pi^+(p)\rangle =
   i f_\pi (p_+)^{n+1} \int_0^1\!du\, (2u-1)^n \phi_\pi(u,\mu)\,.
\eeq
Apart from the flavour content and the extra $\gamma_5$ these are the same
leading twist operators that enter the OPE for the unpolarized deep inelastic
scattering. The difference to the latter case is, however, that in the present
situation we are not restricted to forward matrix elements and hence cannot
neglect the mixing of these operators with the operators containing total
derivatives
\beq{lt-3}
  {\cal O}_{n-k,k} = (i\partial_+)^k \bar d (0)\gamma_+\gamma_5\,
  \big( i\! \stackrel{\leftrightarrow}{D}_+\!\big)^{n-k} u(0)\,.
\eeq
Note that the operators with less total derivatives can only mix with the
operators with more total derivatives, but not the other way around. This implies
that the mixing matrix is triangular and, therefore, the entries on the diagonal
are true anomalous dimensions. Moreover, they are the same anomalous dimensions
that appear in deep inelastic scattering (an extra $\gamma_5$  has no effect in
the flavor non-singlet sector)
\beq{an-dim}
    \gamma^{(0)}_n =  C_F\left(1-\frac{2}{(n+1)(n+2)}
     + 4\sum_{m=2}^{n+1} \frac1{m}\right)\,,
\eeq
so that
\beq{an-dim2}
   \langle P|{\cal O}_{n,0}(\mu)|P\rangle =
    \langle P|{\cal O}_{n,0}(\mu_0)|P\rangle
    \left(\frac{\alpha_s(\mu)}{\alpha_s(\mu_0)}
         \right)^{\gamma^{(0)}_n/\beta_0},
        \qquad \beta_0 = (11/3)N_c -(2/3) N_f\,.
\eeq
Knowing the eigenvalues is helpful but
we also need to find the eigenvectors of the mixing
matrix, which correspond to the multiplicatively renormalizable operators.
Finding such eigenvectors is of course equivalent to finding the
eigenfunctions of the ER-BL kernel.

The problem is solved by the observation that conformal operators
in QCD with different values of the conformal spin cannot mix under
renormalization to leading order. The reason why conformal symmetry
``works'' in this case is that leading-order operator renormalization
is driven by counterterms that are tree-level to this accuracy, and
therefore retain all the symmetry properties of the classical theory.
As explained in Sect.\ 2 the Callan-Symanzik equation that governs
the dependence on the renormalization scale is nothing else  but
the Ward identity for the dilatation operator.

In our case the relevant conformal operators are
essentially ${\mathbb Q}^{1,1}(x)$, given in Eq.~(\ref{co-19}),
correcting for the flavor and Dirac structure:%
\footnote{We hope that using the same notation
for conformal operators built of quark fields with
different flavor and Dirac structure will not cause confusion;
the choice is always  clear from the context.}
\beq{lt-geg}
  {}{\mathbb Q} _n^{1,1}(x) = (i\partial_+)^n \left[\bar d(x)
 \,\gamma_+\,\gamma_5\, C^{3/2}_n
\left(\stackrel{\leftrightarrow}{D}_+/\partial_+\right)
u(x)\right].
\eeq
Because of the flavor structure there can be no mixing with operators
built of two gluon fields, cf.\ Eq.\ (\ref{co-20}), and all
operators built of three and more fields have higher twist so that
there will be no mixing with them as well. We conclude that the
operators in Eq.\ (\ref{lt-geg}) are the only relevant ones, and, therefore,
they must be multiplicatively renormalized. The beauty of the
argument is that to arrive to this conclusion we do not need to know the
explicit form of the mixing matrix, alias the ER-BL kernel.
Comparing Eqs.\ (\ref{lt-OPE}) and (\ref{lt-geg}), we conclude that the
Gegenbauer moments of the pion distribution amplitude
are given by reduced matrix elements of conformal operators
\bea{lt-gm}
\int_0^1\!du\, C_n^{3/2}(2u-1) \phi_\pi(u,\mu) =
  \langle\!\langle {\mathbb Q} _n^{1,1}\rangle\!\rangle,
\qquad
 \langle 0| {\mathbb Q} _n^{1,1}(0)|\pi^+(p)\rangle = i f_\pi p_+^{n+1}
  \langle\!\langle {\mathbb Q} _n^{1,1}\rangle\!\rangle
\eea
and they must be renormalized multiplicatively, with the anomalous
dimensions given in Eq.\ (\ref{an-dim}).

Last but not least, note that the Gegenbauer polynomials appearing on
the l.h.s.\ of Eq.\ (\ref{lt-geg}) are orthonormal with the weight function
$u(1-u)$ and form a complete set of functions on the interval $0<u<1$ \cite{BE}:
\beq{norm1}
     \int_0^1\! du\, u(1-u)\, C_n^{3/2}(2u-1) \, C_m^{3/2}(2u-1) =
        \delta_{mn}\, \frac{(n+1)(n+2)}{4(2n+3)}\,.
\eeq
As a consequence,  the distribution amplitude $\phi_\pi(u,\mu)$ can be
expanded in a series over Gegenbauer polynomials with multiplicatively
renormalizable coefficients
\bea{lt-pion}
     \phi_\pi(u,\mu) &=& 6u(1-u)\sum_{n=0}^\infty \phi_n(\mu) \,
      C_n^{3/2}(2u-1)\,,
\nonumber\\
      \phi_n(\mu) &=& \frac{2(2n+3)}{3(n+1)(n+2)}
      \langle\!\langle {\mathbb Q} _n^{1,1}(\mu)\rangle\!\rangle,
    \qquad
      \phi_n(\mu) = \phi_n(\mu_0)
      \left(\frac{\alpha_s(\mu)}{\alpha_s(\mu_0)}
         \right)^{\gamma^{(0)}_n/\beta_0}.
\eea
This is the final result \cite{ER80,LB80}. The normalization in Eq.\
(\ref{lt-2}) is chosen such that $\int du\, \phi_\pi(u) =1$ which implies
$\phi_0 =1$. The other coefficients are genuine nonperturbative
parameters (at a certain reference scale) that encode the structure of
the pion at small transverse separations between the constituents.

We note in passing  that the orthogonality of Gegenbauer polynomials allows one
to write the ER-BL kernel in terms of the anomalous dimension:
\beq{BLham}
  V_0(u,v) = -\frac{\alpha_s}{4\pi} u(1-u) \sum_{n=0}^\infty
  \frac{4(2n+3)}{(n+1)(n+2)}\,\gamma^{(0)}_n\, C^{3/2}_n(2u-1) C^{3/2}_n(2v-1)\,.
\eeq
This representation is valid for $V_0$ acting on arbitrary polynomials
on the interval $0< u< 1$. In Sect.\ 4 we will develop a powerful
approach to the solution of three-particle evolution equations which is
based on this kind of representations for the evolution kernels.

At this place, several comments are in order.
\begin{itemize}
 \item{} To put it simple, the conformal symmetry allows one to resolve the mixing with
         operators involving total derivatives.
\vspace*{-0.1cm}
\item{} Whether this is enough to achieve multiplicative
         renormalization, depends on the problem. We have profited
         from working with flavor non-singlet operators. In the
         flavor singlet sector (e.g., for the distribution amplitude of
         $\eta'$-meson) one  has to take into account two-gluon
         operators. This will result in a $2\times2$ mixing matrix for each
         conformal spin. For the higher-twist conformal operators
         given in Eq.\ (\ref{co-19}) one also has to take into account
         the mixing with quark-antiquark gluon operators of the same
         twist and conformal spin so that the situation gets complicated.
         We will return to this problem in the next sections.
\vspace*{-0.1cm}
\item{}  In order to make use of the completeness condition in
         Eq.\ (\ref{norm1}) it is important that the region of integration
         over the momentum fraction $u$ in Eq.\ (\ref{lt-2}) is precisely
         $0<u<1$. This is always the case for distribution amplitudes,
         but can be different in other applications, e.g., for
         generalized parton distributions
         \cite{MuePHD,MueRobGeyDitHor94,Ji96Rad96}. Whenever the integration
         regions do not match, it is not possible to invert the moments and
         write the physical
         amplitude as a sum over multiplicatively renormalizable
         contributions, although the conformal operators are the same.
\vspace*{-0.1cm}
\item{}  The expansion in Eq.\ (\ref{lt-pion}) is formal, in the sense that
         its convergence properties cannot be derived from the symmetry
         considerations. In QCD, however, the anomalous
         dimensions of conformal operators are ordered and rise
         logarithmically with the conformal spin. This implies that
         contributions of operators with high spin
         are suppressed at large scales and asymptotically
         $\phi_\pi(u,\mu) \to 6u(1-u)$ at $\mu \to \infty$. This
         expression is called asymptotic pion distribution amplitude.
         Whether the pion distribution amplitude at hadronic scales
         is close to its asymptotic form, has to be decided by
         experiment. There exist strong indications \cite{Sav97}
         that contributions of higher-spin operators are indeed small
         compared to the asymptotic term so that the conformal expansion
         is well-behaved at already low scales. It is worth while to
         note that convergence of the conformal expansion of
         distribution amplitudes is used implicitly in the proofs of QCD
         factorization in exclusive reactions.
\vspace*{-0.1cm}
\item{}  The validity of Eq.\ (\ref{lt-pion}) is not
         spoiled by quark masses, although the mass terms certainly
         break the conformal symmetry already at the classical level.
         The reason for this is that although
         {\it finite parts} of the matrix elements of the operators
         (\ref{lt-geg}) and therefore the coefficients $\phi_n(\mu)$
         depend on the quark masses (e.g., they are different for $\pi$
         and $K$ mesons), the {\it infinite}, {UV divergent}
         contributions to the matrix elements are removed by the
         mass-independent counterterms, so that the renormalization
         group behavior is not affected.
\end{itemize}
The expansion (\ref{lt-pion}) is generic and is valid for all leading twist
quark-antiquark meson distribution amplitudes, albeit with different coefficients
and anomalous dimensions. As a further example, consider distribution amplitudes
of a vector particle, e.g., the $\rho$-meson \cite{CZreport,BB96}. There exist
two independent distribution amplitudes corresponding to $\rho$-mesons  with
longitudinal $\lambda=0$ and transverse $\lambda=\pm 1$ polarizations:
\bea{lt-rho}
 \langle 0| \bar d(0) \gamma_+ u(\alpha) |\rho^+(P,\lambda)\rangle &=&
 f_\rho m_\rho e^{(\lambda)}_{+}
\int_0^1 du\,{\rm e}^{-iu \alpha p_+} \phi_{\rho}(u,\mu)\,,
\nonumber\\
\langle 0| \bar d(0) \sigma_{\perp+} u(\alpha) |\rho^+(P,\lambda)\rangle &=&
i f_\rho^T e^{(\lambda)}_{\perp}p_+
\int_0^1 du\,{\rm e}^{-iu \alpha p_+} \phi^T_{\rho}(u,\mu)\,,
\eea
where $e^{(\lambda)}_{\mu}$ is the polarization vector and
$f_\rho$ and $f^T_\rho$ are the corresponding decay constants.

Since $\gamma_\perp$ commutes with the quark spin projection
operators (\ref{co-16}), both $\gamma_+$ and $\sigma_{\perp+}$
project onto twist-two operators
built of two ``plus'' quark fields, so that the conformal
operators only differ by this substitution. Repeating the
above derivation, we end up  the same expansions
\bea{lt-rho2}
     \phi_\rho(u,\mu) &=& 6u(1-u)\sum_{n=0}^\infty
     \phi_n(\mu_0)
      \left(\frac{\alpha_s(\mu)}{\alpha_s(\mu_0)}
         \right)^{\gamma^{(0)}_n/\beta_0}\,
      C_n^{3/2}(2u-1)\,,
\nonumber\\
     \phi^T_\rho(u,\mu) &=& 6u(1-u)\sum_{n=0}^\infty
      \phi^T_n(\mu_0)
      \left(\frac{\alpha_s(\mu)}{\alpha_s(\mu_0)}
         \right)^{(\gamma^{T(0)}_n-\gamma^{T(0)}_0)/\beta_0}\!\!
      C_n^{3/2}(2u-1)\,,
\eea
where in the second case $\gamma^{T(0)}_0$ is subtracted in order to
compensate for the scale dependence of $f^T_\rho$. The anomalous
dimensions $\gamma^{(0)}_n$ are the same as for the pion while
$\gamma^{T(0)}_n$ are equal to \cite{SV81}:
\beq{an-dim-tensor}
  \gamma^{T(0)}_n = C_F\left(1
     + 4\sum_{m=2}^{n+1} \frac1{m}\right)\,.
\eeq

{}From all these examples we see that the task accomplished by the conformal
symmetry is the separation of variables. A very close analogy is provided by
nonrelativistic quantum mechanics in a spherically symmetric potential. As well
known,  the $O(3)$  symmetry allows for the separation of angular and radial
degrees of freedom by going over to the basis of states with given orbital
angular momentum $l$. For fixed $l$ one is left with a one-dimensional
Schr\"odinger equation for the radial coordinate, whereas the angular dependence
is included in spherical harmonics which are the irreducible representations of
the group of rotations. In our case, the conformal symmetry invites to go over to
the basis of states with given conformal spin, conformal partial waves. {}For
each spin the dependence of the wave function on transverse coordinates is given
by a simple renormalization group equation and the dependence on longitudinal
coordinates (momentum fractions) is included in orthogonal polynomials which form
an irreducible representation of the collinear conformal group, $SL(2,{\mathbb
R})$.

The group-theoretical content of this problem becomes especially transparent
using the basis of one-particle conformal states defined in Eq.~(\ref{coh-55}). A
generic multi-particle distribution amplitude can be defined as the matrix
element of a multi-local operator
\bea{dada}
 \langle 0| \Phi_1(\alpha_1)\ldots  \Phi_m(\alpha_m)|p\rangle
   &=& \int_0^\infty\!\! dp_1\ldots \!\int_0^\infty\!\! dp_m\,
   \delta(\sum p_i-p) \,\, \exp[-{i(\alpha_1 p_1 +\ldots + \alpha_n
    p_m)}]\!
\nonumber\\
&&\times
\sum_{k_1,\ldots,k_m=0}^\infty
    \langle0|\varphi_{k_1}\ldots\varphi_{k_m}|p\rangle
 \,\,
     \langle p_1|j_1,k_1\rangle\ldots \langle p_m|j_m,k_m\rangle,
\eea
where we used the momentum conservation, the fact that
all states have positive energy, hence $p_k \equiv (p_+)_k>0$,
and inserted the expansion (\ref{coh-NN}).
The matrix elements $\langle0|\varphi_{k_1}\ldots\varphi_{k_m}|p\rangle$
are just constants; call them $\omega_{k_1,\ldots k_m}$.
The basis vectors $\langle p_i|j_i k_i\rangle$ (\ref{coh-55})
are homogeneous functions; hence one can redefine them in terms of
momentum fractions $u_i = p_i/p$ and
the whole expression becomes, up to the overall normalization
\bea{dddd}
 \langle 0| \Phi_1(\alpha_1)\ldots  \Phi_m(\alpha_m)|p\rangle &\sim&
  \int [du]\,   {\rm e}^{ip(\alpha_1 u_1 +\ldots + \alpha_n u_m)}
   \phi(u_1,\ldots u_m)\,,
\nonumber\\
   \phi(u_1,\ldots,u_m) &=& \sum_{k_1,\ldots,k_m=0}^\infty \omega_{k_1,\ldots k_m}
    \,\langle u_1|j_1,k_1\rangle\ldots \langle u_m|j_m,k_m\rangle.
\eea
where
$
   \int [du] = \int_0^1\!\! du_1\ldots \!\int_0^1\!\! du_m\,
   \delta(\sum u_i-1)
$.
What we would like to see, on the other hand, is the expansion
of the distribution amplitude in terms of the irreducible representations
of $SL(2,{\mathbb R})$ with given conformal spin $j$
\beq{DAexpand}
  \phi(u_i) = \sum_{j=j_1+\ldots+j_m}^\infty \sum_{R} \omega'(j,R)
   \langle u_i|j,0\rangle_R
\eeq
where the summation over $R$  goes over all (degenerate) representations
with the same spin. Going over from Eq.\ (\ref{dddd}) to Eq.\  (\ref{DAexpand})
is the classical problem of spin summation, with the only difference
that the spins ``live'' on the hyperboloid since the algebra
of the collinear conformal group $SL(2,{\mathbb R})$
(\ref{SL2R2}) coincides with the algebra of the
Minkowski group in 2+1 dimensions, $O(2,1)$. Restricting oneself to
infinitesimal group transformations one can say that the conformal
spin $j$ corresponds to the ``mass'' and $|j_0|$ to the ``energy'' of
a particle in this internal space. Using this analogy it becomes
intuitively obvious that the lowest conformal spin (``mass'') for a
multi-particle state is just the sum of spins (``masses'') of the
constituents; this state is nondegenerate and is given by the product
of one particle states ``at rest'', that is with the lowest
values of the ``energies'' $|j_0| = j$, alias $k=0$, in our
notation. This lowest state defines what is called the asymptotic
distribution amplitude in the general case \cite{BF90}
\beq{asDA}
   \phi_{\rm as}(u_i) = \langle u_i|j=j_1+\ldots+j_m,0\rangle
   = \frac{\Gamma(2j_1+\ldots+2j_m)}{\Gamma(2j_1)\ldots \Gamma(2j_m)}
     u_1^{2j_1-1}u_2^{2j_2-1}\ldots u_m^{2j_m-1}\,,
\eeq
where the normalization is chosen such that $\int [du_i]\,\phi_{\rm as}(u_i)=1$.
The constructions of states with higher spin reduces to finding the
corresponding Clebsch-Gordan coefficients
\beq{Clebsch}
 \langle u_1,\ldots,u_m|j,k\rangle = \sum_{k_1+\ldots +k_m = k}
  C^{j,k}_{j_1k_1,\ldots,j_m k_m}
 \langle u_1|j_1k_1\rangle\ldots  \langle u_m|j_mk_m\rangle
\eeq
In particular, for two particles
\bea{Clebsch2}
   \langle u_1u_2|j,k\rangle &=& (u_1+u_2)^k \sum_{k_1+k_2=j+k-j_1-j_2}(-1)^{k_1}
    \left(
      \begin{array}{c}
       k_1+k_2 \\ k_1
      \end{array}
 \right)
    \langle u_1|j_1,k_1\rangle  \langle u_2|j_2,k_2\rangle
\nonumber\\
&\sim& (u_1+u_2)^k \phi_{\rm as}(u_1,u_2) P^{2j_1-1,2j_2-1}_{j-j_1-j_2}
\left(\frac{u_1-u_2}{u_1+u_2}\right),
\eea
cf.~Eq.\ (\ref{co-11}). Note that the ``raising'' operator ${\JJ}_+ = {
\JJ}_{1,+} +\ldots+ {\JJ}_{m,+}$ acts as a multiplication operator on the
distribution amplitudes,
 ${\JJ_+}\, \phi(u_i) \sim (u_1+\ldots+u_m) \phi(u_i)$, and the prefactor
is equal to unity because of the phase-space condition $\sum u_i =1$. Hence
only the lowest ``energy'' state with $k=0$ is physically relevant.
In the general situation with more than two particles it can be shown
that higher spin states are obtained by multiplication of the asymptotic
distribution amplitude by a polynomial of degree $n=j-j_1-j_2$ and the
polynomials corresponding to different conformal spins are mutually
orthogonal with the weight function $\phi_{\rm as}(u_i)$ (\ref{asDA}).
A convenient complete set of orthogonal polynomials for three particles
is defined in Appendix A. Note that the whole construction is equivalent
to finding of the corresponding local conformal operators.

\subsection{\it Higher Twists}

Contributions to physical cross sections that are suppressed by a power of the large
momentum are generically referred to as higher twists. The most important question
to be addressed in connection with higher twist contributions is always to prove
that  QCD factorization can be extended to include power-suppressed
corrections. If this is the case, the second question is to obtain a general classification
of the relevant hadronic quantities in terms of the independent nonperturbative
 parameters. Since the number of parameters usually prolifiterates compared to the
leading twist, a practical task is to establish their hierarchy and build
selfconsistent approximations   with minimum nonperturbative input.

Conformal partial wave expansion proves to be indispensable for the construction of higher-twist
distribution amplitudes. At present, this program is completed for pseudoscalar \cite{BF90,Ball99} and
vector mesons \cite{BBKT98,BB99}, for the real photon \cite{BBK89,BBK03} and partially also for
the nucleon \cite{BFMS00}.  The summary
of results goes beyond the tasks of this review. Instead, we will try to explain basic ideas
and techniques beyond such calculations, and also indicate existing problems.
Our first example will be the construction of  selfconsistent approximations to pion
distribution amplitudes of twist-three. This is the simplest case which is, on the other
hand, sufficiently general; applications to twist-four and to other hadrons follow the same
scheme. Second, we discuss a recent work \cite{Andersen00,GG03}
concerning convergence properties of the conformal expansion for higher twist distributions
using the concept of renormalons \cite{Beneke99,BBr01}.
Finally, we consider meson mass corrections and in particular the so-called
Wandzura-Wilczek contributions to distribution amplitudes, and how they can be included in
the conformal expansion. To this end twist-three distributions of the $\rho$ meson are
considered.

\subsubsection{\it Pion Distribution Amplitudes: Twist-three}

Pion distribution amplitudes of twist-three are defined as matrix elements of nonlocal twist-three
operators. One finds two two-particle distributions
\bea{ht-1}
 \langle 0| \bar d(0) i\gamma_5 u(\alpha) |\pi^+(p)\rangle &=&
 \frac{f_\pi m_\pi^2}{m_u+m_d}
\int_0^1 du\,{\rm e}^{-iu \alpha p_+} \phi_p(u,\mu)\,,
\nonumber\\
\langle 0| \bar d(0) \sigma_{\mu\nu}\gamma_5 u(\alpha) |\pi^+(p)\rangle &=&
\frac{i \alpha f_\pi m_\pi^2}{6(m_u+m_d)}\big(p_\mu n_\nu-p_\nu n_\mu\big)
\int_0^1 du\,{\rm e}^{-iu \alpha p_+} \phi_\sigma(u,\mu)\,,
\eea
and one three-particle distribution
\bea{ht-2}
\langle0|\bar d(\alpha_2) \sigma_{\mu\nu}\gamma_5 gG_{\alpha\beta}(\alpha_3) u(\alpha_1)
|\pi^+(p)\rangle
&=&
i \Big[p_\alpha(p_\mu g_{\nu\beta}-p_\nu g_{\mu\beta}) -
\big(\alpha\leftrightarrow\beta\big)\Big]
\nonumber\\
&&{}\times
\int [du_i]\, {\rm e}^{-ip_+(\alpha_1 u_1+\alpha_2 u_2+\alpha_3 u_3)}\,
   \phi_{3\pi}(u_i,\mu)\,.
\eea
The normalization is chosen such that $\int\! du\, \phi_p(u) = \int\! du\, \phi_\sigma(u) = 1$.
We will use a shorthand notation
\beq{ht-3}
R = {f_\pi m_\pi^2}/({m_u+m_d}) \simeq -\langle \bar u u\rangle -\langle\bar d d\rangle\,,
\eeq
where $ \langle \bar u u\rangle \simeq \langle\bar d d\rangle $ is the quark condensate.
We will tacitly imply the chiral limit and neglect quark mass corrections in what follows.

The three functions $\phi_p(u)$, $\phi_\sigma(u)$ and $\phi_{3\pi}(u_i)$  are
related to each other by equations of motion~\cite{Gorsky87,BF90}. One way to see
this is to observe that relevant matrix elements of chiral-odd operators (with
even number of $\gamma$-matrices) correspond to one ``plus'' component of the
quark field and one ``minus'' component. This also explains the twist assignment,
since going over from a ``plus'' to a ``minus'' component adds one unit of twist.
As well known in context of the light-cone quantization \cite{KS70}, ``minus''
components of the spinor fields do not present independent degrees of freedom,
but can be eliminated using the Dirac equation. A convenient way to derive these
relations is to make use of the operator identities \cite{BB89,Ball99}
\bea{ht-4}
   \partial_\mu \Big\{\bar d(-x)\sigma_{\mu\nu}\gamma_5 u(x)\Big\} &=&
       i \frac{\partial}{\partial x^\nu} \Big\{\bar d(-x)\gamma_5 u(x)\Big\}
     + \int_{-1}^1 \! vdv\, \bar d(-x) x^\rho gG_{\rho\nu}(vx) \gamma_5 u(x)
\nonumber\\
 &&{}+ i \int_{-1}^1 \! dv\, \bar d(-x) x_\rho gG^{\rho\mu}(vx)\sigma_{\mu\nu} \gamma_5 u(x)\,,
\nonumber\\
    \frac{\partial}{\partial x^\nu} \Big\{\bar d(-x)\sigma_{\mu\nu}\gamma_5 u(x)\Big\} &=&
       i  \partial_\nu \Big\{\bar d(-x)\gamma_5 u(x)\Big\}
     + \int_{-1}^1 \! dv\, \bar d(-x) x^\rho gG_{\rho\nu}(vx) \gamma_5 u(x)
\nonumber\\
 &&{}+ i \int_{-1}^1 \! vdv\, \bar d(-x) x_\rho gG^{\rho\mu}(vx)\sigma_{\mu\nu} \gamma_5 u(x)
\eea
for the total translation and dilatation of the relevant nonlocal operators;
 $\partial_\nu$ stands for the derivative with respect to the total translation, for example
\beq{ht-6}
\partial_\nu[\bar d(-x)\gamma_5 u(x)] =\frac{\partial}{\partial y^\nu}
[\bar d(-x+y)\gamma_5 u(x+y)]\Big|_{y\to 0}\,.
\eeq
Taking in Eq.\ (\ref{ht-4}) the light-cone limit $x^2\to 0$ and the matrix elements
between vacuum and the one-pion state, one obtains the system of two differential equations
connecting the three functions $\phi_p$, $\phi_\sigma$ and $\phi_{3\pi}$, see \cite{BF90,Ball99}
for details.

Our aim is to build a selfconsistent description of the distribution amplitudes in terms of a minimal
number  of nonperturbative parameters. To be precise, by a "selfconsistent" description we mean
the following:
\begin{itemize}
\item{} The constraints that follow from equations of motion have to be fulfilled identically.
              That is, the relations between $\phi_p$, $\phi_\sigma$ and $\phi_{3\pi}$ that follow from
               the exact QCD operator identities (\ref{ht-4}) have to be satisfied.
\vspace*{-0.1cm}
\item{} The structure of the distribution amplitudes has to be preserved by the evolution.
              For example, if we choose a model for the distribution amplitude at a certain scale
              as a polynomial, the degree of this polynomial remains the same at all scales.
              We also require that the renormalization group equations for the relevant parameters
              take the simplest possible form.
\end{itemize}
Both tasks are achieved if the distribution amplitudes are expanded in
contributions of increasing conformal spin. In the previous section  we have seen
already that conformal expansion simplifies the evolution equations: To one-loop
accuracy, the renormalization of relevant operators is driven by counterterms
that are tree-level and retain the symmetries of the classical Lagrangian; it
follows that operators with different conformal spin cannot mix with each other.
{}For the present case  there is a complication: The three-particle
representations of the collinear conformal group are degenerate, there exist
multiple operators of the same conformal spin and the symmetry does not prevent
their mixing. The mixing matrix is, therefore, not completely diagonalized but
rather separated  in smaller blocks. More importantly, QCD equations of motion do
not receive any quantum corrections whatsoever and, therefore, respect all
symmetries of the classical theory. As a consequence, if two operators are
related by the equations of motion, their matrix elements must have the same
transformation properties under conformal transformations and, in particular, the
same conformal spin. This implies that the relations between $\phi_p$,
$\phi_\sigma$ and $\phi_{3\pi}$ that follow from the operator identities in
Eqs.~(\ref{ht-4}) can be solved order by order in the conformal expansion.

After these general remarks, let us proceed with the explicit construction.  We
start with the three-particle distribution (\ref{ht-2}). It is easy to see that
the twist-3 part of interest can be separated by the Lorentz projection $
\bar d
\sigma_{+\perp}\gamma_5 gG_{+\perp} u$ and corresponds to the maximum possible
conformal spins of the three constituent fields: $j_{\bar d} =1$, $j_u =1$, and
$j_g =3/2$. The nonlocal composite operator in Eq.\ (\ref{ht-2}) can, therefore,
be expanded in contributions with conformal spin $j = j_{\bar d} + j_u + j_g +n =
7/2 +n$, $n=0,1,2,\ldots$. The contribution with the lowest spin $j=7/2$ can be
referred to as the asymptotic distribution amplitude and higher spin
contributions are given by the corresponding orthogonal polynomials. Using the
master-formula in Eq.\ (\ref{asDA}) and taking into account that the distribution
$\phi_{3\pi}(u_i)$ has to be symmetric under the interchange of $u_1$ and $u_2$
because of G-parity, we obtain \cite{BF90}
\bea{phi3pi}
 \phi_{3\pi}(u_i) &=& 360 u_1 u_2 u_3^2\Big[\omega^{7/2} +\omega^{9/2}\frac12
                                    (7u_3-3)
                                 + \omega^{11/2}_1 (2-4u_1u_2-8u_3+8u_3^2)
\nonumber\\
&&{}\hspace*{3cm}+\omega^{11/2}_2  (3 u_1 u_2 -2 u_3 +3 u_3^2)+\ldots \Big].
\eea
Here $\omega^j_i$ are scale-dependent nonperturbative
coefficients; the superscript refers to the conformal
spin. Conformal symmetry implies that $\omega^{7/2}$ and $\omega^{9/2}$ are renormalized
multiplicatively (to leading order), while $\omega^{11/2}_1$ and $\omega^{11/2}_2$
can mix with each other. These properties are indeed confirmed by the explicit calculation,
the anomalous dimensions and the mixing matrix for $\omega^{11/2}_1$ and
$\omega^{11/2}_2$ can be found in \cite{BF90}. The  expansion can easily be extended
to include  higher conformal spins using the conformal basis defined in Appendix A.

Going over to the two-particle distribution amplitudes  (\ref{ht-1}) we have, first of all,
to separate contributions with different (anti)quark  spin projections. To this end
we define two auxiliary distribution amplitudes \cite{BF90}
\bea{ht-aux}
  \langle 0| \bar d(0) i\gamma_-\gamma_+\gamma_5 u(\alpha) |\pi^+(p)\rangle &=&
R \int_0^1 du\,{\rm e}^{-iu a p_+} \phi_{\downarrow\uparrow}(u,\mu)\,,
\nonumber\\
\langle 0| \bar d(0) i\gamma_+\gamma_-\gamma_5 u(\alpha) |\pi^+(p)\rangle &=&
R \int_0^1 du\,{\rm e}^{-iu a p_+} \phi_{\uparrow\downarrow}(u,\mu)\,.
\eea
It is easy to see that, e.g., $\phi_{\downarrow\uparrow}(u)$ corresponds to the
spin projections $s_{\bar d}= -1/2$, $s_u=+1/2$ so that $j_{\bar d}=1/2$ and
$j_u=1$. It follows that the conformal expansion of
$\phi_{\downarrow\uparrow}(u)$ and $\phi_{\uparrow\downarrow}(u)$ goes over
Jacobi polynomials (\ref{co-12})
\bea{downup1}
  R\, \phi_{\downarrow\uparrow}(u) &\!\!\!=\!\!\!& 2 u\Big[ \kappa_{\downarrow\uparrow}^{3/2}
       + \kappa_{\downarrow\uparrow}^{5/2} P_1^{(0,1)}(\xi)
      +\kappa_{\downarrow\uparrow}^{7/2} P_2^{(0,1)}(\xi)+
       \kappa_{\downarrow\uparrow}^{9/2} P_3^{(0,1)}(\xi) +
         \kappa_{\downarrow\uparrow}^{11/2} P_4^{(0,1)}(\xi)+\ldots\Big],
\nonumber\\
  R\, \phi_{\uparrow\downarrow}(u) &\!\!\!=\!\!\!& 
        2 \bar u\Big[ \kappa_{\uparrow\downarrow}^{3/2}
        + \kappa_{\uparrow\downarrow}^{5/2} P_1^{(1,0)}(\xi)
      +\kappa_{\uparrow\downarrow}^{7/2} P_2^{(1,0)}(\xi)
       +  \kappa_{\uparrow\downarrow}^{9/2} P_3^{(1,0)}(\xi) +
         \kappa_{\uparrow\downarrow}^{11/2} P_4^{(1,0)}(\xi)+\ldots\Big],
\eea
with  nonperturbative coefficients $\kappa^j$.
Here and below in this section we use a shorthand notation
\beq{xiii}
               \xi = 2u-1\,.
\eeq
The superscripts $j=3/2, 5/2, \ldots$
refer to the conformal spin.
On the other hand, obviously
\bea{ht-psig}
 && \phi_p(u) = \frac12\big[\phi_{\uparrow\downarrow}(u)+
                       \phi_{\downarrow\uparrow}(u)\big]\,,\qquad
\frac{d}{du} \phi_\sigma(u) = 6\big  [\phi_{\uparrow\downarrow}(u)-
                        \phi_{\downarrow\uparrow}(u)\big]\,,
\eea
so that the knowledge of $\phi_{\downarrow\uparrow}(u)$ and $\phi_{\uparrow\downarrow}(u)$
is sufficient to determine both distributions $\phi_p(u)$ and $\phi_\sigma(u)$%
\footnote{The integration constant in the second of equations in Eq.\ (\ref{ht-psig}) is fixed by the
normalization condition $\int du\, \phi_\sigma(u) =1$, so there is no ambiguity.}.

Finally, we can use equations of motion or, equivalently, the operator identities
(\ref{ht-4}) in order to relate the expansion coefficients $\kappa^j$ and
$\omega^j_i$. After simple algebra (see \cite{BF90}) one obtains
\bea{ht-EOM}
  &  \kappa_{\downarrow\uparrow}^{3/2} = \kappa_{\uparrow\downarrow}^{3/2} = R\,,
 \qquad
    \kappa_{\downarrow\uparrow}^{5/2} = - \kappa_{\uparrow\downarrow}^{5/2} =
    0\,,\qquad
    \kappa_{\downarrow\uparrow}^{7/2} =
\kappa_{\uparrow\downarrow}^{7/2}= 30\, \omega^{7/2}\,,
&
\nonumber\\[3mm]
& \kappa_{\downarrow\uparrow}^{9/2} =
 - \kappa_{\uparrow\downarrow}^{9/2} = 30\, \omega^{9/2}\,,\qquad
    \kappa_{\downarrow\uparrow}^{11/2} = \kappa_{\uparrow\downarrow}^{11/2} =
           (3/2)\big[ 4\,\omega_1^{11/2}-\omega_2^{11/2}\big]\,,
&
\eea
etc. We see that everything fits together, equations of motion only relate
contributions with the same conformal spin, as expected. In particular, the
conformal spin of the quark condensate contribution $R=-2\langle \bar \psi
\psi\rangle = -\langle\bar\psi\gamma_-\gamma_+\psi\rangle
-\langle\bar\psi\gamma_+\gamma_-\psi\rangle $ is indeed $j=1/2+1 = 3/2$. Note
that $\kappa^{11/2}$ is {\it not} multiplicatively renormalized, but rather is
given by a certain combination of the two existing operators with spin 11/2. The
reason for this is that the quark-antiquark twist-3 operators entering
(\ref{ht-aux}) mix with the quark-antiquark-gluon operator (\ref{ht-2}) of the
same twist.

The conventional  distribution amplitudes defined in Eq.\ (\ref{ht-1}) can easily be restored
using Eq.\ (\ref{ht-psig}). One obtains \cite{BF90}
\bea{ht-psig2}
 R\, \phi_p(u) &=& R + 30\omega^{7/2} C_2^{1/2}(\xi)
                + \frac32\left[4\omega_1^{11/2}-\omega_2^{11/2}-2 \omega^{9/2}\right]\,
                    C_4^{1/2}(\xi)\,,
\nonumber\\
R\,\phi_\sigma(u) &=& 6u(1-u)\Big\{
                       R+ \Big[5\omega^{7/2}- \frac12\omega^{9/2}\Big] C_2^{3/2}(\xi)
                       + \frac{1}{10}\left[4\omega_1^{11/2}-\omega_2^{11/2}\right]
                       C_4^{3/2}(\xi)\Big\}\,,
\eea
where $C_n^{1/2}(\xi)$ and $C_n^{3/2}(\xi)$ are Gegenbauer polynomials. Note that also
in this case the expansion is organized in a natural way in terms of orthogonal polynomials.
This is no more a conformal expansion, however. The coefficient in front of each polynomial
is equal to the sum of two contributions with neighboring conformal spins. In particular,
the coefficient in front of $C_4^{3/2}$ in the second equation in (\ref{ht-psig2}) will be
modified once contributions of higher spin 13/2 are taken into account. For this reason,
and also to avoid superficial cancellations between $\phi_p(u)$ and $\phi_\sigma(u)$, the
distribution amplitudes $\phi_{\downarrow\uparrow}(u)$ and
$\phi_{\uparrow\downarrow}(u)$ are preferable for the description of
twist-three effects in exclusive reactions with pions.

\subsubsection{\it Renormalon Model of Higher-Twist Distribution Amplitudes}

The next problem to address is the convergence property of the conformal expansion
for higher-twist hadron distribution amplitudes. Assuming that the anomalous dimensions of
higher-twist operators increase logarithmically with the conformal spin, similar to the
leading twist, and repeating the argumentation of Sec.~3.1, one is led to the conclusion that
the conformal expansion must converge starting at a certain high renormalization scale.
{}From the practical point of view this argument is not very assuring, however, since
the suppression of higher spin operators by the QCD anomalous dimensions is rather weak,
at least in perturbation theory. For leading twist, the better argument for the convergence
of the conformal expansion at already the scales of order several GeV comes from experiment:
The CLEO data on the $\gamma \gamma^\ast \pi$ transition form factor
\cite{Sav97} strongly favor the
pion distribution amplitude that is not very different from the asymptotic form.
For higher twist, a similar direct experimental verification is impossible.
However, if convergence of the conformal expansion for leading twist
distributions (at a given scale) is taken for granted, then one can derive a reasonable
upper bound for the contributions of higher conformal spin operators of twist 4
using the technique of renormalons \cite{Beneke99,BBr01}.

The basic idea is that twist-four contributions to physical amplitudes are in
one-to-one correspondence  to the contributions of low momenta in perturbative Feynman diagrams.
Consider the following example:
\bea{ren-1}
   \lefteqn{
    \langle 0| \bar d (0) \not\! x \,\gamma_5 d(x)|\pi^+(p)\rangle
            }
\\
 &=& i f_\pi (px) \int_0^1\! du\, {\rm e}^{-iupx}
 \left[ \left(1+\sum_{k=1}^\infty c_k(u,\mu^2x^2)\alpha_s^k(\mu^2)\right)\otimes\phi_\pi(u,\mu^2)
        + x^2 g_1(u,\mu^2) + {\cal O}(x^4)\right]
\nonumber
\eea
where $x^2\Lambda_{\rm QCD}^2 \ll 1$, $\otimes$ stands for the convolution,
 and the gauge factor is assumed along the straight line
connecting the quark and the antiquark. The expansion on the r.h.s.\ of Eq.\ (\ref{ren-1}) is
nothing else but the operator product expansion and the contributions with the increasing
powers of $x^2$ correspond to the increasing twist. The relevant matrix element
of the leading twist operator defines the usual pion distribution amplitude $\phi_\pi(u)$
and it is multiplied by the coefficient function that can be calculated order by order
in perturbation theory. $g_1(u,\mu^2)$ is usually referred to, somewhat inaccurately, as the
two-particle pion distribution amplitude of twist-four. Its physical interpretation is to
describe the $k_\perp^2$ distribution of the quark and the antiquark in the pion,
and it can be expressed in terms of three-particle
quark-antiquark-gluon twist-four distribution amplitudes
using equations of motion, similar as for the twist-three distributions considered above.
Using these relations, and expanding the relevant three-particle distributions in
contributions of conformal operators one can work out the conformal expansion for
$g_1(u)$, see \cite{BF90} for the derivation. Taking into account the
contributions with the lowest and next-to-lowest conformal spin, the result reads%
\footnote{In notation of Ref.~\cite{BF90} $g^{j=3} = 5/6 \delta^2$, $g^{j=4} =
 1/2 \epsilon \delta^2$.}
\bea{ren-2}
 g_1(u,\mu) &=&   g^{j=3}(\mu)\,  u^2 \bar u^2
 \\
&+&\! g^{j=4}(\mu)
    \!\left[ \bar u u(2\!+\!13 \bar u u) + 10 u^3 \ln u
     \left( 2\!-\!3u \!+\! \frac65 u^2\right) + 10 \bar u^3 \ln \bar u
    \left(2\!-\!3\bar u \!+\! \frac65 \bar u^2\right)\right]\!+\ldots
\nonumber
\eea
where $g^{j=3}(\mu)$ and  $g^{j=4}(\mu)$ are the reduced matrix elements of
conformal twist-four operators. They are renormalized multiplicatively, with
known anomalous dimensions. Note that the $j=4$ and higher spin contributions are
not given by polynomials in the momentum fraction in this case, which implies
that twist-four two-particle conformal operators are not given by polynomials in
the covariant derivatives (i.e., they are non-local), cf. \cite{DGI82,BF90}. The
question that we want to address now is whether the approximation in Eq.\
(\ref{ren-2}) is sufficient, or higher spin operators can play an important
r\^ole.

Let us assume for a moment that the scale separation in Eq.\ (\ref{ren-1}) is done with a momentum
cutoff, that is contributions of high loop momenta $|k| >\mu $ are included in the coefficient functions
$c_k(u,\mu^2 x^2)$ and contributions of low momenta $|k| < \mu$ in the matrix elements.
The coefficient functions can be expanded in the Laurent series in $\mu^2 x^2 \ll 1$ and
apart from the usual logarithmic terms $\sim \ln^k x^2 \mu^2$ will also contain the power
corrections, schematically
\beq{ren-3}
  \left(1+\sum_{k=1}^\infty c_k(u,\mu^2x^2)\alpha_s^k(\mu^2)\right) \to
  \left(1+\sum_{k=1}^\infty c_k(u,\ln \mu^2x^2)\alpha_s^k(\mu^2)\right)
 - d(u) \mu^2 x^2 + {\cal O}(\mu^4 x^4)\,,
\eeq
where the terms $\sim \mu^2 x^2$ are also accompanied by a series in $\ln \mu^2x^2$ that we do not
show for brevity. As well known, the logarithmic scale dependence of the coefficients
is compensated by the scale dependence of matrix elements, in our case the leading twist pion
distribution amplitude, so that the product of the leading twist coefficient function and
the relevant operator matrix element does not depend on the factorization scale
{\it to logarithmic accuracy}\,. The crucial observation is that independence of the
result on the factorization scale {\it to power accuracy} involves cancellations
between contributions of different twists. Technically, this cancellation proceeds as follows.
As mentioned above,  $g_1(u,\mu^2)$ is determined by reduced matrix elements of twist-four
operators that have not only the usual logarithmic, but also the quadratic UV divergence.
If, as we have assumed,  the matrix elements are calculated with an explicit UV cutoff,
the results for the matrix elements will be, schematically
\beq{ren-4}
     g_1(u,\mu^2x^2) \to \mu^2 \,d(u)\otimes \phi_\pi(u) + \Lambda_{\rm QCD}^2\, g_1(u,\ln \mu^2x^2)\,,
\eeq
with {\it the same} function $d(u)$ as in Eq.\ (\ref{ren-3}). The result is that
power-suppressed infrared contributions to the coefficient functions cancel
against ultraviolet quadratically divergent contributions to the matrix elements
of higher twist operators. If the factorization is done using dimensional
regularization (and MS-subtraction) instead of the momentum cutoff, then
power-like scale dependence does not arise, but the perturbative series becomes
factorially divergent in high orders. Divergences of the perturbative series in
high orders are called renormalons and they can be due to both the infrared and
the ultraviolet regions. In this language, one observes that infrared renormalons
in the coefficient functions cancel against the ultraviolet renormalons in the
matrix elements of higher twist operators.

Without going into details, we note that the function $d(u)$ can be calculated in perturbation
theory and
$\mu^2 d(u)\otimes \phi_\pi(u)$ can be taken as an estimate (``renormalon model'')
for the twist-four distribution
amplitude. 
The leading-order result reads \cite{Andersen00,GG03}
\bea{ren-5}
   g_1^{\rm ren}(u) &=& \mu^2 \Bigg\{\bar u\! \int_{\bar u}^1\!\frac{dv}{v^2}
       \left[ 1+\left(\frac{v}{\bar u}-1\right) \ln \left(1-\frac{\bar u}{v}\right)\right]\,
                                          \phi_\pi (\bar v)
\nonumber\\&&{}
       +  u \!\int_{ u}^1\!\frac{dv}{v^2}
       \left[ 1+\left(\frac{v}{ u}-1\right) \ln \left(1-\frac{ u}{v}\right)\right]\,
                                          \phi_\pi ( v)\Bigg\},
\eea
where the scale $\mu^2$ has to be of the order of $\Lambda_{\rm QCD}^2$ but otherwise is arbitrary.
This expression corresponds to taking into account contributions of all conformal operators with
arbitrary spin, assuming ``ultraviolet dominance'' \cite{BBr01,Braun97}
of the corresponding matrix elements and neglecting the (logarithmic) effects of the
anomalous dimensions altogether. Because the anomalous dimensions suppress contributions
with higher conformal spins, the renormalon model is likely to overestimate them,
and thus can be viewed  as an upper bound for higher spin contributions.

\begin{figure}[tb]
\begin{center}
\begin{minipage}[t]{8 cm}
\epsfig{file=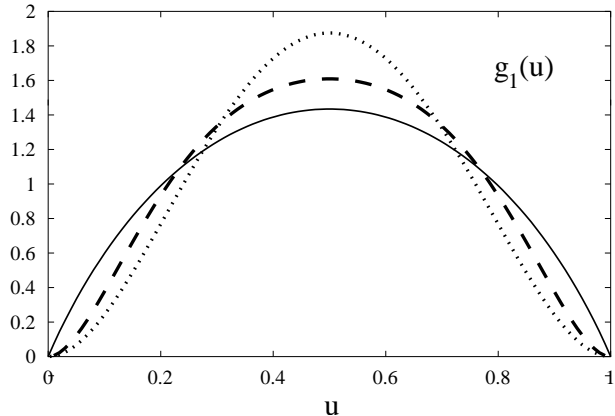,scale=0.7}
\end{minipage}
\caption{The twist-four two-particle pion distribution amplitude $g_1(u)$ in the renormalon
model (\ref{ren-5}) (solid line) compared with contributions of the first two orders
in the conformal partial wave expansion (\ref{ren-2}) (dashed line) with the coefficients
estimated using QCD sum rules \cite{BF90}. The asymptotic distribution amplitude
corresponding to the first term in Eq.\ (\ref{ren-2}) is shown by the dotted curve.
The normalization is adjusted so that $\int_0^1 du\, g_1(u) =1$.
\label{fig:g1}}
\end{center}
\end{figure}

The comparison of the renormalon model (\ref{ren-5}) with the contributions of the lowest two
conformal partial waves (\ref{ren-2}) is shown in Fig.~\ref{fig:g1}, where we used the
asymptotic pion distribution amplitude $\phi_\pi(u) = 6u\bar u$ for the evaluation of
Eq.\ (\ref{ren-5}) and adjusted the overall normalization.
%
%
%
We see that the global shape of $g_1(u)$ in both models is very similar, and the only
essential difference concerns the behavior at the end points: In the renormalon model
$g_1(u) \sim u(1-u)$ for $u\to0,1$, whereas each term in the conformal
expansion has the $\sim u^2(1-u)^2$ behavior. The difference means that the conformal
expansion is not converging uniformly at the end points, in the renormalon model
approximation. As mentioned above, contributions of higher spin
operators alias the end-point behavior of the renormalon ansatz will be tamed by taking
into account the  anomalous dimensions of higher twist operators, so that what we see
in Fig.~\ref{fig:g1} has to be considered as an upper bound. Phenomenological consequences
of this different behavior still have to be investigated, but in any case the renormalon
approach provides a viable alternative to the uses of conformal expansion truncated
at the first few orders.

\subsubsection{\it Wandzura-Wilczek Contributions and Meson Mass Corrections}

Last but not least, we want to consider meson mass corrections and discuss how they
can be included in the conformal expansion.
To this end we will consider a subset of the distribution amplitudes of the
$\rho$-meson, defined by the matrix elements of nonlocal operators with even
number of  $\gamma$-matrices in between the quark and the antiquark operators.
This is a close analogue to the example of pion twist-three distributions considered
above, and the difference is, first, that $\rho$ meson has spin one,
and, second, it has a mass which is non-negligible in comparison to the
mass scale of quark-gluon correlations. Following \cite{BBKT98} we define
\bea{ww-1}
\langle 0| \bar d(0) \sigma_{\mu\nu} u(\alpha) |\rho^+(P,\lambda)\rangle &=&
i f_\rho^T\Bigg[
(e^{(\lambda)}_{\perp\mu}p_\nu - e^{(\lambda)}_{\perp\nu}p_\mu)
\int_0^1 du\,{\rm e}^{-iu \alpha p_+} \phi_\perp(u,\mu)
\nonumber\\
&&{}+(p_\mu n_\nu-p_\nu n_\mu)
\frac{e^{(\lambda)}_+}{p_+^2} m_\rho^2
\int_0^1 du\,{\rm e}^{-iu \alpha p_+} h^{(t)}_\parallel(u,\mu)
\nonumber\\
&&{}+\frac{1}{2p_+}
(e^{(\lambda)}_{\perp\mu}n_\nu - e^{(\lambda)}_{\perp\nu}n_\mu) m_\rho^2
\int_0^1 du\,{\rm e}^{-iu \alpha p_+} h_3(u,\mu)\Bigg]\,,
\nonumber\\
\langle 0| \bar d(0)  u(\alpha) |\rho^+(P,\lambda)\rangle &=&
\frac{i\alpha}{2 } f_\rho^T m_\rho^2 e^{(\lambda)}_+
\int_0^1 du\,{\rm e}^{-iu \alpha p_+} h^{(s)}_\parallel(u,\mu)\,,
\eea
where $P^2=m_\rho^2$,
   $   p_\mu = P_\mu - n_\mu {m_\rho^2}/({2p_+})$, $p^2=0$,
and the polarization vector $e^{(\lambda)}_\mu$ of the $\rho$-meson is decomposed
into projections onto the two light-like vectors $n_\mu$, $\bar n_\mu = p_\mu/p_+ $ and
the orthogonal plane:
\beq{ww-3}
  e^{(\lambda)}_\mu = e^{(\lambda)}_+ \bar n_\mu + e^{(\lambda)}_-   n_\mu
  + e^{(\lambda)}_{\perp\mu}\,.
\eeq
The function $\phi_\perp(u)$ is the leading twist distribution amplitude of the transversely
polarized $\rho$-meson, $ h^{(t)}_\parallel(u) $ and $ h^{(s)}_\parallel(u) $ are twist-three
distributions for the longitudinally polarized $\rho$-meson, and  $h_3(u)$ is the twist-four
distribution, for transverse polarization. In addition, there exist one three-particle twist-three
distribution for longitudinal polarization, and several three-particle (and four-particle)
twist-four distributions for transversely polarized $\rho$ mesons. Explicit expressions
can be found in \cite{BBKT98,BB99}, we will not need them in what follows.

The distributions $ h^{(t)}_\parallel(u) $ and $ h^{(s)}_\parallel(u) $ are very
similar to the tensor $\phi_\sigma(u)$ and the pseudoscalar $\phi_p(u) $
twist-three distributions  of the pion that we have just constructed. In full
analogy with the latter, $ h^{(t)}_\parallel(u) $ and $ h^{(s)}_\parallel(u) $
contain a ``genuine" twist-three contribution that can be related to the existing
three-particle distribution using equations of motion and following the scheme
described above. Note that these relations are between the two-particle and the
three-particle distributions corresponding to the same (longitudinal)
polarization. A peculiar new feature is, however,  that in addition to this
``genuine" twist-three part,  $h^{(t)}_\parallel(u) $ and $ h^{(s)}_\parallel(u)
$ contain one more  contribution that can be expressed through the distribution
amplitude $\phi_T(u)$ with a lower (leading) twist-two and different (transverse)
polarization. Such contributions have first been found in \cite{WW77} for the
structure function $g_2(x,Q^2)$ of the deep-inelastic lepton-hadron scattering
from transversely polarized nucleons, and are generically referred to as
Wandzura-Wilczek contributions. We can write
\bea{ww-4}
  h^{(t)}_\parallel(u) = h^{(t)WW}_\parallel(u) + h^{(t) qGq}_\parallel(u)\,,
&\qquad &
h^{(s)}_\parallel(u) = h^{(s)WW}_\parallel(u) + h^{(s) qGq}_\parallel(u)\,,
\eea
 and will concentrate on the Wandzura-Wilczek parts; the construction of  ``genuine"
twist-three contributions is fully analogous  to the pion case.

In order to derive the Wandzura-Wilczek contributions one starts with the same
operator identities (\ref{ht-4}) (omitting the $\gamma_5$-matrix), takes the
appropriate matrix elements and neglects  contributions of $\bar q q G$
operators. The relations that arise in this way have a unique solution
\cite{BBKT98}
\bea{ww-5}
 h^{(t)WW}_\parallel(u) = (2u-1)
 \left(\int_0^u\!dv\, \frac{\phi_\perp(v)}{\bar v}
       - \int_u^1\!dv\, \frac{\phi_\perp(v)}{ v}\right),
\nonumber\\
h^{(s)WW}_\parallel(u) =  2
 \left(\bar u \int_0^u\!dv\, \frac{\phi_\perp(v)}{\bar v}
        + u \int_u^1\!dv\, \frac{\phi_\perp(v)}{ v}\right).
\eea
These are the analogues of the Wandzura-Wilczek contributions to the nucleon structure
functions $g_T(x,Q^2)$ and $h_L(x,Q^2)$ \cite{JJ91}.

Our goal is to understand the structure of these contributions in the context of
the conformal expansion. Similar as for the pion, we need to go over to auxiliary
amplitudes with fixed spin projections:
\bea{ww-6}
  \langle 0| \bar d(0) \gamma_+\gamma_-  u(\alpha) |\rho^+(P,\lambda)\rangle &=&
f_\rho^T m_\rho^2 \frac{e^{(\lambda)}_+}{p_+}
 \int_0^1 du\,{\rm e}^{-iu \alpha p_+} h_{\uparrow\downarrow}(u)\,,
\nonumber\\
\langle 0| \bar d(0) \gamma_-\gamma_+  u(\alpha) |\rho^+(P,\lambda)\rangle &=&
f_\rho^T m_\rho^2 \frac{e^{(\lambda)}_+}{p_+}
 \int_0^1 du\,{\rm e}^{-iu \alpha p_+} h_{\downarrow\uparrow}(u)\,,
\eea
which are related to $ h^{(t)}_\parallel(u)$ and $ h^{(s)}_\parallel(u)$ by
\bea{ww-7}
h_{\uparrow\downarrow}(u)= \phantom{-}h^{(t)}_\parallel(u)
     + \frac12 \frac{d h^{(s)}_\parallel(u)}{du},
&\qquad&
h_{\downarrow\uparrow}(u)= -h^{(t)}_\parallel(u)
     + \frac12 \frac{d h^{(s)}_\parallel(u)}{du}.
\eea
The conformal expansion of $ h_{\uparrow\downarrow}(u)$ and
$h_{\downarrow\uparrow}(u)$ is straightforward and is given by
\bea{ww-8}
 h_{\uparrow\downarrow}(u) = 2\bar u \sum_{n=0}^\infty
                           h_{\uparrow\downarrow}^{n+3/2} P^{(1,0)}_n(\xi)\,,
&\qquad &
h_{\downarrow\uparrow}(u) = 2\bar u \sum_{n=0}^\infty
                           h_{\downarrow\uparrow}^{n+3/2} P^{(1,0)}_n(\xi)\,.
\eea
On the other hand
\beq{ww-9}
  \phi_\perp (u) = 6 u\bar u \sum_{n=0}^\infty \phi_\perp^{n+2} C^{3/2}_n(\xi)\,.
\eeq
In the both cases the superscripts ($j= n+3/2$ and $j=n+2$)  refer to the
conformal spin. Substituting these expansions in Eq.~(\ref{ww-5}),  after some
algebra one obtains \cite{BBKT98}
\bea{ww-10}
 H_{WW}^{n+3/2} = \frac{3(n+1)}{2n+3} \phi_\perp^{n+2}\,,
&\qquad&
h_{WW}^{n+3/2} = \frac{3(n+1)}{2n+1} \phi_\perp^{n+1}\,,
\qquad (n=0,2,4,\ldots)\,,
\nonumber\\
H_{WW}^{n+3/2} = -\frac{3(n+1)}{2n+1} \phi_\perp^{n+1}\,,
&\qquad&
h_{WW}^{n+3/2} = \frac{3(n+1)}{2n+3} \phi_\perp^{n+2}\,,
\qquad (n=1,3,5,\ldots)\,,
\eea
where $H_{WW}^{n+3/2}$ and $h_{WW}^{n+3/2}$ are the Wandzura-Wilczek parts of
the G-parity even and odd combinations of the coefficients:
\beq{ww-11}
 H^{n+3/2} = \frac12 \left[h_{\uparrow\downarrow}^{n+3/2}
                                  -(-1)^n h_{\downarrow\uparrow}^{n+3/2}\right]\,,
\qquad
h^{n+3/2} = \frac12 \left[h_{\uparrow\downarrow}^{n+3/2}
                                  +(-1)^n h_{\downarrow\uparrow}^{n+3/2}\right]\,.
\eeq
We see that the conformal spins on the l.h.s.\ and the r.h.s.\ of Eq.~(\ref{ww-10}) do not
coincide which calls for an explanation.

The basic idea is the following \cite{BBKT98}. The Wandzura-Wilczek relations (\ref{ww-5})
are in fact consequence of the Lorentz symmetry. Working out such relations essentially
corresponds to reexpressing matrix elements of conformal operators between vacuum and
the longitudinally polarized $\rho$-meson in terms of the similar matrix elements
involving the transversely polarized meson, or vice versa. In our context it is important that
the transversely polarized state is obtained (in the $\rho$-meson rest frame) from
the longitudinally polarized state by the spin rotation which does not commute
with the generators of the collinear conformal group. This rotation gives rise to
the shift in conformal spin and exactly explains the mismatch appearing in Eq.\ (\ref{ww-10}).
The moral is that the conformal symmetry is realized for the Wandzura-Wilczek contributions
as well, but to see this one has to supplement the conformal classification of operators
by conformal transformation properties of the meson states.

It is convenient to work in the helicity basis: $\rho$-mesons with $\lambda=\pm
1$ correspond to the two states with transverse polarization, $\lambda=0$ stands
for the longitudinal polarization. Introducing conformal eigenstates $\JJ^2
|j,k\rangle = j(j-1) |j,k\rangle$, $\JJ_0 |j,k\rangle =
-(j+k)|j,k\rangle$, cf. (\ref{coh-5}), we
can separate different terms in Eq.\ (\ref{ww-9}) as
\beq{ww-111}
   \phi_\perp^{n+2} \propto \langle n+2,0|\rho(P,\lambda=\pm 1)\rangle\,,
\eeq
where the proportionality factor is irrelevant for what follows. In the $\rho$-meson rest
frame the $\lambda = +1$ state is related to the $\lambda=0$ state by the spin rotation
\beq{ww-12}
    |\rho(P=m_\rho,\lambda= + 1)\rangle \propto (M_{23}+i M_{31})\, |\rho(P=m_\rho,\lambda= 0)\rangle,
\eeq
where $M_{23}+i M_{31}$ is the appropriate step-up operator of ordinary angular
momentum. A fast-moving $\rho$-meson $|\rho(P,\lambda=\pm 0)\rangle$ is then
obtained by the Lorentz boost in the appropriate direction $P_\mu
=(P_0,0,0,P_3)$:
\beq{ww-13}
  |\rho(P,\lambda)\rangle = {\cal U}(\omega)\, |\rho(P=m_\rho,\lambda)\rangle
\eeq
with ${\cal U}(\omega) =  {\rm e}^{-i \omega M^{03}}$, $ \mbox{\rm tanh}(\omega)
= P_3/P_0$. We can thus write
\beq{ww-14}
   \phi_\perp^{n+2} \propto \langle \Psi |\rho(P,\lambda= 0)\rangle\,,
\eeq
where
\beq{ww-15}
    |\Psi\rangle = {\cal U}(\omega) \,(M_{23}+i M_{31}) \,{\cal U}^{-1}(\omega) |n+2,0\rangle\,.
\eeq
Working out the commutation relations one obtains \cite{BBKT98}
\beq{ww-16}
  |\Psi\rangle  = c_1 |n+3/2,0\rangle
                + c_2 |n+3/2,1\rangle
                + c_3 |n+5/2,0\rangle,
\eeq
where $c_k$ are certain numerical coefficients. Eq.~(\ref{ww-16}) shows that the contributions
of $\phi_\perp^{n+2}$ to the distribution amplitude of the longitudinal $\rho$-meson
correspond to the conformal spins $j=n+3/2$ and $j=n+5/2$, which explains the pattern
appearing in Eq.\ (\ref{ww-10}). This derivation is valid also for chiral-even distribution
amplitudes, see \cite{BBKT98}.

The structure of meson mass corrections to twist-four distribution amplitudes is much more
complicated \cite{BB99} as it turns out that they have several sources.
First, mass corrections are generated
by the contributions of leading twist operators in which case the conditions of symmetry
and zero traces imply a certain Lorentz structure that multiplies the relevant
reduced matrix elements. Such corrections are the direct analogue of the Nachtmann's correction
to deep inelastic scattering. This result is not complete, however,
 because in exclusive processes one has to take into account higher-twist operators containing
total derivatives, and vacuum-to-meson matrix elements of such operators reduce
to powers of the meson mass times reduced matrix elements of leading twist
operators. The operators containing the square of the total derivative times the
leading twist operator are the simplest ones and can be taken into account in a
systematic way, see \cite{BB99}. The major complication arises because of
operators that are given by (or can be reduced to) the {\it divergence}\, of the
leading twist conformal operators considered in Sec.~3.1. As well known
\cite{FGPG72}, conformal operators are divergenless in the conformal-invariant
theory, so that in QCD their divergence must be proportional to the strong
coupling and involve the gluon field. Indeed, one obtains that this divergence
can be expressed in terms of contributions of quark-antiquark-gluon operators
\cite{BB89,BB99,BBK03}, schematically
\beq{ww-18}
    \partial_{\alpha_1} {\mathbb O}_n^{\alpha_1,\alpha_2,\ldots,\alpha_{n}} =
   \sum \,\bar q gG q~\mbox{\rm operators}.
\eeq
Sandwiching Eq.\ (\ref{ww-18}) between vacuum and the meson state, one
obtains nontrivial relations between reduced matrix elements of
quark-antiquark-gluon operators on the r.h.s., and the meson mass
squared times the leading twist matrix elements on the l.h.s. This
implies that the meson mass effects and the effects of quark-gluon
correlations in mesons cannot be separated from one another in a
meaningful way: The approximation of taking into account the meson mass
corrections and neglecting contributions of quark-gluon operators
appears to be not self-consistent, see \cite{BB99} for a more detailed
discussion. The similar problem is present for photon distribution
amplitudes \cite{BB89,BBK03} in which case one would like to separate
contributions of higher-twist quark-photon operators (contact terms)
from the quark-gluon ones, and also for processes involving off-forward
(generalized) parton distributions if one tries to separate higher-twist
corrections that are proportional to either the nucleon mass squared, or
the four-momentum transfer $t=(P_2-P_1)^2$ \cite{BelMue01}. Here
$P_1^\mu$ and $P_2^\mu$ are the momenta of the initial and the final
state nucleon, respectively.

\section{Hamiltonian Approach to QCD Evolution Equations}

\subsection{\it Conformal Symmetry and Hermiticity}

We have seen in Sect.~3.1 that the kernel (\ref{BLkernel}) of the ER-BL
equation for the scale dependence of pion distribution amplitudes can be
written in terms of the anomalous dimension (\ref{BLham}) and the
characteristic polynomials of local conformal operators (Gegenbauer
polynomials) which form a complete basis and are orthogonal with the
weight function $u(1-u)$. We have also demonstrated in Sect.~2.4.2 that
the orthogonality relation follows from conformal symmetry and,
in particular, is a direct consequence of special conformal invariance.

In this section we want to elaborate on the possibility to use such
representations, first suggested in \cite{BukFroKurLip85}, and rewrite
Eq.~(\ref{BLham}) in an abstract operator form such that the evolution equation
(\ref{ERBL}) becomes a spectral problem for a one-dimensional Hamiltonian which
is manifestly conformally invariant: it commutes with the $SL(2,{\mathbb R})$
generators  and is self-adjoint (hermitian) with respect to the conformal scalar
product defined in (\ref{co-13}) and (\ref{sc-prod}). This reformulation proves
to be very convenient for the analysis of more complicated, three-particle
evolution equations which we consider at the next step.

\begin{figure}[t]
\centerline{{\epsfysize3cm \epsfbox{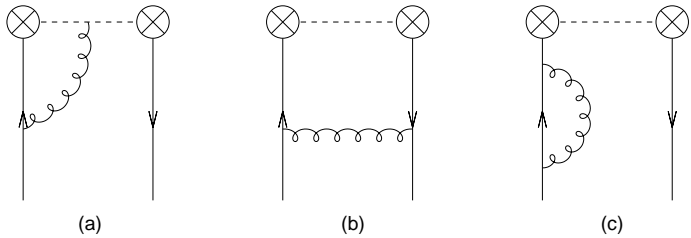}}}
\caption[]{Examples of a `vertex' correction (a), 'exchange'
diagram (b) and self-energy insertion (c) contributing to the renormalization of
quark-antiquark operators in Feynman gauge. Path-ordered gauge factors are shown
by the dashed lines. The set of all diagrams includes possible permutations.}
\label{diag-2}
\end{figure}

To begin with, let us rederive the ER-BL equation in a somewhat different language,
as an evolution equation for the light-ray operator built from quark and antiquark fields
\bea{O-2}
{\cal Q}(\alpha_1,\alpha_2)= \bar \psi(\alpha_1)
\gamma_+ [\alpha_1,\alpha_2]\,\psi(\alpha_2)\,,
\eea
cf. Eq.~(\ref{co-14}). This operator serves as a generating function for
local quark-antiquark twist-two operators. In order to
avoid complications due to mixing with gluon operators, we will tacitly assume that the
quark fields have different flavor. The renormalization group equation for the
operator (\ref{O-2}) can be written as \cite{BB89}
\bea{H-EQ}
\left(\mu\frac{\partial}{\partial\mu}+\beta(g)\frac{\partial}{\partial g}\right)
{\cal Q}(\alpha_1,\alpha_2)=-\frac{\alpha_s C_F}{4\pi}\left[\mathbb{H} \cdot
{\cal Q}\right](\alpha_1,\alpha_2)\,,
\eea
where $\mathbb{H}$ is some integral operator acting on the light-cone coordinates
of quark fields. Expanding the both sides of this equation in powers of
$\alpha_1-\alpha_2$ one can reproduce the evolution equations for local twist-two
operators. One of the reasons why we prefer to work with the nonlocal operator
(\ref{O-2}) is that the conformal symmetry becomes manifest in the coordinate
space. To the one-loop accuracy the evolution kernel $\mathbb{H}$ is given by the
Feynman diagrams shown in Fig.~\ref{diag-2}
%
%
 and can be decomposed in three pieces:
\bea{H-ker}
\mathbb{H}= 2 \mathcal{H}_{\rm v}^{(12)} -2\mathcal{H}_{\rm e}^{(12)}+1\,.
\eea
The first contribution, denoted by  $\mathcal{H}_{\rm v}^{(12)}$, describes a
gluon exchange between the quark (antiquark) and the Wilson line, as
displayed in
Fig.~\ref{diag-2}a.
It is given in the Feynman gauge by
\bea{H-v}
[\mathcal{H}_{\rm v}^{(12)}\cdot
{\cal Q}](\alpha_1,\alpha_2)=-\int_0^1\frac{du}{u}(1-u)
\left\{ {\cal Q}(\alpha_{12}^u,\alpha_2)+{\cal Q}(\alpha_1,\alpha_{21}^u)-2{\cal Q}(\alpha_1,\alpha_2)
 \right\}\,,
\eea
where $\alpha_{jk}^u\equiv \alpha_j(1-u)+\alpha_ku$. The second term,
$\mathcal{H}_{\rm e}^{(12)}$, corresponds to a gluon exchange between the
quark and the antiquark, see Fig.~\ref{diag-2}b.
It admits the following representation
\bea{H-e}
[\mathcal{H}_{\rm e}^{(12)}\cdot
{\cal Q}](\alpha_1,\alpha_2)=\int_0^1[du]\,{\cal Q}(\alpha_{12}^{u_1},\alpha_{21}^{u_2})\,,
\eea
where the integration measure is defined as $[du]=du_1 du_2
du_3\,\delta(u_1+u_2+u_3-1)$. Finally, the extra ``+1'' takes into
account the self-energy insertions, Fig.~\ref{diag-2}c.
Notice that the action of the evolution kernel $\mathbb{H}$
on the nonlocal quark-antiquark operator ${\cal Q}(\alpha_1,\alpha_2)$
reduces to a displacement of the quarks along the light-cone in the direction of each
other.

Expanding the nonlocal operator ${\cal Q}(\alpha_1,\alpha_2)$ at short distances
over the local composite operators
\bea{O-exp}
{\cal Q}(\alpha_1,\alpha_2)=\sum_{N\ge 0}
\sum_{k_1+k_2=N}\frac{ \alpha_1 ^{k_1}}{k_1!}
\frac{\alpha_2^{k_2}}{k_2!} {\cal Q}_{k_1,k_2}(0)\,,
&\qquad&
{\cal Q}_{k_1,k_2}(0) = \bar \psi(0)
\stackrel{\leftarrow}{D}_+{\!\!\!\!}^{k_1}\gamma_+
\stackrel{\rightarrow}{D}_+{\!\!\!\!}^{k_2}\,\psi(0)\,,
\nonumber
\eea
one finds that the evolution kernel (\ref{H-ker}) preserves the total number of
derivatives $N$ and, therefore, the evolution equation (\ref{H-EQ}) takes the
matrix form. This is a consequence of Poincar\'e invariance.
For given $N$ the local operators ${\cal Q}_{k_1,k_2}(0)$ mix
under renormalization and their mixing matrix has the size $(N+1)$. As
we have seen in Sect.~3.1, the form of this matrix is severely
constrained by the conformal symmetry so that only the diagonal entries
(anomalous dimensions) remain undetermined.

{} For the time being, we want to stay with the description in terms of
nonlocal operators, however.
According to Eq.\ (\ref{SL2R2}), the action
of the collinear conformal group on the two-particle operator
${\cal Q}(\alpha_1,\alpha_2)$ is given by the differential operators
\bea{2-particle}
\JJ_a {\cal Q}(\alpha_1,\alpha_2)= (\JJ_{1,a}+\JJ_{2,a}){\cal Q}(\alpha_1,\alpha_2)\,,
\eea
where $a=+,-,0$ and $\JJ_{k,a}$ is the differential operator (\ref{SL2R2}) acting
on the light-cone coordinate $\alpha_k$ of the $k$th quark,  $k=1,2$.
 Using Eqs.~(\ref{H-v}) and (\ref{H-e}) it is straightforward to verify the
following relations:
\bea{103}
 [ \mathbb{H} \cdot \JJ_a \,{\cal Q}](\alpha_1,\alpha_2)= \JJ_a \,[\mathbb{H} \cdot {\cal Q}](\alpha_1,\alpha_2)\,.
\eea
Hence the evolution kernel $\mathbb{H}$, acting on the Hilbert space of nonlocal
light-cone operators ${\cal Q}(\alpha_1,\alpha_2)$ that admit the short-distance
expansion, commutes with the generators $\JJ_+$, $\JJ_-$ and $\JJ_0$
(\ref{O-exp})
\bea{H-comm-J}
 [ \mathbb{H}, \JJ_+ ] = [ \mathbb{H}, \JJ_- ] = [ \mathbb{H}, \JJ_0 ] = 0\,.
\eea
This is nothing else but the statement of invariance
under the projective transformations on
the light-cone. Eq.~(\ref{H-comm-J}) implies that  $\mathbb{H}$
must be a certain function  of the two-particle Casimir operator
\bea{Casimir}
{\JJ}^2_{12}=\JJ_0^2+\frac12\left(\JJ_+\JJ_- +
\JJ_-\JJ_+\right)=-\partial_{\alpha_1}\partial_{\alpha_2}
\alpha_{12}^2 \,,
\eea
where in the last equality we used that the conformal spins of the quark and
the antiquark are
equal to one for the leading twist, $j_q=j_{\bar q}=1$, cf. (\ref{co-18}).

To find the explicit form of the dependence of $\mathbb{H}$ on ${\JJ}_{12}^2$, it is
enough to compare their action on the homogeneous translation invariant
polynomials $\alpha_{12}^n\equiv (\alpha_1-\alpha_2)^n$ with $n\ge 0$. These
polynomials diagonalize the operator ${\JJ}_{12}^2$ and are annihilated by $\JJ_-$
\bea{J-eig}
{\JJ}^2_{12} \alpha_{12}^n = (n+2)(n+1) \alpha_{12}^n\,,\qquad \JJ_-\alpha_{12}^n =
0\,.
\eea
As a consequence, they are the highest weights of the irreducible representation
of the conformal group, cf. Eqs.\ (\ref{co-9}) and (\ref{co-10}).
Substituting ${\cal Q}(\alpha_1,\alpha_2) \to
\alpha_{12}^n$, we obtain from Eqs.~(\ref{H-v}), (\ref{H-e}) and (\ref{Casimir})
\bea{H-act}
\mathcal{H}_{\rm v}^{(12)}\, \alpha_{12}^n = 2\left[\Psi(n+2)-\Psi(2)\right]\, \alpha_{12}^n
\,,\qquad
\mathcal{H}_{\rm e}^{(12)}\, \alpha_{12}^n = 1/[(n+2)(n+1)]\, \alpha_{12}^n\,,
\eea
where $\Psi(x)=d\ln\Gamma(x)/dx$ is the Euler $\psi-$function. This can be cast
in an operator form introducing the two-particle spin operator $J_{12}$ as a
formal solution to the operator relation
\bea{104}
{\JJ}^2_{12}=J_{12} (J_{12}-1)\,.
\eea
{}From Eq.\ (\ref{J-eig}) it follows that the eigenvalues of $J_{12}$ are
equal to $j_{12}=n+2$ and specify the total conformal spin
of the quark-antiquark state $j_{12} = n+ j_q+j_{\bar q}$, in agreement
with the general rule for the sum of two conformal $SL(2,\mathbb{R})$ spins:
\bea{sum-spins}
[j_1] \otimes [j_2] = \bigoplus_{n\ge 0} [j_1+j_2+n]\,.
\eea
Using the expressions in (\ref{J-eig}) and (\ref{H-act})
we  obtain
\beq{H-op-rel}
\mathcal{H}_{\rm v}^{(12)} =
2\left[\Psi(J_{12})-\Psi(2)\right]\,,\qquad
\mathcal{H}_{\rm e}^{(12)} = 1/[J_{12}(J_{12}-1)]=1/{\JJ}^2_{12} \,
\eeq
and, finally,
\beq{Hfull}
 \mathbb{H} = 4\left[\Psi(J_{12})-\Psi(2)\right] -2/[J_{12}(J_{12}-1)]+1\,.
\eeq
Diagonalization of the renormalization group equation
(\ref{H-EQ}) requires the solution of the  one-dimensional Schr\"odinger equation
\bea{H-EQ2}
 \frac{\alpha_s C_F}{4\pi}{\mathbb H}(J_{12}) \, {\cal O}(\alpha_1,\alpha_2) =
     \gamma\, {\cal O}(\alpha_1,\alpha_2) \,,\qquad
  \gamma =\frac{\alpha_s} {4\pi} \gamma^{(0)}+\ldots
\eea
Since $\mathcal{H}_{\rm v}^{(12)}$ and
$\mathcal{H}_{\rm e}^{(12)}$ are real functions of the two-particle spin
operator $J_{12}$ which is  self-adjoint on the $SL(2,\mathbb{R})$
representation space, the evolution Hamiltonian $\mathbb{H}$ inherits the same
property. The conformal invariance manifests itself through the fact that the
Hamiltonian commutes with the generators of the conformal group and, as a
consequence, in our case it becomes diagonal for the states carrying a
definite conformal spin.

Solutions to (\ref{H-EQ2}) we already know. They are given by
 homogeneous polynomials $(\alpha_1-\alpha_2)^n$,
see (\ref{H-act}), which are nothing else but the characteristic polynomials
of the conformal operators $\mathbb{O}^{1,1}_n$ in the adjoint representation,
cf. Eq.~(\ref{co-10})
 \bea{H-EQ22}
     {\mathbb H}(J_{12}) \, \widetilde{\mathbb P}_n (\alpha_1,\alpha_2) =
     (\gamma^{(0)}_n/C_F) \, \widetilde{\mathbb P}_n (\alpha_1,\alpha_2)\,,
\eea
and the anomalous dimension is equal to
$\gamma^{(0)}_n=C_F\, \mathbb{H}( J_{12}\to n+2)$,
in agreement with Eq.~(\ref{an-dim}).

The major advantage of the representation of the ER-BL evolution kernel
in the operator form (\ref{Hfull}) is that it is  covariant, i.e.,
retains its form if going over to a different representation of the conformal group.
{}For example, one may prefer to have an equation for the characteristic
polynomials of conformal operators in the ``direct'' rather than adjoint
representation. This equation can easily be derived
applying ${\mathbb P}_n(\partial_{\alpha_1},
\partial_{\alpha_2})$ to the both sides of the evolution equation (\ref{H-EQ})
and making use of Eqs.~(\ref{cf1}) and (\ref{cf2}). One finds that the
characteristic polynomial satisfies the Schr\"odinger equation
\bea{Sch-eq}
\mathbb{H}(\widetilde J_{12})\, {\mathbb P}_n(\xxi_1,\xxi_2) =
(\gamma^{(0)}_n/C_F)\, {\mathbb P}_n(\xxi_1,\xxi_2)\, ,
\eea
where $\mathbb{H}$ is given by the same expression as before, Eq.~(\ref{Hfull}),
and the only difference is that the two-particle spin
operator has to be taken in the adjoint representation, Eq.~(\ref{adjoint})
\bea{tilde-J2}
\widetilde J_{12}(\widetilde J_{12}-1)=\widetilde \JJ_0^2+\frac12(\widetilde
\JJ_+\widetilde \JJ_-+\widetilde \JJ_-\widetilde
\JJ_+)=-(\partial_{\xxi_1}-\partial_{\xxi_2})^2 \xxi_1 \xxi_2\,.
\eea
As before $\widetilde \JJ_+=\widetilde \JJ_{1,+}+\widetilde \JJ_{2,+}$, etc. One
can verify by explicit calculation that $\widetilde J_{12}$ and, as a
consequence, the Hamiltonian $\mathbb{H}(\widetilde J_{12})$ are self-adjoint
operators on the Hilbert space endowed with the scalar product (\ref{sc-prod})
with $j_1=j_2=1$. As a consequence, the eigenfunctions ${\mathbb
P}_n(\xxi_1,\xxi_2)$ form an orthonormal basis and the corresponding eigenvalues
$\gamma^{(0)}_n$ take real values. Notice that the form of  ${\mathbb
P}_n(\xxi_1,\xxi_2)$ is uniquely fixed by the conformal invariance
\bea{J-dual}
\JJ^2_{12}\, {\mathbb P}_n(\xxi_1,\xxi_2)=(n+2)(n+1){\mathbb P}_n(\xxi_1,\xxi_2)\,,\qquad
\widetilde \JJ_-\, {\mathbb
P}_n(\xxi_1,\xxi_2)=0\,,
\eea
cf.~Eq.\ (\ref{J-eig}), where the second, highest weight condition ensures that
the  polynomials are irreducible with respect to multiplication by a power of
$(\xxi_1+\xxi_2)$. The solutions to Eq.\ (\ref{J-dual}) are given in Eq.\
(\ref{co-11}).

Using the completeness condition $\sum |n\rangle \langle n| =1$ with
$\langle\xxi_1\xxi_2|n\rangle \sim {\mathbb P}_n(\xxi_1,\xxi_2) $,
 one can expand the Hamiltonian over its eigenstates. This leads to the
ER-BL kernel (\ref{BLham}):
\bea{H-dec}
 \mathbb{H}(\xxi_1',\xxi_2';\xxi_1,\xxi_2)= \xxi_1\xxi_2\sum_{n= 0}^\infty
{\mathbb P}_n(\xxi_1',\xxi_2') \,(\gamma^{(0)}_n/C_F)\, {\mathbb
P}_n(\xxi_1,\xxi_2)\,.
\eea

\subsection{\it Baryon Distribution Amplitudes}

The Hamiltonian formalism
is very powerful for the analysis of three-parton systems. We will concentrate
on two examples -- the nucleon distribution amplitude, $\phi_N(\xxi_i)$, and the
distribution amplitude of the $\Delta-$resonance state with helicity$-3/2$,
$\phi_\Delta^{3/2}(\xxi_i)$, both are leading twist.
 They are defined \cite{earlybaryon,CZreport} through the
matrix elements of nonlocal three-quark operators of definite chirality
$(q^{\uparrow (\downarrow)}=\frac12 \left (1\pm \gamma_5\right) q)$:
\begin{eqnarray}
\lefteqn{
\langle 0|\epsilon^{ijk}(u_i^\uparrow(\alpha_1)\,C\!\!\not\!nu_j^\downarrow(\alpha_2))
       \not\!nd_k^\uparrow(\alpha_3)|P(p)\rangle =
}
\nonumber\\&&{}\hspace*{2.5cm}=
    - \frac{f_N}{2} 
p_+ \not\!n N^\uparrow(p) \int \![d\xxi]\,
{\rm e}^{-i p_+ \sum_{i=1}^3\alpha_i\xxi_i}
\phi_N(\xxi_i)
\label{Nucleon}
\end{eqnarray}
and
\begin{eqnarray}
\lefteqn{\hspace*{-1cm}
 \langle 0|\epsilon^{ijk}
(u_i^\uparrow(\alpha_1)\,C\sigma_{\mu+} u_j^\uparrow(\alpha_2))
       (\bar\Delta^\mu\!\!\not\!n)\,u_k^\uparrow(\alpha_3)
|\Delta(p)\rangle =
}
\nonumber\\&&{}\hspace*{2.0cm}=
   {i} f_\Delta^{3/2} 
p_+ \bar\Delta^\uparrow_\nu \!\not\!n \Delta^{\uparrow,\nu}\!\!
\int \![d\xxi]\,
{\rm e}^{-i p_+ \sum_{i=1}^3\alpha_i\xxi_i}\phi_\Delta^{3/2}(\xxi_i)\, .
\label{repr3/2}
\end{eqnarray}
Here $i,j,k$ are color indices, $C$ is the charge conjugation matrix,
$|P(p)\rangle $ and $|\Delta(p)\rangle$ are the proton and $\Delta-$resonance
state described by the spin$-1/2$ spinor $N(p)$ and spin$-3/2$ Rarita-Schwinger
field $\Delta^\mu_\alpha(p)$, respectively.

The scale dependence of $\phi_\Delta^{3/2}(\xxi_i,\mu^2)$ and
$\phi_N(\xxi_i,\mu^2)$  is governed by the renormalization group equations for
the three-quark operators with three spinor indices~\cite{BDKM},
\bea{B3/2}
B^{3/2}_{abc}(\alpha_1,\alpha_2,\alpha_3) &=&
\varepsilon^{ijk}(\!\not\!n \psi_{i}^\uparrow(\alpha_1))_a
(\!\not\!n \psi_{j}^\uparrow(\alpha_2))_b (\!\not\!n
\psi_{k}^\uparrow(\alpha_3))_c\,,
\\
B^{1/2}_{abc}(\alpha_1,\alpha_2,\alpha_3) &=&
\varepsilon^{ijk}(\!\not\!n \psi_{i}^\uparrow(\alpha_1))_a
(\!\not\!n \psi_{j}^\downarrow(\alpha_2))_b (\!\not\!n
\psi_{k}^\uparrow(\alpha_3))_c\,,
\label{B1/2}
\eea
respectively. The superscript $1/2$ or $3/2$ denotes helicity. The nonlocal operators
(\ref{B3/2}) and (\ref{B1/2}) have to be interpreted as generating functions for the
renormalized local operators
\bea{Blocal}
B(\alpha_1,\alpha_2,\alpha_3)=\sum_{N\ge 0} \sum_{k_1+k_2+k_3=N}
\frac{\alpha_1^{k_1}}{k_1!}
                \frac{\alpha_2^{k_2}}{k_2!}
                \frac{\alpha_3^{k_3}}{k_3!}
 (nD)^{k_1} \psi(0) (nD)^{k_2} \psi(0) (nD)^{k_3} \psi(0)\,,
\eea
whose reduced matrix elements define moments $\int [d\xxi]\,
\xxi_1^{k_1}
\xxi_2^{k_2}\xxi_3^{k_3}
\phi(\xxi_i)$ of the distribution amplitudes.

All three-quark local operators in (\ref{Blocal}) with the same total number of
derivatives $N=k_1+k_2+k_3$ get mixed by the evolution, so that for given $N$ the
size of the mixing matrix is $N(N+1)/2$. As before, conformal symmetry allows one
to resolve the mixing with operators containing total derivatives. This is done
in the standard way, introducing the basis of three-particle conformal operators
\bea{O3quark}
\mathbb{O}^{1,1,1}_{N,q}(0)=
\mathbb{P}_{N,q}(\partial_{\alpha_1},\partial_{\alpha_2},\partial_{\alpha_3})
B(\alpha_1,\alpha_2,\alpha_3)\bigg|_{\alpha_i=0}\,,
\eea
which are enumerated by the subscript $q$ and satisfy the defining equations
\bea{conf-const}
{\widetilde \JJ}^2\, \mathbb{P}_{N,q}(\xxi_1,\xxi_2,\xxi_3)
         =(N+3)(N+2)\mathbb{P}_{N,q}(\xxi_1,\xxi_2,\xxi_3)\,,
\quad \quad
\widetilde \JJ_-\, \mathbb{P}_{N,q}(\xxi_1,\xxi_2,\xxi_3)=0\,.
\eea
Here $\widetilde \JJ^2$ is the three-particle Casimir operator in the adjoint
representation. In the present case ($j_1=j_2=j_3=1$) it can be written as a sum
of two-particle Casimir operators
 ${\widetilde \JJ}^2={\widetilde \JJ}_{12}^2+{\widetilde \JJ}_{23}^2+
{\widetilde \JJ}_{31}^2$. The highest
weight condition $\widetilde \JJ_-\, \mathbb{P}_{N,q}=0$ can be replaced by
$\widetilde \JJ_0\, \mathbb{P}_{N,q}= (N+3)\mathbb{P}_{N,q}$.

Note that three-particle conformal operators with different conformal
spin $N+3$ are orthogonal with respect to the conformal scalar product
\bea{sc-prod-3}
\langle \mathbb{P}_{N,q}|\mathbb{P}_{N'q'}\rangle \equiv \int_0^1 [d\xxi]\,
\xxi_1^{2j_1-1}\xxi_2^{2j_2-1}
\xxi_3^{2j_3-1}\,
\mathbb{P}_{N'q'}(\xxi_1,\xxi_2,\xxi_3)\mathbb{P}_{N,q}(\xxi_1,\xxi_2,\xxi_3)
\sim \delta_{NN'}\,.
\eea
This orthogonality relation can easily be established generalizing
the analysis in Sect.~2.4.2 to the correlation function of two three-particle
conformal operators.

In difference to the situation that one encounters in the analysis of
leading-twist meson distribution amplitudes, the constraints (\ref{conf-const})
(or, equivalently, the orthogonality relation (\ref{sc-prod-3})) do not fix the
characteristic polynomials of conformal operators uniquely, cf.~Sect.~2.2 and
Eqs.~(\ref{co-90}), (\ref{co-91}). Hence we need an extra index $q$ to
distinguish between different conformal operators with the same conformal spin
$j= N+j_1+j_2+j_3=N+3$ alias  the same number of derivatives $N$, cf.~Eq.\
(\ref{co-91}). Existence of multiple three-particle conformal operators with the
same spin can easily be understood as the multiplicity of the corresponding
irreducible component in the sum of three quark spins
\bea{105}
[j_1]\otimes[j_2]\otimes[j_3]=\bigoplus_{N\ge 0} \, [j_1+j_2+j_3+N]\,.
\eea
Applying the summation rule for two conformal spins in Eq.~(\ref{sum-spins}),
$[j_1]\oplus[j_2] = \sum [j_{12}=j_1+j_2+n]$, we find that the total  spin
$j=N+3$ component has a nontrivial multiplicity $(N+1)$ corresponding to possible
values of the two-particle spin $0\le j_{12}\le N$, in agreement with our
discussion in Sect.~2.2.  As the result, for a given $N$  the conformal
constraints (\ref{conf-const}) generate $(N+1)$ linear independent solutions for
the characteristic polynomials $\mathbb{P}_{N,q}(\xxi_1,\xxi_2,\xxi_3)$. The
corresponding conformal operators have the same spin $N+3$ and, therefore,
nothing prevents them from mixing with each other. It follows that the mixing
matrix is not diagonalized but its size is reduced to $(N+1)$. In other words,
the leading twist baryon distribution amplitudes have to be written as a double
sum~\cite{earlybaryon}
\bea{expansion}
\phi(\xxi_1,\xxi_2,\xxi_3;\mu^2)= 120 \xxi_1 \xxi_2 \xxi_3
\sum_{N=0}^\infty\sum_{q=0}^N \phi_{N,q}(\mu_0)\,\mathbb{P}_{N,q}(\xxi_1,\xxi_2,\xxi_3)
\left (\frac{
\alpha_s(\mu)}{\alpha_s(\mu_0)}
\right )^{\gamma_{N,q}/\beta_0},
\eea
where $\mathbb{P}_{N,q}(\xxi_1,\xxi_2,\xxi_3)$ are certain polynomials that are
constrained, but not fixed uniquely by Eqs.~(\ref{conf-const}) (or, equivalently,
by the orthogonality relation (\ref{sc-prod-3})) and $\gamma_{N,q}$ are the
corresponding anomalous dimensions. The prefactor $120\xxi_1\xxi_2\xxi_2$ in
Eq.~(\ref{expansion}) is the asymptotic distribution amplitude,
cf.~Eq.~(\ref{asDA}). It is nothing else but the weight function in the conformal
scalar product and is fixed by the symmetry. On the other hand, $\gamma_{N,q}$
and $\mathbb{P}_{N,q}(\xxi_i)$ have to be found by the explicit diagonalization
of the (reduced) mixing matrix
\cite{barmix}, where mixing with the operators containing total derivatives can
be ignored, i.e., by taking forward matrix elements. Progressing to large values
of $N$ one obtains  rather complicated matrices which are not symmetric and have
no obvious structure. For many years the only method to study the evolution of
baryon distribution amplitudes was based on numerical evaluation of such matrices
written in terms of contributions of symmetrized Appell's polynomials, see e.g.,
\cite{Stefanis},  which correspond to one possible choice (\ref{co-91}) of the
three-particle conformal operators. It was believed that the r\^ole of conformal
symmetry is exhausted by this representation and the problem has no further
symmetry; hence, no analytic approach seemed feasible.

We will demonstrate that this pessimistic conclusion is not warranted. It turns
out that all important features of the spectrum of anomalous dimensions and
eigenfunctions of three-quark states can be completely understood  in analytic
form. The first step is to find an orthonormal basis of conformal operators, such
that any two operators with different spin $N$ and/or  different $q$ are
orthogonal with respect to the scalar product (\ref{sc-prod-3}), $\langle
\mathbb{P}_{N,q}|\mathbb{P}_{N,q'}\rangle ={\cal N}\delta_{qq'}$. Expanding the
solutions of the evolution equations in this basis, one obtains the reduced
mixing matrix
\bea{picture}
&& {}
\nonumber
\\[-3mm]
&&
\underbrace{\left(
\begin{array}{c}
\mbox{\rm full~mixing matrix~of}\\
\mbox{\rm composite~ operators}\\
\end{array}
\right)}_{\mbox{$N(N+1)/2$}}\quad
\Longrightarrow\quad
\underbrace{\left(
\begin{array}{c}
{\rm  hermitian~kernels}\\
{\rm in~conformal~basis}
\end{array}
\right)}_{\mbox{$N+1$}}
\eea
which has the same size $N+1$ as in the standard approach, but is {\it
symmetric}. This symmetry is a direct consequence of the underlying
$SL(2,{\mathbb R})$ symmetry of the evolution equations and explains, e.g., why
all anomalous dimensions of three-quark operators are real numbers. Most
importantly, in this form new (hidden) symmetries can be observed and we will
find that the evolution equation for the distribution amplitude of the
$\Delta-$resonance is completely integrable \cite{BDM98,BDKM}. Last but not
least, for symmetric matrices one can use a simple perturbation theory to take
into account contributions that are suppressed by a certain parameter, and this
proves to be crucial in many further applications.

\setlength{\unitlength}{0.7mm}
\begin{figure}[t]
\vspace{7.3cm}
\begin{picture}(120,160)(0,1)
\mbox{\epsfxsize16.0cm\epsfbox{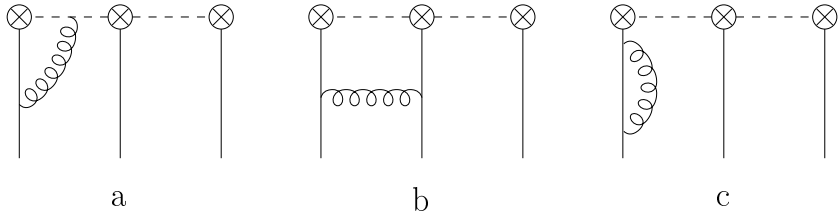}}
\end{picture}
\vspace*{-16.0cm}
\caption{Examples of a `vertex' correction (a), 'exchange'
diagram (b) and self-energy insertion (c) contributing to the renormalization of
three-quark operators in Feynman gauge. Path-ordered gauge factors are shown by
the dashed lines. The set of all diagrams includes possible permutations. }
\label{figure1}
\end{figure}
We begin now with the explicit
construction.
The nonlocal operators $B_{3/2}$ (\ref{B3/2}) and $B_{1/2}$
(\ref{B1/2}) do not mix with each other since
they belong to different representations of the Lorentz group. To one-loop
accuracy the corresponding renormalization group equation can be written
as~\cite{BukFroKurLip85,BB89}
\begin{equation}
 \left\{\mu\,\frac{\partial}{\partial \mu}
         +\beta(g)\,\frac{\partial}{\partial
         g}\right\}B^\lambda(\alpha_1,\alpha_2,\alpha_3)
  = \frac{\alpha_s}{4\pi}\left[{\mathbb H}_\lambda\cdot B^\lambda\right]
     (\alpha_1,\alpha_2,\alpha_3),
\label{RG}
\end{equation}
where the evolution kernel ${\mathbb H}_\lambda$ depends on the helicity of the
three-quark operator and corresponds
to contributions of the Feynman diagrams shown in Fig.~\ref{figure1}. Apart from
the different color factor these diagrams coincide with those defining the
two-particle evolution kernels $\mathcal{H}_{\rm v}$ (\ref{H-v}) and
$\mathcal{H}_{\rm e}$ (\ref{H-e}), so that
\bea{H-lambda}
{\mathbb H}_{\lambda}= (1+1/N_c)\mathcal{H}_{\lambda}+ 3C_F/2  \,,
\eea
where the last term is due to the self-energies, Fig.~\ref{figure1}c,
the color factor follows from the identity
$\epsilon_{ijk}t^a_{ii'}t^a_{jj'}=-\varepsilon_{i'j'k}(1+1/N_c)/2$ and
the Hamiltonians $\mathcal{H}_{\lambda}$ are given by the sum of
two-particle kernels
\bea{H-helicity}
\mathcal{H}_{\lambda=3/2}=
\mathcal{H}_{\rm v}^{(12)}+\mathcal{H}_{\rm v}^{(23)}+\mathcal{H}_{\rm v}^{(13)}
\,,\qquad
\mathcal{H}_{\lambda=1/2}=\mathcal{H}_{\lambda=3/2}
-\mathcal{H}_{\rm e}^{(12)}-\mathcal{H}_{\rm e}^{(23)}\,,
\eea
where the superscripts are introduced to show that $\mathcal{H}_{\rm
v}^{(ik)}$ and $\mathcal{H}_{\rm e}^{(ik)}$ act on the light-cone
coordinates of the quarks `$i$' and `$k$'. Note that the difference between the two
Hamiltonians for $\lambda=3/2$ and $\lambda=1/2$ in
Eqs.~(\ref{H-helicity}) is due to the fact that the gluon exchange
diagram in Fig.~\ref{figure1}b vanishes unless the
two quarks have opposite chirality.

Since the two-particle kernels $\mathcal{H}_{\rm v}^{(ik)}$ and
$\mathcal{H}_{\rm e}^{(ik)}$, by construction, only depend on the corresponding
spin operators $J_{ik}$, the Hamiltonians $\mathcal{H}_{\lambda}$ are explicitly
$SL(2,{\mathbb R})$ invariant and commute with the three-particle generators
$\JJ_a=\JJ_{1,a}+\JJ_{2,a}+\JJ_{3,a}$ (with $a=+,-,0$):
\bea{H3-com}
[ \mathcal{H}_\lambda, \JJ^2 ] = [ \mathcal{H}_\lambda, \JJ_+ ]
 = [ \mathcal{H}_\lambda, \JJ_- ] =
[ \mathcal{H}_\lambda, \JJ_0 ] = 0\,.
\eea
Finding solution to the renormalization group equation (\ref{RG}) is equivalent
to solving the Schr\"odinger equation for the characteristic polynomials of
multiplicatively renormalizable operators
\bea{Sch-3-eq}
\mathcal{H}_\lambda(\widetilde J_{12},\widetilde J_{23},\widetilde J_{13})\cdot
\mathbb{P}_{N,q}^\lambda(\xxi_1,\xxi_2,\xxi_3)= E^\lambda_{N,q}
\mathbb{P}_{N,q}^\lambda(\xxi_1,\xxi_2,\xxi_3)\,,
\eea
and the anomalous dimensions are given in terms of the ``energies'' as
\beq{restore}
 \gamma^\lambda_{N,q} = (1+1/N_c) E^\lambda_{N,q} + 3C_F/2\,.
\eeq
Since the Hamiltonian commutes with the three-particle $SL(2,{\mathbb R})$
generators, we enforce additional conditions (\ref{conf-const}) on the
eigenfunctions, i.e., require that they are characteristic
polynomials of conformal operators.
In addition, since the Hamiltonian is hermitian, all its eigenvalues
$E^\lambda_{N,q}$ are real numbers and the eigenfunctions corresponding
to different $E^\lambda_{N,q}$, alias different $q$, have to be
orthogonal with respect to the conformal scalar product
(\ref{sc-prod-3}).

The Schr\"odinder equation~(\ref{Sch-3-eq}) is written in operator form, and in
order to obtain the general solution it is convenient to expand the
eigenfunctions $\mathbb{P}_{N,q}(\xxi_i)$ over an orthonormal basis of functions
satisfying the same conformal constraints, Eq.~(\ref{conf-const}). A suitable
basis can be constructed as follows \cite{BDM98,BDKM,BKM01}. We define the set of
polynomials $\mathbb{Y}^{(12)3}_{N,n}(\xxi_1,\xxi_2,\xxi_3)$ with
$n=0,1,\ldots,N$ by requiring that in addition to (\ref{conf-const}) they
diagonalize the two-particle Casimir operator for the $(12)$-quark pair
(\ref{tilde-J2}), ${\widetilde \JJ}_{12}^2=\widetilde J_{12}(\widetilde
J_{12}-1)$:
\bea{basis2}
{\widetilde \JJ}_{12}^2\,\mathbb{Y}^{(12)3}_{N,n}(\xxi_i)
 = j_{12}(j_{12}-1)\, \mathbb{Y}^{(12)3}_{N,n}(\xxi_i),
\qquad (j_{12}= n+2\,,\quad 0 \leq n \leq N)\,.
\eea
The solutions to the combined equations (\ref{conf-const}) and (\ref{tilde-J2})
are given by products of Gegenbauer and Jacobi polynomials and are presented
for the general case of arbitrary conformal spins $j_1,j_2,j_3$ in
Appendix~\ref{app:a}. For our present purposes it turns out to be
convenient to use a different normalization, so we choose
\bea{dif-norm}
\mathbb{Y}^{(12)3}_{N,n}(\xxi_i)=\frac12(N+n+4)(n+2)\,
{Y}^{(12)3}_{J=N+3,j=n+2}(\xxi_i)\,,
\eea
where the functions $Y^{(12)3}_{J,j}(\xxi_i)$ are given in Eq.~(\ref{YJi123}).
The functions $\mathbb{Y}^{(12)3}_{N,n}(\xxi_i)$ depend on the pair of integers
$N$ and $n$ which are related to the total conformal spin of the three-quark
operator $j_1+j_2+j_3+N=3+N$ and the conformal spin of the $(12)-$pair
$j_{12}=j_1+j_2+n=2+n$ with $j_1=j_2=j_3=1$. In what follows we often drop the
subscript `$N$' if it is clear from the context. The particular choice of a quark
pair in Eq.~(\ref{basis2}) is of course arbitrary and we might use, e.g.,
${\widetilde \JJ}_{23}^2$ to construct a different basis of functions
$\mathbb{Y}^{1(23)}_n(\xxi_i)$. Here, the superscript indicates the order in
which the tensor product of three $SL(2)$ representations has been decomposed
into the irreducible components. The two set of basis functions
$\mathbb{Y}^{1(23)}_n(\xxi_i)$ and $\mathbb{Y}^{(12)3}_n(\xxi_i)$ are related to
each other through the Racah $6j-$symbols of the $SL(2)$ group (see Appendix~B in
Ref.~\cite{BKM01}).

One convenient property of the functions $\mathbb{Y}^{(12)3}_n(\xxi_i)$ is that
the action of the Casimir operators on them turns out to be
 rather simple. By construction,
$\mathbb{Y}^{(12)3}_{N,n}(\xxi_i)$ diagonalize the total three-particle Casimir
operator $\widetilde{{\JJ}}_{}^2$ and the Casimir operator for the $(12)-$quark
pair $\widetilde{{\JJ}}_{12}^2$, whereas the remaining two two-particle Casimir
operators turn out to be three-diagonal:
\begin{eqnarray}
\widetilde{{\JJ}}_{23}^2\, \mathbb{Y}^{(12)3}_n(\xxi_i) &=& f_n\!\left[ \phantom{-}
\frac{1}{(n+1)}\mathbb{Y}^{(12)3}_{n-1}(\xxi_i)+ \frac{2n+3}{(n+2)(n+1)}\mathbb{Y}^{(12)3}_n(\xxi_i)
                                +\frac{1}{(n+2)}\mathbb{Y}^{(12)3}_{n+1}(\xxi_i)
\right]\!,
\nonumber\\
\widetilde{{\JJ}}_{31}^2\, \mathbb{Y}^{(12)3}_n(\xxi_i)&= & f_n\!\left[
-\frac{1}{(n+1)}\mathbb{Y}^{(12)3}_{n-1}(\xxi_i)+\frac{2n+3}{(n+2)(n+1)}\mathbb{Y}^{(12)3}_n(\xxi_i)
                                -\frac{1}{(n+2)}\mathbb{Y}^{(12)3}_{n+1}(\xxi_i)
\right]\!,
\nonumber\\{}
\label{threediagonal}
\end{eqnarray}
where $f_n$ is defined in Eq.~(\ref{u-coeff}) below.
This property turns out to be crucial for simplification of the evolution
equation.

The most important property, however, is that
$\mathbb{Y}^{(12)3}_{N,n}(\xxi_i)$ form an orthogonal basis on the space of
characteristic polynomials
\bea{basis-33}
\langle \mathbb{Y}^{(12)3}_{N,n}|\mathbb{Y}^{(12)3}_{N',n'}\rangle
 \sim \delta_{NN'}\delta_{nn'}\,,
\eea
the reason being that ${\widetilde \JJ}_{12}^2$ is a self-adjoint operator with
respect to the scalar product (\ref{sc-prod-3}) so that its eigenstates with
different $n$ have to be mutually orthogonal.
As the result, the Hamiltonians $\mathcal{H}_{\lambda=3/2}$ and
$\mathcal{H}_{\lambda=1/2}$ (\ref{H-helicity}) can be written in the conformal basis
as
\bea{106}
  \left[\mathcal{H}_{\lambda}\right]_{kn} =
   \langle \mathbb{Y}^{1(23)}_k|\mathcal{H}_{\lambda}|\mathbb{Y}^{1(23)}_n\rangle
/(\|\mathbb{Y}^{1(23)}_k\|\cdot\|\mathbb{Y}^{1(23)}_n\|)\,,
\eea
where $\left[\mathcal{H}_{\lambda}\right]_{nk}=
\left[\mathcal{H}_{\lambda}\right]_{kn}$ are
hermitian (symmetric) matrices of the size $N+1$ which
can be expressed in terms of the Racah
$6j-$symbols of the $SL(2,{\mathbb R})$ group.

We look now for the solutions $\mathbb{P}_{N,q}$
of the Schr\"odinger equation (\ref{Sch-3-eq}), expanding them in this basis
\beq{u-coeff}
\mathbb{P}_{N,q}(\xxi_1,\xxi_2,\xxi_3)=\sum_{k=0}^N  {i^k}\frac{c_k(q)}{f_k} \,
\mathbb{Y}^{(12)3}_{N,k}(\xxi_1,\xxi_2,\xxi_3)\,,
\quad\
f_k=\frac{(k\!+\!1)(k\!+\!2)}{2(2k\!+\!3)} (N\!-\!k\!+\!1)(N\!+\!k\!+\!4)\,,
\eeq
where $c_k(q)$ are the expansion coefficients and the factor $i^k/f_k$ was
inserted for later convenience.
This gives a matrix equation for $c_n(q)$
\bea{Matrix-H}
\sum_{k=0}^{N}i^{n-k} \left[\mathcal{H}_{\lambda}\right]_{nk}
c_k(q)/f_k = E^\lambda_{N,q} \,c_n(q)/f_n\,,
\eea
where it is tacitly assumed that $c_n(q)$ also depend on the total spin $N$ and
helicity $\lambda$. The parameter $q$ enumerates different solutions to Eq.\
(\ref{Matrix-H}). As was already mentioned, for given $N$ one expects to find
$N+1$ different eigenstates. Using explicit expressions for the matrices
$\left[\mathcal{H}_{\lambda}\right]_{kn}$ is becomes straightforward to solve
Eq.\ (\ref{Matrix-H}) and obtain the spectrum of anomalous dimensions
$\gamma^{\lambda=1/2}_{N,q}$ and $\gamma^{\lambda=3/2}_{N,q}$ for fixed $N$. The
results of numerical calculations for $0\le N \le 30$ are shown in
Fig.~\ref{figure2}. We notice that the both spectra exhibit remarkable
regularity. The spectrum of $\mathcal{H}_{1/2}$ is very similar to that of
$\mathcal{H}_{3/2}$ except the few lowest eigenvalues for each $N$, which are
separated from the rest of the spectrum by a gap that remains finite at large
$N$. As we will show in the next Section, these properties are a manifestation of
a hidden symmetry of the Hamiltonian $\mathcal{H}_{3/2}$.

\subsection{\it Complete Integrability}

As was explained in the previous Section, the QCD evolution equations for
three-quark operators have the form of the Schr\"odinger equations
(\ref{Sch-3-eq}) describing a three-particle system with three degrees of freedom
corresponding to quark momentum fractions $\xxi_i$. In this way, the scale
dependence of baryon distribution amplitudes in QCD corresponds to a  {\it
one-}dimensional quantum mechanical 3-body problem with very peculiar
Hamiltonians, (\ref{H-lambda}), (\ref{H-helicity}) and (\ref{H-op-rel}),
determined by the underlying QCD dynamics. The conformal symmetry allows us to
trade two degrees of freedom for two quantum numbers corresponding to the total
conformal spin $\widetilde{{\JJ}}^2$ and its projection $\widetilde \JJ_0$,
Eq.~(\ref{conf-const}). The remaining degree of freedom is described in the
conformal basis by the set of coefficients $c_n$ introduced in
Eqs.~(\ref{u-coeff}) and (\ref{Matrix-H}). Thus, thanks to conformal invariance,
the original 3-body Schr\"odinger equation (\ref{Sch-3-eq}) is reduced to a
(complicated) one-body problem (\ref{Matrix-H}) which is not possible to solve
analytically for arbitrary $N$, unless it has some additional symmetry.

\begin{figure}[t]
\begin{picture}(100,100)
\put(275,192){\scalebox{0.9}[0.9]{\includegraphics*[400,500][0,0]{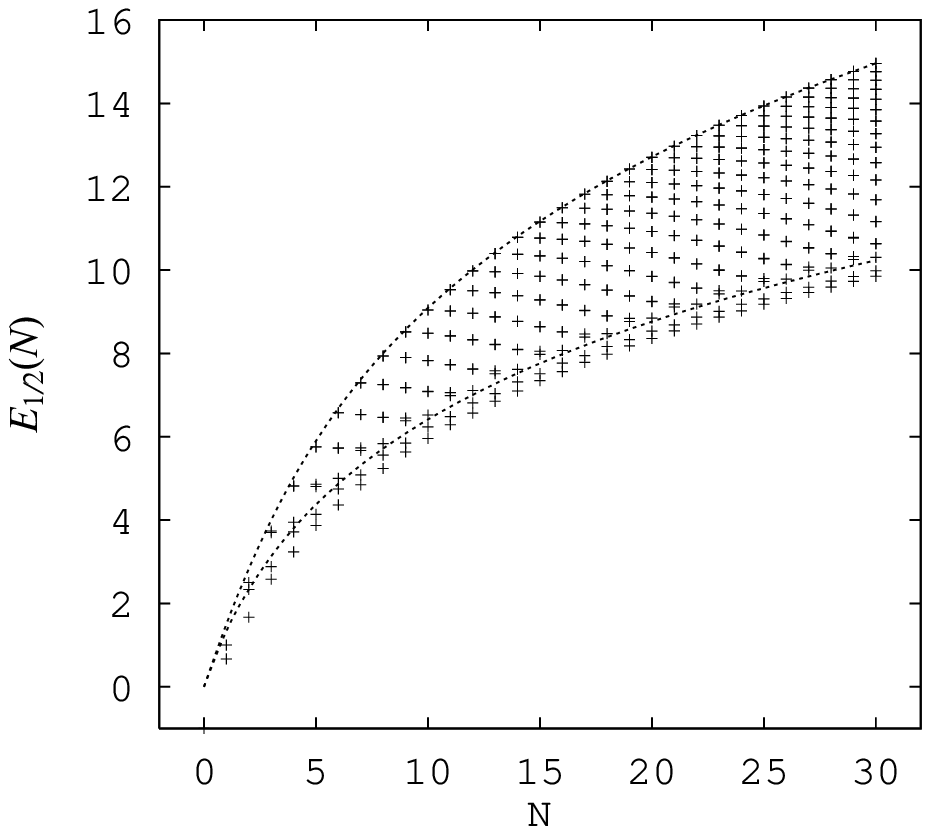}}}
\put(152,101){\scalebox{0.9}[0.9]{\includegraphics*[400,300][0,0]{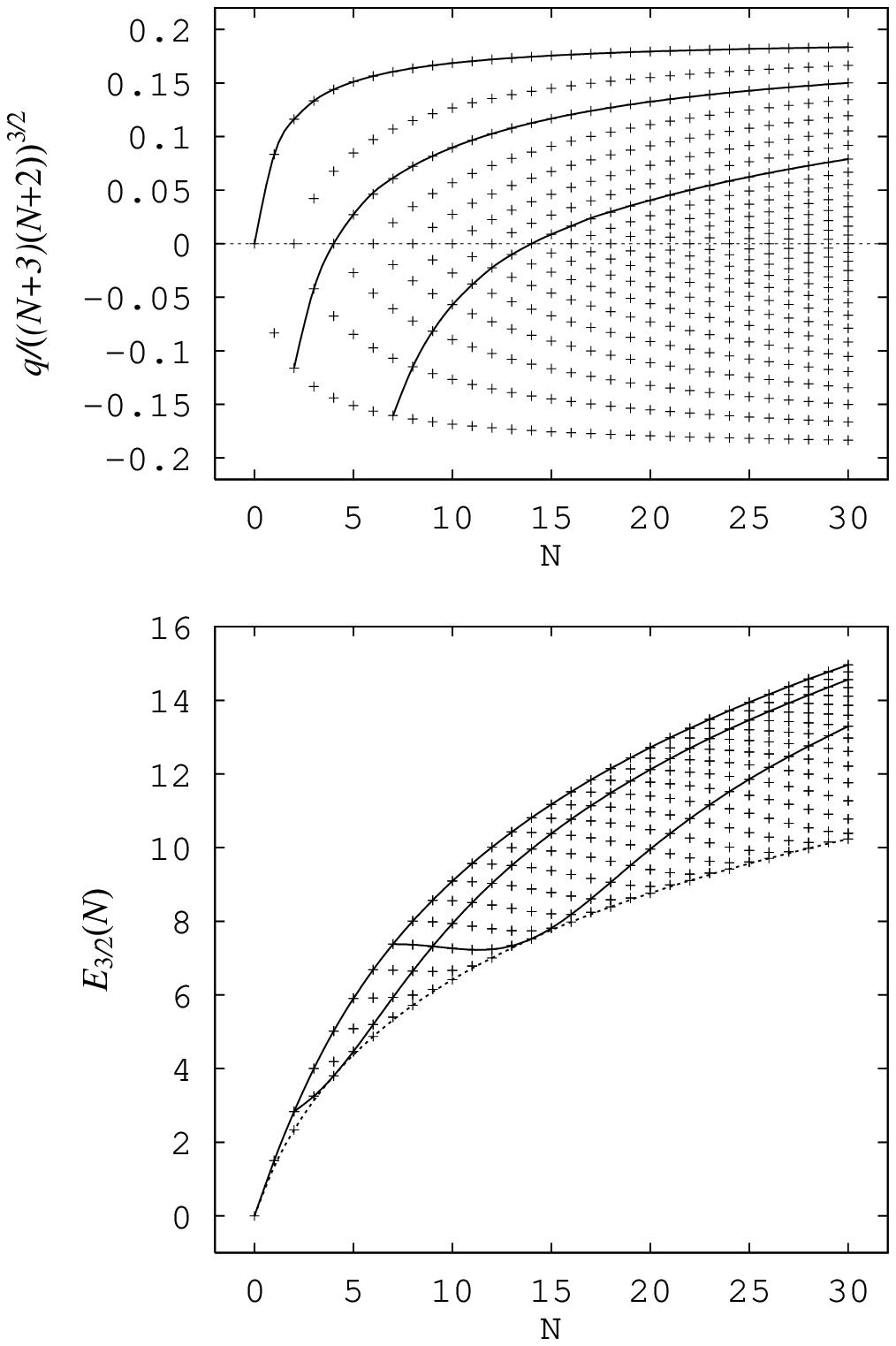}}}
\end{picture}
\vspace*{0.5cm}
\caption[]{The spectrum
of eigenvalues
for the helicity-3/2 (left) and helicity-1/2 (right) Hamiltonians  ${\cal
H}_{3/2}(N)$ and ${\cal H}_{1/2}(N)$, respectively. The lines of the largest and
the smallest ${E}_{3/2}$ are indicated on the plot for ${E}_{1/2}$ by dots for
comparison. The solid curves in the left panel show three different trajectories
for the anomalous dimensions (see text and Fig.~\ref{figure3}).}
\label{figure2}
\end{figure}

It turns out that the Hamiltonian ${\cal H}_{3/2}$ (but not ${\cal H}_{1/2}$)
possesses such additional `hidden' symmetry. Namely, one can construct the
operator
\bea{Q-def}
 {\mathbb Q} = \frac{i}{2}[{\widetilde{\JJ}}_{12}^2,{\widetilde{\JJ}}_{23}^2]
= i \left(\partial_{\xxi_1}-\partial_{\xxi_2}\right)
                        \left(\partial_{\xxi_2}-\partial_{\xxi_3}\right)
                        \left(\partial_{\xxi_3}-\partial_{\xxi_1}\right)
   \xxi_1 \xxi_2 \xxi_3 \,,
\eea
that commutes with ${\cal H}_{3/2}$ and with the $SL(2)$ generators:
\begin{eqnarray}
  [{\mathbb Q}, \widetilde{\JJ}_\alpha] = 0\,,
\quad\quad
[{\mathbb Q}, {\cal H}_{3/2}] =0\,.
\label{Q3}
\end{eqnarray}
Thus, the operator ${\mathbb Q}$ is the integral of motion and its eigenvalues
$q$ specify uniquely the spectrum of the anomalous dimensions of the three-quark
operators of helicity $\lambda=3/2$. The first relation in Eq.\ (\ref{Q3})
follows readily from the definition of the charge ${\mathbb Q}$. The second
relation can be checked using the commutation relations between ${\mathbb Q}$ and
two-particle Hamiltonians, Eqs.~(\ref{H-helicity}) and (\ref{H-op-rel})
\begin{equation}
 [{\cal H}^{(12)}_{\rm v},{\mathbb Q}] =i({\widetilde{\JJ}}_{23}^2-{\widetilde{\JJ}}_{31}^2)\,, \qquad
 [{\cal H}^{(23)}_{\rm v},{\mathbb Q}] =i({\widetilde{\JJ}}_{31}^2-{\widetilde{\JJ}}_{12}^2)\,, \qquad
 [{\cal H}^{(31)}_{\rm v},{\mathbb Q}] =i({\widetilde{\JJ}}_{12}^2-{\widetilde{\JJ}}_{23}^2)\,.
\label{Q3com}
\end{equation}
To prove these operator identities, apply the both sides to the function
$\mathbb{Y}^{(12)3}_n(\xxi_i)$ belonging to the conformal basis, use
Eqs.~(\ref{threediagonal}), and take into account that the operator ${\mathbb
Q}$, Eq.~(\ref{Q-def}), is represented in the conformal basis by a $(N+1)\times
(N+1)$ matrix with only two nonzero subleading diagonals
\bea{Q-matrix}
{\mathbb Q}\, \mathbb{Y}^{(12)3}_n(x_i) = if_n\left[ \mathbb{Y}^{(12)3}_{n+1}
(x_i)-\mathbb{Y}^{(12)3}_{n-1} (x_i)
\right].
\eea
Existence of a nontrivial integral of motion ${\mathbb Q}$ implies that for $\lambda=3/2$ the
Schr\"odinger equation (\ref{Matrix-H}) is completely integrable. This allows us
to calculate the corresponding spectrum of the anomalous dimensions analytically
by applying a powerful technique of integrable models.

Remarkably enough the Hamiltonian ${\cal H}_{3/2}$ is well-known in the theory of
integrable models as a generalization of the celebrated Heisenberg spin magnet
describing the nearest-neighbor interaction between spin$-1/2$ operators (Pauli
matrices) with the Hamiltonian
\bea{Heis-1/2}
H_{\rm XXX}=-\frac12\sum_{n}\vec{\sigma}_n \cdot \vec\sigma_{n+1}\,.
\eea
This model has been studied in 1931 by Bethe as a model of one-dimensional metal
and solved by the method which has become well-known as the Bethe Ansatz. The
Hamiltonian (\ref{Heis-1/2}) possesses the set of integrals of motion and the
corresponding Schr\"odinger equation is completely integrable for
arbitrary number of spins. Later it was
recognized that the Heisenberg spin magnet can be generalized to spins belonging
to higher-dimensional representations of the $SU(2)$ and the $SL(2,\mathbb{R})$
groups \cite{TTF}. Although the Hamiltonian becomes much more involved, it
inherits the property of complete integrability and its spectrum can be studied
by the Bethe Ansatz technique.

In application to the QCD evolution equations, one encounters the integrable
generalization of the Heisenberg magnet for the $SL(2,\mathbb{R})$ spins. Indeed,
an inspection shows that the $SL(2)$ generators (\ref{adjoint}) for quarks with
conformal spin $j_k=1$ can be interpreted as the usual spin $s=-1$ operators. In
this way, we may consider the Hamiltonian ${\cal H}_{3/2}$ as describing the
system of three interacting spins each acting on its internal space labeled by
the coordinates $\xxi_k$. These spins carry the index of the corresponding
particles and form a one-dimensional spin chain with three sites which coincides
identically with the one-dimensional homogeneous  Heisenberg spin magnet of
noncompact spin $s=-1$. This model has been solved using the Bethe Ansatz
technique and the detailed results can be found in Refs.~\cite{K97}. In the rest
of this section we explain the exact solution for the Hamiltonian ${\cal
H}_{3/2}$ and discuss the main features of the spectrum.

\begin{figure}[t]
\vspace{0.2cm}
\begin{picture}(100,100)
\put(175,104){\scalebox{0.95}[0.95]{\includegraphics*[400,500][0,300]{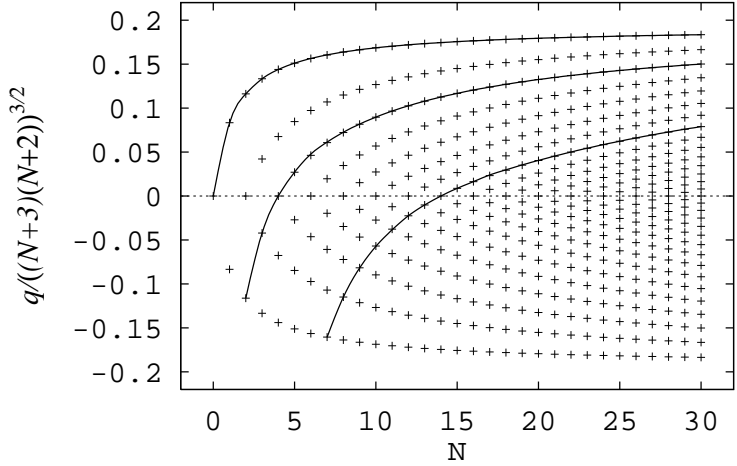}}}
\end{picture}
\hfill
\parbox[b]{7cm}{
\caption[]{
The spectrum of eigenvalues for the conserved charge ${\mathbb Q}$.
}
\label{figure3}
\vspace{4.0cm}
}
\vspace*{-1cm}
\end{figure}

Complete integrability of the Schr\"odinger equation for $\lambda=3/2$ implies
that ${\cal H}_{3/2}$ is a (complicated) function of two and only two mutually
commuting operators ${\mathbb Q}$ and ${\widetilde{\JJ}}^2$. Therefore, instead
of solving the Schr\"odinger equation (\ref{Sch-3-eq}) directly, one can solve
much simpler equations (\ref{conf-const}) supplemented by the additional
constraint
\bea{Q3psi}
 {\mathbb Q}\,\mathbb{P}_{N,q}(\xxi_i) \equiv i
\left(\partial_{\xxi_1}-\partial_{\xxi_2}\right)
                        \left(\partial_{\xxi_2}-\partial_{\xxi_3}\right)
                        \left(\partial_{\xxi_3}-\partial_{\xxi_1}\right)
   \xxi_1 \xxi_2 \xxi_3\, \mathbb{P}_{N,q}(\xxi_i) = q\, \mathbb{P}_{N,q}(\xxi_i)
\eea
and find the spectrum of the Hamiltonian ${\cal H}_{3/2}={\cal
H}_{3/2}({\widetilde{\JJ}}^2,Q)$ by replacing the operators by their
corresponding eigenvalues. Expanding $\mathbb{P}_{N,q}(\xxi_i) $ over the
conformal basis, Eq.~(\ref{u-coeff}), and making use of Eq.~(\ref{Q-matrix}), one
finds that the equation ${\mathbb Q}\,\mathbb{P} = q\,\mathbb{P}$ is equivalent
to the following three-terms recurrence relation for the coefficients $c_n$,
$(n=0,\ldots,N)$:
\begin{equation}
q \,c_n = f_n\left( c_{n+1}+c_{n-1} \right),\qquad c_{-1}=c_{N+1}=0.
\label{masterrec}
\end{equation}
Since the operator ${\mathbb Q}$ is self-adjoint with respect to the scalar
product (\ref{sc-prod-3}), its eigenstates $\mathbb{P}_{N,q}(\xxi_i)$ are
orthogonal to each other for different set of quantum numbers. Their norm depends
on the overall normalization of $c_n$. It is convenient to choose it in such a
way that the eigenstates $\mathbb{P}_{N,q}(\xxi_i)$ form an orthonormal set of
the states
\bea{ortho}
\langle\mathbb{P}_{N,q}|\mathbb{P}_{N',q'}\rangle = \delta_{NN'} \delta_{qq'}\,.
\eea

The recurrence relations (\ref{masterrec}) represent the system of $N+1$ linear
homogeneous equations on the coefficients $c_n$. Solution of this system is
equivalent to diagonalization of a $(N+1)\times(N+1)$ matrix with only two
subleading diagonals nonzero. The consistency condition for this system leads to
$N+1$ quantized values of $q$, which have the properties of roots of orthogonal
polynomials. That is, they are real, simple and for different $N$ the sets of
quantized $q$ are interlaced, see Fig.~\ref{figure3}. The set of coefficients
$c_n\equiv c_n(N,q)$ defines uniquely the characteristic polynomial of the
three-quark operator, Eq.~(\ref{u-coeff}). The corresponding exact value of the
anomalous dimension is given by
\begin{equation}
E^{3/2}_{N,q}=4\,{\rm Re}\frac {\sum_{n=0}^{N}i^n\, c_n(q)\,
\left[\Psi(n+2)-\Psi(2)\right]
\,(2n+3)}
{\sum_{n=0}^{N}i^n\, c_n(q)\, (2n+3)}\,.
\label{energy1}
\end{equation}
The recurrence relations (\ref{masterrec}) combined with the expression for the
anomalous dimensions (\ref{energy1}) and the characteristic polynomials
(\ref{u-coeff}) allow us to determine the spectrum of local conformal
three-quark operators of helicity$-3/2$, (\ref{O3quark}). One finds from
(\ref{masterrec}) that $c_n(-q)=(-1)^n c_n(q)$ and, a consequence, the energy
levels corresponding to nonzero values of quantized $q$ are double degenerate
\bea{degeneracy}
E^{3/2}_{N,q}=E^{3/2}_{N,-q}\,.
\eea
The resulting spectrum of the conserved charge $q$ and the ``energy''
$E^{3/2}_{N,q}$ for $0\le N\le 30$ is shown in Fig.~\ref{figure3} and
Fig.~\ref{figure2}, respectively.

As follows from (\ref{repr3/2}), conformal moments of the distribution amplitude
of the $\Delta$-resonance with helicity $\lambda=3/2$ are related to the
(reduced) matrix elements of conformal three-quark operators
\bea{Phi-P}
\int_0^1[d\xxi]\, \phi_{3/2}(\xxi_1,\xxi_2,\xxi_3;\mu)\mathbb{P}_{N,q}(\xxi_1,\xxi_2,\xxi_3)=
\langle 0|\mathbb{O}^{1,1,1}_{N,q}(\mu_0)|P \rangle
    \left(\frac{\alpha_s(\mu)}{\alpha_s(\mu_0)}
         \right)^{\gamma^{3/2}_{N,q}/\beta_0}\,.
\eea
Thanks to the orthogonality condition (\ref{ortho}) this relation can be inverted
to obtain the desired expansion of the distribution amplitude
envisaged in Eq.~(\ref{expansion}).

{}For fixed $N$ the dominant contribution to the sum in (\ref{expansion}) comes
from the conformal operators with the lowest anomalous dimension. For
\textit{even} $N$ such operators are those with $q=0$. The recurrence relation
(\ref{masterrec}) can be easily solved in this case and leads to $c_{2n}=(-1)^n
c_0$ and $c_{2n+1}=0$. Using (\ref{energy1}) then yields
\begin{equation}
   E^{3/2}_{N,q=0} = 4\Psi(N+3)+4\gamma_E-6\,.
\label{e0}
\end{equation}
The corresponding curve is shown in Fig.~\ref{figure2} (left) by dots. The
corresponding eigenfunction (\ref{u-coeff}) is a completely symmetric real
function of $\xxi_i$ and is given (up to an overall normalization factor) by
\cite{BDKM}
\begin{eqnarray}
\lefteqn{ 
\xxi_1 \xxi_2 \xxi_3\, \mathbb{P}_{N,q=0}(\xxi_1,\xxi_2,\xxi_3) }
\nonumber\\&=&
     \xxi_1(1-\xxi_1)C_{N+1}^{3/2}(1-2\xxi_1)
    +\xxi_2(1-\xxi_2)C_{N+1}^{3/2}(1-2\xxi_2)
    +\xxi_3(1-\xxi_3)C_{N+1}^{3/2}(1-2\xxi_3),
\label{ground32}
\end{eqnarray}
where $\xxi_1+\xxi_2+\xxi_3=1$.

The expressions for $E^{3/2}_{N,q}$ and $\mathbb{P}_{N,q}$ for the
general case $q\neq 0$ cannot be written in a closed form.
A significant simplification occurs for large $N$, however.
In this limit one can solve the recurrence relations (\ref{masterrec}) by the WKB
method and develop the systematical expansion of the spectrum in powers of $1/N$.
One finds that for $N\gg 1$ the energies and conserved charges $q$
occupy the bands
\bea{band}
4\ln N \le E^{3/2}_{N,q} \le 6\ln N\,,\qquad 0 \le q^2 \le N^3/27\,.
\eea
The dependence of $E^{3/2}_{N,q}$ on the charge $q$ (the ``dispersion
curve'') takes the form
\bea{E-asym}
E^{3/2}_{N,q}=2\ln 2 -6 +6\gamma_{\rm E} + 2{\rm Re}\,\sum_{k=1}^3
\Psi(1+i\lambda_k)+\mathcal{O}(1/N^6)
\eea
where $\gamma_{\rm E}=0.57722\ldots$ is the Euler number and $\lambda_k$ are
defined by the three  roots of
the cubic equation
$2\lambda_k^3-\lambda_k-q=0$.

Possible values of the charge $q$ satisfy the quantization conditions
\bea{LO-quant}
q N^{-2}\ln N -{\rm arg}\, \Gamma(1+iqN^{-2}) + \mathcal{O}(1/N) =
\frac{\pi}6(N-2\ell),
\eea
where $\ell$ is a nonnegative integer. Solving (\ref{LO-quant}) one obtains the
quantized values of the charge $q=q(N,\ell)$. For given $\ell$ the values of $q$
form a  trajectory which depends analytically on $N$. Each trajectory $q(\ell,N)$
is mapped into the corresponding trajectory for $E^{3/2}_{N,\ell}$ as shown in
Fig.~\ref{figure2}. The $\ell-$th trajectory starts at $N=\ell$, approaches the
`Fermi surface' (\ref{e0}) at $N=2\ell$, gets repelled from it and monotonously
grows to infinity at large $N$. The corresponding asymptotic expression for the
energy reads (with $\eta=\sqrt{(N+3)(N+2)}$) \cite{K97}:
\begin{equation}
E^{3/2}_{N,\ell} = 6\ln\frac{\eta {\rm e}^{\gamma_E-1}}{\sqrt{3}}   -\frac{3(2\ell+1)}{\eta}
- \frac{30\ell^2+ 30\ell  -7}{6\eta^2}
- \frac{464\ell^3+696\ell^2-802\ell -517}{72\eta^3} +{\cal O}(1/\eta^4)
\label{upperE}
\end{equation}
at large $N\gg \ell$, and~\cite{BDM98,BDKM}
\begin{eqnarray}
E^{3/2}_{N,\ell}= 4\ln(N+3)+4\gamma_{\rm E}-6+\frac{\pi^2\zeta(3)}
{18\ln^2(\eta\,{\rm e}^{\gamma_E})} (N-2\ell)^2
\label{middleE}
\end{eqnarray}
in the vicinity of $N=2\ell$.
These relations
define the asymptotic expansion for the energy levels of the Hamiltonian ${\cal
H}_{3/2}$ parameterized by the integer $\ell$ in the upper and the lower parts of the band
(\ref{band}), respectively.
The corresponding level spacing is equal to
\begin{equation}
\delta E^{3/2}_{N,\ell} \stackrel{q\to 0}{=}
    \mathcal{O}\left(1/{\ln^2N}\right)\,,
\qquad
\delta E^{3/2}_{N,\ell} \stackrel{q\to N^3/\sqrt{27}}{=}
  \mathcal{O}\left(1/{N}\right)\,.
\label{levelspacing}
\end{equation}

To summarize, the
evolution equation for the distribution amplitude
$\phi^{3/2}_\Delta$ is exactly solvable. The physical interpretation of
integrability is that we are able to identify a new `hidden' quantum
number $q$, which distinguishes components in the $\Delta$-resonance
with different scale dependence, Eq.~(\ref{expansion}). In this case,
the coefficients $c_n$ in the expansion (\ref{u-coeff}) of the
eigenfunctions of the evolution equation over the
complete set of orthogonal conformal polynomials can be calculated
using a simple three-term recurrence relation (\ref{masterrec}).
The corresponding anomalous dimensions are given in terms
of $c_n$ by Eq.~(\ref{energy1}). For large $N$ the spectrum is
described to a high accuracy by the WKB formulae (\ref{upperE}) and (\ref{middleE}).

The most interesting result concerns the structure of the component of
$\phi^{3/2}_\Delta$ with the smallest anomalous dimension for each $N$. It
represents the leading contribution to the distribution amplitude
(\ref{expansion}) in the limit $\mu^2\to\infty$.
\begin{figure}[t]
\centerline{\epsfxsize7cm\epsfysize6.2cm\epsfbox{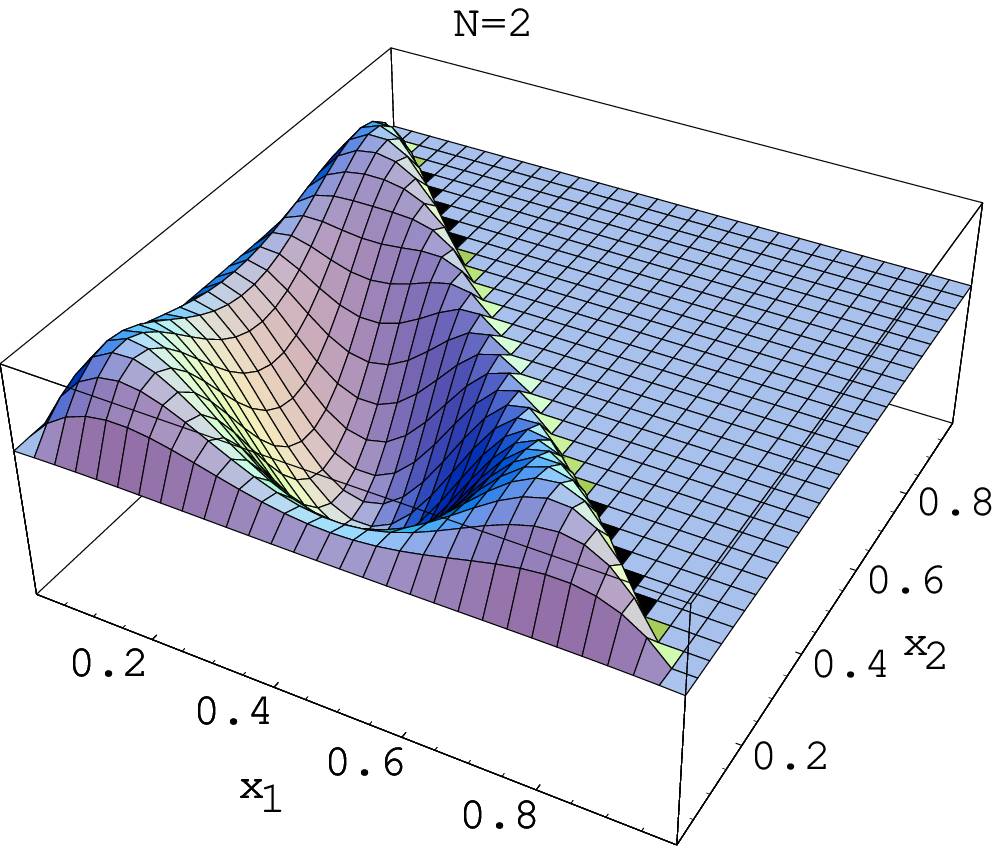}~~~
            \epsfxsize7cm\epsfysize6.2cm\epsfbox{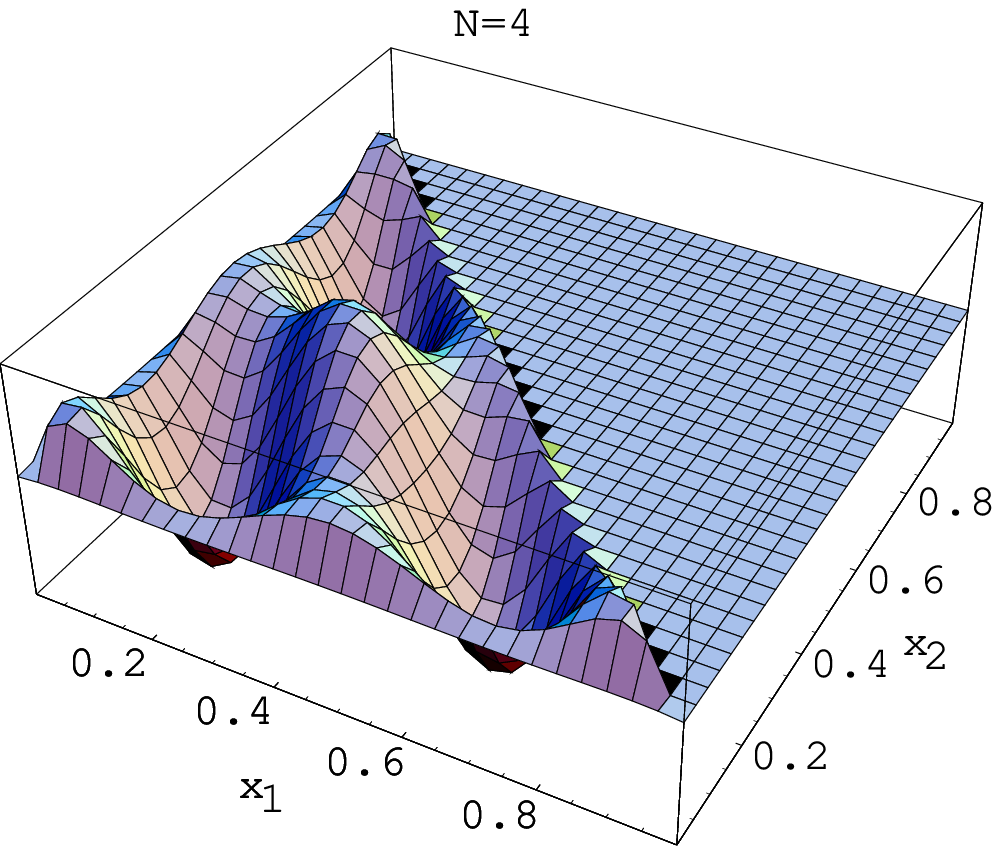}
            }
\caption[]{\small Contributions to the $\lambda=3/2$ distribution amplitude
$\phi_\Delta^{3/2}(\xxi_i) $ with lowest anomalous dimensions for $N=2$ and
$N=4$. The normalization is arbitrary.}
\label{fig:WF0}
\end{figure}
The corresponding polynomials are known exactly and are given in
Eq.~(\ref{ground32}) (see Fig.~\ref{fig:WF0}). The physical interpretation of
such `ground states' is most transparent in coordinate space.  Up to contribution
of operators with total derivatives, one can represent the three-quark `ground
state' in a concise form as the nonlocal light-cone operator \cite{ABH}
\bea{Blow32}
B_{3/2}^{(q=0)}(\alpha_1,\alpha_2,\alpha_3)=\frac12 \sum_{a,b=1,2,3
\atop
a\neq b} \varepsilon^{ijk}\int_0^1\!dv\, \!\not\!n \psi_{i}^\uparrow (\alpha_a)
   \!\not\!n \psi_{j}^\uparrow(v\alpha_a+(1-v)\alpha_b)
   \!\not\!n \psi_{k}^\uparrow (\alpha_b)\,.
\eea
The Taylor expansion
at short distances, $\alpha_{12}, \alpha_{32} \to 0$, generates the series of local
multiplicatively renormalizable three-quark local operators with the lowest
anomalous dimension for each even $N$:
\begin{equation}
B_{3/2}^{(q=0)}(\alpha_1,\alpha_2,\alpha_3)=\sum_{N={\rm even}} \frac{
\alpha_{12}^N+\alpha_{23}^N+\alpha_{31}^N}{(N+1)!}\
 \mathbb{P}_{N,q=0}^{3/2}(\partial_1,\partial_2,\partial_3)\
B(\alpha_1,\alpha_2,\alpha_3)\bigg|_{
\alpha_1=\alpha_2=\alpha_3=0
\atop
\partial_1+\partial_2+\partial_3=0} +\ldots\, .
\end{equation}
Note the  integration in Eq.\ (\ref{Blow32}) with unit weight over the position of the
quark in the middle that goes in between  the light-cone positions
 of the other two quarks, up to permutations.  If renormalization of the
operator is interpreted as interaction, this unit weight
can in turn be interpreted as the statement that the quark in the middle is effectively
`free': In the `ground state' with the lowest `energy', the interaction of the
quark in the middle with its right and left neighbors exactly compensate each
other.

\subsection{\it Further Developments}

Let us now consider the evolution equation for helicity$-1/2$ distribution
amplitude $\phi_N(\xxi_i)$. We recall that the corresponding Hamiltonian differs
from that for $\phi^{3/2}_\Delta$ by the additional contributions of  gluon
exchange between the quarks with opposite helicity, see Eq.~(\ref{H-helicity})
and Fig.~\ref{figure1}
\bea{pert}
\mathcal{H}_{1/2}=\mathcal{H}_{3/2}+V\,,\qquad
V=-1/{\widetilde{{\JJ}}_{12}^2}-1/{\widetilde{{\JJ}}_{23}^2}\,.
\eea
The added terms destroy the complete integrability  and hence the Schr\"odinder
equation for $\mathcal{H}_{1/2}$ cannot be exactly solved. The numerical
calculation for $0\le N\le 30$ gives the spectrum in Fig.~\ref{figure2}. As seen
from this figure, the spectra of ${\cal H}_{1/2}$ and ${\cal H}_{3/2}$ are very
similar in the upper part, for larger eigenvalues, and at the same time the two
lowest levels of the ${\cal H}_{1/2}$ Hamiltonian appear to be special and `dive'
considerably below the line of lowest eigenvalues of ${\cal H}_{3/2}$, given by
Eq.~(\ref{e0}).

The comparison  of the spectra in Fig.~\ref{figure2} suggests to consider the
operator $V$ in Eq.\ (\ref{pert}) as a perturbation. As usual in quantum
mechanics, for perturbation theory to be applicable the perturbation should be
smaller than the level spacing. To verify whether this is the case, one has to
examine the matrix elements of $V$ over the exact eigenstates of
$\mathcal{H}_{3/2}$. The explicit calculation shows that the matrix $\langle
\mathbb{P}_{N,q}|V|\mathbb{P}_{N,q'}\rangle$ is strongly peaked at the diagonal $q=q'$ and decreases
rapidly with $|q-q'|$. In the lower ($q\to 0$) and the upper
($q\to N^3/\sqrt{27})$ part of the spectrum it scales for large $N$ as
\begin{equation}
\langle \mathbb{P}_{q',N}  | {V} | \mathbb{P}_{q,N}  \rangle
 \stackrel{q,q'\to 0}{=}
   \mathcal{O}\left(1/{\ln N }\right)\,,
\qquad
\langle \mathbb{P}_{q',N}  | {V} | \mathbb{P}_{q,N}  \rangle
\stackrel{q,q'\to N^3/\sqrt{27}}{=}
 \mathcal{O}\left(1/{N^2}\right)\,,
\label{levelperturbation}
\end{equation}
respectively.
Comparing (\ref{levelperturbation}) with (\ref{levelspacing})
 we conclude that the perturbation theory  is
justified for large $N$ for the upper part of the spectrum.
A simple calculation then leads to
\bea{upper-PT}
E^{1/2}_{N,\ell}-E^{3/2}_{N,\ell} \simeq
\langle \mathbb{P}_{q,N}  | {V} | \mathbb{P}_{q,N}  \rangle
= - \frac{6}{\eta^2}\left[1+\frac{2(2\ell+1)}{\eta}\right]
+\mathcal{O}(1/\eta^4)\,,
\eea
where $\eta = \sqrt{(N+3)(N+2)}$.

On the contrary, in the lower part of the spectrum several (of order $\sim\ln N$)
lowest energy eigenstates are affected strongly and their mixing cannot treated
perturbatively. It turns out that the effective Hamiltonian describing the
interactions between the lowest energy levels can be brought to the form of
(generalized) Kroning-Penney model of a single particle in a periodic
$\delta-$function potential. Using the well-known solution to this model,
one can show that the two lowest eigenvalues (hence also the anomalous
dimensions) $E^{1/2}_\pm(N)$ decouple from the rest of the spectrum and in the
limit $N\to\infty$ are separated from the other eigenvalues by a finite constant
--- the `mass gap',
\bea{gap}
E^{1/2}_\pm(N)-E^{3/2}_{N,q=0} \simeq -0.3\,,
\eea
where $E^{3/2}_{N,q=0}$ is defined in Eq.\ (\ref{e0}). The corresponding contributions
to the distribution amplitude are given for $N\to\infty$ by
\begin{equation}
 \xxi_1 \xxi_2 \xxi_3\, \mathbb{P}^{1/2,\pm}_{N}(\xxi_1,\xxi_2,\xxi_3) =
  \xxi_1 \xxi_2 \xxi_3 \left[P^{(3,1)}_N(2\xxi_3-1)\pm P^{(3,1)}_N(2\xxi_1-1)\right],
\label{12a}
\end{equation}
where $P^{(3,1)}_N$ are the Jacobi polynomials. Notice that the expression in the
square brackets does not depend on the momentum fraction of the quark with
helicity opposite to that of the parent baryon. As a consequence, going over from
the conformal three-quark operators defined by the characteristic polynomials
(\ref{12a}) to the generating nonlocal light-cone operator
$B(\alpha_1,\alpha_2,\alpha_3)$, one finds that Eq.\ (\ref{12a}) corresponds, in the
same sense as Eq.~(\ref{Blow32}), to the following contribution
\begin{eqnarray}
\lefteqn{
B^\pm(\alpha_1,\alpha_2,\alpha_3)=
}
\nonumber\\ &=&
\varepsilon^{ijk}\left[ (\!\not\!n \psi_{i}^\uparrow
   \!\not\!n \psi_{j}^\downarrow)(\alpha_1)
   \!\not\!n \psi_{k}^\uparrow (\alpha_3)\delta(\alpha_2-\alpha_1)
\pm
   \!\not\!n \psi_{i}^\uparrow(\alpha_1)
   (\!\not\!n \psi_{j}^\downarrow)
   \!\not\!n \psi_{k}^\uparrow)(\alpha_3)\delta(\alpha_2-\alpha_3)\right].
\label{Blow12}
\end{eqnarray}
Formation of the mass gap in the spectrum of anomalous dimensions is, therefore,
naturally interpreted as due to binding of the quarks with opposite helicity into
scalar diquarks.

Note that while the expression (\ref{ground32})  for the eigenfunction is
exact, the result (\ref{12a}) is only valid in the asymptotic $\ln N\to\infty$
limit. In the coordinate space picture, the restriction to large $N$ is
translated to the condition that the light-cone separation between the same
helicity quarks is very large to allow for the formation of a diquark. In the
momentum space, the result means that at sufficiently large normalization scale
$\mu^2$ the quark carrying a very large momentum fraction is more often with the
same helicity as of the parent baryon. This observation seems to be in
qualitative agreement with phenomenological models of baryon distribution
amplitudes derived from QCD sum rules~\cite{CZreport,CZ84,FZOZ88}.

We have demonstrated that the Hamiltonian approach offers a convenient technique
for finding the scale dependence of the baryon distribution amplitudes. The power
of this approach is largely due to the fact that it is conformally covariant,
with the Hamiltonians being self-adjoint operators on the space of characteristic
polynomials endowed with the conformal scalar product (\ref{sc-prod-3}). They
only depend on the two-particle conformal spins but the explicit form of the
dependence is not fixed by the symmetry. Finally, the Hamiltonian
$\mathcal{H}_{3/2}$ possesses a hidden symmetry which allows one to solve the
corresponding Schr\"odinger equation exactly and identify a new quantum number
$q$ that distinguishes different components of the helicity$-3/2$ baryon
distribution amplitude. Although the Hamiltonian $\mathcal{H}_{1/2}$ does not
possess the same symmetry due to presence of additional terms, its complete
integrability is broken ``softly''. Its energy spectrum has many features common
with the spectrum of $\mathcal{H}_{3/2}$ and only a few lowest energy levels are
affected. We would like to stress that integrability of $\mathcal{H}_{3/2}$ is
intrinsically tied to a peculiar feature of anomalous dimensions in gauge
theories, which involve Euler $\Psi$-functions and are rising logarithmically
with the spin (dimension) \cite{K87}. {}For large $J_{12}$ one finds from
(\ref{H-op-rel})
that the both Hamiltonians $\mathcal{H}_{3/2}$ and $\mathcal{H}_{1/2}$ have the
same {\it universal\/} form
\bea{blabla2}
\mathcal{H}_{\lambda} = 2\ln J_{12} + 2\ln J_{23} + 2\ln J_{31}
+\mathcal{O}(1/J_{12}^2,1/J_{23}^2,1/J_{31}^2)\,.
\eea
The integrability of $\mathcal{H}_{3/2}$ imposes severe
restrictions on the form of subleading in $1/J_{ik}$ terms.

The same approach can be used for studies of the evolution of
quark-antiquark-gluon and three-gluon operators. These are interesting, e.g., in
connection with the structure function $g_2(x,Q^2)$ of polarized deep inelastic
scattering, which is attracting increasing interest as it provides a direct
measurement of three-parton correlations. The scale-dependence of $g_2(x,Q^2)$
can be traced to the renormalization-group equations for the
quark-antiquark-gluon operator \cite{BB89,ABH}
\bea{Spm}
   S^\pm (\alpha_1,\alpha_2,\alpha_3) =
   \frac12\bar \psi(\alpha_1)[ig\widetilde G_{\perp +}(\alpha_2)\pm
   gG_{\perp +}(\alpha_2)\gamma_5]\!\not\!n\psi(\alpha_3)\,,
\eea
where $\widetilde G_{\mu \nu}=\frac12\varepsilon_{\mu\nu\rho\lambda}
G^{a,\rho\lambda}t^a$, $G_{\mu \nu}= G_{\mu\nu}^a t^a$. In the
flavor singlet sector one also has to include the
three-gluon operator
\bea{Odual}
  \widetilde O (\alpha_1,\alpha_2,\alpha_3)
 = \frac{ig}{2} f^{abc}G^a_{+\lambda}(\alpha_1)
   \widetilde G^b_{+\perp}(\alpha_2)G^c_{+\lambda}(\alpha_3)\,
\eea
which mixes with the C-even part of the flavor singlet $qGq$ operator
(\ref{Spm}).

To begin with, consider the flavor non-singlet sector. In this case the
three-gluon operator drops out and the evolution of $S^\pm$ simplifies
dramatically in the limit of large number of colors, $N_c\to\infty$. The
corresponding Hamiltonian reads, e.g., for $S^+$ \cite{BDM98}
\beq{qGq-1}
   {\mathbb H}_{S^+} = N_c \Big[V_{qG}(J_{12})+ U_{Gq}(J_{23})\Big]
+{\mathcal O}\left(1/{N_c}\right)\,,
\eeq
where
\bea{qGq-2}
  V_{qG}(J) &=& \Psi\!\!\left(\!\! J+\frac{3}{2}\!\right)\! +
\Psi\!\!\left(\!\! J-\frac{3}{2}\!\right)\! -2\Psi(1)-\frac{3}{4}\,,
\nonumber\\
U_{Gq}(J) &=& \Psi\!\!\left(\!\! J+\frac{1}{2}\!\right)\! +
\Psi\!\!\left(\!\! J-\frac{1}{2}\!\right)\! -2\Psi(1)-\frac{3}{4}\,,
\eea
and it turns out to be equivalent to the Hamiltonian of the completely
integrable {\it open} Heisenberg magnet \cite{BDM98}. The corresponding conserved
charge is equal to
\beq{qGq-charge}
   {\mathbb Q}_{S^+} = \{{ \JJ}_{12}^2,{ \JJ}_{23}^2\}
  -\frac12{ \JJ}_{12}^2 - \frac92{ \JJ}_{23}^2\,,
\eeq
where $\{,\}$ stands for an anticommutator. Properties of this Hamiltonian were
studied in detail in Refs.\ \cite{Belitsky99,DKM00,Belitsky00} and the
corrections $\mathcal{O}(1/N_c^2)$ to the spectrum were calculated in Ref.\
\cite{BKM00}. The most interesting result is that the lowest energy, albeit the
lowest anomalous dimension in the spectrum of quark-gluon operators can be
exactly found, and the corresponding eigenfunction coincides to leading
logarithmic accuracy with the structure function $g_2$ itself. Thus, a simple
evolution equation arises~\cite{ABH}.

The study of the flavor singlet sector  is much more
challenging \cite{BKM01}.  Ignoring quarks for a moment, the
evolution kernel of three-gluon operators
becomes the Hamiltonian of {\it closed\/} Heisenberg magnet
with three ``gluonic'' sites of the spin $j_g=3/2$, whose integrability is
``softly'' broken by additional terms \cite{Belitsky00}.
This can be studied along the same lines as the three-quark operators,
although the algebra becomes more involved because of higher spins.
The most interesting part appears to be the mixing between quark-gluon
and three-gluon operators which for large $N$ can be interpreted as
describing the interaction between open and closed Heisenberg magnets.
It turns out that this mixing has rather peculiar features, which we
cannot discuss in detail in this review, but the outcome is that
the mixing can to a large extent be reduced to a few participating levels.
Identifying important degrees of freedom (at least for large $N$) one
can write down an approximate two-channel evolution equation for
the structure function $g_2$ in terms of transverse spin
densities \cite{BKM01}. This approximate evolution equation looks as
follows.

To  leading logarithmic accuracy, the structure function $g_2(x,Q^2)$ is
expressed as
\bea{g2final}
 g_2^{\rm LL}(x,Q^2) &=& g^{\rm WW}_2(x,Q^2)+
\frac{1}{2}\sum_q e^2_q
      \int_{x}^1 \frac{dy}{y} \,\Delta q^+_T(y,Q^2)\,,
\eea
where $g^{\rm WW}_2$ is the Wandzura-Wilczek
contribution, expressible in terms of the twist-two structure function
$g_1$. The transverse quark distribution
$\Delta q^+_T(x,Q^2)=\Delta q_T(x,Q^2)-\Delta q_T(-x,Q^2)$ is
defined as
\bea{Tq}
\lefteqn{\hspace*{-2cm}
\int^\alpha_{-\alpha} \!dv\,\langle p,s|
\left[(\alpha+v)\,S^+(\alpha,v,-\alpha)+(\alpha-v)S^-(\alpha,v,-\alpha)
\right]|p,s\rangle=}
\nonumber\\&&{}\hspace*{4cm}=
-i s_\perp
\!\int_{-1}^1\!
 \!dx\, {\rm e}^{2ix \alpha p_+}\,\, \Delta q_T(x,\mu^2)\,.
\eea
Here $|p,s\rangle$ is the polarized nucleon state with momentum $p_\mu$ and
polarization vector $s_\mu$. In the leading logarithmic approximation,
the gluon contribution only appears due to the QCD evolution, in
particular
\bea{stand}
&&Q^2\frac{d}{d Q^2}\Delta q^{+}_T(x;Q^2) =
\frac{\alpha_s}{4\pi}\int_{x}^1
\frac{dy}{y} \left[P^T_{qq}(x/y)\Delta q^{+}_T(y;Q^2)+
P^T_{qg}(x/y)\Delta g_T(y;Q^2)
\right],
\nonumber\\
&&Q^2\frac{d}{d Q^2}\Delta g_T(x;Q^2)=\frac{\alpha_s}{4\pi}\int_{x}^1
\frac{dy}{y} P^T_{gg}(x/y)\Delta g_T(y;Q^2)\,.
\eea
The gluon transverse spin density $\Delta g_T$ is defined
in terms of the three-gluon operators
\cite{BKM01a}
\bea{Tg}
 \lefteqn{\hspace*{-2cm}\int^\alpha_{-\alpha}\!\!\!\! \!dv\,\langle p,s|
\Big[(\alpha+v)\widetilde O(v,\alpha,-\alpha)
+2\alpha\, \widetilde O(\alpha,v,-\alpha) + (\alpha-v)\widetilde
O(\alpha,-\alpha,v)\Big] |p,s\rangle =} 
\nonumber\\&&{}\hspace*{4cm}=
2 s_\perp p_+\!
\!\int_{-1}^1\!\!\!\!
 \!dx\,\textrm{e}^{2ix \alpha p_+}\! x \Delta g_T(x,\mu^2)
\eea
with $x$ having the meaning of the Bjorken variable. The evolution kernels are
given by
\bea{otvet2}
 P^T_{qq}(x) &=&
\left[\frac{4C_F}{1-x}\right]_+
      +\delta(1-x)\left[C_F+\frac{1}{N_c}\left(2-\frac{\pi^2}{3}\right)\right]
      - 2C_F\,,
\nonumber\\
 P^T_{gg}(x) &=& \left[\frac{4 N_c}{1-x}\right]_+
 + \delta(1-x) \left[ N_c\lr{\frac{\pi^2}3-\frac13}\,-\frac23{N_f}\right]
\nonumber\\
 &&{}
+
  N_c\lr{\frac{\pi^2}3-2}+ N_c\ln\frac{1-x}{x}\lr{\frac{2\pi^2}{3}-6}\,,
\nonumber\\
 P^T_{qg}(x) &=& 
      4 N_f \left[ x - 2(1-x)^2\ln(1-x)\right]\,.
\eea
Here the first two expressions are accurate up to corrections of order
$\mathcal{O}(1-x)$ for $x\to 1$ and the third one  has the accuracy
$\mathcal{O}((1-x)^3)$. Since quark and gluon distributions in the nucleon
strongly decrease for $x\to 1$, the contribution of large $x$ to the splitting
functions is the most important one numerically.

\subsection{\it The Regge Limit of QCD}

In this section we shall describe another example in which conformal symmetry
plays a crucial r\^ole --- finding the asymptotic behavior of QCD scattering
amplitude in the high-energy (Regge) limit. In the BFKL
approach~\cite{BFKL,Lipatov85,L-review} partial waves of the scattering amplitude
satisfy a Bethe-Salpeter-like equation which involves the effective Hamiltonian
$\mathcal{H}_{\rm BFKL}$ acting on two-dimensional plane of transverse
coordinates. It turns out that this Hamiltonian has many features in common with
the evolution kernels considered in the previous Section. It is invariant under
conformal transformation on the two-dimensional (Euclidean) transverse plane and
is also related to integrable Heisenberg magnets.
In distinction with the previous case,
the conformal transformations act on the plane (and not on the line) and the
corresponding generators form the $SL(2,\mathbb{C})$ algebra (and not the
$SL(2,\mathbb{R})$ algebra). The origin of the $SL(2,\mathbb{C})$ symmetry and
integrability properties of the Hamiltonian  $\mathcal{H}_{\rm BFKL}$ remains
unclear.
However, they allow us to solve the
Schr\"odinger equation for this Hamiltonian  and obtain the high-energy
asymptotics of the scattering amplitudes.

Consider the elastic scattering of two colorless hadronic states in the Regge
limit, when their center-of-mass energy $s$ is much larger than their masses and
momentum transferred in the $t-$channel, $s\gg -t,\, m^2$. For the sake of
simplicity and in order for perturbative QCD to be applicable, one may choose
these states to be heavy onia, i.e. the bound states of heavy quark and heavy
antiquark with small transverse size $\sim 1/m\ll 1/\Lambda_{\rm QCD}$. To the
lowest order in the coupling constant the elastic scattering of two onia with the
momenta $p_1^\mu \simeq \sqrt{s/2} n^\mu$ and $p_2^\mu \simeq \sqrt{s/2}\bar
n^\mu$ occurs through the exchange by two gluons with momenta  $k^\mu$ and
$q^\mu-k^\mu$. Using the Sudakov parametrization $k^\mu = \alpha_1 p_1^\mu +
\alpha_2 p_2^\mu + k_\perp^\mu$ the four-dimensional integration over the gluon
momenta becomes $d^4k = (s/2)d\alpha_1d\alpha_2d^2k_\perp$. It can be shown that
at high energies the dominant contribution to the scattering amplitude
$\mathcal{A}(s,t)$ comes from the momentum region such that $k_\mu^2\approx
-k_\perp^2$ and $(k-q)^2\approx -(k-q)_\perp^2$ and the integrations over the
longitudinal momentum fractions $\alpha_1$ and $\alpha_2$ get factorised,
see~\cite{L-review}. As a consequence, the two-gluon scattering amplitude can be
written in the so-called impact factor representation
\bea{2-gluon-amp}
\mathcal{A}(s,t)= is \alpha_s^2
\int\frac{d^2k_\perp}{(2\pi)^2}\frac1{k_\perp^2(k-q)_\perp^2}
\Phi_1(k_\perp,q_\perp-k_\perp)\Phi_2(k_\perp,q_\perp-k_\perp)\,,
\eea
where the functions $\Phi_{1,2}(k_\perp,q_\perp-k_\perp)$ do not depend on the
energy $s$ and describe the coupling of two gluons to the onia states with momenta
$p_1$ and $p_2$, respectively.
Gauge
invariance implies that these functions vanish if momenta of one of the gluons
vanishes
$\Phi_{1,2}(k_\perp,q_\perp-k_\perp)\to 0$
for $k_\perp \to 0$ or $q_\perp-k_\perp \to 0$. The same condition ensures that
the amplitude (\ref{2-gluon-amp}) is infrared finite.
The two-gluon exchange amplitude (\ref{2-gluon-amp}) is of order
$\mathcal{A}(s,t)\sim s^1$ which corresponds to the constant cross section.
Going over to higher orders in perturbation theory,
one finds that the scattering amplitude receives corrections enhanced by
logarithms of the energy $\sim \alpha_s^2(\alpha_s \ln s)^m[1+\mathcal{O}(\alpha_s)]$.
The resummation of such corrections in the leading logarithmic approximation (LLA)
in energy gives rise to the famous result~\cite{BFKL,Lipatov85}
\beq{BFKL}
\mathcal{A}(s,t)\sim i\,s^{1+\frac{\alpha_s N_c}{\pi}4\ln 2}\,,
\eeq
known as the BFKL pomeron.

The resummation of terms $\sim (\alpha_s\ln s)^m$ is done by expanding the scattering
amplitude over the partial waves with the complex angular momentum
\bea{omega}
\mathcal{A}(s,t)=is \alpha_s^2 \int_{\delta-i\infty}^{\delta+i\infty} \frac{d\omega}{2\pi
i}\, s^\omega \widetilde{\mathcal{A}}(\omega,t)\,,
\eea
where the  integration contour goes  to the right of all singularities of
$\widetilde{\mathcal{A}}(\omega,t)$ on the complex $\omega-$plane.
In this way the expansion over $(\alpha_s\ln s)^m$ is traded for the expansion over
$(\alpha_s/\omega)^m$ with the coefficients that depend on $t$, and the
asymptotical behavior in Eq.\ (\ref{BFKL}) arises because the
resummed expression for $\widetilde{\mathcal{A}}(\alpha_s/\omega,t)$ develops
a nontrivial singularity for positive $\omega$ in the sum of all orders in $\alpha_s$.

To leading logarithmic accuracy the partial wave
$\widetilde{\mathcal{A}}(\alpha_s/\omega,t)$ is given by
\beq{2-vev}
\widetilde{\mathcal{A}}(\omega,t)
=\int\frac{d^2k_\perp}{(2\pi)^2}\int\frac{d^2k_\perp'}{(2\pi)^2}
\Phi_1(k_\perp,q_\perp-k_\perp)
T_\omega(k_\perp,k_\perp';q_\perp)
\Phi_2(k_\perp',q_\perp-k_\perp')\,,
\eeq
where $T_\omega(k_\perp,k_\perp';q_\perp)$ is the partial wave amplitude for the
scattering of two off-shell gluons with the virtualities $k_\perp^2$,
$(q_\perp-k_\perp)^2$, $k_\perp'{}^2$, $(q_\perp-k_\perp)'{}^2$ and vacuum
quantum numbers in the $t-$channel. Comparing Eq.~(\ref{2-vev}) with
(\ref{2-gluon-amp}) one finds the function $T_\omega$ which, in leading order, is
given by the product of two gluon propagators:
\bea{T0}
T_\omega(k_\perp,k_\perp';q_\perp)=\frac1{\omega}\left[
T^{(0)}(k_\perp,k_\perp';q_\perp) +\mathcal{O}(\alpha_s/\omega)\right],
\quad
T^{(0)}(k_\perp,k_\perp';q_\perp)=\frac{(2\pi)^2\delta(k_\perp-k_\perp')}
{k_\perp^2(q_\perp-k_\perp)^2}\,,
\eea
A nontrivial analysis which cannot repeat
here, see \cite{BFKL,Lipatov85,L-review},  
shows that the resummed
partial wave $T_\omega$ satisfies a Bethe-Salpeter like
equation
\bea{BS}
\omega\, T_\omega(k_\perp,k_\perp';q_\perp) = T^{(0)}(k_\perp,k_\perp';q_\perp)
+\frac{\alpha_s N_c}{\pi}\left[\mathcal{H}_{\rm BFKL}\otimes
T_\omega\right](k_\perp,k_\perp';q_\perp)\,,
\eea
where the integral operator $\mathcal{H}_{\rm BFKL}$ describes the interaction between
gluons which are propagating in the $t-$channel. It acts only on their two-dimensional
transverse momenta:
\bea{ker}
\lefteqn{\hspace*{-1cm}
\left[\mathcal{H}_{\rm BFKL}\otimes
T_\omega\right](k_\perp,k_\perp';q_\perp)=}
\nonumber\\
&=&\frac1{2\pi}\int\frac{d^2l_\perp}{(k-l)_\perp^2}
\bigg\{\left[k^2_\perp(q-l)_\perp^2+(q-k)_\perp^2l_\perp^2-(l-k)_\perp^2q_\perp^2
\right]T_\omega(l_\perp,k_\perp';q_\perp)
\nonumber
\\
&&{}-\left[\frac{k^2_\perp}{l^2_\perp+(k-l)_\perp^2}+\frac{(q-k)_\perp^2}
{(q-l)_\perp^2+(k-l)_\perp^2}\right]
k_\perp^2(q-k)_\perp^2T_\omega(k_\perp,k_\perp';q_\perp)\bigg\}.
\eea
Solving Eq.~(\ref{BS}) by iterations corresponds to the perturbation theory in
$(\alpha_s/\omega)$.
Instead, in order to identify the singularities of the partial wave in the complex $\omega-$plane one
has to solve Eq.\ (\ref{BS}) exactly. To this end it is convenient to go over
from the momentum to the impact parameter representation:
\bea{Four}
&&\Phi_a(k,q-k)=\int d^2 b_1 d^2 b_2\, \textrm{e}^{-i(kb_1)-i((q-k)b_2)}
\Phi_a(b_1,b_2)\,,
\\
&&T_\omega(k,k';q)\delta^{(2)}(q-q')=\int\prod_{k=1,2} d^2 b_k\, d^2b_k'\,
\textrm{e}^{i(kb_1)+i((q-k)b_2)-i(k'b_1')-i((q'-k')b_2')}
T_\omega(b_1,b_2;b_1',b_2')\,, \nonumber
\eea
where it is implied that all vectors belong to two-dimensional plane of
transverse coordinates (impact parameters) and we use the same notation for
functions in the momentum and the coordinate space.

The rewriting of (\ref{2-vev}) in the impact parameter space reads
\bea{A-Four}
\widetilde{\mathcal{A}}(\omega,t)&=&\int d^2 b_0 \, \textrm{e}^{i(qb_0)} \int
\prod_{k=1,2} d^2 b_k\, d^2b_k'\,
\Phi_1(b_1-b_0,b_2-b_0)T_\omega(b_1,b_2;b_1',b_2') )\Phi_2(b_1',b_2')
\nonumber
\\
&\equiv& \int d^2 b_0 \, \textrm{e}^{i(qb_0)} \langle \Phi(b_0)|
\mathbb{T}_\omega |\Phi(0)\rangle\,,
\eea
where
in the second line  the notation $\mathbb{T}_\omega$ was introduced for the operator
whose kernel is given by $T_\omega(b_1,b_2;b_1',b_2')$, defined by (\ref{ker}), (\ref{Four}).
As the result, the Bethe-Salpeter equation (\ref{BS}) takes a
simple operator form
\beq{BS-coor}
\omega\mathbb{T}_\omega=\mathbb{T}_\omega^{(0)} +\frac{\alpha_s N_c}{\pi}\,
\mathbb{H}_{\rm BFKL}\,\mathbb{T}_\omega\,.
\eeq
Solution of this equation can formally be written as
\beq{sol}
\mathbb{T}_\omega=\left(\omega-\frac{\alpha_s N_c}{\pi}\,
\mathbb{H}_{\rm BFKL}\right)^{-1}\mathbb{T}_\omega^{(0)}\, ,
\eeq
where from one concludes that the singularities of $\mathbb{T}_\omega$ in the
$\omega$-plane are determined by the eigenvalues of the BFKL operator
\beq{H-bfkl}
\left[\mathbb{H}_{\rm BFKL}\cdot \psi_\alpha\right](b_1,b_2)
=  E_{\alpha}\; \psi_\alpha(b_1,b_2)\,,
\eeq
with $\alpha$ numerating the solutions. The high-energy behavior of the
scattering amplitude is governed by the right-most singularity of
$\mathbb{T}_\omega$, which corresponds to the maximal eigenvalue ${\rm
max}_\alpha E_{\alpha}$. Eq.~(\ref{H-bfkl}) has the form of a Schr\"odinger
equation for the system of two interacting particles on the two-dimensional
plane. Such particles can be identified as reggeized gluons and the eigenstates
$\psi_\alpha(b_1,b_2)$ have the meaning of the wave function of the color-singlet
compound states built from two reggeized gluons.

The BFKL operator $\mathbb{H}_{\rm BFKL}$ has a number of remarkable properties
which allow one to solve the Schr\"odinger equation (\ref{H-bfkl})
exactly~\cite{Lipatov85,L-review,Lipatov90}. First of all, $\mathbb{H}_{\rm
BFKL}$ splits into the sum of two operators acting on the holomorphic and the
antiholomorphic coordinates  $b_{1,2}\to (z_{1,2},\bar z_{1,2})$, Eq.~(\ref{z}),
on the transverse plane:
\beq{HH}
 \mathbb{H}_{\rm BFKL}
 =\mathcal{H}_2 +\overline{\mathcal{H}}_2\,,\qquad
  \mathcal{H}_2 = \partial_{z_1}^{-1} \ln (z_{12})\, \partial_{z_1} +
  \partial_{z_2}^{-1} \ln (z_{12}) \,\partial_{z_2}
  + \ln (\partial_{z_1}\partial_{z_2})-2\Psi(1)\,,
\eeq
where $z_{12}=z_1-z_2$ and $\overline{\mathcal{H}}_2$ is given by the similar
expression in the $\bar z-$sector. Consider the generators of the holomorphic
$SL(2,\mathbb{C})$ transformations (\ref{laws}) corresponding to the field with
$h=\bar h=0$%
\footnote{We would like to stress that these quantum numbers correspond to
a scalar field in a two-dimensional field theory and are different from
transverse components of a physical gluon field. Because of this the
$SL(2,\mathbb{C})$ symmetry of $\mathbb{H}_{\rm BFKL}$ does not follow
immediately from the conformal symmetry of the QCD Lagrangian.}
\beq{c-gen}
\JJ_{k,-}=-\partial_{z_k}\,,\qquad \JJ_{k,0}=z_k\partial_{z_k}\,,\qquad
\JJ_{k,+}=z_k^2\partial_{z_k}\,,
\eeq
and the corresponding antiholomorphic  generators $\bar \JJ_{k,-}$, $\bar
\JJ_{k,0}$ and $\bar
\JJ_{k,+}$ given by similar expressions with $z_k$ replaced by $\bar z_k$,
 with $k=1,2$ enumerating particles. By inspection one finds
that $\mathbb{H}_{\rm BFKL}$ commutes with all two-particle generators
\beq{?-1}
 [\mathbb{H}_{\rm BFKL},\JJ_{1,a}+\JJ_{2,a}] =
 [\mathbb{H}_{\rm BFKL},\bar \JJ_{1,a}+\bar \JJ_{2,a}] = 0\,.
\eeq
with $a=+,-,0$ and, therefore, is invariant under the $SL(2,\mathbb{C})$
transformations of the transverse plane. This implies that $\mathbb{H}_{\rm
BFKL}$ only depends on the two-particle Casimir operators of the
$SL(2,\mathbb{C})$ group
\beq{SL2C-Casimir}
{\JJ}_{12}^2=-(z_1-z_2)^2\partial_{z_1}\partial_{z_2}\,,\qquad
\bar{{\JJ}}_{12}^2=-(\bar z_1-\bar z_2)^2\partial_{\bar z_1}
\partial_{\bar z_2}\,,
\eeq
and 
leads to $\mathcal{H}_2=\mathcal{H}_2({\JJ}_{12}^2)$ and $
\overline{\mathcal{H}}_2=\overline{\mathcal{H}}_2(\bar{{\JJ}}_{12}^2)$.
As a consequence, solutions of the Schr\"odinger equation (\ref{H-bfkl}) have to be eigenstates
of  the Casimir operators
\beq{eig-SL2C}
{\JJ}_{12}^2\psi_{n,\nu}=h(h-1) \psi_{n,\nu}\,,\qquad
\bar {\JJ}_{12}^2\psi_{n,\nu}=\bar h(\bar h-1) \psi_{n,\nu}\, .
\eeq
Here the pair of complex conformal spins
\beq{h}
h=\frac{1+n}2+i\nu \quad\mbox{and}\quad
\bar h=\frac{1-n}2+i\nu\,,
\eeq
with a nonnegative integer $n$ and real $\nu$ specify the irreducible (principal
series) representation of the $SL(2,\mathbb{C})$ group to which $\psi_{n,\nu}$
belongs to. The solutions to Eqs.\ (\ref{eig-SL2C}) read
\beq{wf-2}
\psi_{n,\nu}(b_1,b_2)
=\left(\frac{z_{12}}{z_{10}z_{20}}\right)^{(1+n)/2+i\nu}
\left(\frac{\bar z_{12}}{\bar z_{10}\bar z_{20}}\right)^{(1-n)/2+i\nu}\,,
\eeq
where $z_{jk}=z_j-z_k$ and $b_0=(z_0,\bar z_0)$ is the collective
coordinate, reflecting the invariance of $\mathbb{H}_{\rm BFKL}$ under
translations. We recall that the solution (\ref{wf-2}) defines the wave
function of the color-singlet compound state, which is built from two
reggeized gluons with the coordinates $b_1=(z_1,\bar z_1)$ and
$b_2=(z_2,\bar z_2)$. The integer $n$ fixes the two-dimensional Lorentz
spin of the state, real valued $\nu$ gives the scaling dimension
$\ell=1+2i\nu$, and the two-dimensional vector $b_0$ sets up the
center-of-mass coordinate of the state.

We conclude that, similar to the two-quark evolution equations discussed
in Sect.~4.1, the solution to the Schr\"odinger equation for the
compound states of reggeized gluons is uniquely fixed by conformal
invariance. We would like to stress the following important differences
between the two cases. The conformal $SL(2,\mathbb{R})$ symmetry of the
evolution equations is a consequence of the symmetry of the QCD
Lagrangian, but the origin of the $SL(2,\mathbb{C})$ symmetry of the
BFKL operator remains unclear. In the former case, the conformal group
acts along the line defined by the light-cone direction and conformal
spins of operators take positive (half)integer values. In the latter
case the conformal group acts independently along two ``orthogonal''
$z-$ and $\bar z-$directions on the two-dimensional Euclidean plane and
conformal spins of gluonic states take complex values specified in Eq.\
(\ref{h}).

To find the eigenvalues in Eq.\  (\ref{H-bfkl}), one substitutes
the wave function (\ref{wf-2}) into the Schr\"odinger equation (\ref{H-bfkl})
and uses the explicit form of the BFKL kernel
(\ref{ker}) leading to~\cite{BFKL,Lipatov85,L-review}
\beq{2-en}
E_{n,\nu}=2\Psi(1)-\Psi\left(\frac{n+1}2+i\nu\right)
-\Psi\left(\frac{n+1}2-i\nu\right)\,.
\eeq
Its maximal value $\textrm{max}\,E_{n,\nu}=4\ln 2$ corresponds to $n=\nu=0$, or
equivalently $h=\bar h=1/2$. It defines the position of the right-most
singularity of the partial wave amplitude and is translated into the BFKL
behavior of the scattering amplitude in the leading logarithmic approximation,
Eq.~(\ref{BFKL}). Due to accumulation of the energy levels $E_{n,\nu}$ around
$n=\nu=0$, the corresponding Regge singularity is not a pole but a square-root
cut so that (\ref{BFKL}) is modified by additional $(\alpha_s\ln
s)^{-1/2}-$factor.

Using (\ref{2-en}) one can reconstruct the operator form of the BFKL kernel
$\mathbb{H}_{\rm BFKL}$ on the representation space of the principal series of
the $SL(2,\mathbb{C})$ group
\beq{H-oper}
\mathbb{H}_{\rm BFKL}=\frac12\left[H(J_{12}) + H(\bar J_{12})\right]
\,,\qquad H(j)=2\Psi(1)-\Psi(j)-\Psi(1-j)\,,
\eeq
where, as before, the two-particle spins are defined as
${\JJ}_{12}^2=J_{12}(J_{12}-1)$ and $\bar{\JJ}_{12}^2=\bar J_{12}(\bar
J_{12}-1)$.
Notice that we already encountered similar Hamiltonians in Sect.~4.1 and later found in Sect.~4.3
that they give rise to complete integrability for the three-particle evolution
equations  for the helicity$-3/2$ baryon distribution amplitudes.
It turns out that the BFKL kernel (\ref{H-oper}) has {\it the same\/} hidden
symmetry. In order to see this, we have to go over to the states containing more than
two particles. In the Regge limit, this amounts to considering color-singlet compound
states built from three and more reggeized gluons, or briefly
$N-$reggeon states. In this terminology, the BFKL pomeron is a
particular example of the two-reggeon state.
In spite of the fact that the contribution of $N$-reggeon compound states is
suppressed by the factor $\alpha_s^{N-2}$ as compared with the leading term,
they have to be taken into account at very high energy and are expected \cite{BKP} to tame
a power rise with energy of the BFKL solution (\ref{BFKL}), which is in conflict with unitarity
constraints.

The contribution of the $N-$reggeon state to the scattering amplitude is given by
the expression similar to (\ref{A-Four}) in terms of the impact factors
$\Phi_{1,2}(b_1,\ldots,b_N)$ that take into account the coupling of $N$ gluons to
the onia states and the kernel $T_\omega(b_1,\ldots,b_N;b_1',\ldots,b_N')$
describing propagation of $N$ interacting reggeons in the $t-$channel. The
corresponding operator $\mathbb{T}_\omega$ acts on the two-dimensional
coordinates of $N$ reggeized gluons and satisfies the Bethe-Salpeter equation
(\ref{sol}), in which $\mathbb{T}_\omega^{(0)}$ corresponds to the product of $N$
gluon propagators and $\mathbb{H}_{\rm BFKL}$ is replaced by the $N-$reggeon
effective Hamiltonian
\beq{HN}
\mathbb{H}_N=-\frac1{2N_c}\sum_{1\le j< k\le N} t^a_j t^a_k
\left[H(J_{jk}) + H(\bar J_{jk})\right],
\eeq
where the sum goes over all pairs $(j,k)$ of reggeized gluons. Each term in the
sum has the color factor given by the product of the color charge of interacting
gluons. It is multiplied by the pair-wise Hamiltonian (\ref{H-oper}),
depending on the two-particle conformal spins
$J_{jk}(J_{jk}-1)=-z_{jk}^2\partial_{z_j}\partial_{z_k}$ and $\bar J_{jk}(\bar
J_{jk}-1)=-\bar z_{jk}^2\partial_{\bar z_j}\partial_{\bar z_k}$. The $N$-reggeon
states satisfy the Bartels-Kwiecinski-Praszalowicz equation
\bea{BKP}
[\mathbb{H}_N\cdot \psi_\alpha](b_1,\ldots,b_N) = E_\alpha
\psi_\alpha(b_1,\ldots,b_N)\,,&~~~&
\sum_{j=1}^N t^a_j\,\psi_\alpha(b_1,\ldots,b_N)=0\,,
\eea
which is a generalization of the Schr\"odinger equation (\ref{H-bfkl}) to
$N$-particle system with pair-wise interaction \cite{BKP}. Here the second relation
ensures that the total color charge of the state equals zero.

At $N=2$ and $N=3$ the color factor in (\ref{BKP}) can be reduced to a number
$t^a_1t^a_2=-N_c$ and $t^a_jt^a_k=-N_c/2$, respectively.
For $N\ge 4$ several color structures exist and finding the general solution to the
$N$-particle Schr\"odinger equation (\ref{BKP}) becomes an
extremely difficult task.
A considerable simplification occurs in the large-$N_c$ limit. In this case the relevant
Feynman diagrams have a topology of cylinder so that the interaction takes place
between nearest neighbors only, leading to $t^a_jt^a_k=-\delta_{j,k+1}N_c/2$. Thus, in
the multi-color limit the Hamiltonian (\ref{HN}) takes the form
\beq{H-multi}
\mathbb{H}_N\stackrel{N_c\to\infty}{=}\frac1{4}\sum_{1\le k \le N}
\left[H(J_{k,k+1}) + H(\bar J_{k,k+1})\right]\,,
\eeq
where $J_{N+1,N}=J_{1,N}$ and similar for $\bar J_{N+1,N}$ \cite{Lipatov90,Lipatov94}. Notice that
(\ref{H-multi}) is exact for $N=2$ and $N=3$.

Since the Hamiltonian (\ref{H-multi}) depends on the two-particle spins, it
commutes with the total conformal $SL(2,\mathbb{C})$ spin of the system
$\JJ_a=\sum_{k=1}^N \JJ_{k,a}$ and $\bar \JJ_a=\sum_{k=1}^N \bar \JJ_{k,a}$. This
allows one to impose the conformal constraints on its eigenstates
\beq{sl2c-const}
{\JJ}^2\, \psi_{n,\nu} =h(h-1) \psi_{n,\nu}\,,\qquad
\bar {\JJ}^2\, \psi_{n,\nu} =\bar h(\bar h-1) \psi_{n,\nu}\,,
\eeq
where the conformal spins $h$ and $\bar h$ are defined in Eq.\ (\ref{h}). For $N\ge 3$
these conditions do not fix the eigenstates uniquely. Similar to the situation
with the evolution equations for three quark operators,  additional constraints
follow from the analysis of symmetry properties of the Hamiltonian.

It turns out that the Hamiltonian (\ref{H-multi}) possesses a set of integrals of
motion $\mathbb{Q}_k$ and $\bar\mathbb{Q}_k$ (with $k=2,\ldots,N$)
\cite{Lipatov94,FK95}
\beq{Qk}
\mathbb{Q}_k = i^k \sum_{1\le j_1 < \ldots < j_k \le N} z_{j_1j_2} \ldots z_{j_{k-1} j_k}
z_{j_k j_1}\partial_{z_{j_1}}\ldots\partial_{z_{j_{k-1}}}\partial_{z_{j_k}}\,,
\eeq
$\bar \mathbb{Q}_k$ are given by the same expression with $z$ replaced by $\bar
z$. It is easy to verify that $\mathbb{Q}_2={\JJ}^2$ and $\bar
\mathbb{Q}_2=\bar
{\JJ}^2$. The holomorphic charges satisfy the following commutation relations
\beq{Q-rel}
 [ \mathbb{Q}_k,\mathbb{Q}_n] = [\mathbb{Q}_k, \JJ_a] = [\mathbb{Q}_k, \mathbb{H}_N] =0
\eeq
and similar relations hold in the antiholomorphic sector. The existence of the
integrals of motion implies that the Schr\"odinger equation (\ref{BKP}) is
completely integrable in the multi-color limit. Hence the Hamiltonian depends
only on the total set of conserved charges $\mathbb{Q}_2,\ldots,\mathbb{Q}_N$ and
their counterparts in the antiholomorphic sector. This allows one to replace
the eigenvalue problem (\ref{BKP}) by a simpler one of finding simultaneous
eigenstates for the set of conserved charges
\beq{Q-eq}
\mathbb{Q}_k\, \psi_{n,\nu,q}(b_1,\ldots,b_N) =
q_k\,  \psi_{n,\nu,q}(b_1,\ldots,b_N)\quad\mbox{with}\quad  k=2,\ldots,N\, ,
\eeq
supplemented by the dependence of the energy $E(q_2,\ldots,q_N)$ on the
integrals of motion. The quantization conditions for the
charges $q_k$ follow from the requirement that $\psi_{n,\nu,q}(b_i)$ is a
single-valued function on the two-dimensional plane, normalizable with respect to
the $SL(2,\mathbb{C})$ invariant scalar product.

We have seen that the evolution kernel for helicity$-3/2$ baryon operator
coincides with the Hamiltonian of the Heisenberg magnet with the spin operators
being the generators of the $SL(2,\mathbb{R})$ group. The $N$-reggeon
Hamiltonian (\ref{H-multi}) has the similar hidden symmetry in the multi-color
limit. Remarkably enough, for arbitrary $N$ this Hamiltonian coincides with the
Hamiltonian of the Heisenberg magnet with $N$ sites and the spin operators
belonging to the principal series representation of the $SL(2,\mathbb{C})$ group
\cite{FK95}. As follows from Eq.\ (\ref{c-gen}), ${\JJ}_k^2=0$ so that the single
particle spin equals zero. These features allow one to apply powerful methods of
integrable models to solve to the spectral problem (\ref{H-multi}). The current
situation
is the following. For $N=3$ the solution was
found in Refs.~\cite{JW,BLV}. For $N=4$ the spectrum was recently calculated by two
groups Refs.~\cite{KKM1,KKM2} and \cite{DeVega} with somewhat different results.
The spectrum of $5\le N\le 8$ states and the generalization to arbitrary $N$ were
considered in Ref.~\cite{KKM2}. We refer an interested reader to these papers for further
details.

\section{Conformal Symmetry Beyond the Leading Logarithms}

Beyond the leading logarithmic approximation conformal symmetry is certainly
broken by the trace anomaly of the energy-momentum tensor. In the present section
we analyse this breaking in detail and establish the following schematic
structure of the perturbative series for a generic quantity ${\cal Q}$
\bea{DefForConLim}
{\cal Q} = {\cal Q}^{\rm con} + \frac{\beta(g)}{g} \Delta{\cal Q}\,
,\quad\mbox{where}\quad
 \Delta{\cal Q} = \mbox{power\ series in\ } \alpha_s\,.
\eea
Here ${\cal Q}^{\rm con}$ is the result in the formal conformal limit, obtained
by setting the $\beta$-function to zero by hand. It has full symmetry of a
conformally invariant theory. The extra term $\Delta{\cal Q}$ can be
perturbatively evaluated in a power series with respect to $\alpha_s$ and
vanishes to leading order (LO). Note that the evaluation of the leading
${\mathcal O}(\alpha_s)$ contribution to $\Delta{\cal Q}$ requires only little
effort, since it is sufficient to calculate the $N_f$ proportional terms in
$\beta_0=11N_c/3 - 2N_f/3$ via quark bubble insertions.

The possibility of the separation of conformally symmetric and
$\beta$-proportional contributions is by no means trivial. In
particular, since the anomalous dimensions are scheme-dependent beyond
LO, ${\cal Q}^{\rm con}$ appears to be scheme-dependent as well and one
might expect that the representation in (\ref{DefForConLim}) can only be
valid in a certain scheme, if at all. We will indeed find that
perturbative corrections in general violate conformal Ward identities
and destroy the symmetry. The most important result is, however, that
the $n$-loop anomalous contribution to the dilatation and the
$n\!-\!1$-loop special conformal anomaly are related to each other and
they can {\it both} be removed simultaneously by a finite
renormalization. Neglecting the $\beta$-function, the conformal
covariance can, therefore, be restored in a special, conformal scheme.

This general result can be used in two different ways. In the conformal
scheme, mixing matrices involving operators with total derivatives
(alias full ER-BL kernels) are completely fixed (to all orders in
$\alpha_s$) by the diagonal entries. This implies that, e.g. two-loop
off-forward anomalous dimensions (or two-loop ER-BL kernel) in an {\it
arbitrary} scheme, say the modified MS ($\overline{\rm MS}$), can be
restored from their diagonal (forward) parts, making a finite
renormalization from the conformal to the $\overline{\rm MS}$ scheme.
This transition requires a calculation of the one-loop contribution to
the special conformal anomaly and is considerably simpler than the full
two-loop calculation. Another option is to abandon the MS-like schemes
completely and do the whole analysis in the conformal scheme. We will
give examples for both strategies.

The presentation is organized as follows.
In Sect.~\ref{SubSec-CreRel} we consider the so-called Crewther
relation in which case the above-mentioned difficulties are absent.
This relation has been tested to ${\mathcal O}(\alpha_s^3)$ by explicit
calculations and provides a beautiful illustration of the power of the
conformal symmetry.
In Sect.~\ref{SubSec-ConCovRenSch} we consider conformal Ward identities
in an interacting theory and show that the conformal covariance can be
restored (up to corrections in $\beta$) in a special renormalization scheme ---
we call it  conformal subtraction (CS) scheme --- that differs from a
MS-like scheme by a finite renormalization. In particular the conformal
operator product expansion, derived in Sect.~\ref{SubSec-COPE}, holds true
in the CS scheme beyond the leading order.
In Sect.~\ref{SubSec-AnoDimNLO} the leading order special conformal
anomaly is calculated and we demonstrate that
using this anomaly and imposing the conformal constraints one can
derive the NLO expressions for the matrices of anomalous dimensions of
{all} twist-two operators in a MS scheme and reconstruct the
corresponding evolution kernels.
{}Finally, in Sect.~\ref{SubSec-ConPreTwoPhoPro} we consider
the application of the conformal operator product expansion to
light-cone dominated two-photon processes.

\subsection{\it The Crewther Relation}
\label{SubSec-CreRel}

The Crewther relation \cite{Cre72} (see also \cite{AdlCalGroJac72})
predates QCD and provides  the relation between three quantities:
the Bjorken sum rule for the structure function of polarized deep inelastic
scattering
\bea{Bjo-SumRul}
\int_0^1 dx \left(g_1^{p}-g_1^{n} \right)(x,Q^2)
=  \frac{1}{6} \left| \frac{g_A}{g_V} \right| C_{\rm Bj}(\alpha_s(Q^2))  +
\mathcal{O}(1/Q^2)\,,
\eea
the isovector part of the ratio of the total cross-sections
\bea{R-rat}
R(s) = \frac{\sigma(e^+ e^- \to\mbox{hadrons})}{\sigma(e^+ e^- \to \mu^+ \mu^-)}
\eea
at large center-of-mass energy $s$ and the axial anomaly constant, which
governs the pion decay $\pi^0 \to \gamma \gamma$. It was later extended
to QCD in Ref.~\cite{Cre97}. In its present form, the
so-called generalized Crewther relation reads
\bea{Rel-Cre-gen}
C_{\rm Bj}(\alpha_s) D(\alpha_s)   = 1 +
\frac{\beta(g)}{g} \left[ {\rm power\ series\ in\ }\alpha_s \right]\,,
\eea
where $D=D_{\rm NS}$ is the flavor non-singlet part of the Adler $D$ function
\cite{Adl74}. It is related to the $R-$ratio by the dispersion relation
\bea{Rel-RAdl}
D(\alpha_s,Q^2) = Q^2 \frac{d}{d Q^2} \int_{4 m_\pi^2}^\infty ds \frac{Q^2
R(s)}{s(s+Q^2)}\,.
\eea
The Crewther relation (\ref{Rel-Cre-gen}) relates the perturbative expansion of
two coefficient functions and it is valid up to corrections $\mathcal{O}(1/Q^2)$.
In comparison with its original form \cite{Cre72}, Eq.\ (\ref{Rel-Cre-gen}) contains
an additional term proportional to the $\beta-$function%
\footnote{In pre-QCD times it was
believed that the conformal symmetry is only softly broken, i.e., the trace of
the energy momentum tensor would provide only power corrections.}. The Crewther
relation can be extended to the flavor singlet sector in which case it relates
the coefficient function entering the Gross-Llewellyn Smith sum rule, $C_{\rm
GLS}(\alpha_s)$, and the singlet part of the Adler $D-$function.
The validity of the Crewther relations in both the
flavor singlet and non-singlet sector has been verified  \cite{BroKat93} using
the available QCD results for $C-$ and $D-$functions up to the order
$\mathcal{O}(\alpha_s^3)$, \cite{LarVer91,GorKatLar91} (see also
\cite{BroLu95}--\cite{Rat96}).

\begin{figure}[t]
\unitlength1pt
\vspace{1.2cm}
\begin{minipage}[t]{4cm}
\begin{center}
\mbox{
\begin{picture}(100,60)
\put(0,0){\insertfig{4.3}{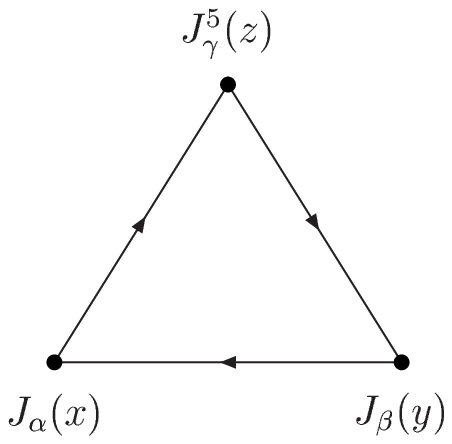}}
\put(135,18){\insertfig{3.6}{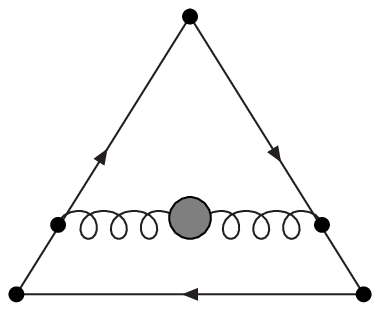}}
\put(55,0){\small (a)}
\put(180,0){\small (b)}
\end{picture}
}
\end{center}
\end{minipage}
\hfill
\parbox[b]{7cm}{
\caption{\label{Fig-DiaTri}
The VVA correlator (\ref{Def-AVV}) at tree level (a).
In (b) we display one contribution out of six to the
r.h.s.\ of the CWIs (\ref{VVA-CWI}), where the gray blob denotes the
renormalized trace anomaly.}
\vspace{0.7cm}
}
\end{figure}

To derive Eq.~(\ref{Rel-Cre-gen}), consider the three-point
correlation function
\bea{Def-AVV}
C_{\alpha\beta\gamma}(x-y,z-y;\alpha_s) =
\langle 0|
{\rm T} J_{\alpha}(x) J_{\beta}(y) J^5_{\gamma}(z) |0 \rangle \, ,
\eea
where $J_{\alpha}$ is the electromagnetic current and $J^5_{\gamma}$ is the
isovector axial-vector current. Neglecting quark masses the both currents are
conserved in QCD and have zero anomalous dimensions so that their scaling
dimensions are not altered by the interaction, $\ell=\ell^{\rm can} = 3$.

If the conformal symmetry in QCD were exact, the three-point correlation function
(\ref{Def-AVV}) would be uniquely fixed up to normalization \cite{Sch71} (see
Sect.~2.4.2). Moreover, since the scaling dimensions of the currents do not depend on
$\alpha_s$, $C_{\alpha\beta\gamma}$ would be given
by the lowest-order triangle diagram in Fig.~\ref{Fig-DiaTri}(a), up to
an overall normalization. Notice, however, that the
normalization of Eq.~(\ref{Def-AVV}) is protected by the $U(1)$ axial anomaly
condition
\bea{Nor-Rel-Cre}
\frac{\partial}{\partial z_\gamma} C_{\alpha\beta\gamma}(x-y,z-y;\alpha_s) =
\frac{1}{2\pi^2}\,
\epsilon_{\alpha\beta\mu\nu}
\frac{\partial}{\partial x_\mu} \frac{\partial}{\partial y_\nu}
\delta^{(4)}(x-z ) \delta^{(4)}(y-z ) \,.
\eea
According to the Adler-Bardeen theorem \cite{AdlBar69}, the r.h.s.\ of this
relation does not receive radiative corrections and, therefore, in the conformal
limit, $C_{\alpha\beta\gamma}$ coincides with the lowest-order result that we
denote as $\Delta_{\alpha\beta\gamma}$ (see Fig.\ \ref{Fig-DiaTri}(a))
\bea{Cre-Sol-ConInvThe}
C_{\alpha\beta\gamma}(x-y,z-y) = \Delta_{\alpha\beta\gamma}(x-y,z-y)\,.
\eea
In QCD, Eq.~(\ref{Cre-Sol-ConInvThe}) is modified because of the non-vanishing
$\beta-$function. To understand the modifications, note that the currents
$J_{\alpha}$ and $J_{\mu}^5$ are transformed under conformal transformations as
primary fields of spin$-1$ and scaling dimension $\ell=3$. Hence
$C_{\alpha\beta\gamma}$ obeys the conformal Ward identities (\ref{CWI-D}) and
(\ref{CWI-K}):
\bea{VVA-CWI}
\left[{\cal D}_x +{\cal D}_y+{\cal D}_z\right]
C_{\alpha\beta\gamma}(x-y,z-y) &\!\!\! =&\!\!\! \frac{\beta(g)}{2g}
\langle 0|
{\rm T} J_{\alpha}(x) J_{\beta}(y) J^5_{\gamma}(z)
\int d^d\! w\,  [G_{\mu\nu}^a G^{a\mu\nu} ](w)|0 \rangle \, , \\
\left[{\cal K}^{\mu}_x +{\cal K}^{\mu}_y +{\cal K}^{\mu}_z\right]
C_{\alpha\beta\gamma}(x-y,z-y) &\!\!\! =&\!\!\! \frac{\beta(g)}{2g}
\langle 0|
{\rm T} J_{\alpha}(x) J_{\beta}(y) J^5_{\gamma}(z)
\int d^d\! w\, 2 w^{\mu} [G_{\mu\nu}^aG^{a\mu\nu} ](w)|0 \rangle
\nonumber
\,.
\eea
Here we introduced a shorthand notation for the differential operators
that enter in the \ l.h.s.\ of (\ref{CWI-D}) and (\ref{CWI-K}):
\bea{DefDandK}
{\cal D}_x  = \lsc + x\cdot\partial_x \quad\mbox{and}\quad
{\cal K}_x^\mu = 2 x^\mu\, x\cdot\partial_x -x^2 \partial^\mu_x+2 \ell\, x^\mu
   - 2 x_\nu\Sigma^{\mu\nu}
\eea
with $\ell=3$ and $\Sigma_{\mu\nu}$ being the operator of spin rotations for a
vector field, Eq.~(\ref{spin1}). $[G_{\mu\nu}^aG^{a\mu\nu}]$ denotes the operator
renormalized in a MS-like scheme.

{}For $\beta(g)=0$ one can verify that the solution to the conformal Ward
identities (\ref{VVA-CWI}), supplemented by the additional condition
(\ref{Nor-Rel-Cre}) is given by Eq.\ (\ref{Cre-Sol-ConInvThe}). For $\beta(g)\neq
0$ the anomalous terms entering the r.h.s.\ of Eqs.~(\ref{VVA-CWI}) are described
by diagrams such as in Fig.\ \ref{Fig-DiaTri}(b). They modify the general
solution to the system of two inhomogeneous differential equations
(\ref{VVA-CWI}) by terms proportional to the $\beta-$function
\bea{VVA--Sol}
C_{\alpha\beta\gamma}(x-y,z-y;\alpha_s) =
\Delta_{\alpha\beta\gamma}(x-y,z-y) +
\frac{\beta(g)}{g}
\overline{\Delta}_{\alpha\beta\gamma}(x-y,z-y;\alpha_s)\, ,
\eea
where $ \overline{\Delta}_{\alpha\beta\gamma}=\mathcal{O}(\alpha_s)$. Since this
addition has still to satisfy the anomaly condition (\ref{Nor-Rel-Cre}),
$\overline{\Delta}_{\alpha\beta\gamma}$ must have a vanishing derivative with respect
to $z_\gamma$ and also respect the vector current conservation.

On the other hand,  the correlation function (\ref{VVA--Sol}) can be calculated
at short distances using the operator product expansion. The idea
\cite{Cre72,AdlCalGroJac72,Cre97} is to consider the non-symmetric limit with
$|x-y| \ll |z-y|$, so that one has to expand the product of two electromagnetic
currents $J_{\alpha}(x) J_{\beta}(y)$ as the first step. Choosing $y=0$ for
simplicity we obtain
\bea{ShoDisExpV}
{\rm T} J_{\alpha}(x) J_{\beta}(0) \simeq C^{\rm R}_{\alpha\beta}(x) I + C^{{\rm
K}}_{\alpha\beta\mu}(x)  J^{5\mu}(0) + \cdots,
\eea
where the coefficient functions $C^{\rm R}_{\alpha\beta}(x)$ and $C^{{\rm
K}}_{\alpha\beta\mu}(x)$ define the perturbative expansion of $C_{\rm
Bj}(\alpha_s)$ and $D(\alpha_s)$, Eqs.~(\ref{Bjo-SumRul}) and (\ref{Rel-RAdl}),
respectively. Inserting this expansion into the
correlation function (\ref{Def-AVV}) we find that the contribution
of the unit operator in (\ref{ShoDisExpV}) vanishes. Making use of the operator
product expansion for $z\to 0$ we obtain
\bea{ShoDisExpAA}
{\rm T} J^5_{\mu}(0) J^5_{\gamma}(z) \simeq C^{{\rm R}^\prime}_{\mu\gamma}(z)  I
+ \cdots\, ,
\eea
so that finally
\bea{Res-Cre}
C_{\alpha\beta\gamma}(x,z) \simeq
 C^{{\rm K}\ \mu}_{\alpha\beta}(x) C^{R'}_{\mu\gamma}(z)\,,\qquad x,z
\to 0\,,
 \qquad  x \ll z\,.
\eea
Notice that $C^{{\rm R}^\prime}_{\mu\gamma}(z)$
coincides with the isovector part of the correlator $\langle 0|
\textrm{T} J_{\alpha}(x) J_{\beta}(0)|0\rangle=C^{\rm R}_{\alpha\beta}(z) $ and,
therefore, it contributes to the flavor non-singlet part of the $D-$function,
Eq.~(\ref{Rel-Cre-gen}). Equating the conformal prediction
(\ref{VVA--Sol}) and
the OPE result (\ref{Res-Cre}) one obtains nontrivial constraints on the
perturbative expansions of the coefficient functions $C^{{\rm
R}^\prime}_{\mu\gamma}(z)$ and $C^{{\rm K}\ \mu}_{\alpha\beta}(x)$.
Going over to the momentum space one finds \cite{Cre97} that
Eq.~(\ref{VVA--Sol}) and Eq.~(\ref{Res-Cre}) give rise to the expressions entering
the\ r.h.s.\ and the \ l.h.s. of the Crewther relation (\ref{Rel-Cre-gen}),
respectively.  Repeating the
similar analysis in the opposite limit $z\ll x$ with $x,z\to 0$ one can
derive the Crewther relation in the flavor singlet sector.
Since $\overline{\Delta}_{\alpha\beta\gamma}=\mathcal{O}(\alpha_s)$, the
perturbative series on the r.h.s.\ of Eq.~(\ref{Rel-Cre-gen}) starts at order
$\alpha^2_s$.

It is worth mentioning that within the so-called Brodsky-Lepage-Mackenzie
scale setting prescription \cite{BroLepMac83} in which the terms in $\beta$
are absorbed into the redefinition of the scale of the running coupling,
the generalized Crewther relation (\ref{Rel-Cre-gen}) takes the same
form as in the conformal invariant theory \cite{BroGabKatLu95}:
\bea{Rel-Cre-gen-BLM}
C(Q)  D(\overline{Q}^\ast) =
\left[1-\hat \alpha_C(Q)\right]\left[1+\hat \alpha_D(\overline{Q}^\ast)\right]
=1 \, ,
\eea
where $\hat\alpha_C$ and $\hat\alpha_D$ are the corresponding effective charges
satisfying a geometrical commensurate {\em single} scale relation.
Eq.~(\ref{Rel-Cre-gen-BLM}) has been verified using the known
$\mathcal{O}(\alpha_s^3)$ results, but a general proof has not been given so far.

\subsection{\it Conformally Covariant Renormalization Scheme}
\label{SubSec-ConCovRenSch}

The case of the Crewther relation is special in that all relevant
operators are protected by exact symmetries and are not renormalized. In
the general situation the transformation properties of conformal
operators can easily be spoiled by renormalization and this is indeed
what happens in the usual $\overline{\rm MS}$ scheme. For the sake of
simplicity we consider the conformal tower of twist-two quark-antiquark
operators in the flavor non-singlet sector
\bea{RedConOpe}
{\ccal O}_{nl} = (i \partial_+)^{l-n}\, {\mathbb Q}_n^{1,1}(0)\, ,
\eea
where ${\mathbb Q}_n^{1,1}(0)$ is defined in Eq.~(\ref{co-19}) and $l
\ge n$ is the total number of derivatives. For a moment let us take the
formal conformal limit and set the $\beta-$function to zero. Conformal Ward
identities (CWIs) for the Green function made of the renormalized operator
$[{\ccal O}_{nl}]$ (here and below $[\ldots]$ stands for the renormalized
quantity in a MS-like scheme) and arbitrary number $N$ of fundamental fields
$\langle[{\ccal O}_{nl}] {\cal X}_N\rangle\equiv
\langle 0|[{\ccal O}_{nl}]\Phi(x_1)\ldots \Phi(x_N)|0\rangle$ have the
following general form (cf. Eqs.~(\ref{DCWI-QCD}) and
(\ref{SCCWI-QCD})):
\bea{CWI-confOpe}
\sum_{i=1}^N {\cal D}_i\, \langle [{\ccal O}_{nl}]  {\cal X}_N  \rangle &=&
    -\sum_{m=0}^{n}\left[ \lsc_l^{\rm can}\delta_{nm}+
    \hat \gamma_{nm}\right] \langle [{\ccal O}_{ml}] {\cal X}_N \rangle,
\\
\label{CWI-confOpe-sc}
\sum_{i=1}^N {\cal K}_i^- \,\langle [{\ccal O}_{nl}]  {\cal X}_N \rangle &=&
    i\sum_{m=0}^{n} \left[ a(n,l)\delta_{nm} +\hat\gamma^c_{nm}(l)\right]
    \langle [{\ccal O}_{ml-1}] {\cal X}_N \rangle,
\eea
where the gauge-fixing and ghost terms are not displayed, since they do not
contribute to physical quantities. Here ${\cal D}_i$ and ${\cal K}_i^-= \bar
n_\mu {\cal K}_i^\mu$ denote the differential operators (\ref{DefDandK}) acting
on the primary fields in the monomial ${\cal X}_N$, with the scaling dimensions
of fields $\ell_\Phi=
\ell_\Phi^{\rm can} +
\gamma_\Phi$. The terms $\sim \delta_{nm}$ on the r.h.s.\ of Eqs.~(\ref{CWI-confOpe}),
(\ref{CWI-confOpe-sc}) originate from the same differential operators acting on
${\ccal O}_{nl}$ which has the Lorentz spin $l+1$ and the canonical dimension
$\ell_l^{\rm can}=2\lsc_\psi^{\rm can}+l= 3+l$. The coefficient $a(n,l)$ is equal
to
\bea{Def-aMat}
a(n,l)= -2(l-n) (l-n+ 2j_n -1) = -2(l-n) (n+l+3) = 2(n-l) (\lsc_n^{\rm can}+l)
\,,
\eea
where in the first expression $j_n=(\lsc_n^{\rm can}+n+1)/2$ is the (canonical)
conformal spin, cf. Eq.~(\ref{traOnk}).

The remaining entries are the anomalous-dimension matrix $\hat\gamma =
\left\{ \gamma_{nm}\right\}$ and the so-called special-conformal anomaly
matrix $\hat\gamma^c(l) = \left\{\gamma^c_{nm}\right\}$, cf.\ Eqs.\
(\ref{def-delaAno}) and (\ref{def-scAno}). They both are induced by the
trace anomaly and have the standard perturbative expansion
\bea{AnoDim-PerExp}
\hat\gamma(\alpha_s)= \frac{\alpha_s}{2\pi}\hat\gamma^{(0)} +
\frac{\alpha_s^2}{(2\pi)^2}\hat\gamma^{(1)} + \mathcal{O}(\alpha_s^3)\,
\quad\mbox{and}\quad
\hat\gamma^c(\alpha_s)= \frac{\alpha_s}{2\pi}\hat\gamma^{c(0)} +
\mathcal{O}(\alpha_s^2)\, .
\eea
We recall that the dilatation Ward identity (\ref{CWI-confOpe}) is nothing else
but the Callan-Symanzik renormalization group equation and what it tells us is
that conformal operators in general can mix with the operators with the same
dimension and lower conformal spin. Due to Poincar\'e invariance, the both
matrices $\hat\gamma$ and $\hat\gamma^c$ are in general triangular and, in
addition, $\hat\gamma$ is independent on the Lorentz spin $l+1$. At the same
time, since the special conformal transformations do not commute with the
Poincar\'e transformations ($[{\bf K}_\alpha,{\bf M}_{\beta\gamma}]\neq 0$ and
$[{\bf K}_\alpha,{\bf P}_{\beta}]\neq 0$), both $a(n,l)$ and $\hat\gamma^c(l)$
depend on $l$. The conformal invariance is broken if the conformal anomaly
matrices $\hat\gamma$ and $\hat\gamma^c(l)$ are not diagonal. For instance, if
the special conformal anomaly matrix $\hat\gamma^c(l)$ has nonzero off-diagonal
terms, then the operator $[{\ccal O}_{nn}]$ is not annihilated by ${\cal K}_-$
and its variation is expressed in terms of the operators $[{\ccal O}_{mn-1}]$
with $m \le n-1$.

The crucial point is that the two conformal anomalies, $\hat\gamma$ and
$\hat\gamma^c(l)$, are not independent. The relation between the two matrices is
imposed by the collinear conformal algebra. Using
\bea{Emp-ConComRel}
\sum_{i=1}^N{\cal P}_{i+} \langle \left[{\ccal O}_{nl}\right] {\cal X}_N \rangle =
i \langle \left[{\ccal O}_{nl+1}\right] {\cal X}_N \rangle\, ,
\qquad
\sum_{i=1}^N{\cal M}_{i-+} \langle \left[{\ccal O}_{nl}\right] {\cal X}_N \rangle =
-(l+1) \langle \left[{\ccal O}_{nl}\right] {\cal X}_N \rangle\, ,
\eea
and applying the identities
\bea{conCon}
[{\cal D},{\cal K}_-] = {\cal K}_-\,,  \quad
[{\cal K_-},{\cal P}_+] = -2({\cal D}+{\cal M}_{-+})\,
\eea
to $\langle[{\ccal O}_{nl}] {\cal X}_N\rangle$ one obtains the desired
constraints. They consist of the commutator relation
\bea{conf-constr-KD1}
\left[\hat a(l)+\hat{\gamma}^c(l;\alpha_s)
,\hat{\gamma}(\alpha_s)\right]=0\,   \ \mbox{where}\ \
\hat a(l) = \left\{a(n,l) \delta_{nm}\right\}\,
\eea
and hence
\bea{conf-constr-KD}
2(n-m)(n+m+3)\gamma_{nm}(\alpha_s)=
\left[\hat{\gamma}(\alpha_s),
\hat{\gamma}^c(l;\alpha_s)\right]_{nm} ,\ \ n> m\, ,
\eea
between the anomalous-dimension and special-conformal anomaly matrices,
and the relation
\bea{conf-constr-KP}
\hat{\gamma}^c(l+1;\alpha_s)-\hat{\gamma}^c(l;\alpha_s)=
-2\hat{\gamma}(\alpha_s)\, ,
\eea
which determines the spin dependence of the special conformal anomaly.
Eq.~(\ref{conf-constr-KP}) ensures that the $l$ dependence in the commutator
relation (\ref{conf-constr-KD}) is spurious and cancels out.

The constraint (\ref{conf-constr-KD1}) immediately tells us that the anomalous
dimension matrix is diagonal in LO, since its perturbative expansion gives
\bea{conf-constr-KD-LO}
\left[\hat{a}(l),\hat{\gamma}^{(0)}\right]_{nm}=
2(n-m)(n+m+3)\gamma^{(0)}_{nm} =0\,, \qquad n>m\,.
\eea
Beyond this approximation one realizes from Eq.\ (\ref{conf-constr-KD}) that the
special conformal anomaly induces the off-diagonal entries of the anomalous
dimension matrix, starting $\mathcal{O}(\alpha_s^2)$.

It is instructive to take a closer look at what is happening to order
${\cal O}(\alpha_s)$. To this accuracy $\gamma_{nm}^{(0)}=\gamma_n^{(0)}
\delta_{nm}$ and the dilatation covariance is preserved. In contrast,
the covariance with respect to the special conformal transformation is
lost, i.e., the special conformal anomaly matrix $\hat\gamma^{c(0)}$ is
non-diagonal in a general renormalization scheme. The reason for this is
that in distinction with one-loop anomalous dimension $\gamma_n^{(0)}$,
which is induced by UV-divergent parts of Feynman diagrams, the special
conformal anomaly receives contribution also from UV finite parts.
Notice, however, that the off-diagonal entries of $\hat\gamma^{c(0)}(l)$
can be removed and, as a consequence, the special conformal covariance
can be restored by performing a finite renormalization
\bea{FinRenConOpe}
 {\ccal O}_{nl}^{\mbox{\tiny CS}}&=&
    [{\ccal O}_{nl}] + \frac{\alpha_s}{2\pi}
        \sum_{m=0}^{n-2}
    \frac{\gamma^{c(0)}_{mn}}{a(m,n)}
    [{\ccal O}_{ml}] + O(\alpha_s^2).
\eea
This transformation defines the {conformal subtraction} (CS) scheme. Indeed,
inserting Eq.~(\ref{FinRenConOpe}) in (\ref{CWI-confOpe-sc}) and taking into
account the identity $a(n,l)= a(n,m)+ a(m,l)$ we find that the operators ${\ccal
O}_{nl}^{\mbox{\tiny CS}}$ do not mix under infinitesimal special conformal
transformations and satisfy the Ward identity
\bea{CWI-confOpe-LO-true-sc}
\sum_{i=1}^N {\cal K}_i^- \,
\langle {\ccal O}_{nl}^{\mbox{\tiny CS}} {\cal X}_N  \rangle =
    i\left[
    a(n,l)+ \frac{\alpha_s}{2\pi} \gamma^{c(0)}_{nn}(l)\right]
    \langle {\ccal O}_{nl-1}^{\mbox{\tiny CS}} {\cal X}_N \rangle+ O(\alpha_s^2). 
\eea
On the other hand, the finite renormalization (\ref{FinRenConOpe}) does
not affect the dilatation Ward identity (\ref{CWI-confOpe}) to order
$\alpha_s$:
\bea{CWI-confOpe-LO-true}
\sum_{i=1}^N {\cal D}_i\, \langle {\ccal O}_{nl}^{\mbox{\tiny CS}} {\cal X}_N  \rangle =
    -\left[
    \lsc_l^{\rm can}+\frac{\alpha_s}{2\pi}\gamma^{(0)}_{n}
    \right]
\langle {\ccal O}_{nl}^{\mbox{\tiny CS}} {\cal X}_N \rangle + O(\alpha_s^2),
\eea
As we will show below in Eqs.\ (\ref{spconam})  and (\ref{CWI-scBre2}), the
diagonal matrix elements $\gamma^{c(0)}_{nn}(l)$ are related to the anomalous
dimensions $\gamma^{(0)}_n$ as $\gamma^{c(0)}_{nn}(l)=2(n-l)\gamma^{(0)}_n$.
Therefore, in both the dilatation and the special conformal Ward identities,
Eqs.~(\ref{CWI-confOpe-LO-true}) and (\ref{CWI-confOpe-LO-true-sc}), the
conformal symmetry {\em breaking} terms, containing one-loop anomalous dimension,
can be {\em absorbed} through the redefinition of the scaling dimension
\bea{ShiCanDimOPe}
 \lsc_n^{\rm can} &\Rightarrow& \lsc_n(\alpha_s) =
\lsc_n^{\rm can}+\gamma_n(\alpha_s)\,,
\\
\nonumber
 a(n,l)&\Rightarrow&  a(n,l) + 2 (n-l) \gamma_n(\alpha_s) = 2
(n-l)(\lsc_n(\alpha_s)+l) \,.
\eea
Thus, the conformal properties of the renormalized operators (\ref{FinRenConOpe})
are restored at leading order. As in a free theory, the operators
(\ref{FinRenConOpe}) form the irreducible conformal tower (\ref{co-4}) but their
conformal spin is modified by the anomalous dimension term
$j_n=(\lsc_n(\alpha_s)+n+1)/2$.

Beyond the leading order, the commutator relation (\ref{conf-constr-KD})
tells that the special conformal anomaly gives rise to off-diagonal
entries in the anomalous dimension matrix. Inserting the decomposition
of the anomalous dimension matrix in the diagonal and off-diagonal parts
\bea{Dec-AnoDim}
\hat{\gamma} = \hat{\gamma}^{\rm D} + \hat{\gamma}^{\rm ND}\,
,\quad \mbox{with}\quad
\hat{\gamma}^{\rm D} = \bigg\{ \gamma_n \delta_{nm}\bigg\}
\quad \mbox{and}\quad
\hat{\gamma}^{\rm ND} = \bigg\{ \gamma_{nm}\, \theta(n > m)\bigg\} \, ,
\eea
into Eq.\ (\ref{conf-constr-KD}) leads to a recurrence relation for
$\hat\gamma^{\rm ND}$, which can be solved perturbatively
\bea{adcodt}
\hat{\gamma}^{\rm ND} = - {{\cal G} \over
                          \hat{1} + {\cal G}} \hat{\gamma}^{\rm D}
= -{\cal G} \hat{\gamma}^{\rm D} +
       {\cal G}^2 \hat{\gamma}^{\rm D}  -  \cdots, \ \ \mbox{with}\ \
{\cal G}\hat{A}:= \left\{
     \frac{\theta(n>m)}{a(n,m)}
\left[\hat{\gamma}^c(l)
,\hat{A}\right]_{nm}
\right\}\,
\eea
by an expansion in powers of $\hat{\gamma}^{\rm c}(\alpha_s)$
\cite{Mue94}. The $l$ dependence appearing here on the r.h.s.\ is
spurious.

The conformally covariant operators ${\ccal O}_{nl}^{\mbox{\tiny CS}}$ are
obtained from those in a MS-like scheme by the rotation 
\bea{transformation}
{\ccal O}_{nl}^{\mbox{\tiny CS}}=\sum_{m=0}^{n} B^{-1}_{nm} [{\ccal O}_{ml}]\,,
\eea
where the matrix $\hat{B}=\{B_{nm}\}$ is defined by the requirement that
it diagonalizes the anomalous dimension matrix $\hat \gamma$:
$\hat{\gamma}^{\rm D} = \hat{B}^{-1}\hat{\gamma}\hat{B}$. It is given by
the following expansion \cite{Mue94}
\begin{eqnarray}
\label{sol-transformation-1}
\hat{B}=
    \frac{\hat 1}{{\hat 1}-{\cal L}\hat{\gamma}^{\rm ND}}
    ={\hat 1}+{\cal L}\hat{\gamma}^{\rm ND}+
     {\cal L}\left(\hat{\gamma}^{\rm ND}{\cal L}\hat{\gamma}^{\rm
    ND}\right)+\cdots \ \ \mbox{with} \ \
{\cal L}\hat{A}:= \left\{
                  -\theta(n>m){A_{nm}\over \gamma_n-\gamma_{m}} \right\} \,.
\eea
Inserting Eq.\ (\ref{adcodt}) into Eq.\ (\ref{sol-transformation-1}),
one finds after some algebra that the diagonal anomalous dimension
matrix cancels out, so that the rotation matrix only depends on the
special conformal anomaly \cite{Mue97a}:
\bea{blefdt-1}
\hat{B} = {\hat{1} \over \hat{1}+{\cal J}\hat{\gamma}^c}
   = \hat{1}-{\cal J}\hat{\gamma}^c + {\cal J}(\hat{\gamma}^c
            {\cal J}\hat{\gamma}^c)-\cdots, \ \ \mbox{with}\ \
{\cal J}\hat{A}:=
                \left\{ \theta(n>m) {A_{nm}\over a(n,m)}\right\}\,  .
\eea
It remains to be shown that the operator ${\ccal O}_{nl}^{\mbox{\tiny CS}}$ is
covariant with respect to the special conformal transformations. To this end we
apply the transformation (\ref{transformation}) to the special conformal Ward
identity (\ref{CWI-confOpe-sc}). Hence ${\ccal O}_{nl}^{\mbox{\tiny CS}}$ appears
on the l.h.s.\ while the r.h.s.\ reads
\bea{DiaSpeConAno}
\hat{B}^{-1} \left[\hat{a}(l) + \hat{\gamma}^c(l) \right]\hat{B}  =
 \hat{a}(l) + \hat{B}^{-1} [\hat{a}(l),\hat{B}] +
\hat{B}^{-1} \hat{\gamma}^c(l)\hat{B}
= \left\{ 2 (n-l) (n+l+3+\gamma_n) \delta_{nm} \right\}
\, .
\eea
Here the identity $[\hat a(l),\hat{B}]_{nm} =
-\{\hat\gamma^c(n)\hat B\}_{nm}$ (see Eq.\ (A5) of Ref.\
\cite{Mue97a})
has been used. This proves that the special conformal anomaly
(\ref{DiaSpeConAno}) in the CS scheme is indeed diagonal and has the
form (\ref{ShiCanDimOPe}). The conformal spin of the operators is
modified by the anomalous dimensions and the operator with the highest
conformal weight is annulled by the collinear special conformal
transformation.

We conclude that a CS scheme exists in which {covariance} of conformal operators
holds true in the formal conformal limit. Beyond the conformal limit the
constraint (\ref{conf-constr-KD}) is modified by a $\beta$ proportional term
while the condition (\ref{conf-constr-KP}) remains unchanged. Besides the
commutator relation (\ref{conCon}) and the conformal Ward identities (see
(\ref{SCCWI-QCD})), the derivation of the full constraint requires the
renormalization group equation for the trace anomaly. The derivation is tricky
and all details for the scalar theory in dimensional regularization are given in
\cite{Mue91a}. The analogous calculation for the gauge field theories leads as
expected to the absence of any explicit gauge dependence in the commutator
relation \cite{MuePHD,BelMue98a,BelMue98c}:
\bea{conf-constr-KD-full}
\left[\hat{a}(l)+\hat{\gamma}^c(l)
+2{\beta(g)\over g}\hat{b}(l)
,\hat{\gamma}\right]=0\, ,
\qquad \hat b(l)= \bigg\{ b_{nm}(l,3/2) \bigg\} \, .
\eea
The matrix  $b_{nm}(l,\nu) $ appearing here is defined by
the expansion of
\beq{Def-DefEqu-Mat-b}
\left[2 l-(2u-1)\frac{d}{du}\right] C_n^\nu(2u-1)
= \sum_{n=0}^m  b_{nm}(l,\nu) C_m^\nu(2u-1)
\eeq
in terms of Gegenbauer polynomials and reads
\bea{Def-Mat-b}
b_{nm}(l,\nu) =
2\theta(n\ge m) \left\{(l+m+\nu)\delta_{nm} -
\left[1+(-1)^{n-m} \right](m+\nu)\right\}\, .
\eea

The solution to the constraint (\ref{conf-constr-KD-full}), replacing
$\hat{\gamma}^c(l)$ by $\hat{\gamma}^c(l)+2\beta(g)\hat{b}(l)/g$ in Eq.\
(\ref{adcodt}), is now the off-diagonal part of $\hat\gamma$ in the {\em full
perturbative theory}. Knowing the special conformal anomaly at $n$th order in
$\alpha_s$ allows us to predict the off-diagonal entries to order $n+1$. In NLO
they read
\bea{gamND-NLO}
\gamma^{{\rm ND}(1)}_{nm} = \frac{\gamma_n^{(0)}-\gamma_m^{(0)}}{a(n,m)}
 \left(\gamma^{c(0)}_{nm} - \beta_0 b_{nm} \right)\quad\mbox{with}\quad
\beta_0= \frac{11}{3} N_c - \frac{2}{3} N_f \, ,
\eea
while the evaluation of $\gamma^{c(0)}_{nm}$ remains.

In the next Section we explain the calculation of the LO special conformal
anomaly, give the results for both the flavor non-singlet and singlet sectors,
and also consider the reconstruction of the NLO evolution kernels based on the
relation in Eq.~(\ref{gamND-NLO}).

\subsection{\it Evaluation of Conformal Anomalies and Evolution Kernels}
\label{SubSec-AnoDimNLO}
%
\begin{figure}[t]
\unitlength1pt
\vspace{-2.cm}
\begin{center}
\mbox{
\begin{picture}(0,140)(250,0)
\put(0,14){\insertfig{9}{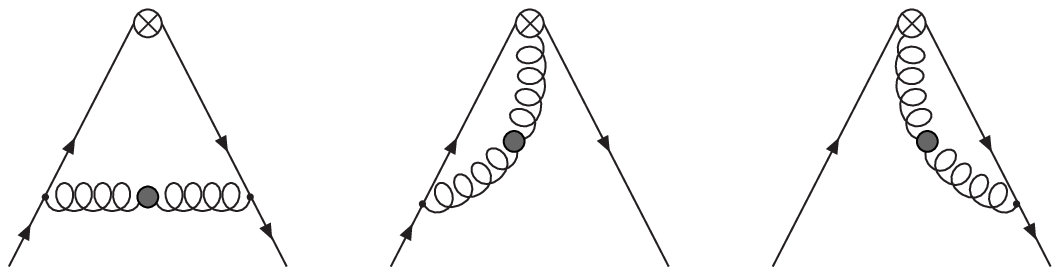}}
\put(30,0){$(a)$}
\put(125,0){$(b)$}
\put(215,0){$(c)$}
\put(300,10){\insertfig{6}{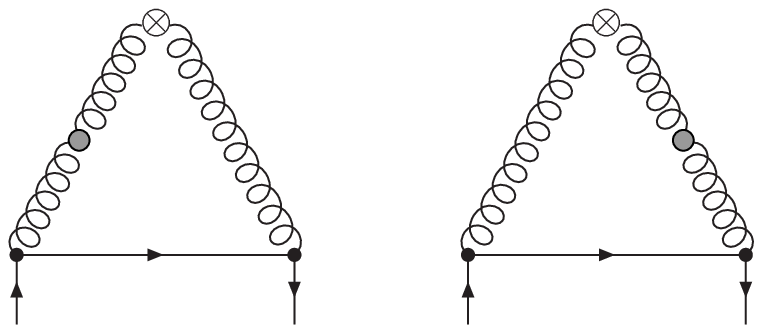}}
\put(330,0){$(d)$}
\put(430,0){$(e)$}
\end{picture}
}
\end{center}
\caption{\label{FigQua}
The special conformal anomaly at one-loop order in the quark-quark $(a,b,c)$ and
gluon-quark $(d,e)$ channels. Here the symbol $\otimes$ and the gray blob denote
the operator insertions of the conformal operator ${\ccal O}_{nl}$ and $\int dx\,
2x^- \Delta(x)$, respectively. }
\end{figure}

As the simplest example, we consider the conformal anomalies for flavor
non-singlet quark-antiquark operators
\beq{CWI-QED-LO}
\left[{\cal D}_y + {\cal D}_z\right] \langle [{\ccal O}_{nl}] \psi(y)
\bar\psi(z)\rangle
= -(l+3 )\langle [{\ccal O}_{nl}] \psi(y) \bar\psi(z) \rangle -i\!
\int\!\!d^dx \langle 0|[{\ccal O}_{nl}] \Delta(x) \psi(y) \bar\psi(z)
\rangle,
\eeq
\beq{CWI-QED-LO-sc}
\left[{\cal K}^-_y + {\cal K}^-_z\right]
\langle [{\ccal O}_{nl}] \psi(y) \bar\psi(z) \rangle
= i\, a(n,l)
\langle [{\ccal O}_{nl-1}] \psi(y) \bar\psi(z) \rangle
 -i\! \int\!\!d^dx \, 2x^-\,\langle [{\ccal O}_{nl}] \Delta(x) \psi(y) \bar\psi(z)
\rangle,
\eeq
where $\langle\ldots\rangle$ stands for the vacuum expectation value of the
T-product and $[{\ccal O}]$ is the renormalized operator insertion in a MS-like
scheme. For this case it is enough to consider the abelian gauge theory, the QCD
result can be restored by adding the color factor $C_F$. It proves to be
advantageous to choose the Landau gauge $\xi\to 0$ in which the ``quark''
anomalous dimension is zero, $\gamma_\psi=0$. Within these specifications we
have, cf. (\ref{TraceAnomaly})
\beq{Def-traano-LO-QED}
\Delta (x) = (d\!-\!4)\frac14 G_{\mu\nu}G^{\mu\nu}(x)
 -\frac{d\!-\!4}{2}\bar\psi(x) i\!\not\stackrel{\leftrightarrow}{D}\psi(x)
 -(d\!-\!2)\frac{1}{\xi}\partial_\mu(A^\mu(x)\partial_\nu A^\nu(x))
+ {\rm higher\ orders\ in\ }\alpha_s \,.
\eeq
The last, gauge-fixing term in $\Delta(x)$ does not induce UV divergencies and
can be omitted.

{}For the dilatation Ward identity (\ref{CWI-QED-LO}) we only have to consider
the first term in $\Delta(x)$, the insertion of $\frac{i}{4}\int
\!d^dx\,(G_{\mu\nu})^2(x)$ in an internal gluon line, see
Fig.~\ref{FigQua}(a)--(c). This is simply $i (g_{\mu\nu} k^2 - k_\mu k_\nu)$, so
that the effective gluon propagator with this insertion is just the usual gluon
propagator (in Landau gauge)
\bea{Con-OA}
 D_{\alpha}^{\ \mu}(k)\,i \left(g_{\mu\nu} k^2 - k_\mu k_\nu\right) D_{\
\beta}^{\nu}(k)
=  D_{\alpha \beta}(k)
 \quad\mbox{with}\quad
D_{\alpha\beta}(k)= \frac{-i}{k^2} \left({g_{\alpha\beta}  - k_\alpha
k_\beta/k^2}\right)\,.
\eea
Hence we are left with the usual one-loop diagrams and their divergent part is
equal to $-(\alpha_s/4\pi\epsilon) \gamma^{(0)}_n {\ccal O}_{nl}$, where
$\gamma^{(0)}_n$ is the one-loop anomalous dimension of ${\ccal O}_{nl}$. In the
notations of App.~B, $Z^{[1](0)}_n =
\gamma^{(0)}_n/2$. Inserting this result in (\ref{CWI-QED-LO}) completes
the calculation, reproducing Eq.~(\ref{CWI-confOpe}) to the desired accuracy.

In the case of the special conformal Ward identity (\ref{CWI-QED-LO-sc}) we have
two contributions. The insertion of  $i\int \!d^dx\, 2 x^-\psi(x)
i\!\not\!\stackrel{\leftrightarrow}{D}\psi(x) $ is a EOM operator and it can be
treated by partial integration in the functional integral. For the relevant
(divergent) part we have
\beq{EOM-LO-QED}
i\int\!\!d^dx\,2x^-\langle [{\ccal O}_{nl}]
\bar\psi(x) i\!\not\!\stackrel{\leftrightarrow}{D}\psi(x) \psi(y) \bar\psi(z)
\rangle
= -
\int\!\! d^dx\, 2 x^- \langle\left(\frac{\delta [{\ccal O}_{nl}]}{\delta
\psi(x)} \psi(x)
+ \bar\psi(x) \frac{\delta [{\ccal O}_{nl}]}{\delta \bar\psi(x)}\right)
\psi(y) \bar\psi(z)
\rangle.
\eeq
Using the explicit expression for ${\ccal O}_{nl}$ this can be brought to the
form
\bea{ActofEOM}
\lefteqn{\hspace*{-0.0cm}
\int d^dx\, 2 x^-\, \left(\frac{\delta [{\ccal O}_{nl}]}{\delta
\psi(x)} \psi(x)
+ \bar\psi(x) \frac{\delta [{\ccal O}_{nl}]}{\delta \bar\psi(x)}\right) =}
\nonumber\\
&&\hspace*{2.0cm}= \left(1+ \frac{\alpha_s}{4\pi\epsilon} \gamma^{(0)}_{n}
\right)
  \left(i \partial_x+ i\partial_y \right)^l
C_n^{3/2}\left(\frac{\partial_x \!-\! \partial_y}{\partial_x+\partial_y}\right)
\Bigg\{\bar \psi(y)\, 2
 (x^-+y^-) \gamma_+ \psi(x)\Bigg\}\Bigg|_{x=y=0} \, ,
\nonumber
\eea
where $\partial_x = D^x_+$ etc. and the prefactor comes from the renormalization
of ${\ccal O}_{nl}$. This is a local operator with the dimension by one unit less
than ${\ccal O}_{nl}$, so it can be expanded in conformal operators ${\ccal
O}_{ml-1}$ with $m\le n$. To obtain the coefficients note that the multiplication
by $x^-$ corresponds in the momentum space to the derivative with respect to the
incoming momenta, schematically $2 i\left(\partial_p +
\partial_q\right)\big[ C^{3/2}_n\big(\frac{p-q}{p+q}\big) (p+q)^l\big]$, and
this is the same differential operator that appears in the definition of the
matrix $\hat b (l)$ in Eq.~(\ref{Def-DefEqu-Mat-b}). Consequently, we have the
expansion
\bea{ActofEOM2}
\int d^dx\, 2 x^-\,  \left(\frac{\delta [{\ccal O}_{nl}]}{\delta
\psi(x)} \psi(x)
+ \bar\psi(x) \frac{\delta [{\ccal O}_{nl}]}{\delta \bar\psi(x)}\right) = 2i
\left(1+ \frac{\alpha_s}{4\pi\epsilon} \gamma^{(0)}_{n} \right)  \sum_{m=0}^{n}
b_{nm}(l) {\ccal O}_{ml-1}\,,
\eea
where ${\ccal O}_{ml-1}$ on the r.h.s.\ are not renormalized  and possess the
UV divergencies $-(\alpha_s/4\pi\epsilon)\gamma^{(0)}_{m}$ that have to be
subtracted. We find
\bea{EOM-LO-QED-2}
\lim_{\epsilon\to 0}
\epsilon\,
i\int\!\!d^dx\,2x^-\langle [{\ccal O}_{nl}]
\bar\psi(x) i\!\not\stackrel{\leftrightarrow}{D}\psi(x) \psi(y) \bar\psi(z)
\rangle
= - \frac{\alpha_s}{2\pi}\,  i\sum_{m=0}^n [\hat \gamma^{(0)},\hat b(l) ]_{nm} \,
\langle [{\ccal O}_{ml-1}]  \psi(y) \bar\psi(z)  \rangle
\, ,
\eea
with the $\hat b$-matrix given in Eq.~(\ref{Def-Mat-b}) (for $\nu=3/2)$.

The second part of the special conformal anomaly is induced by the insertion of
the operator $\frac{i}{4}\!\int \!\!d^dx\,2x^-(G_{\mu\nu})^2(x)$, see Fig.~\ref{FigQua}
(a)--(c). In difference to the case of the dilation anomaly (\ref{Con-OA}), the
operator insertion now contains a derivative with respect to the loop momentum.
The part of the effective gluon propagator that induces a UV divergence can
schematically be written as \cite{BelMue98c}
\bea{Con-OAmin}
2\bar n \cdot\! \stackrel{\leftarrow}{\partial}_k D_{\alpha\beta}(k) -
D_{\alpha\beta}(k)\,  2\bar n \cdot\! \stackrel{\rightarrow}{\partial}_k
\,.
\eea
In the case of Fig.~\ref{FigQua} (a) we can use momentum conservation on the both
``quark-gluon-quark'' vertices to replace these derivatives by those with respect
to the external momenta. The result for this diagram is thus proportional to the
corresponding part of the anomalous dimension and the derivatives give rise to
the same $\hat b$--matrix as above. In the case of Feynman diagrams in
Figs.~\ref{FigQua}(b) and (c) there is a similar contribution, so that the whole
anomalous dimension of ${\ccal O}_{nl}$ is restored. However, the derivative with
respect to the loop momentum also acts in this case directly on the operator
insertion ${\ccal O}_{nl}$ and in this way additional terms arise that will be
denoted as $w_{nm}$. More precisely, we define
\beq{OAmin-LO-QED}
\lim_{\epsilon\to 0}
\epsilon\,
i\int\!\!d^dx\,2x^-\langle [{\ccal O}_{nl}]
\frac12[G_{\mu\nu}^2](x) \psi(y) \bar\psi(z)
\rangle
= \frac{\alpha_s}{2\pi}\, i\sum_{m=0}^n
\left\{- \hat \gamma^{(0)} \hat b(l)
+ \hat{w} \right\}_{nm} \,
\langle [{\ccal O}_{ml-1}]  \psi(y) \bar\psi(z) \rangle.
\eeq
Inserting Eqs.~(\ref{EOM-LO-QED-2}) and (\ref{OAmin-LO-QED}) in
Eq.~(\ref{CWI-QED-LO-sc}) yields the special conformal anomaly
\beq{spconam}
\hat{\gamma}^{c(0)}(l)= -\hat{b}(l) \hat{\gamma}^{(0)}+\hat{w}\,.
\eeq
The $w$-term is $l$-independent and requires an explicit calculation. To this end
it is convenient to go over from the matrix $w_{nm}$ to the kernel $[w(u,v)]_+$
defined as
\bea{ConMom-wKer}
\int_0^1\! du\, C_n^{3/2}(2u-1) \left[w(u,v)\right]_+ =
\sum_{m=0}^n w_{nm} C_m^{3/2}(2v-1)\,.
\eea
The result for the kernel can be obtained by using non-local light-ray operators
instead of ${\ccal O}_{nl}$ (see for instance Eq.~(\ref{co-14})) and reads
\bea{Def-wKer}
\left[w(u,v)\right]_+ &\!\!\! =\!\!\! & w(u,v) -
\delta(u-v) \int_0^1\! dt\, w(t,v) + \frac{d}{du}\delta(u-v)
\int_0^1\! dt\, (t-v) w(t,v)\,
\\
w(u,v) &\!\!\! =\!\!\! & -C_F \theta(v-u) \frac{u}{v} \frac{2}{(u-v)^2} +
\left\{ {u\to 1-u \atop v\to 1-v} \right\}\, ,
\nonumber
\eea
where the ``+'' subtraction is extended in such a way that the second order pole,
appearing at $u=v$, is regularized, cf.~Eq.~(\ref{BLkernel}). The $w$-term is
absent in a scalar theory \cite{Mue94} and is responsible for the deviation of
the explicit NLO calculations in QCD of the off-forward evolution kernels in the
$\overline{\rm MS}$-scheme \cite{DitRad84Sar84Kat85MikRad85} from the
expectations based on the conformal operator product expansion
\cite{BroFriLepSac80,CraDobTod85,BroDamFriLep86}. Forming moments with respect to
Gegenbauer polynomials (\ref{ConMom-wKer}) one obtains \cite{Mue94}:
\bea{CWI-scBre2}
w_{nm} &=&
- C_F \theta(n>m)\left(1+ (-1)^{n-m}\right)
                           (2m+3)
\\&&{}\times
\left[
4A_{nm} + {a(n,m) \over (m+1)(m+2)} \left\{A_{nm}-\psi(n+2)+\psi(1)\right\}
\right]
\nonumber
\eea
where $A_{nm}=
    \psi\left({n+m+4\over 2}\right)-\psi\left({n-m\over 2}\right)
       +2\psi\left(n-m\right)-\psi\left(n+2\right)-\psi(1)$.
This form of the special conformal anomaly matrix and the explicit expression for
the $[w(u,v)]_+$ kernel are valid for all twist-two quark operators, irrespective
of their Dirac structure. Of course, the anomalous dimensions for tensor
operators differ from those of (axial-)vector ones.

Let us add that the calculation of conformal anomalies can be systematically
extended to higher loops. In abelian theory this can be achieved by introducing
renormalized operator products such as
\bea{Def-ZA-QED}
[{\cal O}_A^- {\ccal O}_{nl}] =  [{\cal O}_A^-] [{\ccal O}_{nl}] -
\sum_{m=0}^n \left\{ Z^-_{A} \right\}_{nm} [{\ccal O}_{ml-1}]\,,
\eea
where ${\cal O}_A^- = \int d^dx\,x^-\,G^a_{\mu\nu}(x) G^{a\mu\nu}(x)$ and
$Z^-_{A}$ are the corresponding additive renormalization constants. In QCD
additional counterterms appear due to peculiarity of the coupling constant
renormalization for non-abelian gauge fields, cf. Eq.~(\ref{ren-A}). Within the
definition (\ref{Def-ZA-QED}) the quark-quark entry  in the special conformal
anomaly $\hat\gamma^{c(0)}$  reads in Landau gauge
\bea{Def-SpeConAnoQQ}
\hat\gamma^{c(0)}=
2\, [\hat Z^{[1](0)},\hat b ] + \hat Z^{[1](0)-}_{A}\, , \ \mbox{with}\ \
\hat Z^{[1](0)} = \frac{1}{2}\hat\gamma^{(0)}\, .
\eea
Obviously, the renormalization constant $\hat Z^{[1](0)-}_{A}= - 2\hat
Z^{[1](0)}\hat b +\hat w$ appears on the r.h.s.\ of Eq.\ (\ref{OAmin-LO-QED}).

We are now in a position to discuss the anomalous dimensions in NLO. The special
conformal anomaly (\ref{spconam})  inserted in the conformal prediction
(\ref{gamND-NLO}) yields together with the forward anomalous dimensions the full
anomalous dimension matrix to NLO accuracy
\bea{AnoDimND-NonSin}
\hat\gamma^{(1)} = \hat\gamma^{{\rm D}(1)} + \hat\gamma^{{\rm ND}(1)}
\quad\mbox{with}\quad
\hat\gamma^{{\rm ND}(1)} =
-\left[ \hat \gamma^{(0)}, \hat d \right] \left(\beta_0 \hat 1 +
    \hat\gamma^{(0)}  \right)+
\left[\hat \gamma^{(0)},  \hat g
\right]\, ,
\eea
where we have introduced the matrices $\hat d$ and $\hat g$ defined as $ d_{nm}=
b_{nm}/a(n,m)$ and $g_{nm}=w_{nm}/a(n,m)$. The corresponding evolution kernel
$V^{(1)}(u,v)$ is defined by its conformal moments as
\bea{Def-ConMom}
\int_{0}^1\! du\, C_n^{3/2}(2u-1)\,  V^{(1)}(u,v)
= -\frac{1}{2} \sum_{m=0}^n \gamma^{(1)}_{nm}\,C_m^{3/2}(2v-1)\, ,
\eea
and for the present case it was calculated explicitly in
\cite{DitRad84Sar84Kat85MikRad85}. For completeness, we now present the method
for the construction of this kernel from the anomalous dimensions
(\ref{AnoDimND-NonSin}). This method is general and can be used in other cases as
well.

The matrix multiplication in Eq.~(\ref{AnoDimND-NonSin}), e.g., $\hat
\gamma^{(0)}\hat g$, corresponds in the momentum fraction representation
to the convolution of the kernels $ -2(V^{0}\otimes \hat g)(u,v) \equiv -2
\int_0^1\! dt\, V^{(0)}(u,t) g (t,v)$, so that all we need to do is to find the
kernels $g(u,v)$ and $\dot V(u,v)$ that give rise to the conformal moments
$g_{nm}$ and $[ \hat \gamma^{(0)}, \hat d]_{nm}$, respectively. The precise
definition is
\bea{Sum-gKer}
g(u,v) &=& \sum_{n=0}^\infty \sum_{m=0}^n \frac{4(2n+3)}{(n+1)(n+2)} (1-u) u
C_n^{3/2}(2u-1) g_{nm} C_m^{3/2}(2v-1)
\eea
and similar for $\dot V(u,v)$, with the replacement $g_{nm} \to [ \hat
\gamma^{(0)}, \hat d]_{nm}$.

{}First, we construct the $g(u,v)$ kernel. Since the Gegenbauer polynomials obey
the  equation
\bea{EigValEq-GegC}
\frac{d^2}{du^2} (1-u)u C_n^{3/2}(2u-1) = -(n+1)(n+2) C_n^{3/2}(2u-1)\ ,
\eea
the multiplication with $a(n,m)/2 = (n+1)(n+2)- (m+1)(m+2)$ in the conformal
moment space corresponds to a second order differential operator in the momentum
fraction representation. Thus, from the definition $a(n,m) g_{nm} = w_{nm}$ it
follows that $g(u,v)$ satisfies the inhomogeneous second order differential
equation
\bea{DifEq-gKer}
\frac{d^2}{dv^2}(1-v)v g(u,v) - (1-u)u \frac{d^2}{du^2} g(u,v)  =
\frac{1}{2} w(u,v)\quad\mbox{for}\quad u\not= v\,.
\eea
The homogeneous solution of this equation is diagonal with respect to Gegenbauer
polynomials and is not of interest for our purpose. The particular
integral  which
contains the conformal moments $g_{nm}$ for $n> m$, reads
\bea{Def-gKer}
g(u,v) & =& - C_F\Bigg[\theta(v-u) \frac{\ln\left(1-\frac{u}{v}\right)}{v-u} +
 \left\{{u\to 1-u \atop v\to 1-v} \right\} \Bigg]_+\, ,
\eea
where the ``+'' subtraction is defined in Eq.~(\ref{+}).

Second, we construct $\dot V(u,v)$. It turns out that the matrix elements
$d_{nm}=b_{nm}/a(n,m)$ occur in the expansion of the Gegenbauer polynomials with
a shifted index $\nu=3/2+\epsilon$ in terms of those with index $\nu=3/2$ at
first order in $\epsilon$:
\bea{Exp-GegBau}
C^{3/2+\epsilon}_n (2u-1) = C^{3/2}_n (2u-1) -2\epsilon \sum_{m=0}^n d_{nm}
C^{3/2}_m (2u-1)  + {\cal O}(\epsilon^2)\, ,
\eea
where we are only interested in $d_{nm}$ for $n>m$. To show this, use the
defining differential equation for Gegenbauer polynomials
\beq{Def-DifEqGegBau}
\frac{d^2}{du^2} (1\!-\!u)u C_n^{3/2+\epsilon}(2u-1)
-\epsilon \frac{d}{du} (2u-1) C_n^{3/2+\epsilon}(2u-1)  = -(n+1)(n+2+2\epsilon
)C_n^{3/2+\epsilon}(2u-1),
\eeq
and the definition (\ref{Def-DefEqu-Mat-b}) for the $b$-matrix. The desired
relation $d_{nm}= b_{nm}/a(n,m)$ then follows upon the expansion in $\epsilon$.
Thus, we look for the kernel $V(u,v|\epsilon)$ with eigenfunctions
$C_n^{3/2+\epsilon}$ and eigenvalues $\gamma^{(0)}_n$ for $\epsilon=0$. This
kernel must be symmetric with respect to the weight function
$[(1-u)u]^{1+\epsilon}$, i.e. $[(1-v)v]^{1+\epsilon} V(u,v|\epsilon) =
[(1-u)u]^{1+\epsilon} V(v,u|\epsilon)$ and can easily be deduced from the LO
evolution kernel (\ref{BLkernel}) by including an additional factor
$(u/v)^\epsilon$:
\bea{Def-KerV}
V(u,v|\epsilon) = C_F \left[\theta(v-u) \left(\frac{u}{v}\right)^{1+\epsilon}
\left(1+ \frac{1}{v-u} \right) +
 \left\{{u\to 1-u \atop v\to 1-v} \right\}\right]_+ \, .
\eea
The derivative of $V(u,v|\epsilon)$ with respect to $\epsilon$ leads to the
dotted kernel, which is therefore given by a logarithmical modification of
$V^{(0)}(u,v)$:
\bea{Def-dotV}
\dot V(u,v) = \frac{\partial}{\partial \epsilon} V(u,v|\epsilon)\Big|_{\epsilon=0}
=C_F \left[\theta(v-u) \frac{u}{v}\left(1+ \frac{1}{v-u} \right) \ln\frac{u}{v} +
 \left\{{u\to 1-u \atop v\to 1-v} \right\}\right]_+ \, .
\eea

Collecting everything, we find the following structure of the evolution kernel at
NLO
\bea{def-ND-kernel-NonSin}
V(u, v;\alpha_s) &\!\!\! = &\!\!\!
\frac{\alpha_s}{2\pi} V^{(0)} (u, v) +  \frac{\alpha_s^2}{(2\pi)^2} V^{(1)} (u, v) +
O(\alpha_s^3)\, ,
\\
V^{(1)} (u, v)
 &\!\!\! = &\!\!\!  D(u, v) + G(u, v)-
\left\{ \dot V \otimes V^{(0)} - \frac{\beta_0}{2} \dot V
+ g\otimes V^{(0)} -  V^{(0)}\otimes g
\right\} (u, v)\,,
\nonumber
\eea
The first two terms $D(u, v)$ and $G(u, v)$ only possess diagonal Gegenbauer
moments
and are distinguished by color factors. They can be restored, at least in
principle, through an integral transformation of the two-loop DGLAP kernel
\cite{BelMue98a} where, to avoid double counting, one has to subtract the
diagonal contributions that have already been included in the last term in
(\ref{def-ND-kernel-NonSin}).
The $D$-kernel can be represented as a convolution of the two separate diagonal
parts of the LO kernels (\ref{BLkernel}), while the $F$-kernel is related to the
crossed-ladder two-loop diagram and is rather complicated. It is given by the
rational function $(u/v)[1+1/(v-u)] $ which enters the LO kernel
(\ref{BLkernel}), decorated with Spence functions and double logs. The
explanation that the crossed-ladder diagram is diagonal is rather simple: it
contains no subdivergence and hence only a single $1/\epsilon$ pole.
Consequently, only tree-level counterterms are required that preserve conformal
symmetry. The explicit expressions can be found in Ref.~\cite{BelMueFre99}.

The NLO evolution kernel for the tensor quark-antiquark operator can be
constructed along the similar lines, see \cite{BelMueFre99,BelFreMue00}. It has
the same structure (\ref{def-ND-kernel-NonSin}) with $g(u,v)$ defined in
Eq.~(\ref{Def-gKer}), but the LO kernel is of course different:
\bea{Def-TraKerV}
V^{T(0)}(u,v) = C_F \left[\theta(v-u) \frac{u}{v}\frac{1}{v-u} +
 \left\{{u\to 1-u \atop v\to 1-v} \right\}\right]_+ - \frac{C_F}{2}  \delta(u-v)\, .
\eea
The corresponding dotted kernel $\dot V^{T}$ is given by the logarithmical
modification of $V^{T(0)}$, in full analogy to Eq.~(\ref{Def-dotV}). It turns out
that $G^T$ can easily be read off from the known result for $G(u,v)$ in the
vector case by a simple replacement of the corresponding LO structures, e.g., $
\frac{u}{v}\left(1+ \frac{1}{v-u} \right) \to \frac{u}{v}\left(\frac{1}{v-u} \right) \,.
$ {}For the remaining diagonal kernel $D^T(u, v)$ one just uses an ansatz, as a
linear combination of diagonal LO kernels and their convolutions, and finds the
corresponding coefficients from the known diagonal anomalous dimensions or,
alternatively, the corresponding DGLAP evolution kernel to NLO.

The similar procedure allows one to calculate the evolution kernel for gluon
tensor twist-two operators \cite{BelMue00,BelFreMue00}, where in addition
supersymmetric conditions are used to reconstruct the corresponding gluonic $G(u,
v)$ kernel. Note that the quark and gluon tensor operators belong to different
spin representations of the Lorentz group and, thus, do not mix with each other
under renormalization.

\subsubsection{\it The Flavor Singlet Sector}

\begin{figure}[t]
\unitlength1pt
\begin{center}
\mbox{
\begin{picture}(0,140)(250,0)
\put(0,150){\rotatebox{270}{\insertfig{5.7}{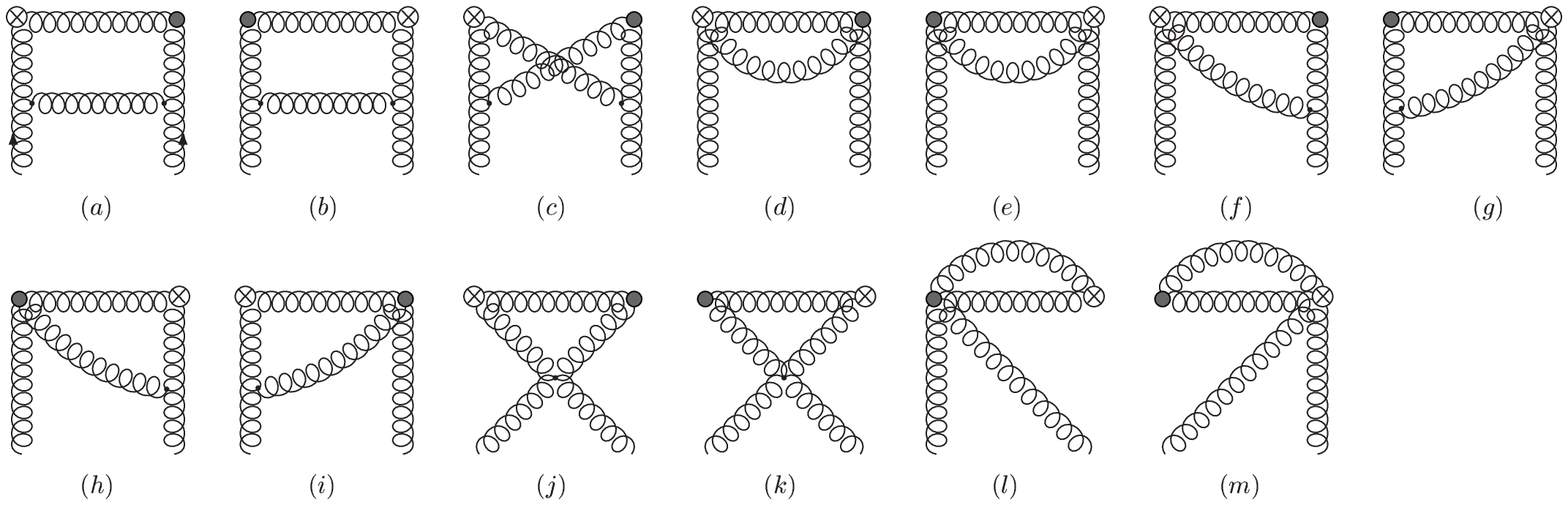}}}
\end{picture}
}
\end{center}
\caption{\label{FigGlu}
The renormalization constant $\hat Z_A^-$ at one-loop order. The notations are
the same as in Fig.~\ref{FigQua}. }
\end{figure}
The technique based on the evaluation of conformal anomalies can be extended to
the flavor singlet sector and indeed it has been employed to calculate the
corresponding anomalous dimension matrices \cite{BelMue98a,BelMue98c} and
evolution kernels at NLO \cite{BelMueFre99,BelMue99c,BelFreMue99}. The formalism
becomes more cumbersome, since the quark (\ref{co-19}) and the gluon operators
(\ref{co-20}) with the same spin mix with each other under renormalization. For
convenience we introduce the two-dimensional vector
\bea{Def-OpeSinSec}
\mbox{{\boldmath ${\ccal {\bf O}} $}}_{nl} =
(i \partial_+)^{l-n} \left( {  {\mathbb Q}_n^{1,1} \atop
 {\mathbb G}_{n-1}^{3/2,3/2} } \right)
\quad\mbox{with}\quad l \ge n,
\eea
 and the  $2\times 2$ matrices:
\bea{Def-SinAnoDim}
\mbox{{\boldmath $\hat a$}}
= \left( \hat a\ \ 0 \atop 0\ \ \hat a \right) ,
\qquad
\mbox{{\boldmath $\hat b$}}
= \left( \hat b\ \ 0 \atop 0\ \  \hat b \right),
\qquad
 \mbox{{\boldmath $\hat\gamma$}}
= \left( {{}^{QQ}\!\hat\gamma\ \  {}^{QG}\!\hat\gamma \atop {}^{GQ}\!\hat\gamma\
\  {}^{GG}\!\hat\gamma  } \right),
\qquad
\mbox{{\boldmath $\hat\gamma$}}^c
= \left( {{}^{QQ}\!\hat\gamma^c\ \  {}^{QG}\!\hat\gamma^c \atop
{}^{GQ}\!\hat\gamma^c\ \  {}^{GG}\!\hat\gamma^c  } \right).
\eea
Although the superscript of the corresponding Gegenbauer polynomials is different
for quark and gluon operators in Eq.~(\ref{Def-OpeSinSec}), both of them have the
same conformal spin. Hence the nonvanishing entries in the matrices {\boldmath
$\hat a$} and {\boldmath $\hat b$} are the same in the quark-quark and
gluon-gluon channels. They are given in Eqs.~(\ref{Def-aMat}) and
(\ref{Def-Mat-b}), respectively. The following construction applies for the both
existing sets of flavor singlet operators which can be distinguished according to
vector ($V$) and axial-vector ($A$) Dirac structure of the quark operators.

The constraints for the conformal anomalies in the flavor singlet sector have
been derived in \cite{BelMue98a,BelMue98c}. To achieve the same form as in the
non-singlet sector, Eqs.~(\ref{conf-constr-KP}) and (\ref{conf-constr-KD-full}),
\bea{conf-constr-KD-full-Sin}
\left[\mbox{{\boldmath $\hat a$}}(l)+\mbox{{\boldmath $\hat\gamma$}}^c(l)
+2{\beta\over g}
\mbox{{\boldmath $\hat b$}}(l), \mbox{{\boldmath $\hat\gamma$}} \right] =0
\qquad \mbox{and} \qquad
\mbox{{\boldmath $\hat\gamma$}}^c(l+1)-\mbox{{\boldmath $\hat\gamma$}}^c(l)
= -2\mbox{{\boldmath $\hat\gamma$}}
\eea
we redefined here the original special conformal anomaly matrix (\ref{def-scAno})
in the gluon-gluon channel by a shift ${}^{GG}\!\hat\gamma^c
\to {}^{GG}\!\hat\gamma^c +  \hat b\beta(g)/2g$. Consequently, the special conformal
anomaly in LO reads (in Landau gauge)
\bea{Def-SinSpeConAnoDim}
\mbox{{\boldmath $\hat\gamma$}}^{c(0)} =
\left( { 2\, [{}^{QQ}\!\hat Z^{[1](0)},\hat b ]
+ {}^{QQ}\!\hat Z^{[1](0)-}_{A}\ \ -2\hat b\, {}^{QG}\!\hat Z^{[1](0)} \atop 2\,
{}^{GQ}\!\hat Z^{[1](0)} \hat b + {}^{GQ}\!\hat Z^{[1](0)-}_{A}\qquad
 {}^{GG}\!\hat Z^{[1](0)-}_{A} + 2\beta_0 \hat b  } \right)\, .
\eea
The matrix of renormalization constants {\boldmath $ Z^{-}_{A}$} enters
in the renormalized operator product
$[{\cal O}_A^-] [\mbox{\boldmath $ {\ccal {\bf O}}$}_{nl}]$ where ${\cal O}_A^- =
\!\int\! d^dx\,x^-G^a_{\mu\nu}(x) G^{a\mu\nu}(x)$, see Eq.~(\ref{ren-A}). It has
been calculated at LO for both vector and axial-vector cases in
\cite{BelMue98a,BelMue98c}. At this order the result in the flavor singlet
quark-quark channel coincides with the non-singlet result in
Eq.~(\ref{CWI-scBre2}). Since in the quark-gluon mixing channel there is no gluon
loop, the operator insertion $[{\cal O}_A^-]$ cannot couple, and, consequently, $
{}^{QG}\!\hat Z^{-}_{A}$ is vanishing. It remains to calculate the anomalies in
the gluon-quark, cf.\ Fig.~\ref{FigQua}(d,e), and gluon-gluon channels, cf.\
Fig.~\ref{FigGlu}. The result reads
\bea{Res-SpeConAnoSin}
 \mbox{\boldmath $\hat\gamma$}^{c(0)}
= - \mbox{\boldmath $\hat b$}\, \!\mbox{\boldmath $\hat\gamma$}^{(0)} +
\mbox{\boldmath $\hat w$} \, ,
\eea
where the explicit expressions for ${}^{AB}\!\gamma^{(0)}_{n}$ and ${}^{AB}\!
w_{nm}$ can be found in Refs.\ \cite{BelMue98c,BelMue00}. Here we only note that
${}^{AB}\! w_{nm}$ are universal, i.e., the same for $V$ and $A$ operators, and
are given by the conformal moments of the corresponding ${}^{AB}\! w(u,v)$
kernels. The quark-quark kernel ${}^{QQ}\! w(u,v) = [w(u,v)]_+$ is defined in
Eq.\ (\ref{Def-wKer}), ${}^{QG}\! w(u,v)=0$ and the remaining two kernels read
\bea{Def-wKerSin}
\label{GQ-w}
{^{GQ}\!w} (u, v) &\!\!\!=\!\!\!& 2 C_F
\left\{ \frac{\theta ( v - u )}{v} - \left( { u \to 1-u \atop v \to 1-v } \right)
\right\}\, ,
\\
\label{GG-w}
{^{GG}\!w} (u, v) &\!\!\!=\!\!\!& - 2 C_A
\left\{
\left[ \frac{u^2}{v^2} \frac{\theta ( v - u )}{( v - u )^2} \right]_+
- \frac{1}{v^2} \theta ( v - u ) + \frac{1}{v} \delta ( u - v ) + \left( { u \to
1-u \atop v \to 1-v } \right)
\right\}\, .
\eea

The constraints (\ref{conf-constr-KD-full-Sin}) immediately yield the
off-diagonal entries of the anomalous dimension matrix at NLO. It has the same
structure (\ref{AnoDimND-NonSin}) as in the flavor non-singlet sector
\bea{AnoDimND-sin}
\mbox{\boldmath $\hat\gamma$}^{(1)} = \mbox{\boldmath $\hat\gamma$}^{{\rm D}(1)} +
\mbox{\boldmath $\hat\gamma$}^{{\rm ND}(1)}
\quad\mbox{with}\quad
\mbox{\boldmath $\hat\gamma$}^{{\rm ND}(1)} =
-\left[
\mbox{\boldmath $ \hat \gamma $}^{(0)}, \mbox{\boldmath $ \hat d$}
\right] \left(\beta_0  \mbox{\boldmath $ \hat 1 $} +
    \mbox{\boldmath $\hat\gamma$}^{(0)}  \right)+
\left[\mbox{\boldmath $ \hat \gamma $}^{(0)},  \mbox{\boldmath $ \hat g $}
\right]\, ,
\eea
where we introduced the matrices $\mbox{\boldmath $ d $}_{nm}=
\mbox{\boldmath $b $}_{nm}/a(n,m)$ and $\mbox{\boldmath $ g $}_{nm}=
\mbox{\boldmath $w $}_{nm}/a(n,m)$. These functions are the same for the
vector and axial-vector operators while the corresponding forward anomalous
dimensions are different (and known to NLO). This result implies the following
structure of the evolution kernels
\bea{def-ND-kernel-Sin}
\mbox{\boldmath$V$}^{(1)} (u, v)
= \mbox{\boldmath${ D}$}(u, v) + \mbox{\boldmath${G}$}(u, v)-
\left\{
\mbox{\boldmath$\dot V$} \otimes
\mbox{\boldmath$V$}^{(0)} - \frac{\beta_0}{2}\, \mbox{\boldmath$\dot V$}
+
\mbox{\boldmath$g$} \otimes \mbox{\boldmath$V$}^{(0)}
-\mbox{\boldmath$V$}^{(0)} \otimes  \mbox{\boldmath$g$}
\right\} (u, v)\, ,
\eea
where the off-diagonal conformal moments of the kernels $\mbox{\boldmath$\dot V$}
$ and $\mbox{\boldmath$g$} $ coincide with the matrix elements of
$\mbox{\boldmath$[\hat\gamma^{(0)},\hat d]$} $ and
\mbox{\boldmath$\hat g$}. Similar to the flavor non-singlet sector the
\mbox{\boldmath$g$} kernel can be reconstructed from
$\mbox{\boldmath$w$} $, Eqs.\ (\ref{GQ-w}) and (\ref{GG-w}), as the solution of a
certain inhomogeneous second order differential equation. The result
reads
\bea{set-g-kernels}
\mbox{\boldmath$g$} (u, v)
= \theta(v - u)
\left(
\begin{array}{cc}
- C_F \Big[\frac{ \ln \left( 1 - \frac{u}{v} \right) }{v - u} \Big]_+
& 0 \\
C_F \frac{u}{v} & - C_A\Big[ \frac{ \ln \left( 1 - \frac{u}{v} \right) }{v - u}
\Big]_+
\end{array}
\right)
\pm
\left\{ u \to 1-u \atop v \to 1-v \right\} \, ,
\eea
where the $+ (-)$ sign is to be taken in the diagonal (off-diagonal) entries,
respectively. In the axial-vector case the dotted kernels are constructed from
the LO evolution kernels by the logarithmical modification displayed in
Eq.~(\ref{Def-dotV}), while in the vector case there are subtleties that induce
extra terms \cite{BelFreMue99}. The main challenge is the construction of the
diagonal $\mbox{\boldmath${G}$}(u, v)$ kernel. The explicit calculation can be
avoided by using six supersymmetric conditions together with the two known
entries in the quark-quark channel for vector and axial-vector channel,
respectively. Here it is crucial that the $\mbox{\boldmath${G}$}(u, v)$ kernel
arises from the $1/\epsilon$ pole of crossed-ladder diagrams, which preserve both
conformal and supersymmetric covariance. This calculation includes a number of
technical details which will be skipped here. The interested reader can find them
in Ref.\ \cite{BelFreMue99}. Finally, the $\mbox{\boldmath${D}$}(u, v)$ kernel
has again a simple representation in terms of known diagonal kernels and can be
fixed by comparison with the known forward anomalous dimensions or DGLAP
evolution kernels.

{}Finally, let us list several consistency checks that have been performed for
the conformal predictions:
\begin{itemize}
\item In the flavor non-singlet sector the conformal prediction
(\ref{AnoDimND-NonSin}) coincides with the explicit NLO calculation
\cite{DitRad84Sar84Kat85MikRad85} of the evolution kernel for the pion
distribution \cite{MuePHD,Mue94}.
\vspace*{-0.1cm}
\item In the singlet sector the $\beta_0$ proportional terms in the
anomalous dimension matrix
 (\ref{AnoDimND-sin}) have been checked by
the calculation of Feynman with quark bubble insertions
\cite{BelMue98a,BelMue98c}.
\vspace*{-0.1cm}
\item  There exist four supersymmetric constraints, arising from the
reduction of QCD to supersymmetric ${\cal N}=1$ Yang-Mills theory, for the
off-diagonal part (\ref{AnoDimND-sin}) of the anomalous dimensions. They are
satisfied in a renormalization scheme that preserves supersymmetry
\cite{BelMueSch98}. As a byproduct it has been shown that two of six constraints
for the LO forward anomalous dimensions derived  in Ref.\ \cite{BukFroKurLip85},
are modified beyond the leading order by off-diagonal entries.
\vspace*{-0.1cm}
\item Using superconformal symmetry \cite{WesZum74a,Fer74,DonSoh74},
which is anomalously broken, one is able to derive four constraints for the
special conformal anomaly. Evaluation of the superconformal anomaly  in LO shows
that these constraints are consistent with the special conformal anomaly
(\ref{Res-SpeConAnoSin}) \cite{BelMue00b}.
\vspace*{-0.1cm}
\item  Rotating the conformal OPE  prediction 
for the hard-scattering amplitudes, valid in the CS scheme, to the $\overline{\rm
MS}$ scheme (see next Section) allows one to resum the conformal partial waves.
In this way one finds a closed expression for this amplitude in the momentum
fraction representation (\ref{Def-ComAmp}) \cite{Mue97a,BelMue97a}. These results
coincide with the diagrammatical evaluation in
Refs.~\cite{ManPilSteVanWei97,JiOsb97JiOsb98}.
\end{itemize}

Let us summarize the results of this Section. The complete set of two-loop
anomalous dimensions and evolution kernels for all twist-two operators has been
obtained using the conformal approach. A complete consistency check for the
special conformal anomaly has been given in the considered order. Together with
the validity of the conformal operator product expansion, shown at NLO, there is
a complete understanding how conformal symmetry is implemented beyond the LO
approximation and its predictive power provides a number of highly nontrivial
results.

\subsection{\it Conformal Operator Product Expansion for Two-Photon Processes}
\label{SubSec-ConPreTwoPhoPro}

The aim of this Section is to illustrate the application of the conformal
operator product expansion (COPE) considered in Sect.~2.3 beyond the leading
order. The relevant physical setup is provided by generic hard exclusive
processes including two photons, which are various cross-channels of the reaction
$\gamma^\ast(q_1) + h(p_1) \to \gamma^{(\ast)}(q_2) + h(p_2)$\,
in the generalized Bjorken kinematics $\nu = p\cdot q\rightarrow
\infty$, $Q^2 = -q^2 \rightarrow \infty$ with $\xi = 1/\omega =
Q^2/p\cdot q$ fixed, where $p=p_1+p_2$ and $q=(q_1+q_2)/2$
\cite{MuePHD,MueRobGeyDitHor94}. Here $h(p_1)$ and $h(p_2)$ are generic hadronic
states with small invariant masses (we also allow for the vacuum state) that can
be neglected to power accuracy in $1/Q^2$. It is convenient to introduce the
so-called skewedness parameter $\eta$ as the second scaling variable
$\eta = {\Delta \cdot q / p\cdot q}$, where  $\Delta=p_2-p_1$.
We assume that $-t= -\Delta^2 \ll Q^2$ and can be neglected as well.

The QCD dynamics in the two-photon processes
in the generalized Bjorken
limit is described by the hadronic tensor
\bea{Def-HadTen}
T_{\mu\nu}(p,\Delta,q) =
   i \int \!d^4 x\, {\rm e}^{i\, q\cdot x}
    \langle h(p_2)|T J_\mu ({x}/{2})
                                  J_\nu\left({-x}/{2}\right)|h(p_1)\rangle,
\eea
which receives a dominant contribution from the region $x^2\sim 1/Q^2$. It can be
evaluated using the operator product expansion of the two currents at short
distances $x^\mu \to 0$, where to power accuracy it is sufficient to retain only
leading twist-two operators. We have demonstrated in the previous Sections that
the conformal covariance of operators can be preserved in QCD, up to corrections
in $\beta(g)$, in the CS scheme. In this scheme, the COPE (\ref{COPE-EleCur}) is
valid to all orders up to corrections proportional to the QCD $\beta-$function.
{}For transverse Lorentz projections of the currents $J_\mu \to J_\perp$ we
obtain, therefore
\bea{cOPE-JJ}
\lefteqn{
T(\omega,\eta,Q^2)=}
\\
&=&\!i\!\!\int\! d^4 x\, \textrm{e}^{iq\cdot x -\frac{i}2\Delta\cdot x}
\sum_{n=0}^\infty C_n\!
\left(\frac1{x^2} \right)^{3-t_n/2}\!
\frac{ (-i x_-)^{n+1}   }{{\rm B}(j_n,j_n)}\!
\int_0^1\!\! du\, [u(1\!-\!u)]^{j_n-1} \langle
h(p_2)|\mathbb{Q}_n^{\mbox{\tiny CS}}(ux_-)|h(p_1)
\rangle \nonumber\\
&&{}+{\mathcal O}(\beta(g)),
\nonumber
\eea
where $\mathbb{Q}_n^{\mbox{\tiny CS}}$ is the renormalized conformal operator
with twist $t_n= 2+\gamma_n(\alpha_s)$ and conformal spin $j_n
=n+2+\frac12\gamma_n(\alpha_s)$ in the
CS scheme%
\footnote{For simplicity we do not include gluon contributions.
and do not display the quark electric charges.}. The hadronic matrix elements,
entering Eq.\ (\ref{cOPE-JJ}), can be parameterized as
\beq{red-mat-el}
\langle h(p_2)|\mathbb{Q}_n^{\mbox{\tiny CS}}(ux_-)|h(p_1) \rangle
=\textrm{e}^{iu\eta (x\cdot p)} \, p_+^{n+1}\, \langle\!\langle
 \mathbb{Q}_n^{\mbox{\tiny CS}}(0) \rangle\!\rangle(\eta,\Delta^2,\mu^2)\,,
\eeq
where we used the identity $(x\Delta)=\eta (xp)$, valid in the Bjorken limit, and
introduced the notation for the reduced matrix elements. The latter depend on the
kinematic invariants $\eta$, $\Delta^2$ and on the normalization scale $\mu^2$.
By construction, the conformal operators have autonomous scale dependence and
satisfy the renormalization group equations
\bea{RGE-MarEle}
\mu \frac{d}{d \mu}
 \langle\!\langle
 \mathbb{Q}_n^{\mbox{\tiny CS}}(0) \rangle\!\rangle = -\gamma_n(\alpha_s(\mu))
 \langle\!\langle
 \mathbb{Q}_n^{\mbox{\tiny CS}}(0) \rangle\!\rangle
\, .
\eea
We note that beyond the conformal limit both Eqs.~(\ref{cOPE-JJ}) and
(\ref{RGE-MarEle}) have to be corrected by terms proportional to $\beta(g)/g\cdot
O(\alpha_s)$. Such terms affect the anomalous dimensions to NLO order and the
Wilson coefficients to the NNLO accuracy. They vanish in the forward limit.

Substituting (\ref{red-mat-el}) into (\ref{cOPE-JJ}) and performing the
$x-$integration, one arrives at
\beq{T-Four}
T(\omega,\eta,Q^2)=\sum_{n=0}^\infty
C_n\left(\frac{\mu^2}{Q^2}\right)^{\gamma_n(\alpha_s)/2}
 \int_0^1 du\,
\frac{\omega^{n+1}[u(1-u)]^{n+1+\gamma_n(\alpha_s)/2}}{[1-\eta\omega
(2u-1)]^{n+1+\gamma_n(\alpha_s)/2}}\,
\langle\!\langle\mathbb{Q}_n^{\mbox{\tiny CS}}(0)
\rangle\!\rangle(\eta,\Delta^2,\mu^2)\,,
\eeq
where $C_n=\widetilde C_n\cdot 2^{n+1}B(n+1,n+2)/B(n+2+\gamma_n/2,n+2+\gamma_n/2
)$; $\widetilde C_n$ are the Wilson coefficients in the polarized deep inelastic
scattering in the flavor non-singlet sector and are known to NNLO \cite{ZijNee94}
\bea{Wil-Coe-DIS}
\widetilde C_n(\alpha_s) = c^{(0)}_n + \frac{\alpha_s}{2\pi} c^{(1)}_n +
\frac{\alpha_s^2}{(2\pi)^2} c^{(2)}_n +
O(\alpha_s^3)\quad \mbox{with}\quad c^{(0)}_n =1\,.
\eea
The integral in
(\ref{T-Four}) can be expressed in terms of the ${}_2F_1-$hypergeometric series.
It depends on the ratio of scaling variables $\eta/\xi=\eta\omega$ and resums the
contribution of the conformal tower of operators with the highest weight
$\mathbb{Q}_n^{\mbox{\tiny CS}}(0)$. Notice that for $|\eta\omega|>1$ the
integrand in (\ref{T-Four}) has a singularity inside the integration region
indicating that the COPE (\ref{T-Four}) is divergent and should be defined by the
analytic continuation from the $|\eta\omega|<1$ region.
It follows that the generalized structure function $T(\omega,\eta,Q^2)$ is an
analytical function on the complex $\omega-$plane with a cut that runs between
the points $\xi=\pm\eta$. To begin with, let us consider the two limiting cases:
$\eta=0$ and $\eta=1$.

The first case, $\eta=0$, corresponds to the forward scattering $\Delta^\mu=0$.
The $u-$integration becomes trivial, the $\xi$-variable coincides with the
Bjorken variable $x_{\rm Bj}$, and the result for $T(\omega,\eta=0,Q^2)$ reduces
to the well-known expansion of the forward Compton amplitude in the unphysical
region $|\omega|<1$. In particular, $\widetilde C_n(\alpha_s)$ and $\gamma_n$
coincide with the Wilson coefficients and the anomalous dimensions in the deep
inelastic scattering. Making use of the dispersion relations, one can identify
the coefficients in front of powers of $\omega$ in the expansion of
$T(\omega,\eta=0,Q^2)$ as the moments of the imaginary part of the Compton
amplitude in the physical region $0< x_{\rm Bj}=1/\omega < 1$.

The second case, $\eta=1$, is realized, e.g., for the photon-to-pion transition
form factor, $\gamma^*(q_1)+\gamma^*(q_2) \to \pi^0(p)$, in which case the matrix
elements are evaluated between the vacuum and the pion state with momentum
$p^\mu=\Delta^\mu=q_1^\mu+q_2^\mu$. Since the both photons are in the initial
state, one has to replace $q_2\to -q_2$. The corresponding hard scale is
$Q^2=-(q_1 - q_2)^2/4=-(q_1^2 + q_2^2)/2$. Due to parity conservation the
hadronic tensor must be of the form
\beq{pion-ff}
T_{\mu\nu}(\omega,\eta=1,Q^2)=i e^2 \epsilon_{\mu\nu\alpha\beta}q_1^\alpha
q_2^\beta F_{\gamma\pi}(\omega,Q)\, ,\qquad \omega= \frac{q_1^2 - q_2^2}{q_1^2 +
q_2^2}\,,
\eeq
where $F_{\gamma\pi}(\omega,Q)$ defines the form factor and $e$ is the electron
charge. In this case it is also sufficient to consider transverse components of
the currents and we can overtake Eq.~(\ref{T-Four}) at $\eta=1$ and conformal
operators defined in the axial-vector sector. Taking care of the appropriate
normalization, the COPE result becomes
\bea{eq:Fpij}
F_{\gamma \pi}(\omega,Q)=
 \frac{\sqrt{2} f_{\pi}}{3 Q^2 } \,\sum_{n=0\atop n\ {\rm even}}^{\infty}{} \;
C_n(\omega| \alpha_s,Q/\mu)\,
\frac{3(n+1)(n+2)}{2(2n+3)}\, \phi_n(\mu) + O(1/Q^4)
 \,,
\eea
where
\beq{DefWilCoe-omega}
C_n(\omega| \alpha_s,Q/\mu)   =
\frac{2^{n+1}\,\widetilde C_n(\alpha_s)B(n+1,n+2)}{B(n+2+\gamma_n/2,n+2+\gamma_n/2)}
\left(\frac{\mu^2}{Q^2}\right)^{\gamma_n(\alpha_s)/2}
\!\!\!\int_0^1\!\! du\,
\frac{\omega^{n}[u(1-u)]^{n+1+\gamma_n(\alpha_s)/2}}{[1-\omega
(2u-1)]^{n+1+\gamma_n(\alpha_s)/2}},
\eeq
$f_{\pi}\simeq 132$ MeV and the scaling variable $|\omega| \le 1$ is called
asymmetry parameter. The nonperturbative parameters $\phi_n(\mu)$ are related to
the reduced matrix elements $\langle\pi(p)|\mathbb{Q}_n^{\mbox{\tiny
CS}}(0)|0\rangle$ and define the partial wave decomposition of the pion
distribution amplitude (\ref{lt-pion}). In the CS scheme $\phi_n(\mu)$ are
multiplicative renormalizable to {\em any} order of perturbation theory, up to
corrections in $\beta-$function. To the NLO accuracy, they obey the
renormalization group equation with inhomogeneous term proportional to $\beta_0$:
\bea{NLO-CS-RGE}
\mu \frac{d}{d\mu} \phi_n(\mu) = - \left[ \frac{\alpha_s(\mu)}{2\pi} \gamma^{(0)}_n +
\frac{\alpha_s^2(\mu)}{(2\pi)^2}
 \gamma^{(1)}_n\right] \phi_n(\mu) -  \frac{\alpha_s^2(\mu)}{(2\pi)^2}\
\beta_0 \sum_{m=0}^{n-2} \Delta^{(0)}_{nm}\, \phi_m(\mu) + O(\alpha_s^3)\, .
\eea
After the consequent expansion with respect to $\alpha_s$, the partial wave
decomposition (\ref{eq:Fpij}) is exact in NLO and after rotation
(\ref{FinRenConOpe}) reproduces the $\overline{\rm MS}$ result \cite{Mue97a}.

To solve (\ref{NLO-CS-RGE}) one introduces a new set of ``renormalization group
improved'' conformal operators that do not mix under renormalization to NLO:
\bea{NewConOpe}
\mu \frac{d}{d\mu} \phi_n^\prime(\mu)(\mu) =  -\gamma_n(\alpha_s(\mu))\phi_n^\prime(\mu)
\quad\mbox{with}\quad
\phi_n^\prime(\mu) =
\sum_{m=0}^n {\cal B}_{nm}^{-1}(\mu) \phi_m(\mu)\, ,
\eea
where $\gamma_n(\alpha_s(\mu))$ is given by the two-loop expression defined in
(\ref{NLO-CS-RGE}) and the mixing matrix ${\cal B}$ satisfies the differential
equation \cite{Mue94,Mue9598}
\bea{Def-MatCalB}
\mu \frac{d}{d\mu} {\cal B}_{nm}(\mu)  =
\left[\hat {\cal B}(\mu), \hat \gamma^{\rm D}(\alpha_s(\mu)) \right]_{nm}  -
\beta_0 \frac{\alpha_s^2(\mu)}{(2\pi)^2}
\left\{\hat \Delta^{(0)}\hat {\cal B}(\mu)\right\}_{nm}
+ O(\alpha_s^3) \,.
\eea
Here $\gamma^{\rm D}_{nm}(\alpha_s)=\delta_{nm}\gamma_n(\alpha_s)$. Since the
pion transition form factor does not depend on the factorization scale $\mu$, the
rotated Wilson coefficients defined as
\bea{newWilCoe}
C^\prime_n(\omega|\alpha_s(\mu),Q/\mu) =
\sum_{m=n}^\infty C_m(\omega|\alpha_s(\mu),Q/\mu) {\cal B}_{mn}(\mu)\,
\eea
satisfy the renormalization group equation
\bea{REG-WilCoe}
\mu \frac{d}{d\mu} C^\prime_n(\omega|\alpha_s(\mu),Q/\mu) =
\left[ \frac{\alpha_s}{2\pi} \gamma_n^{(0)} + \frac{\alpha_s^2}{(2\pi)^2} \gamma_n^{(1)}
+ O(\alpha_s^3)
\right]C^\prime_n(\omega|\alpha_s(\mu),Q/\mu) \, .
\eea
The perturbative solution of this equation to NLO can be cast in the form of Eq.\
(\ref{DefWilCoe-omega}) by modifying the Wilson coefficients \cite{MelMuePas02}
\bea{Def-Coe-FulThe}
\widetilde C_n(\alpha_s) &\Longrightarrow & c_n^{(0)} +
\frac{\alpha_s(\mu)}{2\pi} c_n^{(1)} +
\frac{\alpha^2_s(\mu)}{(2\pi)^2}  c_n^{(2)}
-\frac{\alpha_s(\mu)}{2\pi}
\frac{\beta_0}{2}\ln\left(\frac{Q^2}{\mu^2}\right)
\nonumber \\ & &
\times
\left[ \frac{\alpha_s(\mu)}{2\pi}
 \left(c_n^{(1)}+ c_n^{(0)} \frac{\gamma_n^{(0)}}{4}
\ln\left(\frac{Q^2}{\mu^2}\right)
\right) + c_n^{(0)}  \gamma_n^{(0)}\frac{\partial}{\partial \gamma_n^{(0)}}
\right] + O(\alpha_s^3)\, ,
\eea
where the initial condition is chosen in such a way that the conformal covariance
holds true for $\mu=Q$. Here the derivative in $\gamma_n^{(0)}$ acts on
$C_n(\omega| \alpha_s,Q/\mu)$ in Eq.\ (\ref{DefWilCoe-omega})
 and amounts to replacing $\alpha_s(\mu) \to \int_\mu^Q d\ln \mu'
\alpha_s(\mu')$ in the one-loop expression for the anomalous dimension
$\gamma_n(\alpha_s)$.

\begin{figure}[t]
\unitlength1pt
\begin{center}
\begin{picture}(0,150)(250,0)
\put(85,10){\rotate{$\textstyle  2 Q^2 F_{\gamma \pi}(\omega=\pm 1, Q) \ [{\rm GeV}]$}}
\put(110,0){\insertfig{8}{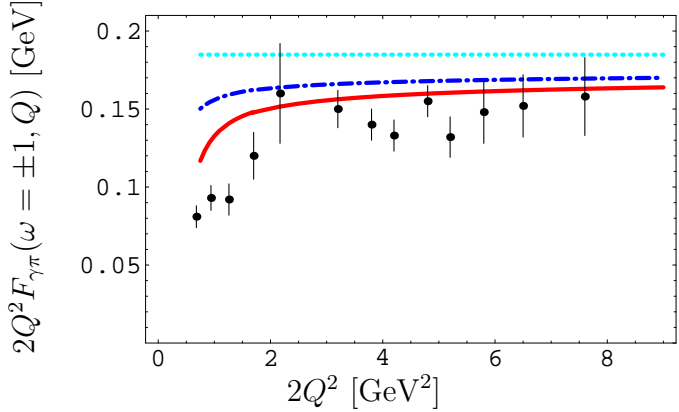}}
\put(190,-8){$2 Q^2\ [{\rm GeV}^2]$}
\end{picture}
\end{center}
\caption{
The perturbative QCD prediction for the scaled pion transition form factor $2
Q^2F_{\gamma\pi}(\omega, Q)$ for one virtual and one real photon ($\omega=1$)
using the asymptotic pion distribution amplitude ($n=0$). The calculations in LO,
NLO, and NNLO are shown by the dotted, dash-dotted, and solid lines,
respectively. We use $\alpha_s^{\overline{\rm MS}}(\mu=M_Z)=0.118$ and set the
scale $\mu^2= 2Q^2=-q_1^2$. The data are taken from
Refs.~\cite{Behetal91,Sav97}.}
\label{FigPre}
\end{figure}

The experimental data are available for the quasi-real photon limit $q_2^2\to 0$
which corresponds to $|\omega | = 1$ and $Q^2 = -q_1^2/2$ \cite{Behetal91,Sav97}
and they are reasonably well described, see Fig.~\ref{FigPre}, by the
contribution of the lowest conformal partial wave alone. This can easily be found
from Eqs.~(\ref{eq:Fpij}) and (\ref{Def-Coe-FulThe}) \footnote{Thanks to current
conservation, the anomalous dimension $\gamma_0$ vanishes. Thus, $C_0(\omega=1)=
3c_0 /2 $,
where $c_0= C_{Bj}$ enters the Bjorken sum rule
 (\ref{Bjo-SumRul}).}
\bea{AsyPre}
&&F_{\gamma \pi}(\omega,Q)= \frac{\sqrt{2} f_{\pi}}{2 Q^2 }
\left\{1-\frac{\alpha_s(\mu)}{\pi} -
 \frac{\alpha_s^2(\mu)}{\pi^2}  \left[3.583 - 2.25
\ln\left(\frac{Q^2}{\mu^2}\right)\right]  +
 O(\alpha_s^3) \right\},
\eea
where we used $N_f=3$. The overall normalization of the LO result is proportional
to the sum of all conformal partial waves $1+\phi_2+\phi_4+\ldots$ so that the
data do not leave much room for higher terms, unless there are sign alternating
contributions.
For now, this is the strongest existing evidence that the pion distribution
amplitude already at relatively low scales is sufficiently close to its
asymptotic form.

It might be possible to measure the pion transition form factor for both virtual
photons for a certain range of $|\omega| < 1$ at the running facilities CLEO,
Barbar and Belle, see e.g. Ref.~\cite{DieKroVog01}. This would help to separate
contributions of different higher conformal partial waves, and also provides a
novel QCD test for exclusive processes that is parameter free
\cite{DieKroVog01,MelMuePas02}:
\begin{eqnarray}
\frac{Q^2}{\omega^{\rm cut}} \int_0^{\omega^{\rm cut}} d\omega
 F_{\gamma \pi}(\omega,Q) =
 \frac{\sqrt{2}f_\pi}{3}\Bigg\{1\!\!\! &-&\!\!\!\frac{\alpha_s(Q)}{\pi} -
3.583 \frac{\alpha_s^2(Q)}{\pi^2}- 20.215\frac{\alpha_s^3(Q)}{\pi^3}
+ O(\alpha_s^4)  \Bigg\},
\nonumber
\end{eqnarray}
where $\omega^{\rm cut} < 0.4$ is required and we set $\mu=Q$.

Returning to the general case $0<\eta<1$ we recall that the COPE expansion
(\ref{T-Four}) diverges in the region $|\eta\omega|>1$. This is the similar
situation as in deep inelastic scattering where the amplitude can only be
restored from the moments by analytic continuation through the Mellin transform.
In the present case we also have to introduce generalized parton distributions
\cite{MuePHD,MueRobGeyDitHor94,Ji96Rad96} given by the Fourier transform of
light-ray operators (\ref{co-14})
\bea{Def-GPDs}
q(x,\eta,\Delta^2,\mu^2) =
\int \frac{d\alpha}{2\pi} {\rm e}^{i \alpha x (n\cdot P)}
\langle h(p_2) |\bar\psi(-\alpha n) \gamma_+ \Gamma
\psi(\alpha n)| h(p_1)\rangle\Big|_{n\cdot\Delta= \eta\, n\cdot p }\, .
\eea
where $\Gamma=\{1\,,\gamma_5\}$ for vector and axial-vector contributions to the
hadronic tensor and the variable $x$ has a partonic interpretation of the
momentum fraction. Using the definition of the conformal operator,
Eq.~(\ref{co-19}), one can identify the reduced hadronic matrix elements
(\ref{red-mat-el}) as conformal moments of the generalized parton distributions
\bea{ConMom-GPD}
\int_{-1}^1\! dx\, \eta^n\, C_n^{3/2}\! \left(\frac{x}{\eta}\right) q(x,\eta,\Delta^2,\mu^2) =
\langle\!\langle\mathbb{Q}_n^{\mbox{\tiny CS}}(0) \rangle\!\rangle(\eta,\Delta^2,\mu^2) \, .
\eea
In the limit of forward scattering $\eta,\,\Delta^2 \to 0$ this relation
coincides with the definition of conventional parton distributions whereas for
the pion transition form factor, $\eta=1$, it reproduces the definition of the
pion distribution amplitude.

The evolution equation for $q(x,\eta,\Delta^2,\mu^2)$ follows from the
renormalization group equation for the light-ray operators
\cite{BB89,GeyRobMueDitHor88,MuePHD,MueRobGeyDitHor94}. In the flavor non-singlet
sector it looks like
\bea{Def-EvoEquGPD}
\mu^2\frac{d}{d\mu^2} q(x,\eta,\Delta^2,\mu^2) =
\int_{-1}^1 \frac{dy}{2|\eta|} \,
V\!\left(\!\frac{\eta+x}{2\eta},\frac{\eta+y}{2\eta};\alpha_s(\mu)\!\right)
q(y,\eta,\Delta^2,\mu^2)\, .
\eea
By construction, at $\eta=1$ this equation has to reproduce the ER-BL evolution
equation (\ref{ERBL}) for the pion distribution amplitude. This implies that for
$v = (1+y)/2$, $u=(1+x)/2$ with $0\le u,v \le 1$ the kernel $V(u,v)$ coincides
with the ER-BL kernel. Using the well-known result for the latter one can restore
the expression for the $V-$kernel entering (\ref{Def-EvoEquGPD})
\cite{MuePHD,MueRobGeyDitHor94}.
Substituting (\ref{ConMom-GPD}) into (\ref{T-Four}) allows for the (formal)
resummation of all conformal partial waves and results into the momentum fraction
representation of the hadronic tensor. In this representation the hadronic tensor
is factorized into a convolution of the hard scattering amplitude
\bea{ForRepHarSca}
\frac{{\cal T} }{\omega} = \sum_{n=0}^\infty
C_n(\omega\eta|\alpha_s,Q/\mu)
%
%
\, C_n^{3/2}\! \left(\frac{x}{\eta}\right)
\eea
and the generalized parton distributions (or generalized distribution amplitudes
for the crossed processes):
\bea{Def-ComAmp}
\xi T(\xi,\eta,\Delta^2,Q^2) \simeq
\int_{-1}^1 \frac{dx}{|\eta|} \left[
{\cal T}\!
\left(\frac{\xi}{\eta},\frac{x}{\eta},\frac{Q}{\mu};\alpha_s(\mu)\right)
\mp \{\xi \to -\xi\}
\right]
q(x,\eta,\Delta^2,\mu^2)
\, ,
\eea
where $- (+)$ is for the (axial-)vector case. Unfortunately, for the CS scheme
exact expressions for the hard scattering amplitudes (\ref{ForRepHarSca}) and the
evolution kernels are not known beyond LO.
To the NLO accuracy they have been evaluated  in the $\overline{\rm MS}$ scheme
by the rotation of the local COPE to this scheme and reconstructing the
corresponding functions in the momentum fraction representation by analytic
continuation \cite{Mue97a,BelMue97a}. After such a resummation we can employ
analytical properties of the Compton amplitude to obtain the result in the
physical region.

\section{Outlook}

To summarize this review, the uses of conformal symmetry in Quantum
Chromodynamics in the last years were developing along two different
lines. The first line of applications is essentially algebraic. The
parton model, by construction, retains all symmetries of the classical
Lagrangian and the QCD corrections in the leading logarithmic
approximation also do. The simplest and the standard application of
conformal symmetry to light-cone dominated processes was (and remains)
the separation of variables corresponding to the transverse and
longitudinal degrees of freedom. This r\^ole is very similar to the
partial wave expansion in nonrelativistic quantum mechanics. Remember
that in the latter case the classification of states in contributions
with given orbital angular momentum remains to be relevant also in the
presence of spin-orbit interaction which breaks the rotational symmetry.
By the same token, the classification in terms of conformal spin in QCD
remains relevant also with higher-order radiative corrections. A deeper
look in the algebraic structure of the evolution equations in QCD
reveals hidden symmetries and interesting connections with completely
integrable systems, with the conformal spins of partons playing the
r\^ole of spins of a generalized Heisenberg magnet. These connections
are intriguing and are not really understood. On the theoretical side,
they may allow for the studies of high-energy QCD using methods
inherited from string theory and possible emerging of a dual string
picture, see \cite{AGMOO}. In this way the strong coupling regime of the
evolution in supersymmetric extensions of QCD can be addressed
\cite{Gubser,BGK}. On the practical side, these developments already led
to an almost complete understanding of the spectra of twist-three
operators, which are relevant for the phenomenology.

The second line of applications has been using constraints imposed by
conformal symmetry, and in particular by conformal Ward identities, in
order to advance perturbative calculations for processes involving
off-forward kinematics to higher orders. As the principal result, there
exists now a complete understanding how conformal symmetry is
implemented beyond the leading logarithmic approximation and its
predictive power provides a number of highly non-trivial results at the
next-to-leading order. It is shown that the covariance of leading twist
conformal operators can be ensured in all orders of perturbation theory
in a special renormalization scheme. This so-called conformal
subtraction scheme differs from a MS-like scheme by a finite
renormalization. The conformal operator product expansion holds true in
this scheme, and indeed it has been checked in NLO by the explicit
calculation. The structure of the conformal constraints is such that,
generally speaking, $n$-loop predictions for off-forward amplitudes can
be obtained from the similar predictions in forward kinematics {\it
and}\ the $n-1$-loop calculation of the special conformal anomaly. The
technique is general and can be applied in a broad context.

There are several other interesting issues that we were not able to
cover, in particular using conformal symmetry in semiclassical
construction of the QCD vacuum, e.g., based on instantons. This is,
however, another subject.

\appendix
\renewcommand{\theequation}{\Alph{section}.\arabic{equation}}
\setcounter{table}{0}
\renewcommand{\thetable}{\Alph{table}}

\section*{Appendices}

\section{The Conformal Basis}
\label{app:a}
\setcounter{equation}{0}

For a generic three-particle operator a ``conformal basis'' can be
constructed as follows \cite{BDKM,BKM01}. The conformal symmetry allows
one to fix the total three-particle conformal spin $J=j_1+j_2+j_3+N$ of
a state. We define a set of functions ${Y}_{Jj}^{(12)3}$ by requiring
that, in addition to a fixed $J$, they also have a definite value of the
conformal spin in the given two--particle channel $(12)$, for
definiteness: $j=j_1+j_2+n$ with $n=0,\dots,N$.

Taken together, these two conditions determine the polynomials
${Y}_{Jj}^{(12)3}$ uniquely and yield the following expression:
\bea{YJi123}
Y_{Jj}^{(12)3}(\xxi_i) &=& (1-\xxi_3)^{j-j_1-j_2}\,
  P_{J-j-j_3}^{(2j_3-1,2j-1)}\lr{1-2\xxi_3}
\,P_{j-j_1-j_2}^{(2j_1-1,2j_2-1)}\lr{\frac{\xxi_2-\xxi_1}{1-\xxi_3}}\,.
\eea
Here $P^{(\xxi,\beta)}_n(x)$ is the Jacobi-polynomial. The basis functions
$Y_{Jj}^{(12)3}(\xxi_i)$ are mutually orthogonal with respect to the conformal
scalar product:
\beq{scalprod}
\int_0^1 [d{\xxi}]\,\,
\xxi_1^{2j_1-1} \xxi_2^{2j_2-1} \xxi_3^{2j_3-1}\,
Y_{Jj}^{(12)3}(\xxi_i) Y_{J'j'}^{(12)3}(\xxi_i) = {\cal
N}_{Jj}\,\delta_{JJ'}\,\delta_{jj'}\,,
\eeq
where
\bea{normY}
{\cal N}_{Jj}
&=&
\frac{\Gamma(j\!+\!j_{1}\!-\!j_{2})\Gamma(j\!-\!j_{1}\!+\!j_{2})}
{\Gamma(j\!-\!j_1\!-\!j_2\!+\!1)\Gamma(j\!+\!j_1\!+\!j_2\!-\!1) (2j\!-\!1)}\,
\frac{\Gamma(J\!-\!j\!+\!j_3)\Gamma(J\!+\!j\!-\!j_3)}
{\Gamma(J\!-\!j\!-\!j_3\!+\!1)\Gamma(J\!+\!j\!+\!j_3\!-\!1)(2J\!-\!1)}\,.
\nonumber\\[-5pt]\eea
The above construction of the conformal basis involves the obvious
ambiguity in what order to couple the spins of partons to the total spin
$J$. Choosing a different two-particle channel, one obtains a different
conformal basis related to the original one through the matrix $\Omega$
of Racah $6j$-symbols of SL$(2,\mathbb{R})$, e.g.,
\bea{Racah}
{Y}_{Jj}^{(31)2}(u_i)=\sum_{j_1\!+\!j_2\leq j'\leq J-j_3}\Omega_{jj'}(J)\,
{Y}_{Jj'}^{(12)3}(u_i)\,.
\eea
The properties of the Racah $6j$-symbols as well as
explicit expressions  in terms of the generalized hypergeometric series
${}_4F_3(1)$ are summarized  in \cite{BKM01}.


\section{Conformal Ward Identities}
\label{App-DerCWI}

To derive the conformal Ward identities (CWIs) we employ
the generating functional
\bea{Def-GenFun}
Z({\bf J}) = \frac{1}{\cal N}
\int {\cal D} \Phi\;  \exp\left\{i [S] + i \int d^d\! x\,  {\bf J}(x)
\Phi(x)
\right\}\quad \mbox{with}\quad
{\cal N} = \int {\cal D} \Phi\;  \exp\left\{i [S] \right\} \, ,
\eea
where $[S] = \int\! d^d\! x\, {\cal L}(x)$ is the renormalized action
and the dimensional regularization is used, i.e., $d= 4-2\epsilon$.
The renormalized connected Green functions result from a functional
derivation with respect to the sources $\bf J$
and one-particle irreducible Green functions, which are not considered
here, are obtained from a Legendre transformation. Replacing the
integration variable $\Phi$ by the new one $\Phi^\prime (\Phi)$, where
the change in the integral measure is given by the Jacobian ${\rm Det}
\left[{\cal D} \Phi^\prime/ {\cal D} \Phi\right]$, does not alter the
generating functional (\ref{Def-GenFun}). Thus, for an infinitesimal
conformal transformation (\ref{trafo})
\bea{Inf-ConVar}
\Phi^\prime(x)
=
\Phi(x) +
\varepsilon \delta_G \Phi(x) \quad\mbox{with}\quad \delta_G \Phi(x)
=
{\cal G}(x,\partial) \Phi(x)\ \ \mbox{for}\ \
G = \{P_\mu,M_{\mu\nu},D,K_\mu\}\, ,
\eea
one finds that the variation of the generating functional
(\ref{Def-GenFun}) has to vanish to order
$\varepsilon$:
\bea{WarIde-1step}
\int {\cal D} \Phi\,
\left[
i\!\int\! d^d\! x\,{\bf J}(x)\left\{ {\cal G}(x,\partial)\Phi(x)\right\}
                +i \delta_G [S] - i \delta_G{\cal N}
        \right]
\exp    \left\{
                i [S]  + i\! \int\! d^d\! x\,  {\bf J}(x) \Phi(x)
        \right\}
\equiv 0\, .
\eea
Here we could safely neglect the Jacobian ${\rm Det} \left[{\bf I} +
\varepsilon {\cal G} \right]$, since it is only a C--number and cancels
out between numerator and denominator, cf.\ Eq.~(\ref{Def-GenFun}). The
variation of the normalization $\delta_G{\cal N} =\int {\cal D} \Phi\,
\delta_G [S] \exp \left\{i [S] \right\}$ is understood as subtraction of
the vacuum expectation value of the operator insertion $\delta_G [S]$
and yields the normal ordering $:\!\!\delta_G [S]\!\!: \equiv \delta_G
[S]-\delta_G{\cal N}$. For simplicity, we do not indicate this operation
here and so Eq.\ (\ref{WarIde-1step}) can be cast in the form
\bea{WarIde-2step}
\int d^d\! x\, {\bf J}(x) \left\{{\cal G}(x,\partial)
\frac{\delta}{\delta{\bf J}(x)} \right\}\; \ln Z({\bf J})
\equiv
- \frac{Z\!\left(\! {\bf J}\big| i \delta_G [S]\right)}{Z({\bf J})}
\, .
\eea
Functional derivatives with respect to ${\bf J}$ give rise to to the
CWIs for connected Green functions. Thanks to Poincar\'e invariance, the
variations $\delta_P^\mu [S]$ and $\delta_M^{\mu\nu} [S]$ vanish
identically, while the conformal variations $\delta_D [S]$ and
$\delta_K^\mu [S] $ contain the trace anomaly, introduced in
Sect.~\ref{SubSec-EneMomTen}. In what follows we calculate $\delta_D
[S]$ and $\delta_K^\mu [S] $ for QCD.

Within the covariant gauge fixing the renormalized QCD Lagrangian reads
\bea{QCD-Lag-Ren}
{\cal L} &=&  Z_\psi \bar\psi i \!\not\!\! D \psi
- \frac{Z_A}{4} G^a_{\mu\nu}G^{a\mu\nu}
- \frac{1}{2\xi} \left( \partial^\mu A^a_\mu \right)^2
+ Z_\omega \ \partial^\mu \bar\omega^a D_\mu^{ab} \omega^b\, .
\eea
Here $\bar\omega$ and $\omega$ are the anti-ghost and ghost fields,
respectively. The renormalized
covariant derivatives in the fundamental and adjoint representation are
$
D_\mu = \partial_\mu  - i  \mu^\epsilon g X A^a_\mu t^a\ \ \mbox{and}\ \
D_\mu^{ab} = \delta^{ab} \partial_\mu  + \mu^\epsilon g X f^{acb} A^c_\mu\,,
$
respectively, and the renormalized field strength tensor is
given by
$
G^a_{\mu\nu} =
\partial_\mu A_\nu^a - \partial_\nu A_\mu^a
+\mu^\epsilon g X f^{abc} A^b_\mu A^c_\nu \, .
$
Here $\mu$ is a mass parameter which plays the role of the
renormalization scale and ensures that the renormalized coupling $g$ is
dimensionless. The $Z_\Phi$ factor renormalizes the corresponding field:
$\Phi^{\rm unr} = \sqrt{Z_\Phi} \Phi$, while $X$ is needed for the
renormalization of the coupling: $g^{\rm unr}= \mu^\epsilon g
X/\sqrt{Z_A}$. We remind that the action is renormalization group
invariant, i.e., the unrenormalized and renormalized actions are equal:
S = [S]. In a MS-like scheme all renormalization factors are given by
Laurent series
\bea{Z-Fac-LauExp}
Z = 1 + \frac{1}{\epsilon} Z^{[1]}(\alpha_s(\mu),\xi(\mu)) +
\frac{1}{\epsilon^2} Z^{[2]}(\alpha_s(\mu),\xi(\mu))  +
O(1/\epsilon^3)\quad \mbox{for}\quad
Z=\left\{Z_\psi, Z_A, Z_\omega, X\right\}\, ,
\eea
where the coefficients $Z^{[i]}(\alpha_s(\mu),\xi(\mu))$ do not
explicitly depend on $\mu$ (see for instance Ref.~\cite{Col84}). Note
that the gauge invariance of the Lagrangian (\ref{QCD-Lag-Ren}) is
broken by the gauge fixing procedure, however, it is invariant under the
following renormalized BRST transformations
\bea{BRST-Tra}
&&\delta^{\rm BRST} \psi =
- i \mu^\epsilon g X Z_\omega\, \omega^a t^a \psi \delta\lambda\, ,
\quad
\delta^{\rm BRST} A^a_\mu =
Z_\omega D_\mu \omega^a \delta\lambda\, , \nonumber\\
&&\delta^{\rm BRST} \omega^a =
\frac{1}{2} \mu^\epsilon  g X Z_\omega f^{abc}
\omega^b\omega^c \delta\lambda\, ,
\quad
\delta^{\rm BRST} \bar{\omega}^a =
\frac{1}{\xi} \partial_\mu A_\mu^a \delta\lambda\, ,
\eea
where $\delta\lambda$ is a  Grassman variable.

The conformal variations of the renormalized action can be calculated using
Eqs.~(\ref{D-variation}) and (\ref{C-variation}):
\bea{ConVar-Act}
\left\{ { \delta_D \atop \delta_K^\mu } \right\} [S] =
\int d^d x \left\{ { 1 \atop 2 x^\mu} \right\} \Delta(x)
\quad\mbox{with}\quad
\Delta(x) =  \Delta_D(x) - \frac{1}{2} \partial_\nu \Delta_K^\nu(x)
\eea
where $\Delta_D(x)$ and the addenda $\Delta_K^\nu(x)$ are defined in
Eqs.~(\ref{Delta-D}) and (\ref{Con-SpeConInv}). As we have mentioned in
Sect.~\ref{SubSec-EneMomTen}, $\Delta_K^\nu(x)$ does not vanish for the
gauge fixing and ghost terms even in four dimensional Minkowski space.
On the other hand, it is possible to achieve that both conformal
variations are expressed in the physical sector in terms of the single
trace anomaly $\Delta_D(x)$. This is done by choosing the scaling
dimensions of physical fields in $d$ dimensions to be equal to the
canonical ones in four dimensions, whereas for the ghost fields we make
the following choice:
\bea{SetScaDimQCDall}
\ell_\psi = 3/2\, , \quad \ell_A=1\, ,\quad
\ell_{\bar{\omega}} = d - 2\, ,
 \  \mbox{and}\  \
\ell_\omega = 0\,.
\eea
That is, $\ell_{\bar{\omega}}$ is simply the canonical dimension in $d$
dimensions and then the choice $\lsc_\omega =0$
is motivated by the BRST transformations (\ref{BRST-Tra}).

Moreover, it turns out that the choice (\ref{SetScaDimQCDall}) allows for a
straightforward decomposition of $\Delta(x)$ in
gauge invariant (type A), BRST exact (type B), and EOM operators:
\bea{TraceAnomaly}
\Delta (x) = -\epsilon\,
\left\{
{\cal O}_A (x) + {\cal O}_B (x)
+ {\mit \Omega}_{\bar \omega} (x)
- {\mit \Omega}_{\bar\psi\psi} (x)
\right\} - (d - 2)\,\partial^\nu {\cal O}_{B\nu}(x)\, .
\eea
Corresponding to their subscript, the operator insertions
\bea{Def-OpeInsAB}
{\cal O}_A (x) = \frac{Z_A}{2} \left( G^a_{\mu\nu} \right)^2\, , \qquad
{\cal O}_B (x) = \frac{\delta^{\rm BRST}}{\delta\lambda}
\bar{\omega}^a\partial_\mu A_\mu^a\, , \ \
{\cal O}_{B\mu} (x) = \frac{\delta^{\rm BRST}}{\delta\lambda}
\bar{\omega}^a A^a_\mu\, ,
\eea
belong to class A and B and the  EOM  operators read
\bea{Def-EOM-QCD}
{\mit \Omega}_A (x) = A^a_\mu \frac{\delta [S]}{\delta A^a_\mu}\, ,
\quad
{\mit \Omega}_{\bar\psi\psi} (x)
= \frac{\delta [S]}{\delta \psi} \psi
+ \bar\psi \frac{\delta [S]}{\delta \bar\psi}\, ,
\quad
{\mit \Omega}_{\bar \omega} (x)
= \bar \omega^a \frac{\delta [S]}{\delta \bar \omega^a} \, .
\eea
The l.h.s.\ of the CWI (\ref{WarIde-2step}) is finite by definition and
so we conclude that the conformal variations (\ref{ConVar-Act}) are
finite operator insertions. This implies that the operators in
Eq.~(\ref{TraceAnomaly}) generate a $1/\epsilon$ pole, which is annulled
by the $\epsilon$ prefactor. In the following we renormalize these
operators.

The general mixing scheme is the following: class A operators need
themselves, class B as well as EOM operators as counterterms, while
class B operators can only mix with themselves or EOM operators
\cite{DixTay74KluZub75JogLee76}. Of course, only operators with the same
quantum numbers and the same or lower twist can appear as counterterms.
The EOM operators are renormalized, since they correspond to a partial
integration in the generating functional (\ref{Def-GenFun}). The
renormalization problem at zero momentum transfer can be completely
solved by means of differential vertex operator insertions, which are
generated by the derivation of renormalized Green--functions with
respect to $g$ and $\xi.$ \cite{Low71}. Thus, they are renormalized and
this allows us to find the integrated and renormalized operator
insertions $\left[{\cal O}_A\right]$ and $\left[{\cal O}_B\right]$
\cite{Nie73,Sar74}:
\bea{def-DIVOI}
[\Delta^g] \equiv g \frac{\partial}{\partial g} [S]
= \left[{\cal O}_A\right]
+ \left[{\cal O}_B\right]
+ {\mit \Omega}_A
+ {\mit \Omega}_{\bar \omega}\, ,
\qquad
[\Delta^\xi] \equiv \xi \frac{\partial}{\partial\xi} [S]
= \frac{1}{2} \Bigl\{ \left[{\cal O}_B\right]
+ {\mit \Omega}_{\bar \omega} \Bigr\}\, ,
\eea
where operators without argument are defined at zero momentum transfer,
for example  $ {\mit \Omega}_\Phi = \int\! d^d\! x\, {\mit \Omega}_\Phi(x)$.

These results together with the general renormalization properties
allow us to express the conformal anomaly in terms of renormalized operator
insertions
\cite{MuePHD,BelMue98a,BelMue98c}:
\bea{trace-anomaly}
\Delta (x)
\!\!\! &=& \!\!\!  \frac{\beta_\epsilon}{g} [ {\cal O}_A (x) ]
+ \left(
\frac{\beta_\epsilon}{g} - \gamma_A
\right)
\Bigl\{
[ {\cal O}_B (x) ]+{\mit \Omega}_{\bar\omega} (x)
\Bigr\}
- \left( \gamma_A - \frac{\beta}{g} \right)
{\mit \Omega}_A (x)  \nonumber\\
&&\!\!- ( \gamma_\psi - \epsilon ) {\mit \Omega}_{\bar\psi\psi} (x)
- 2 \gamma_\omega {\mit \Omega}_{\bar \omega} (x)
- (d - 2) \partial_\mu [ {\cal O}_{B}^\mu (x) ] \, .
\eea
Here, $\gamma_\Phi\equiv \mu \frac{d}{d\mu} \ln \sqrt{Z_\phi}$ denote
the anomalous dimensions and $\beta_\epsilon(g;\epsilon)= - \epsilon g +
\beta(g)$ is the $\beta$ function in $d$ dimensions. Inserting this
result into Eq.~(\ref{WarIde-2step}) and taking derivatives with respect
to the sources ${\bf J}$ we obtain the renormalized CWIs:
\bea{Def-CWI-QCD}
\sum_{i=1}^N {\cal D}_i\,  \langle  {\cal X}_N \rangle \!\!\! &=& \!\!\!
  -\sum_{i=1}^N  \gamma_{\Phi}\, \langle {\cal X}_N  \rangle -
        \frac{\beta}{g} \langle   i [\Delta^g] {\cal X}_N \rangle  - \sigma
                \langle  i [\Delta^\xi] {\cal X}_N \rangle,
\\
\sum_{i=1}^N {\cal K}_i^\mu\, \langle   {\cal X}_N \rangle \!\!\! &=&\!\!\!
         -\sum_{i=1}^N  \gamma_{\Phi}\, 2 x^\mu_i \,
\langle {\cal X}_N \rangle
- \frac{\beta}{g} \langle  i [\Delta^{g\mu} ] {\cal X}_N  \rangle - \sigma
\langle  i [\Delta^{\xi\mu}] {\cal X}_N \rangle  - 4 \langle   i [{\cal
O}^{\mu}_B] {\cal X}_N \rangle \, .
\nonumber
\eea
Here we could safely set $\epsilon$ to zero, $\sigma=-2\gamma_A$ is the
renormalization group coefficient of the gauge fixing parameter and in
analogy to the differential operator vertex insertions (\ref{def-DIVOI})
we defined
\bea{Def-OpeDel-}
[\Delta^{g}_\mu] = \int d^d\!x\, 2x_\mu \left(
\left[{\cal O}_A\right]
+ \left[{\cal O}_B\right]
+ {\mit \Omega}_A
+ {\mit \Omega}_{\bar \omega}
\right)(x)\, , \quad
[\Delta^{\xi}_\mu] = \int d^d\!x\, 2x_\mu \frac{1}{2}\left( \left[{\cal
O}_B\right]
+ {\mit \Omega}_{\bar \omega} \right)(x)\,.
\eea

As the next step we consider the CWIs with one operator insertion
$\mbox{{\boldmath ${\ccal {\bf O}}$}}_{nl}$. These are derived along the same
line from the generating functional
\bea{Def-GenFunInsO}
Z({\bf J}| \mbox{{\boldmath ${\ccal {\bf O}}$}}_{nl}) = \frac{1}{\cal N} \int
{\cal D}
\Phi\;
 \left[\mbox{{\boldmath ${\ccal {\bf O}}$}}_{nl}(\Phi) \right]
\exp\left\{i [S] + i \int d^d\! x\,  {\bf J}(x) \Phi(x)
\right\} \,
\eea
and read for the connected Green functions
\bea{CWI-ComOpe-QCD}
\langle [\mbox{{\boldmath ${\ccal {\bf O}}$}}_{nl}] \left(\delta_G {\cal X}_N
\right) \rangle =
- \langle  \left(\delta_G [\mbox{{\boldmath ${\ccal {\bf O}}$}}_{nl}] \right)
{\cal X}_N \rangle -
\langle [\mbox{{\boldmath ${\ccal {\bf O}}$}}_{nl}] i\left(\delta_G [S]\right)
{\cal X}_N  \rangle \, .
\eea
For simplicity we assume that the composite operators are closed under
renormalization, i.e., we neglect possible counterterms of class B and
EOM operators:
\bea{renOp}
[\mbox{{\boldmath ${\ccal {\bf O}}$}}_{nl}]= \sum_{m=0}^{n} \mbox{{\boldmath
$\hat Z$}}_{nm}(\alpha_s,\epsilon)
\mbox{{\boldmath ${\ccal {\bf O}}$}}_{ml}
\quad \mbox{with} \quad
\mbox{{\boldmath $\hat Z$}}(\alpha_s,\epsilon) = \mbox{{\boldmath $\hat 1$}}
+ \frac{1}{\epsilon}
\mbox{{\boldmath $\hat Z$}}^{[1]}(\alpha_s) + O(1/\epsilon^2)\, .
\eea

The l.h.s.\ of the CWIs (\ref{CWI-ComOpe-QCD}) is finite by definition and so its
r.h.s., too. The conformal variations $\delta_G [\mbox{{\boldmath ${\ccal {\bf
O}}$}}_{nl}]$ of the renormalized operator insertion are the same as at tree
level, except for special conformal transformation, which yields a divergent
expression:
\bea{ConTraRenOpe}
\delta_D \left[\mbox{{\boldmath ${\ccal {\bf O}}$}}_{nl} \right] =
(l+3) \left[\mbox{{\boldmath ${\ccal {\bf O}}$}}_{nl} \right]
\, , \quad
\delta^-_K \left[\mbox{{\boldmath ${\ccal {\bf O}}$}}_{nl} \right] =
-i \sum_{m=0}^n \left\{
\mbox{{\boldmath $\hat Z$}} \mbox{{\boldmath $\hat a$}}(l)
\mbox{{\boldmath $\hat Z^{-1}$}} \right\}_{nm}
\left[\mbox{{\boldmath ${\ccal {\bf O}}$}}_{ml-1} \right]\, .
\eea
The divergencies in $\mbox{{\boldmath $\hat Z$}}
\mbox{{\boldmath $\hat a$}}(l) \mbox{{\boldmath $\hat Z^{-1}$}}$ have
to be annulled by UV-divergencies that arise from the operator product
$[\mbox{{\boldmath ${\ccal {\bf O}}$}}_{nl}] \delta_K^- [S]$. In the case of
dilatation we conclude that $[\mbox{{\boldmath ${\ccal {\bf O}}$}}_{nl}]
\delta_D[S]$ is finite. However, also this operator product will induce
new UV-divergencies that are multiplied by $\epsilon$, see discussion above. The
UV-divergencies of these operator products are caused by the singularities in
$[\mbox{{\boldmath ${\ccal {\bf O}}$}}_{nl}] \Delta(x)$ that are concentrated in
$x=0$, see also Eq.\ (\ref{ConVar-Act}).

The details of the renormalization procedure $[\mbox{{\boldmath ${\ccal {\bf
O}}$}}_{nl}] \Delta(x)$ can be found for Abelian gauge theory in Ref.\
\cite{MuePHD,Mue94} and for QCD in Ref.~\cite{BelMue98c}. Here we only present
the gauge-invariant contribution of $[{\cal O}_A (x)]$,
cf.~(\ref{trace-anomaly}):
\bea{ren-A}
i[{\cal O}_A (x)] [\mbox{\boldmath $ {\ccal {\bf O}}$}_{n l}] = i[{\cal O}_A
(x)\mbox{\boldmath $ {\ccal {\bf O}}$}_{n l}]
\!\!\!&-&\!\!\!
\delta^{(D)} (x) \sum_{m = 0}^{n}
\left\{ \mbox{{\boldmath $ \hat Z_A$}} \right\}_{nm}
[\mbox{\boldmath $ {\ccal {\bf O}}$}_{m l}] -
\frac{i}{2} \partial_+ \delta^{(D)} (x) \sum_{m = 0}^{n}
\left\{ \mbox{{\boldmath $\hat Z_A^-$}} \right\}_{nm}
[\mbox{\boldmath $ {\ccal {\bf O}}$}_{m l-1}]
\nonumber\\
\!\!\!&-&\!\!\!
\left(
g \frac{\partial\ln X}{\partial g}
- 2 \xi \frac{\partial\ln X}{\partial\xi}
\right)
A_\mu^a (x) \frac{\delta}{\delta A_\mu^a (x)} [\mbox{\boldmath $ {\ccal {\bf
O}}$}_{n l}]  + \dots\, ,
\eea
The renormalization is done by an additive subtraction, where counterterms with
higher derivatives, denoted by the ellipses, will not contribute to the CWIs. A
similar formula exists for the renormalization of the product $[{\cal O}_B +
{\mit \Omega}_{\bar \omega}] (x) [\mbox{\boldmath $ {\ccal {\bf O}}$}_{n l}]$,
which is already finite in Landau gauge. These ansatze are motivated by
considering the differential vertex insertions which also relate the
renormalization matrices $\mbox{{\boldmath $\hat Z_A$}}$ and $\mbox{{\boldmath
$\hat Z_B$}}$ to the anomalous dimensions.
The term containing the variation of the composite operators with
respect to the gauge field ensures the correct renormalization group
equation for the composite operators \cite{BelMue98c}. The products of
conformal and EOM operators can easily be treated by partial integration
in the functional integral (\ref{Def-GenFunInsO}). Instead of the MS
prescription, we define
\bea{renEOM}
\langle i [\mbox{\boldmath $ {\ccal {\bf O}}$}_{n l} \Omega_\Phi(x)] {\cal X}_n
\rangle = -
\langle  [\mbox{\boldmath $ {\ccal {\bf O}}$}_{n l} \Phi(x) \frac{\delta}{\delta
\Phi(x)}  {\cal X}_n\rangle \, ,
\eea
where the unrenormalized operator product is related to the renormalized one
as
\bea{renEOM-1}
i [ \mbox{\boldmath $ {\ccal {\bf O}}$}_{n l} ] \Omega_\Phi(x) = i [
\mbox{\boldmath $ {\ccal {\bf O}}$}_{n l} \Omega_\Phi(x)]  -\Phi(x)
\frac{\delta}{\delta \Phi(x)}
[ \mbox{\boldmath $ {\ccal {\bf O}}$}_{n l} ]\,.
\eea

Employing Eqs.~(\ref{CWI-ComOpe-QCD}) and
(\ref{ConTraRenOpe})--(\ref{renEOM-1}) the CWIs can be written after
some algebra as
\begin{eqnarray}
\label{DCWI-QCD}
\sum_{i=1}^N {\cal D}_i\,
\langle [ \mbox{\boldmath $ {\ccal {\bf O}}$}_{n l} ] {\cal X}_N  \rangle
&\!\!\!=& \!\!\!
- \sum_{m = 0}^{n}
 \left\{
( l + 3 ) \mbox{{\boldmath $\hat 1$}} + \mbox{{\boldmath $\hat\gamma$}}
\right\}_{nm}
\langle [\mbox{\boldmath $ {\ccal {\bf O}}$}_{m l} {\cal X}_N  \rangle
- \frac{\beta}{g}
\langle i [ \mbox{\boldmath $ {\ccal {\bf O}}$}_{n l} \Delta^g ] {\cal X}_N
\rangle \\
&&
-  \sigma
\langle i[ \mbox{\boldmath $ {\ccal {\bf O}}$}_{n l} \Delta^\xi ] {\cal X}_N
\rangle\,
- \sum_{i=0}^N \gamma_\Phi \langle [\mbox{\boldmath $ {\ccal {\bf O}}$}_{n l}]
{\cal X}_N \rangle ,
\nonumber\\
\label{SCCWI-QCD}
\sum_{i=1}^N {\cal K}_i^-\, \langle [ \mbox{\boldmath $ {\ccal {\bf O}}$}_{n l} ]
{\cal X}_N  \rangle &\!\!\!=& \!\!\!
i \sum_{m = 0}^{n}
\left\{
\mbox{{\boldmath $\hat a$}} (l) + \mbox{{\boldmath $\hat\gamma^{\rm c}$}}
\right\}_{nm}
\langle [ \mbox{\boldmath $ {\ccal {\bf O}}$}_{m l-1}] {\cal X}_N  \rangle
- \frac{\beta}{g} \langle i [ \mbox{\boldmath $ {\ccal {\bf O}}$}_{n l}
\Delta^g_- ] {\cal X}_N  \rangle
\\
&&
-  \sigma
\langle i[ \mbox{\boldmath $ {\ccal {\bf O}}$}_{n l} \Delta^\xi_- ] {\cal X}_N
\rangle +
2 \langle i[ \mbox{\boldmath $ {\ccal {\bf O}}$}_{n l} \Delta_{\rm ext}^-] {\cal
X}_N  \rangle - \sum_{i=0}^N \gamma_\Phi 2 x_{i}^- \langle [\mbox{\boldmath $
{\ccal {\bf O}}$}_{n l}] {\cal X}_N \rangle \, ,
\nonumber
\end{eqnarray}
where $\Delta_{\rm ext}^- = \int\! d^dx\, 2 \bar n\!\cdot\! x\, \partial_\mu
{\cal O}_B^\mu$.
The finiteness of the anomalous matrices
\begin{eqnarray}
\label{def-delaAno}
\mbox{{\boldmath $ \hat\gamma$}} \!\!\! &=&\!\!\! \lim_{\epsilon\to 0}
\left\{ - \frac{\beta_\epsilon}{g} \mbox{{\boldmath $\hat Z_A$}}
- \left(\frac{\beta_\epsilon}{g}-\gamma_A\right) \mbox{{\boldmath $\hat
Z_B$}}
+ 2 ( \gamma_\psi - \epsilon )
\mbox{{\boldmath $ \hat Z $}} \mbox{{\boldmath $\hat P_Q$}}
\mbox{{\boldmath $\hat Z^{-1}$}}\right\}\, ,\\
\!\!\! &=&\!\!\!
2 \gamma_\psi \mbox{{\boldmath $ P_Q $}}
-2 \left[ \mbox{{\boldmath $\hat Z^{[1]}$}}, \mbox{{\boldmath $\hat P_Q$}}
\right]
+ \mbox{{\boldmath $\hat Z^{[1]}_A $}}
+ \mbox{{\boldmath $\hat Z^{[1]}_B$}}
\, ,
\qquad \mbox{where}\ \ \mbox{{\boldmath $P_Q$}} = \left({ 1\ 0\atop 0\
0 }\right)\, ,
\nonumber
\\
\label{def-scAno}
\mbox{{\boldmath $\hat\gamma^c$}}(l)
\!\!\! &=&\!\!\! \lim_{\epsilon \to 0}
\left\{
\mbox{{\boldmath $\hat Z$}} \left[
\mbox{{\boldmath $\hat a$}}(l) - 2 ( \gamma_\psi - \epsilon )
\mbox{{\boldmath $\hat P_Q$}} \mbox{{\boldmath $\hat b$}}
\right]
\mbox{{\boldmath $\hat Z^{-1}$}}
- \frac{\beta_\epsilon}{g} \mbox{{\boldmath $\hat Z^-_A$}}
- \left( \frac{\beta_\epsilon}{g} - \gamma_A \right) \mbox{{\boldmath $\hat
Z^-_B$}}
- \mbox{{\boldmath $\hat a$}}(l) \right\}
\nonumber\\
\!\!\! &=&\!\!\!
- 2 \gamma_\psi \mbox{{\boldmath $ P_Q \hat b$}}
+ 2 \left[ \mbox{{\boldmath $ \hat Z^{[1]}$}}, \mbox{{\boldmath $ \hat P_Q
\hat b$}} \right]
+ \mbox{{\boldmath $ \hat Z^{[1]-}_A $}}
+ \mbox{{\boldmath $\hat Z^{[1]-}_B $}}
\, ,
\end{eqnarray}
follows from the fact that all other terms in the CWIs (\ref{DCWI-QCD})
and (\ref{SCCWI-QCD}) are finite by definition. In Landau gauge they are
calculable from the counterterms in Eq.\ (\ref{ren-A}) alone, since
\mbox{{\boldmath $\hat Z_B$}} and \mbox{{\boldmath $\hat Z^-_B$}} are
vanishing. Let us add that the definition (\ref{def-delaAno}) coincides
with the usual one for the anomalous dimensions matrix:
\bea{Def-AomDim-FlaSin}
\mbox{{\boldmath $\hat\gamma$}} \equiv \mbox{{\boldmath $\hat\gamma_Z$}}
+ 2\gamma_A
\mbox{{\boldmath $\hat P_G$}} + 2 \gamma_\psi
\mbox{{\boldmath $\hat P_Q$}}  \ \ \mbox{with} \
\mbox{{\boldmath $\hat\gamma_Z$}} =
- \left(\mu \frac{d}{d \mu} \mbox{{\boldmath $\hat Z$}}\right) \,
\mbox{{\boldmath $\hat Z^{-1}$}}
= g \frac{\partial}{\partial g} \mbox{{\boldmath $\hat Z^{[1]}$}}
\ \ \mbox{and}\ \mbox{{\boldmath $P_G$}} = \left({ 0\ 0\atop 0\ 1}\right).
\eea


\end{document}